\documentclass[11pt, a4paper]{article}

\usepackage{amsmath, amsthm, amssymb, amsfonts}
\usepackage{lmodern}
\usepackage[utf8]{inputenc}
\usepackage{mathtools}
\usepackage{hyperref}
\usepackage{graphicx}
\usepackage{subcaption}
\usepackage{dsfont}
\usepackage{fullpage}
\usepackage{color}
\usepackage{soul}
\usepackage[title]{appendix}
\usepackage{tikz}
\usetikzlibrary{patterns}
\usetikzlibrary{shapes.multipart}
\usetikzlibrary{arrows}
\usepackage{braket}
\usepackage{stmaryrd}

\usepackage{comment}

\usepackage{youngtab}

\usepackage{ytableau}

\definecolor{labelkey}{cmyk}{.4,.2,0,0}

\def\Tr{\operatorname{Tr}}

\newcommand{\be}{\begin{equation}}
\newcommand{\ee}{\end{equation}}
\newcommand{\bea}{\begin{eqnarray}}
\newcommand{\eea}{\end{eqnarray}}
\newcommand{\nn}{\nonumber}

\definecolor{violet}{rgb}{0.5,0,0.5}

\def\MM{{\sf{M}}}

\def\BB{{\sf{B}}}

\def\ibig{\mu}
\def\jbig{\nu}
\def\cb{\gamma}

\def\eee{\mathsf{e}}

\def \sgn {{\rm sgn}}

\newcommand{\R}{\ensuremath{\mathbb{R}}}
\newcommand{\C}{\ensuremath{\mathbb{C}}}

\renewcommand{\rho}{\varrho}

\renewcommand{\leq}{\leqslant}
\renewcommand{\geq}{\geqslant}
\renewcommand{\le}{\leqslant}
\renewcommand{\ge}{\geqslant}

\newcommand{\xx}{\mathbf x}
\newcommand{\yy}{\mathbf y}

\newcommand{\Andrea}[2][]{%
  {\color{orange}{#2}}%
  \ifthenelse{\equal{#1}{}}{}{%
    \textcolor{orange}{\textbf{[Andrea: #1]}}%
  }%
}

\DeclareMathOperator{\sech}{sech}
\DeclareMathOperator{\erf}{erf}

\def\ide{\mathbb{I}}

\def\AAA{\mathbb{A}}
\def\BBB{\mathbb{B}}
\def\SSS{\mathbb{S}}
\def\trect{{\bar{\mathcal{T}}}}

\usepackage[backend=bibtex,style=numeric, maxnames=5, maxalphanames=5, sortcites=true, giveninits=true, isbn=false, clearlang=true, doi=false, url=false, sorting=none]{biblatex}
\renewbibmacro{in:}{}
\AtEveryBibitem{\clearlist{language}} 
\DeclareFieldFormat[article]{title}{\mkbibemph{#1}}
 \DeclareFieldFormat[article]{journaltitle}{#1}

\numberwithin{equation}{section}

\usepackage{titlesec}
\titleformat{\section}{\large\bf}{\thesection}{1em}{}
\titleformat{\subsection}[runin]{\bf}{\thesubsection}{1em}{}[.]
\titleformat{\subsubsection}[runin]{\it}{\thesubsubsection}{1em}{}[.]

\def\XX{\vec{\xx}}
\def\YY{\vec{\yy}}

\theoremstyle{definition}

\theoremstyle{definition}

\theoremstyle{definition}

\theoremstyle{definition}

\def\Xint#1{\mathchoice
   {\XXint\displaystyle\textstyle{#1}}%
   {\XXint\textstyle\scriptstyle{#1}}%
   {\XXint\scriptstyle\scriptscriptstyle{#1}}%
   {\XXint\scriptscriptstyle\scriptscriptstyle{#1}}%
   \!\int}
\def\XXint#1#2#3{{\setbox0=\hbox{$#1{#2#3}{\int}$}
     \vcenter{\hbox{$#2#3$}}\kern-.5\wd0}}

\def\dashint{\Xint-}

\graphicspath{{./}}

\usepackage{authblk}
\author[1]{Andrea De Luca and Pierre Le Doussal}
\affil[1]{\normalsize Laboratoire de Physique de l'\'Ecole Normale Sup\'erieure, ENS, Universit\'e PSL, CNRS, Sorbonne Universit\'e, Universit\'e Paris-Cité, 75005 Paris, France}

\title{\bf \large Free energy distribution for two non-crossing
directed polymers on the cylinder}
\date{}
\begin{document}

\maketitle

\begin{abstract}
We consider the partition sum $Z_L(t)$ of
two non-crossing continuum directed polymers of length $t$,
in a common random medium with the geometry of a cylinder of circumference length $L$. In the
limit $t \to +\infty$, the $k$-th cumulant
of (minus) the free energy $\log Z_L(t)$,
grows as $c_k(L) t$, and we compute the amplitudes $c_k(L)$.
To this aim we first study the integer moments
$\langle Z_L(t)^n \rangle$,
whose long-time behavior is
determined by the ground-state energy of the one-dimensional attractive
Lieb--Liniger model in the symmetry sector associated with the Young diagram
$(n,n)$
(a two-component bosonic nested Bethe ansatz). This ground state energy is obtained using
the nested Bethe-Ansatz involving
$2n$ physical rapidities and $n$ auxiliary rapidities.
The analytic
continuation in the replica number $n$, which yields the cumulants $c_k(L)$,
is performed through a non-trivial generalization of the method developed by
Brunet and Derrida for a single polymer on the cylinder.
We obtain fully
explicit formulas for the mean $c_1(L)$ and the variance $c_2(L)$, and
construct a systematic scheme for computing all higher cumulants $c_k(L)$.
Finally, taking $L\to+\infty$ and $n\to0$ with $nL^{1/2}$ fixed, we derive a
universal scaling function which characterizes the large-deviation statistics
of the two-polymer free energy and is expected to be universal within the
one-dimensional two-layer periodic SHE/KPZ class.
The two tails of the large-deviation function are obtained
analytically, revealing different roles of the
non-crossing constraint in favorable and unfavorable
fluctuations of the common disorder.

\end{abstract}

\tableofcontents

\section{Introduction and main results}
\label{sec:intro}

\subsection{Overview}

Directed polymers in a random environment provide one of the paradigmatic models
of disordered statistical mechanics. For a single polymer in $1+1$ dimensions,
the continuum partition function obeys the stochastic heat equation (SHE),
while its logarithm is related by the Cole--Hopf transformation to the
Kardar--Parisi--Zhang (KPZ) equation \cite{kpz}. The replica method maps
integer moments of the partition function to the imaginary-time evolution of
attractively interacting one-dimensional bosons, governed by the integrable
Lieb--Liniger Hamiltonian \cite{kardar1987replica,lieb1963exact}.
This connection has played a central role in exact replica calculations for the continuum polymer on the infinite line
\cite{calabrese2010free,dotsenko2010replica}, whose predictions were
independently obtained and put on a rigorous footing by other methods
\cite{sasamoto2010exact,amir2011probability}.
More generally, the universal 1:2:3 scaling limit (in height fluctuations, space, and time) of one-dimensional models in the KPZ class is described by the KPZ fixed point~\cite{matetski2021kpz}.

The case of a polymer on a cylinder, with periodic transverse coordinate of
circumference $L$, poses a distinct finite-volume problem. Since transverse
KPZ correlations grow on the scale $t^{2/3}$, the finite circumference becomes
relevant on the relaxation time scale $t\sim L^{3/2}$. For
$t\ll L^{3/2}$ the polymer does not feel the periodic geometry and one
recovers the infinite-line KPZ regime, whereas for $t\sim L^{3/2}$ one
encounters a nontrivial crossover described by the periodic KPZ fixed point.
This relaxation regime has been studied extensively for periodic TASEP
\cite{prolhac2016finite,baik2018fluctuations,baik2019multi,
baik2021periodic,prolhac2024kpz}; more recently the periodic KPZ fixed point
has been constructed for general initial conditions in
\cite{baik2026periodic}. Here, as in the works of Brunet and Derrida, we are
instead interested in the opposite regime $t\gg L^{3/2}$: more precisely,
one first takes $t\to+\infty$ at fixed $L$, and subsequently studies the
large-$L$ behavior. In this regime the cumulants of the free energy grow
linearly with time.

In two seminal works, Brunet and Derrida studied the ground-state energy of
the attractive Lieb--Liniger gas as a function of the replica number $n$, and
developed its analytic continuation away from positive integer $n$ in order
to determine the long-time free-energy cumulants of a directed polymer on a
cylinder \cite{brunet2000ground,brunet2000probability}. A crucial insight was
that perturbative expansions and an associated functional equation provide a
representation suitable for this continuation and hence for extracting the
cumulants in the limit $n\to0$. In the subsequent large-$L$ limit, the
resulting scaling function was found to coincide with the universal
large-deviation scaling function previously obtained for the current of the
ASEP on a ring \cite{derrida1998exact,appert1999universal}, providing an
important early manifestation of KPZ universality in finite volume. Brunet
subsequently used the same periodic geometry to study winding-number
fluctuations \cite{brunet2003winding}.

More recently, the Brunet--Derrida approach has been extended to the KPZ
equation on an interval with general boundary slopes
\cite{BarraquandLeDoussal2025KPZInterval}. In that case the direct replica calculation was carried explicitly through
the second cumulant. In the same work, an independent derivation was obtained
by taking an appropriate scaling limit of known exact results for the current
statistics of the open ASEP
\cite{gorissen2012exact,lazarescu2013matrix}. This scaling analysis yields a
functional description of the full cumulant generating function and hence
access to all cumulants \cite{BarraquandLeDoussal2025KPZInterval}; the two approaches agree for the first two cumulants
where they overlap.
For the periodic problem, later probabilistic works have likewise provided
independent expressions for the first two long-time cumulants
\cite{brunet2021high,gu2024kpz}, with the variance appearing in a
Brownian-bridge representation rather different from the Brunet--Derrida
replica formula. These results provide valuable independent checks of the
replica calculations, while the latter give access to the higher cumulants
and to the associated large-deviation scaling functions.

A natural generalization concerns the statistical mechanics of $N$ directed
polymers in a common random environment, constrained not to intersect. Apart
from applications to disordered flux-line arrays and elastic-line systems,
non-intersecting paths also appear in random tilings and dimer models;
they have been investigated
in this context, in particular in connection with the prediction of a glassy super-rough phase
\cite{toner1990superroughening,zeng1999thermodynamics,perret2012superrough,
duits2025gamma}. For fixed ordered initial and final positions, the corresponding $N$-polymer
partition function is the natural many-body analogue of the single-polymer
solution of the SHE. It is defined on the infinite line in the ordered configuration space
$x_1<\cdots<x_N$ and vanishes when two polymer coordinates coincide.
This construction underlies the
multilayer stochastic heat equation introduced by O'Connell and Warren
\cite{oconnell2016multilayer}. Its relation to discrete polymers in the
intermediate-disorder regime, as well as regularity and positivity of the
resulting multilayer partition functions, has been studied in
\cite{corwin2017intermediate,lun2020continuity}.
Related solvable discrete and semi-discrete models are connected with
Whittaker functions, geometric Robinson–Schensted–Knuth (RSK) correspondence, and Macdonald processes
\cite{borodin2014macdonald}. A stationary model of non-intersecting directed
polymers and explicit descriptions of its invariant measures were obtained in
\cite{barraquand2023nonintersecting}; see also recent developments concerning
joint invariant measures in periodic KPZ \cite{corwin2026periodicpitman}.

The replica problem for non-crossing polymers is considerably richer than for
a single polymer. Replicating $N$ mutually avoiding paths $n$ times yields
$Nn$ interacting particles, but the non-crossing constraint selects a
nontrivial representation of the permutation group rather than the completely
symmetric sector of the attractive Bose gas. Such problems can be addressed
using a multicomponent quantum gas and the nested Bethe ansatz. An early
fermionic replica representation of disordered directed-line assemblies was
introduced by Kardar \cite{kardar1987replica}, and its consequences for the
thermodynamics and fluctuations of disordered line lattices were developed by
Emig and Kardar \cite{emig2000thermodynamic,emig2001probability}.
A recent transfer-matrix approach to finite-density mutually avoiding
polymers provides a complementary perspective on these predictions
\cite{mu2026spectral}.
In a bosonic formulation, the connection between non-crossing polymer
observables, nontrivial Young-diagram symmetry sectors and the nested Bethe
ansatz was developed in \cite{de2015crossing,de2016crossing}, followed by
further results for mutually avoiding paths and random-matrix statistics
\cite{de2017mutually}.

The purpose of the present work is to extend the Brunet--Derrida finite-volume
replica approach to $N$ non-crossing directed polymers on a cylinder.
For simplicity, we focus in this paper on the case of two polymers, $N=2$. The
relevant sector of the attractive Lieb--Liniger problem is characterized by
the Young diagram $(n,n)$, and may be formulated through the two-component
bosonic nested Bethe ansatz, with $2n$ physical rapidities and $n$ auxiliary
ones. This should be contrasted with the earlier fermionic formulation of
Emig and Kardar \cite{emig2000thermodynamic,emig2001probability}, which leads
to a different set of Bethe equations and was particularly suited to the
thermodynamic limit in which the number of polymers and the transverse size
are taken large together. Here we instead keep the number of polymers and the
circumference $L$ finite, take the long-time limit, and treat carefully the
analytic continuation in the replica number $n$ around $n=0$. The bosonic
nested Bethe ansatz proves particularly convenient for this purpose, since it
allows us to derive functional relations adapted to this analytic
continuation. We use these relations to study the long-time cumulant
generating function of the two-polymer free energy at finite circumference.
In particular, we obtain exact expressions for its mean and variance and
develop a systematic scheme for computing its higher cumulants. We then study
the large-$L$ limit, in which $n\to0$ and $L\to\infty$ with the appropriate
combination $n\sqrt{L}$ kept fixed. In this regime the cumulant generating
function acquires a non-trivial scaling form, described by a universal
scaling function. This function characterizes, after the appropriate
centering and rescaling, the long-time large-deviation statistics of the
two-polymer free energy, and is expected to be universal within the
one-dimensional two-layer periodic SHE/KPZ class.

At present, in contrast to the case of a single polymer, a route toward a
microscopic confirmation of our replica results for the two-layer problem
appears to be lacking. Although several integrable discrete and semi-discrete
polymer models are known, or conjectured, to converge to the multilayer SHE
under appropriate scaling limits
\cite{corwin2017intermediate,nica2021intermediate,bufetov2018hall},
to our knowledge no analogue of the deformed matrix-product or tilted-generator
approach is presently available that would independently reproduce the
finite-$L$ cumulants obtained here. Even less is known at the level of the
large-$L$ universal regime, where it would be particularly interesting to
identify an integrable periodic discrete model whose scaling limit reproduces
the two-layer scaling function derived in the present work.

The paper is organized as follows. In
Subsec.~\ref{subsec:model} we define the non-crossing polymer partition
functions and the observables considered in this work. Particular care is
required in passing from the line to the circle: while the ordering of the
polymers is unambiguous on the line, on the circle only their cyclic order is
preserved, and different winding and cyclic sectors have to be properly taken
into account. We formulate the corresponding periodic multilayer SHE problem,
including the absorbing boundary conditions at polymer collisions, and define
the long-time cumulant generating function of the free energy. Our main
results, including the finite-$L$ cumulants and the large-$L$ scaling regime,
are summarized in Subsec.~\ref{subsec:mainresults}.

In Sec.~\ref{sec:determinant} we return in detail to the periodic geometry.
We derive the determinantal representation of the non-crossing partition
function on the circle and show how it leads naturally to a free-fermion
formulation, with the appropriate parity-dependent boundary condition around
the circle. After replication and disorder averaging, the problem becomes an
attractive Lieb--Liniger model with $Nn$ particles in a nontrivial symmetry
sector of the permutation group. We then give a detailed derivation of the
corresponding nested Bethe ansatz. In particular, using the representation
theory of the symmetric group and the Young symmetrizer, we identify the
relevant rectangular Young diagram and derive the nested Bethe equations in
the bosonic formulation used throughout the rest of the paper.

We specialize these equations to $N=2$ in Sec.~\ref{sec:replicaN2}. Starting
from the two coupled sets of physical and auxiliary Bethe roots, we introduce
generating functions which encode the roots and allow the Bethe equations to
be rewritten as coupled functional and integral equations. We successively
reduce and decouple these equations, obtain a formulation involving a single
function $B$, and develop the small-$c$ expansion at fixed replica number
$n$. This reformulation is the key step which makes it possible to address
the analytic continuation away from positive integer $n$.

In Sec.~\ref{sec:nExpansion} we develop the expansion around $n=0$. We first
derive equations valid to all orders in the replica number and introduce a
spectral representation adapted to their solution. The resulting recursive
structure provides, in principle, a systematic expansion of the ground-state
energy in powers of $n$. The zeroth and first nontrivial orders are worked out
explicitly, yielding the mean free energy and an exact finite-$L$ expression
for its variance.

Section~\ref{sec:scalingEquations} is devoted to the large-$L$ limit.
Following and extending the Brunet--Derrida analysis for a single polymer
\cite{brunet2000probability}, we write $c=\gamma L/2$ and take
$c\to\infty$ and $n\to0$ with $\epsilon=n\sqrt{c}$ fixed. Starting from the
finite-$c$ functional relations and their spectral representation, we derive
a closed set of coupled scaling equations governing the leading large-$L$
behavior of the two-polymer problem. We also obtain the corresponding scaling
form of the ground-state energy and check its expansion against the finite-$L$
cumulants.

Finally, in Sec.~\ref{sec:WHscaling} we solve the scaling equations by
adapting the Wiener--Hopf method of Brunet and Derrida to the two-layer
problem. The coupled system is reduced to scalar Wiener--Hopf problems and
self-consistency conditions, from which we derive a parametric representation
of the universal large-$L$ scaling function for the two-layer periodic
SHE/KPZ free energy. We obtain its perturbative expansion, and hence the
large-$L$ asymptotics of the higher cumulants, 
and analyze its large negative and positive arguments
to determine both large-deviation tails, the latter
requiring an analytic continuation of the Wiener--Hopf
solution.
The appendices contain complementary derivations and
checks, including the lattice simulations, the ordering
of the symmetry sectors, the relation between row and
column versions of the nested Bethe ansatz, alternative
eliminations of the auxiliary Bethe roots, perturbative
expansions, and the asymptotic analysis of both
large-deviation tails.

\subsection{Model and observables}

\label{subsec:model}

\subsubsection{Non-crossing polymers on the real line}

Let us start with a single directed polymer $x(\tau)$ on the real line with endpoints
$x(0)=y$ and $x(t)=x$. We define its
partition function $z_t(x,y)$ as the solution of the stochastic heat equation (SHE)
\be \label{SHEsingle}
\partial_t z = \frac{1}{2} \partial_x^2 z + \sqrt{\gamma} \eta z
\ee
with initial data $z_0(x,y)=\delta(x-y)$, where $\eta \equiv \eta(x,t)$ is a standard space time white noise, and $\gamma \geq 0$ denotes the strength of the disorder. Using Feynman-Kac it can be written as a path integral
\be \label{pathintegral}
z_t(x,y) = \int_{\substack{x(0) = y \\ x(t) = x}} \mathcal{D}x \; e^{-\int_0^t d\tau \mathcal{L}[x, \dot x, \tau]} \quad , \quad
\mathcal{L}[x, \dot x, \tau] = \frac{1}{2} (\frac{dx(\tau)}{d\tau})^2
- \sqrt{\gamma} \eta(x(\tau),\tau)
\ee
Throughout, the SHE is understood in the It\^o sense,
and the path-integral notation implicitly includes
the corresponding Wick normalization.
\\

Here and below $X = \{ x_1, \ldots, x_N \}$ denotes an (unordered) set of $N$ points on the real line and $\xx = (x_1,x_2,\dots,x_N) \in \mathbb{R}^N$ denotes a $N$-tuple of points on the real line, and similarly for
$Y$ and $\yy$.

\begin{figure}[h]
    \centering
\includegraphics[width=0.75\linewidth]{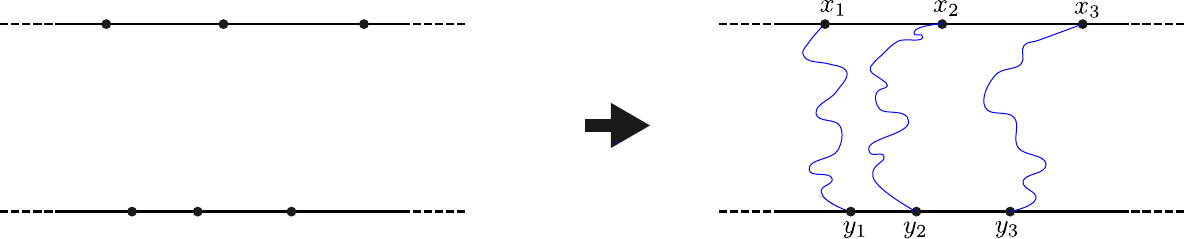}
    \caption{A graphical explanation of Eq.~\eqref{eq:ZtoWeyl}: once the endpoints of the non-crossing paths have been fixed, there is only one way to connect them on the line.}
    \label{fig:line-sketch}
\end{figure}

One considers the partition sum for $N$ non-crossing polymers where we specify the sets of starting points to be $Y$ and the set of ending points to be $X$
\begin{equation}
\label{eq:ZNCline}
    Z_t(X,Y) = \int_{\substack{x_i(0) \in Y \\ x_i(t) \in X }}^{\rm non-crossing} \prod_i \mathcal{D} x_i e^{-\int_0^t d\tau \mathcal{L}[x_i, \dot x_i, \tau]}
\end{equation}
where the path integral is restricted to non-crossing paths, i.e. such that $x_i(\tau) \neq x_j(\tau)$ for all $1 \leq i < j \leq N$ and all $\tau \in [0,t]$.

Crucially, the non-crossing constraint on the real line imposes that the partition sum is restricted to configurations preserving the order: the smallest element in $Y$ is connected with the smallest element in $X$ and so forth. Thus, it is convenient to map the set $X \to \xx$, where $\xx \in \mathbb{R}^N$ is obtained by sorting in ascending order the elements of $X$. More formally, $\xx$ is an element of the Weyl chamber
\begin{equation}
    W_N^\infty= \{ \xx = (x_1,\ldots, x_N) \in \mathbb{R}^N | x_1 < x_2 < \ldots < x_N \}
\end{equation}
In this way, the partition sum $Z_t(X,Y)$ can be expressed as a solution of the multidimensional SHE
\begin{equation}
\label{eq:ZtoWeyl}
Z_t(X,Y) = \mathcal{Z}_t(\xx,\yy) \;, \qquad
\partial_t \mathcal{Z} = \frac{1}{2} \sum_i \partial_{x_i}^2 \mathcal{Z}
+ \sqrt{\gamma} \sum_i \eta(x_i,t) \mathcal{Z} %
\end{equation}
with initial condition $\mathcal{Z}_0(\xx,\yy) = \prod_i \delta(x_i-y_i)$
and absorbing condition at coinciding points at all times
\be
\mathcal{Z}_t(\xx,\yy)|_{x_i=x_j} = 0
\ee
which is equivalent to imposing that $\mathcal{Z}_t(\xx,\yy)$ vanishes at the boundaries of $W_N^\infty$.
Clearly, one must have $\mathcal{Z}_t(\xx,\yy) = 0$ if $\xx$ and $\yy$ are not in the same order since the order of the polymer positions is preserved with time.
This solution can be related to the single-polymer partition function $z_t(x,y)$.
Within the Weyl chamber, i.e. for
$y_1<y_2<\dots<y_N$ and $x_1<x_2<\dots<x_N$ one has the equality
\be
\label{eq:KarlMcGregLine}
\mathcal{Z}_t(\xx,\yy) = \det_{1 \leq i,j \leq N}( z_t(x_i, y_j) )
\ee
This determinantal construction is the stochastic analogue of the Karlin--McGregor formula and provides the multi-layer extension of the stochastic heat equation \cite{karlin1959coincidence,oconnell2016multilayer,de2015crossing}. It also arises as the continuum intermediate-disorder limit of discrete non-intersecting directed polymers \cite{corwin2017intermediate}.

\subsubsection{Non-crossing polymers on the circle}

Let us now assume instead that the polymers live on the circle $\mathcal{C}$ of length $L$, which we write as a submanifold of the complex plane
\begin{equation}
    \mathcal{C} = \{ w \in \mathbb{C} \; | \; 2\pi |w| = L \}
\end{equation}
Thus, the spacetime trajectories $w(\tau)$ live on the cylinder $\mathcal{C} \times [0, t]$. To fix the notation we use the letter $w,v \in \mathcal{C}$ to refer to points on the circle and $x, y \in \mathbb{R}$. We can always map the real line onto the circle using the exponential map
\begin{equation} \label{exponentialmap}
w = {\sf e}(x) := \frac{L}{2\pi} e^{2\pi i x/L}  \;.
\end{equation}
We first consider a single polymer on the circle. Under the exponential map the problem is equivalent to a
polymer on a real line with periodic disorder.
Its partition sum reads
\be
\label{eq:single-polymercyl}
q_t(w,v) = \sum_{m \in \mathbb{Z}} z^{(L)}_t(x, y + L m)
\ee
where $w,v$ are related to $x,y$ through the exponential map and $z_t^{(L)}(x,y)$ is the solution of the SHE \eqref{SHEsingle} with periodic disorder, i.e. $\eta \to \eta^{(L)}$ with $\eta^{(L)}(x+L,t)=\eta^{(L)}(x,t)$,
and with initial condition
$z^{(L)}_0(x, y)=\delta(x-y)$. It can be also written as a path integral
as in \eqref{pathintegral}.
\\

Consider now $N$ non-crossing polymers on the circle. We denote with
$V, W$ two (unordered) ensembles of  $N$ distinct points of the circle $\mathcal{C}$.
Then, we can define the partition sum of $N$ non-crossing polymers on the circle as
\begin{equation}
\label{eq:ZNCcircle}
    Z_{t}^{(L)}(W, V) = \int_{\substack{w_i(0) \in V \\ w_i(t) \in W }}^{\rm non-crossing} \prod_i \mathcal{D} w_i e^{-\int_0^t d\tau \mathcal{L}[w_i, \dot w_i,\tau]}
    \quad , \quad
    \mathcal{L}[w, \dot w, \tau] = \frac{1}{2} \bigl|\frac{dw(\tau)}{d\tau}\bigr|^2
- \sqrt{\gamma} \chi(w(\tau),\tau)
\end{equation}
where $\chi(w, \tau)$ is a standard space-time white noise on the circle $\mathcal{C}$,
and where the path integral is restricted to non-crossing paths, i.e. such that $w_i(\tau) \neq w_j(\tau)$ for all $1 \leq i < j \leq N$ and all $\tau \in [0,t]$. Unlike on the real line, fixing the unordered sets of endpoints on the circle does not uniquely determine the pairing between starting and ending points, since all pairings preserving the cyclic order are allowed. A precise formulation of this cyclic structure, together with the corresponding determinantal representation, is given in Sec.~\ref{sec:determinant}.

In this paper, we are interested in the behavior of $Z_{t}^{(L)}(W, V) $ at large times. In this limit, (minus) the free energy $\sim \log Z_t^{(L)}(W,V)$ is expected to be extensive in the time $t$, so we can write the annealed averages
\begin{equation}
\label{eq:defE}
  E^{(N)}(n, L) := %
   - \lim_{t\to\infty} \frac{1}{t} \log \overline{Z_t(W, V)^n} \;.
\end{equation}
As we will see, this quantity does not depend on the choice of the boundary sets $V, W$, consistently with the fact that this is a bulk contribution to the free energy. For this reason, hereafter, to simplify the notation, we will simply denote $Z(t) = Z_t(W, V)$.

As we show below, for integer $n$, the moments of the polymer partition function
can be mapped, following the replica Bethe-ansatz approach to directed
polymers in random media \cite{kardar1987replica,brunet2000probability},
onto a one-dimensional attractive delta-interacting quantum problem.
Accordingly, $E^{(N)}(n,L)$ can be obtained as the lowest energy of a
Lieb--Liniger Hamiltonian in a particular symmetry sector, selected by the
values of $N$ and $n$.
For $N>1$, this is obtained below using the generalized Bethe ansatz \cite{yang1967some,gaudin1967un,sutherland1968further,gaudin2014bethe,oelkers2006bethe}. Extending the method of Brunet and Derrida~\cite{brunet2000probability} for $N=1$, we are able to perform the analytic continuation in $n$ near $n=0$ to obtain the cumulants of $\log Z(t)$, see next Section. For technical reasons we will later restrict to $N=2$ non-crossing polymers, but in principle the method can be carried out for any $N$. We will thus drop the superscript and below we denote
$E(n,L)= E^{(N)}(n, L)$.

\subsubsection{Observable of main interest}

Our main interest is the asymptotics of the cumulants of (minus)
the free energy $\log Z(t)$,
for infinitely long polymers
\be
c_k(L)  := \lim_{t\to\infty} \frac{1}{t} \kappa_k(\log Z(t)), \label{defck}
\ee
where, for a random variable $X$, $\kappa_k(X)$ denotes the $k$-th cumulant of $X$.
Although $\log Z(t)$ depends on the positions of the endpoints,
we expect that the free energy cumulants per unit length, $c_k(L)$, do not depend on them.

For any finite $t$ the cumulants can be extracted from the
cumulant generating function. Assuming commutation of limits, one can define (minus)
its large-time scaled cumulant generating function (SCGF), for real $n$,
\be
E(n,L)
=
-\lim_{t\to\infty}
\frac{1}{t}
\log \overline{Z(t)^n}.
\ee
It has the formal expansion
\be \label{expansionE}
E(n,L) = - \sum_{k=1}^{\infty} \frac{n^k}{k!} c_k(L)
\ee

Through a Legendre transform one can access the large-deviation form of
the distribution of $\log Z(t)$. Introducing the intensive random variable
\be
H=\frac{\log Z(t)}{t},
\ee
we assume the large-deviation form
\be \label{defPhi}
P\left(\log Z(t)\simeq tH\right)
\asymp
e^{-t\,\Phi_L(H)},
\ee
where $\Phi_L(H)$ is the corresponding rate function and $\asymp$
denotes logarithmic equivalence as $t\to+\infty$.
A saddle-point evaluation then gives
\be \label{legendre1}
E(n,L)
=
-\max_H
\left\{
nH-\Phi_L(H)
\right\}.
\ee
This can be inverted to obtain
\be \label{legendre2}
\Phi_L(H)
=
\max_n
\left\{
E(n,L)+nH
\right\}.
\ee

In this paper our aim is thus to compute $E(n,L)$ for positive integer $n$, then to perform the analytic continuation to real $n$ near $n=0$, and finally obtain the cumulants.
\\

\subsection{Main results}
\label{subsec:mainresults}

Everywhere one denotes, as in \cite{brunet2000probability}
\be
c= \gamma \frac{L}{2}
\ee

\subsubsection{First cumulant}

The first cumulant, i.e. the mean of $\log Z(t)$ per unit time,
can be obtained exactly. It reads
\be \label{c1L}
c_1(L)
=
-\frac{1}{L^2}
\left(
\frac{c^2}{3}+4c+\pi^2
\right)
=
-\frac{\gamma^2}{12}
-\frac{2\gamma}{L}
-\frac{\pi^2}{L^2}.
\ee
For comparison, the Brunet--Derrida result for a single polymer
\cite{brunet2000probability} gives
\be
c_1^{N=1}(L)
=
-\frac{1}{L^2}
\left(
\frac{c^2}{6}+c
\right)
=
-\frac{\gamma^2}{24}
-\frac{\gamma}{2L}.
\ee
Here the Brunet--Derrida result is written in the convention
corresponding to their normalized partition function
$Z/\overline Z$, which is the counterpart of the It\^o/Wick-normalized
partition function used here.

\subsubsection{Small \texorpdfstring{$c=\gamma L/2$}{c=gamma L/2} expansion}

We have obtained the cumulant generating function in an expansion in $c$ for arbitrary $n$.
Up to order $c^5$ it reads
\bea \label{eq:smallcE}
&& E(n,L) = \frac{n \pi^2}{L^2}
- \frac{4 n (n-1)}{L^2} \bigg[
c+\frac{c^2}{12}+\frac{\left(\pi ^4-45\right) c^3 n}{180 \pi
   ^4} \\
   && +\frac{c^4 n \left(2835 (n-4)+\pi ^6 (5 n-6)-63 \pi ^4 (n-1)+945
   \pi ^2\right)}{7560 \pi ^6} \nn \\
   && -\frac{c^5}{226800 \pi
   ^8} n \big(56700 \pi ^2
   (n-4)+14175 (n (17 n-118)+180) \nn
   \\
   && +\pi ^8 ((58-21 n) n-36)+60 \pi ^6 (n-1)
   (5 n-6)-1890 \pi ^4 (n (2 n-7)+2)\big)
   +O\left(c^6\right) \bigg] \nn
\eea
It is interesting to compare with the result of Ref. \cite{brunet2000probability} for a single polymer, $N=1$,
extended up to $O(c^5)$
\begin{equation}
E^{N=1}(n,L)=-\frac{2n(n-1)}{L^2}\left[
\frac{c}{2}+\frac{c^2}{12}+\frac{n c^3}{180}
+\frac{n(5n-6)}{7560}c^4
+\frac{n(21n^2-58n+36)}{226800}c^5
\right]+O(c^6).
\end{equation}

We then obtain the cumulants of $\log Z(t)$ at large $t$ as follows
\be
\overline{(\log Z(t))^k}^c \simeq c_k(L) t \quad , \quad c_k(L) =- \partial_n^k E(n)|_{n=0}
\ee
This leads to the following expansion for the lowest order cumulants
\bea \label{cumsmallc00}
&& L^2 c_1(L) = -\pi ^2-4 c-\frac{c^2}{3}+O\left(c^6\right) \\
&& L^2 c_2(L)=
8 c+\frac{2
   c^2}{3}+\left(-\frac{2}{45}+\frac{2}{\pi ^4}\right)
   c^3+\left(\frac{2}{315}+\frac{12}{\pi ^6}-\frac{1}{\pi ^4}-\frac{1}{15
   \pi ^2}\right) c^4 \nn \\
&& +\left(-\frac{2}{1575}+\frac{90}{\pi
   ^8}-\frac{8}{\pi ^6}-\frac{2}{15 \pi ^4}+\frac{4}{315 \pi ^2}\right)
   c^5+O\left(c^6\right) \nn \\
&&  L^2 c_3(L)=
   \left(\frac{2}{15}-\frac{6}{\pi ^4}\right)
   c^3+\left(-\frac{11}{315}-\frac{45}{\pi ^6}+\frac{3}{\pi
   ^4}+\frac{2}{5 \pi ^2}\right) c^4 \nn  \\
   && +\left(\frac{47}{4725}-\frac{447}{\pi
   ^8}+\frac{30}{\pi ^6}+\frac{9}{5 \pi ^4}-\frac{34}{315 \pi ^2}\right)
   c^5+O\left(c^6\right) \nn \\
&&   L^2 c_4(L)=
   \left(\frac{4}{63}+\frac{36}{\pi ^6}-\frac{4}{5
   \pi ^2}\right) c^4+\left(-\frac{158}{4725}+\frac{810}{\pi
   ^8}-\frac{24}{\pi ^6}-\frac{36}{5 \pi ^4}+\frac{128}{315 \pi
   ^2}\right) c^5+O\left(c^6\right) \nn
\eea
Note that the result for (minus) the average free energy $c_1(L)$ is exact to all orders in $c$, see \eqref{c1L}.

We can compare with the result for a single polymer \cite{brunet2000probability}, extended up to order $O(c^5)$
\begin{align}
L^2 c_1^{N=1}(L) &= -c-\frac{c^2}{6}+O(c^6),
\\
L^2 c_2^{N=1}(L) &= 2c+\frac{c^2}{3}-\frac{c^3}{45}+\frac{c^4}{315}-\frac{c^5}{1575}+O(c^6),
\\
L^2 c_3^{N=1}(L) &= \frac{c^3}{15}-\frac{11c^4}{630}+\frac{47c^5}{9450}+O(c^6),
\\
L^2 c_4^{N=1}(L) &= \frac{2c^4}{63}-\frac{79c^5}{4725}+O(c^6)
\end{align}
Note that at small $L$ one has $c_2(L)\simeq \frac{4 \gamma}{L}$
while for a single polymer $c^{N=1}_2(L)\simeq \frac{\gamma}{L}$.

{\it Remark.} The leading $O(c)$ term in \eqref{eq:smallcE},
$-\frac{4n(n-1)}{L^2}\,c$,
can also be recovered directly from first-order perturbation theory
around the $c=0$ state. For each replica the unperturbed two-polymer
wave function is
$\psi_0(x_1,x_2)\propto
\sin\!\left(\pi(x_1-x_2)/L\right)$, whose one-polymer marginal is
uniform on the circle. Hence, for a fixed pair of distinct replicas
and a fixed choice of one polymer in each replica, the expectation
value of the contact interaction is $1/L$. There are
$n(n-1)/2$ choices of the replica pair and $2\times2=4$ choices of
the polymers, giving
\be
-\gamma\,
\frac{n(n-1)}{2}\,4\,\frac{1}{L}
=
-\frac{4n(n-1)}{L^2}\,c,
\ee
where $c=\gamma L/2$.

\subsubsection{Explicit expression for the variance}

We have developed a method to obtain $E(n)$ in an expansion in powers of $n$
around $n=0$ as in \eqref{expansionE}, for arbitrary $c= \frac{\gamma L}{2}$. This method allows in principle to compute all the cumulants $c_k(L)$ iteratively in $k$. The calculations become very tedious for $k \geq 3$, so we give here only the result for $k=1$ (which was displayed in \eqref{c1L})
and for $k=2$. The result for the variance $c_2(L)$ reads
\be
c_2(L) = \frac{8}{L^2} \bigg[ \frac{3 c}{2}  + \int_0^{+\infty} \frac{d \lambda}{2 \pi^2} e^{-\frac{2 \lambda
   ^2}{c}} \left( \tilde f_0(\lambda) + \tilde f_c(\lambda) \cos \left(\frac{2  \pi  \lambda}{c}\right)
   + \tilde f_s(\lambda) \sin \left(\frac{2  \pi  \lambda}{c}\right)  \right) \bigg]   \label{eq:c2exactIntro}
\ee
where the functions $\tilde f_0,\tilde f_c, \tilde f_s$ read, explicitly
\bea
&& \tilde f_0(\lambda) = 2 \left(c \left(3 \lambda
   ^2+\pi ^2\right)-2 \lambda ^2
   \left(\lambda ^2+\pi
   ^2\right)\right) \coth
   (\lambda )-2 \lambda  \left(c
   \left(\lambda ^2+\pi ^2\right)
   \text{csch}^2(\lambda )+c-2
   \lambda ^2\right) \\
 &&   \tilde f_c(\lambda) =  2 \lambda
   \left(c-2 \lambda ^2+\pi^2\right)+\left(c \left(\pi
   ^2-6 \lambda ^2\right)+4
   (\lambda -\pi ) (\lambda +\pi
   ) \lambda ^2\right) \coth
   (\lambda )+c \lambda  \left(2
   \lambda ^2-\pi ^2\right)
   \text{csch}^2(\lambda ) \nn \\
 &&  \tilde f_s(\lambda) =  \pi
   \left(\left(8 \lambda ^3-6 c
   \lambda \right) \coth (\lambda
   )+3 c \lambda ^2
   \text{csch}^2(\lambda )+c-6
   \lambda ^2\right) \nn
\eea
At small $\lambda$ these functions behave as $\tilde f_0(\lambda) \sim \tilde f_c(\lambda) = O(\lambda)$ and
$\tilde f_s(\lambda)=O(1)$, so the above integrals are convergent.

For comparison, the corresponding result of Brunet--Derrida
\cite{brunet2000probability} for a single polymer follows from the
$n^2$ term of their Eq.~(49). After the change of variables
$\lambda_{\rm DB}=2\lambda/\sqrt{c}$, it can be written in the much simpler form
\be
L^2 c_2^{N=1}(L)
=
8\int_0^{+\infty}d\lambda\,
\lambda^2 e^{-2\lambda^2/c}\coth\lambda .
\label{eq:c2DBexact}
\ee
In this form the Gaussian factor and the argument of the
hyperbolic functions are the same as in \eqref{eq:c2exactIntro}.

{\it Small $c$ expansion}. One performs the small $c$ expansion
separately on the uniform part $\tilde f_0$, and on the oscillating parts $\tilde f_{c,s}$. In the uniform part one first changes $\lambda \to \lambda \sqrt{c}$ and
then expand in $c$. In the oscillating parts one first changes
$\lambda \to c \lambda$ and multiplies by $e^{- \epsilon \lambda}$.
Then one performs the expansion in $c$ and note that each term
can be easily integrated for $\epsilon>0$. Then one takes $\epsilon \to 0$.
Remarkably, the final result is found to reproduce correctly the expansion
\eqref{cumsmallc00}, which we checked up to order $O(c^5)$,
showing that within our method the
expansions in $n$ and $c$ commute.

{\it Large $c$ expansion}. The large-$c$ expansion requires some care,
since the bulk scaling $\lambda\sim\sqrt{c}$ is not uniform near
$\lambda=0$. As detailed in Appendix~\ref{app:largecvariance}, the
additional contribution from the region $\lambda=O(1)$ cancels the
$O(c)$ term of the bulk expansion and produces a $\zeta(3)$ term. One finds
\be
c_2(L)
=
\frac{8}{L^2}
\bigg[
\frac{11}{32}\sqrt{\frac{\pi}{2}}\,c^{3/2}
+
\frac{5\pi^{5/2}}{128\sqrt{2}}\,c^{1/2}
+
\frac{5}{2}\zeta(3)
-
\frac{7\pi^{9/2}}{256\sqrt{2}}\,c^{-1/2}
+
O(c^{-1})
\bigg].
\label{c2largeLintro}
\ee

For comparison, applying the same bulk--boundary decomposition to the
Brunet--Derrida result \eqref{eq:c2DBexact} gives
\be
L^2c_2^{N=1}(L)
=
\sqrt{\frac{\pi}{2}}\,c^{3/2}
+
4\zeta(3)
+
O(c^{-1}).
\label{eq:c2DBlargec}
\ee
Thus in both cases the $\zeta(3)$ term originates from the
$\lambda=O(1)$ region, which is missed by the naive bulk expansion.

Equivalently, recalling that $c=\gamma L/2$, at fixed $\gamma$ this gives
\bea
c_2(L)
&=&
\frac{11\sqrt{\pi}}{16}\,
\frac{\gamma^{3/2}}{L^{1/2}}
+
\frac{5\pi^{5/2}}{32}\,
\frac{\sqrt{\gamma}}{L^{3/2}}
+
\frac{20\zeta(3)}{L^2}
\nn\\
&&
-
\frac{7\pi^{9/2}}{32}\,
\frac{1}{\sqrt{\gamma}\,L^{5/2}}
+
O(L^{-3}),
\label{eq:c2largeLfull}
\eea
whereas for a single polymer
\be
c_2^{N=1}(L)
=
\frac{\sqrt{\pi}}{4}\,
\frac{\gamma^{3/2}}{L^{1/2}}
+
\frac{4\zeta(3)}{L^2}
+
O(L^{-3}).
\label{eq:c2DBlargeL}
\ee

\begin{figure}[t]
\centering
\includegraphics[width=0.75\linewidth]{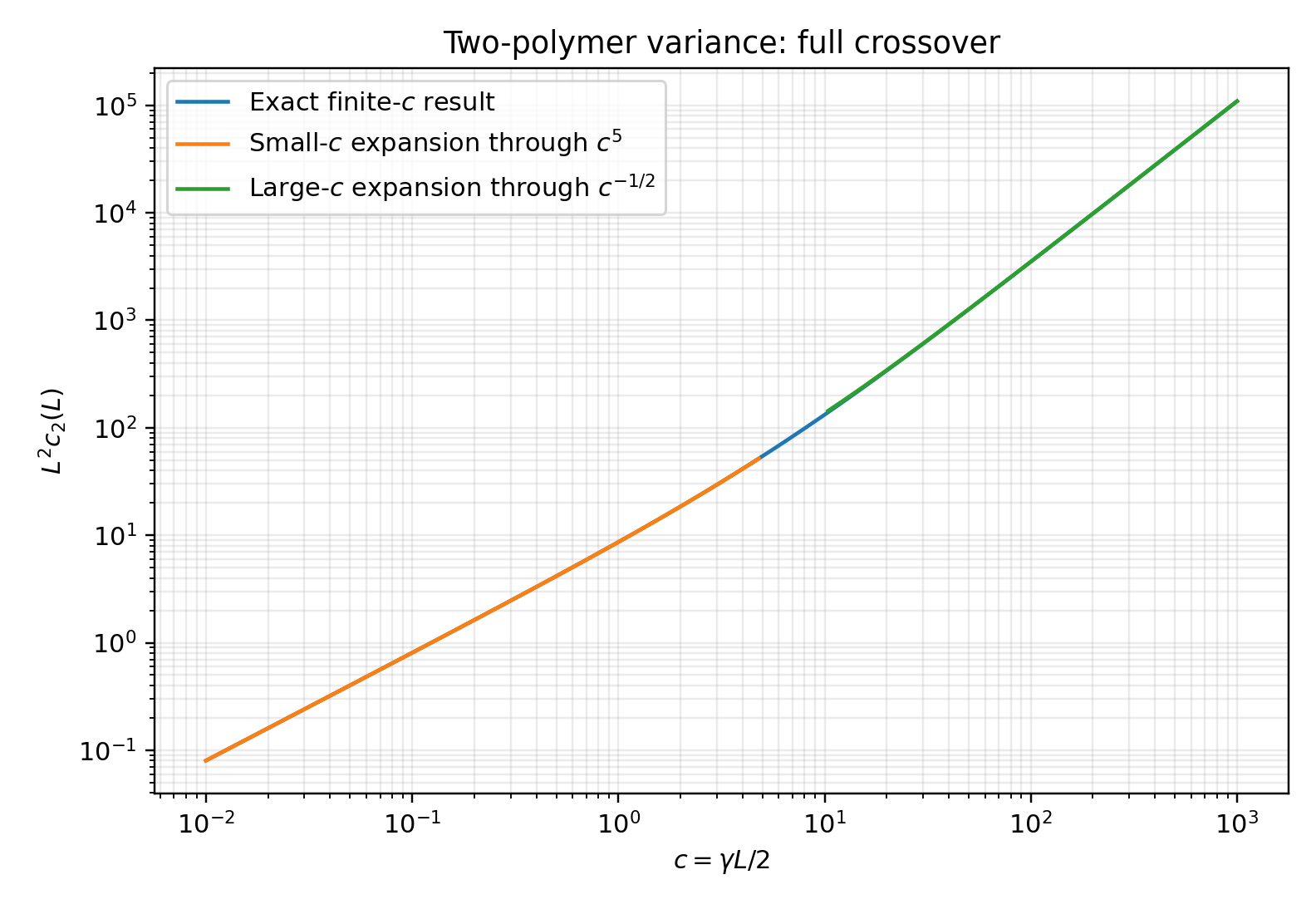}
\caption{Variance amplitude $L^2 c_2(L)$ as a function of
$c=\gamma L/2$. The solid curve shows the exact finite-$c$ result
\eqref{eq:c2exactIntro}. It is compared with the small-$c$ expansion
through order $c^5$ and with the large-$c$ asymptotic expansion through
order $c^{-1/2}$. The logarithmic scales emphasize the crossover from
the weak-coupling behavior $L^2 c_2(L)\simeq 8c$ at small $c$ to the
large-$c$ KPZ regime
$L^2 c_2(L)\simeq \frac{11}{4}\sqrt{\frac{\pi}{2}}\,c^{3/2}$.}
\label{fig:varianceGlobal}
\end{figure}

\begin{figure}[t]
\centering
\includegraphics[width=0.75\linewidth]{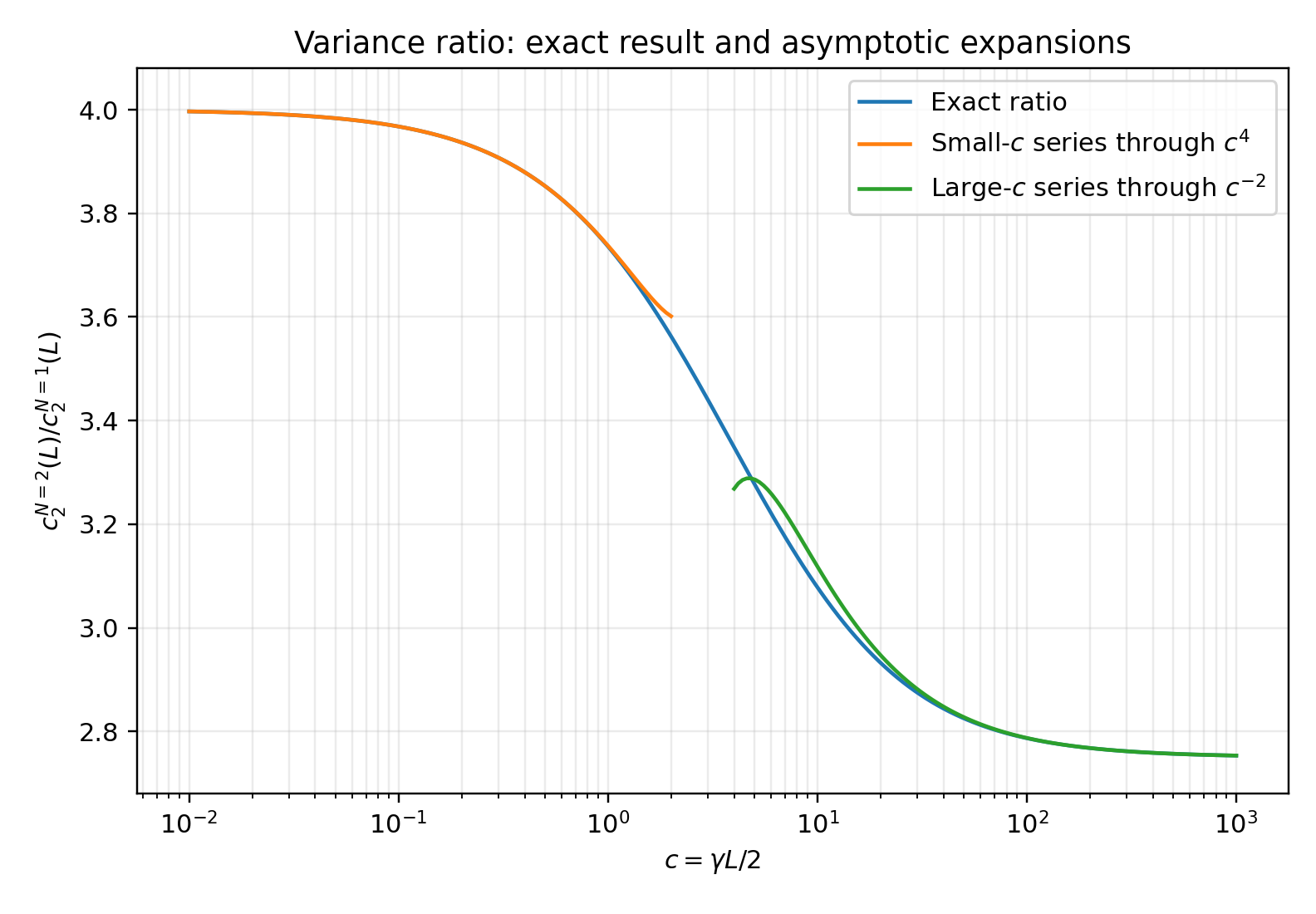}
\caption{Ratio of the two-polymer and one-polymer variance amplitudes,
$R(c)=c_2^{N=2}(L)/c_2^{N=1}(L)$, as a function of
$c=\gamma L/2$. The solid curve shows the exact finite-$c$ ratio obtained
from the corresponding exact variance formulas. It is compared with the
small-$c$ expansion \eqref{eq:varianceRatioSmallC} and the large-$c$
expansion \eqref{eq:varianceRatioLargeC}. The ratio crosses over from
$R(c)\to4$ at weak coupling to $R(c)\to11/4$ in the large-$c$ KPZ regime.}
\label{fig:varianceRatio}
\end{figure}

\begin{figure}[t]
\centering

\begin{subfigure}[t]{0.48\linewidth}
\centering
\includegraphics[width=\linewidth]{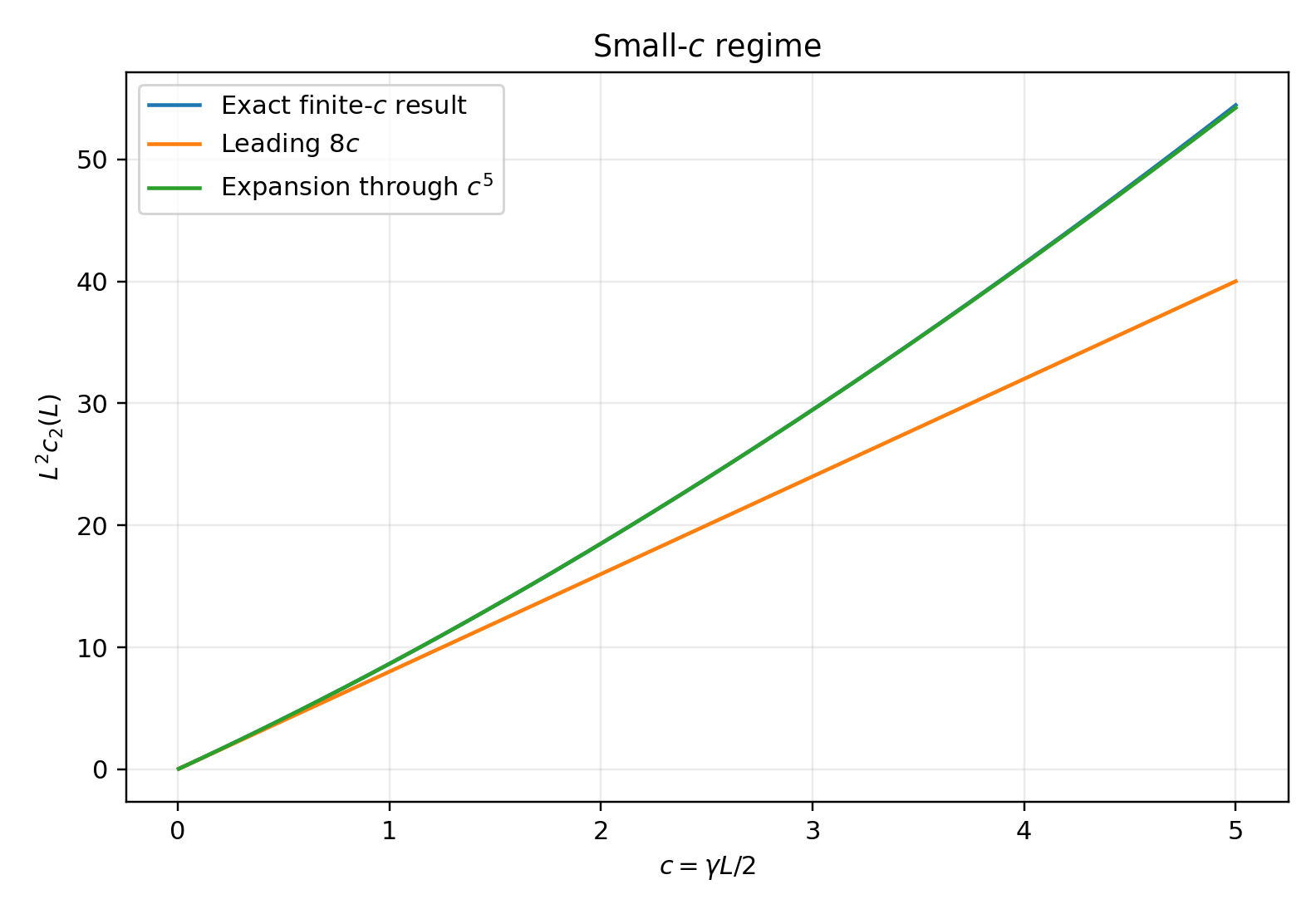}
\caption{Small-$c$ regime. The exact finite-$c$ result for
$L^2c_2(L)$ is compared with the leading behavior
$L^2c_2(L)\simeq 8c$ and with the perturbative expansion through
order $c^5$.}
\label{fig:varianceSmallC}
\end{subfigure}
\hfill
\begin{subfigure}[t]{0.48\linewidth}
\centering
\includegraphics[width=\linewidth]{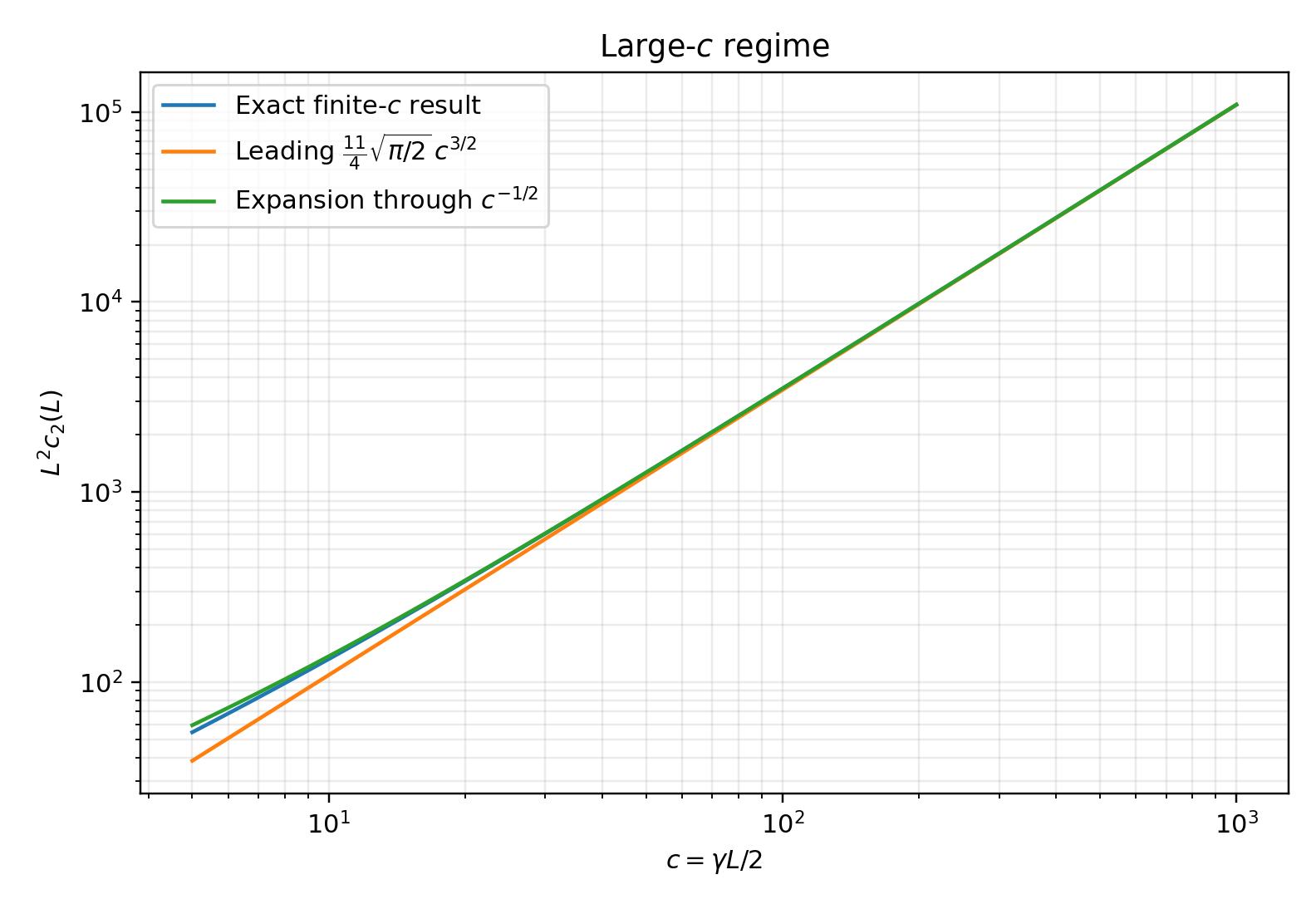}
\caption{Large-$c$ regime. The exact finite-$c$ result is compared
with the leading KPZ behavior
$L^2c_2(L)\simeq \frac{11}{4}\sqrt{\frac{\pi}{2}}\,c^{3/2}$
and with the large-$c$ asymptotic expansion through order
$c^{-1/2}$.}
\label{fig:varianceLargeC}
\end{subfigure}

\caption{Variance amplitude $L^2c_2(L)$ as a function of
$c=\gamma L/2$, with emphasis on the small-$c$ and large-$c$
regimes. The exact expression \eqref{eq:c2exactIntro} interpolates
smoothly between the weak-coupling expansion at small $c$ and the
large-$L$ KPZ scaling regime at large $c$.}
\label{fig:varianceAsymptotics}
\end{figure}

{\it Numerical evaluation}. We have plotted in Fig. \ref{fig:varianceGlobal} the
global form of $L^2 c_2(L)$ versus $c= \gamma L/2$
given by \eqref{eq:c2exactIntro}. The small $c$ and large $c$ behaviors
are plotted in
Fig. \ref{fig:varianceSmallC} and Fig. \ref{fig:varianceLargeC} respectively.
Finally we have plotted in Fig. \ref{fig:varianceRatio} the ratio of the two polymer to the one polymer variances.

\subsubsection{Large-\texorpdfstring{$L$}{L} limit}

\label{susubsec:largeL}

We now consider the large-$L$ limit, equivalently
\be
c=\frac{\gamma L}{2}\to+\infty.
\ee
The relevant replica scaling regime is
\be
n\to0,
\qquad
\epsilon=n\sqrt{c}
\quad\text{fixed}.
\label{eq:mainScalingEpsilon}
\ee
In this regime the ground-state energy takes the scaling form
\be
E(n,L)
=
\frac{4}{L^2}
\left[
\frac{nc^2}{12}
+
\frac{\sqrt{c}}{4}\,
{\cal E}(n\sqrt{c})
\right]
+
o\left(\frac{\sqrt{c}}{L^2}\right).
\label{eq:mainEnergyScaling}
\ee
We refer to ${\cal E}(\epsilon)$ as the large-$L$ scaling function
entering the SCGF.

For comparison, the Brunet--Derrida result for a single polymer
\cite{brunet2000probability}, written in the same variables
$c=\gamma L/2$ and $\epsilon=n\sqrt{c}$, takes the analogous form
\be
E^{N=1}(n,L)
=
\frac{2}{L^2}
\left[
\frac{nc^2}{12}
+
\frac{\sqrt{c}}{4}\,
{\cal E}^{N=1}(n\sqrt{c})
\right]
+
o\left(\frac{\sqrt{c}}{L^2}\right).
\label{eq:mainEnergyScalingDB}
\ee
In terms of the Brunet--Derrida scaling function $G$,
we have defined ${\cal E}^{N=1}(\epsilon)
=
-\frac{1}{\sqrt{\pi}}\,
G(-2\sqrt{\pi}\,\epsilon)$.
Equivalently, $2E^{N=1}(n,L)$ is the SCGF energy of two
single polymers in independent environments. In the normalization
above it has the same leading $nc^2$ term as
\eqref{eq:mainEnergyScaling}; the difference between the
replica-decoupled reference problem and the two non-crossing polymers
is therefore entirely contained, at this order, in
${\cal E}^{N=1}(\epsilon)$ versus ${\cal E}(\epsilon)$.
Let
\be
{\cal E}(\epsilon)
=
\sum_{k\geq1} e_k\epsilon^k.
\label{eq:mainEExpansion}
\ee
Using \eqref{expansionE}, \eqref{eq:mainEnergyScaling}, and
$\epsilon=n\sqrt{c}$, one immediately finds, for $k\geq2$,
\be
L^2c_k(L)
=
-k!\,e_k\,c^{(k+1)/2}
+
o\left(c^{(k+1)/2}\right).
\label{eq:mainCumulantsLargeL}
\ee
Equivalently, since $c=\gamma L/2$,
\be
c_k(L)
\simeq
-k!\,e_k
\left(\frac{\gamma}{2}\right)^{(k+1)/2}
L^{(k-3)/2},
\qquad k\geq2.
\label{eq:mainCumulantsLargeLGamma}
\ee
The first cumulant is special because of the first term in
\eqref{eq:mainEnergyScaling}, and its exact expression was already
given in \eqref{c1L}.

For the second cumulant we recover
\be
L^2c_2(L)
\simeq
\frac{11}{4}\sqrt{\frac{\pi}{2}}\,c^{3/2},
\label{eq:mainc2largec}
\ee
or, equivalently,
\be
c_2(L)
\simeq
\frac{11\sqrt{\pi}}{16}\,
\frac{\gamma^{3/2}}{\sqrt{L}}.
\ee
This agrees with the leading term of the large-$c$ expansion
\eqref{c2largeLintro} of the exact variance \eqref{eq:c2exactIntro}.

For comparison, the leading term of the Brunet--Derrida expansion
\eqref{eq:c2DBlargec}, obtained from the exact formula
\eqref{eq:c2DBexact}, gives
\be
L^2c_2^{N=1}(L)
\simeq
\sqrt{\frac{\pi}{2}}\,c^{3/2},
\qquad
c_2^{N=1}(L)
\simeq
\frac{\sqrt{\pi}}{4}\,
\frac{\gamma^{3/2}}{\sqrt{L}}.
\label{eq:mainc2largecDB}
\ee
Thus the two-polymer variance is enhanced by the factor
\be
\frac{c_2(L)}{c_2^{N=1}(L)}
\underset{L\to\infty}{\longrightarrow}
\frac{11}{4}.
\label{eq:mainc2ratioDB}
\ee

For the third cumulant we obtain
\be
L^2c_3(L)
\simeq
\left[
2+
\left(
\frac{512}{27\sqrt{3}}
-\frac{183}{16}
\right)\pi
\right]c^2
\simeq
0.463044\,c^2.
\label{eq:mainc3largec}
\ee
At the next order,
\bea
L^2c_4(L)
&\simeq&
\frac{\sqrt{\pi}}{1152}
\left[
1728\sqrt{2}
+
\pi
\left(
92745+91746\sqrt{2}-91136\sqrt{6}
\right)
\right]c^{5/2}
\nn\\
&\simeq&
0.167308885\,c^{5/2}.
\label{eq:mainc4largec}
\eea
We also obtain the fifth cumulant explicitly; its expression is given
in \eqref{eq:c5largec}, with
\be
L^2c_5(L)
\simeq
0.0810403524\,c^3.
\ee
For comparison, the corresponding Brunet--Derrida coefficients are
\bea
L^2c_3^{N=1}(L)
&\simeq&
\frac{\pi}{9}
\left(16\sqrt{3}-27\right)c^2
\simeq
0.248818648\,c^2,
\nn\\
L^2c_4^{N=1}(L)
&\simeq&
\pi^{3/2}
\left(
-16\sqrt{6}+18+15\sqrt{2}
\right)c^{5/2}
\simeq
0.118981533\,c^{5/2},
\nn\\
L^2c_5^{N=1}(L)
&\simeq&
0.067091280\,c^3.
\label{eq:mainHigherCumulantsDB}
\eea
Thus the large-$L$ powers $c^{(k+1)/2}$, equivalently
$L^{(k-3)/2}$ in \eqref{eq:mainCumulantsLargeLGamma}, are the same
as for a single polymer, while their amplitudes are modified by the
two-layer constraint.

\paragraph{Parametric representation of the scaling function.}

Beyond its expansion around $\epsilon=0$, the scaling function
${\cal E}(\epsilon)$ admits a closed parametric representation.
Introduce
\be
H_{A,B}(q)
=
2\sqrt{\pi}\,A(1+Bq^2)e^{-q^2},
\label{eq:mainHAB}
\ee
and the odd function
\be
\Omega_{A,B}(q)
=
\frac{1}{2\pi}
\dashint_{\mathbb R} dk\,
\frac{
\log\left[1-2H_{A,B}(k)\right]
}{q-k}.
\label{eq:mainOmegaAB}
\ee
We then define
\be
\hat f(A,B)
=
-\frac12
\int_{\mathbb R}\frac{dq}{2\pi}\,
\log\left[1-2H_{A,B}(q)\right],
\label{eq:mainfhat}
\ee
\be
\hat h(A,B)
=
\frac12
\int_{\mathbb R}\frac{dq}{2\pi}\,
q^2\log\left[1-2H_{A,B}(q)\right],
\label{eq:mainhhat}
\ee
and
\be
\hat g(A,B)
=
AB
\int_{\mathbb R}\frac{dq}{2\pi}\,
\frac{
8\sqrt{\pi}\,q^2e^{-q^2}
}{
1-H_{A,B}(q)
+
\sqrt{1-2H_{A,B}(q)}
\cos\left(\Omega_{A,B}(q)\right)
}.
\label{eq:mainghat}
\ee

For each $A$, let $B_A$ denote the solution of
\be
\hat f(A,B_A)
=
\hat g(A,B_A)
\label{eq:mainBA}
\ee
on the branch continuously connected to the origin,
\be
B_A\to2,
\qquad
A\to0.
\ee
The scaling variable and the scaling function are then given
parametrically by
\be
\epsilon
=
\hat f(A,B_A),
\qquad
{\cal E}(\epsilon)
=
-4\hat h(A,B_A)
.
\label{eq:mainParametricScalingFunction}
\ee
Near the origin one has
\be
A\simeq\frac{\epsilon}{2},
\qquad
B_A\simeq2.
\ee

For comparison, the Brunet--Derrida single-polymer scaling function
appearing in \eqref{eq:mainEnergyScalingDB} is parametrized by a single
auxiliary variable $\xi$,
\be
\epsilon
=
\frac{1}{2\sqrt{\pi}}\,
{\rm Li}_{3/2}(\xi),
\qquad
{\cal E}^{N=1}(\epsilon)
=
\frac{1}{\sqrt{\pi}}\,
{\rm Li}_{5/2}(\xi).
\label{eq:mainParametricScalingFunctionDB}
\ee
Here $\xi$ is the auxiliary parameter denoted $\epsilon$ in
Ref.~\cite{brunet2000probability}, Eqs.~(68)--(69); we use a different
symbol to avoid confusion with our scaling variable
$\epsilon=n\sqrt{c}$.
Thus
\eqref{eq:mainParametricScalingFunction} is the natural two-polymer
generalization of the Brunet--Derrida parametrization. The additional
quantity $B_A$, determined self-consistently by \eqref{eq:mainBA},
encodes the non-trivial two-layer structure. Its derivation from the
large-$L$ scaling equations using a Wiener--Hopf construction is
presented in Sec.~\ref{sec:WHscaling}.

\paragraph{Large-deviation scaling.} It is useful to state explicitly the consequence of
\eqref{eq:mainEnergyScaling} for the large-deviation function
$\Phi_L(H)$ defined in \eqref{defPhi}. In the large-$L$ scaling regime, equivalently large $c=\gamma L/2$ at fixed $\gamma$, we write
\be
H
=
-\frac{c^2}{3L^2}
+
\frac{c}{L^2}\,Y =
-\frac{\gamma^2}{12}
+
\frac{\gamma}{2L}\,Y,
\qquad
c\to+\infty,
\qquad
Y\ {\rm fixed}.
\label{eq:mainHScaling}
\ee
Using the Legendre relation \eqref{legendre2},
the terms proportional to $nc^2$ in
\eqref{eq:mainEnergyScaling} and $nH$ cancel, and one obtains
\be
\Phi_L\left(
-\frac{c^2}{3L^2}
+
\frac{c}{L^2}Y
\right)
=
\frac{\sqrt{c}}{L^2}\,
\Phi(Y)
+
o\left(\frac{\sqrt{c}}{L^2}\right),
\label{eq:mainPhiScaling}
\ee
Equivalently, using $c=\gamma L/2$, the same scaling form can be
written, for fixed $\gamma$ and $L\to+\infty$, as
\be
\Phi_L\left(
-\frac{\gamma^2}{12}
+
\frac{\gamma}{2L}Y
\right)
=
\sqrt{\frac{\gamma}{2}}\,
\frac{1}{L^{3/2}}\,
\Phi(Y)
+
o\left(L^{-3/2}\right).
\label{eq:mainPhiScalingGamma}
\ee
where the scaled rate function is the Legendre transform
\be
\Phi(Y)
=
\max_{\epsilon}
\left[
{\cal E}(\epsilon)+\epsilon Y
\right].
\label{eq:mainScaledRateFunction}
\ee
Following the analogous interpretation for the open KPZ fixed point
in Ref.~\cite{BarraquandLeDoussal2025KPZInterval}, the scaling
\eqref{eq:mainPhiScalingGamma} suggests that, assuming that the
large-$L$ and long-time limits can be exchanged, $\Phi(Y)$ can be
interpreted, up to normalization conventions, as the long-time
large-deviation rate function of the two-layer periodic KPZ fixed
point. This provides a fixed-point interpretation of the universal
scaling function obtained here.

In particular, matching the term linear in $n$ in
\eqref{eq:mainEnergyScaling} with the exact mean free energy
\eqref{c1L} gives ${\cal E}'(0)=4$. Hence the minimum of
$\Phi(Y)$ occurs at $Y=-4$, consistently with the large-$L$
expansion of the mean free energy.

For a single polymer we denote by $\Phi_L^{N=1}$ the corresponding
finite-$L$ large-deviation function, defined by
\be
P\left(\log Z_{N=1}(t)\simeq tH\right)
\asymp
e^{-t\,\Phi_L^{N=1}(H)}.
\label{eq:defPhiN1}
\ee
For comparison, for a single polymer $N=1$ the corresponding
Brunet--Derrida scaling follows from
\eqref{eq:mainEnergyScalingDB}. Writing
\be
H
=
-\frac{c^2}{6L^2}
+
\frac{c}{2L^2}\,Y
=
-\frac{\gamma^2}{24}
+
\frac{\gamma}{4L}\,Y,
\label{eq:mainHScalingDB}
\ee
one finds
\be
\Phi_L^{N=1}\left(
-\frac{c^2}{6L^2}
+
\frac{c}{2L^2}Y
\right)
=
\frac{\sqrt{c}}{2L^2}\,
\Phi^{N=1}(Y)
+
o\left(\frac{\sqrt{c}}{L^2}\right),
\label{eq:mainPhiScalingDB}
\ee
or equivalently
\be
\Phi_L^{N=1}\left(
-\frac{\gamma^2}{24}
+
\frac{\gamma}{4L}Y
\right)
=
\frac{1}{2}\sqrt{\frac{\gamma}{2}}\,
\frac{1}{L^{3/2}}\,
\Phi^{N=1}(Y)
+
o\left(L^{-3/2}\right).
\label{eq:mainPhiScalingDBGamma}
\ee
Here
\be
\Phi^{N=1}(Y)
=
\max_{\epsilon}
\left[
{\cal E}^{N=1}(\epsilon)+\epsilon Y
\right].
\label{eq:mainScaledRateFunctionDB}
\ee

\paragraph{Lower tail.}
We find that the two-polymer scaling function has the large
negative-$\epsilon$ asymptotics
\be
{\cal E}(\epsilon)
\simeq
-\frac{4}{5}(3\pi)^{2/3}|\epsilon|^{5/3},
\qquad
\epsilon\to-\infty.
\label{eq:mainLargeNegativeEpsilon}
\ee
For comparison, the Brunet--Derrida scaling function
\eqref{eq:mainParametricScalingFunctionDB} behaves as
\be
{\cal E}^{N=1}(\epsilon)
\simeq
-\frac{4}{5}\,2^{-2/3}(3\pi)^{2/3}
|\epsilon|^{5/3},
\qquad
\epsilon\to-\infty,
\label{eq:mainLargeNegativeEpsilonDB}
\ee
so that
\be
{\cal E}(\epsilon)
\simeq
2^{2/3}\,
{\cal E}^{N=1}(\epsilon),
\qquad
\epsilon\to-\infty.
\label{eq:mainEComparisonNegative}
\ee

Taking the Legendre transforms
\eqref{eq:mainScaledRateFunction} and
\eqref{eq:mainScaledRateFunctionDB} gives
\be
\Phi(Y)
\simeq
\frac{\sqrt{3}}{20\pi}|Y|^{5/2},
\qquad
Y\to-\infty,
\label{eq:mainScaledDeepLowerTail}
\ee
whereas
\be
\Phi^{N=1}(Y)
\simeq
\frac{\sqrt{3}}{10\pi}|Y|^{5/2}.
\label{eq:mainScaledDeepLowerTailDB}
\ee
Thus
\be
\Phi(Y)
\simeq
\frac12\Phi^{N=1}(Y),
\qquad
Y\to-\infty.
\label{eq:mainScaledPhiComparison}
\ee
The factor $1/2$ has a simple interpretation in terms of the other
natural reference problem. For two unconstrained polymers in the same
environment one has $Z_{\rm b}=(Z^{N=1})^2$, hence
${\cal E}_{\rm b}(\epsilon)=\frac12{\cal E}^{N=1}(2\epsilon)$ and
$\Phi_{\rm b}(Y)=\frac12\Phi^{N=1}(Y)$. Thus
\eqref{eq:mainScaledPhiComparison} shows that, in the deep lower tail,
the non-crossing constraint does not modify the leading
large-deviation cost.

Equivalently, using \eqref{eq:mainHScaling}, the result can be written
directly in terms of (minus) the physical free-energy density as
\be
\Phi_L(H)
\simeq
\frac{\sqrt{c}}{L^2}\,
\frac{\sqrt{3}}{20\pi}
\left|
\frac{L^2}{c}H+\frac{c}{3}
\right|^{5/2},
\label{eq:mainDeepLowerTail}
\ee
in the corresponding large-$L$ deep lower-tail regime. Combining
\eqref{eq:mainPhiScaling}, \eqref{eq:mainPhiScalingDB}, and
\eqref{eq:mainScaledPhiComparison}, one equivalently finds
\be
\Phi_L(H)
\simeq
\Phi_L^{N=1}\left(\frac{H}{2}\right).
\label{eq:mainPhiComparison}
\ee
Thus the deep lower tail has the same $5/2$ exponent as for a single
polymer, with the rescaling $H\to H/2$ reflecting the two-polymer
free energy. This is precisely the physical-rate-function relation
obeyed by the same-environment unconstrained reference problem.
Equivalently, in this deep lower-tail large-$L$ regime, the random
variables $t^{-1}\log Z_{N=2}(t)$ and
$2t^{-1}\log Z_{N=1}(t)$ have the same leading large-deviation rate
function. Physically, a single rare fluctuation of the common
environment depresses both polymer partition sums simultaneously,
rendering the non-crossing constraint irrelevant to the leading cost;
see Sec.~\ref{sec:largeNegativeEpsilon} for a more detailed discussion.

\paragraph{Right tail.}
On the opposite side, an analytic continuation of the scaling SCGF is required, and is performed in Sec.~\ref{sec:largePositiveEpsilon}. It leads to
the following asymptotics at large positive $\epsilon$
\be
{\cal E}(\epsilon)
=
-\frac{\epsilon^3}{3}
+
\frac{\pi^2}{\epsilon}
+
O(\epsilon^{-3}),
\qquad
\epsilon\to+\infty,
\label{eq:mainLargePositiveEpsilon}
\ee
and hence
\be
\Phi(Y)
=
\frac{2}{3}Y^{3/2}
+
\frac{\pi^2}{\sqrt{Y}}
+
O(Y^{-3/2}),
\qquad
Y\to+\infty.
\label{eq:mainScaledRightTail}
\ee
For a single polymer, the Brunet--Derrida scaling function has the same
leading cubic asymptotics, so that
$\Phi^{N=1}(Y)\simeq\Phi(Y)\simeq \frac{2}{3}Y^{3/2}$ at leading
order, although the first correction is exponentially small for
$N=1$ and algebraic for $N=2$. Equivalently,
${\cal E}(\epsilon)/{\cal E}^{N=1}(\epsilon)\to1$ as
$\epsilon\to+\infty$. Taking into account the different physical scalings in
\eqref{eq:mainPhiScaling} and \eqref{eq:mainPhiScalingDB}, this gives
\be
\Phi_L(H)
\simeq
2\,\Phi_L^{N=1}\left(\frac{H}{2}\right)
\label{eq:mainRightTailPhysicalComparison}
\ee
in the corresponding large-$L$ deep right-tail regime. These two
relations have the same physical interpretation: relative to the
replica-decoupled reference problem, the optimal favorable
fluctuation asymptotically splits into two spatially separated
corridors, each carrying one polymer and half of the total excess
free energy, so that the two disorder costs add independently. The
analytic continuation and comparison with the single-polymer problem
are discussed in Sec.~\ref{sec:largePositiveEpsilon}, while the
optimal-fluctuation argument is given in
App.~\ref{app:rightTailOF}.

Thus the two tails approach two different natural reference problems:
the common-environment unconstrained problem on the left, and the
independent-environment problem on the right.

\vspace{.5cm}

\begin{figure}[t]
\centering
\includegraphics[width=0.98\linewidth]{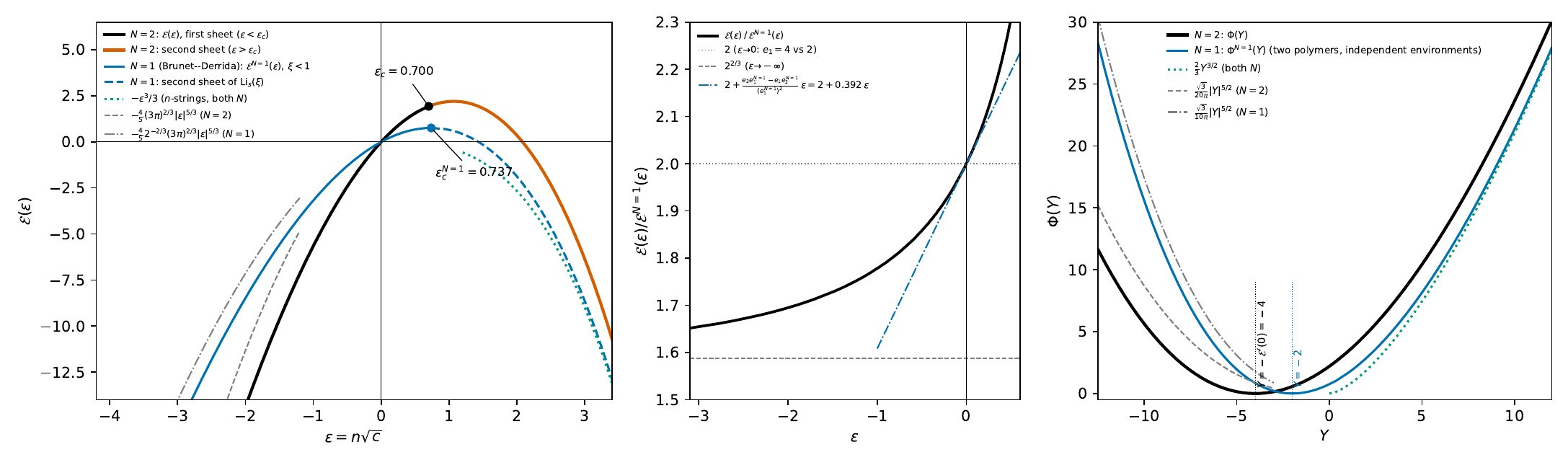}
\caption{The large-$L$ scaling function of two non-crossing polymers compared with the one-polymer result. Left:
${\cal E}(\epsilon)$ from the parametric representation \eqref{eq:mainParametricScalingFunction}, on its principal
sheet (black) and on the second sheet beyond the branch point $\epsilon_c=0.700$ (red), together with the
Brunet--Derrida function ${\cal E}^{N=1}(\epsilon)$, which in the normalization of \eqref{eq:mainEnergyScalingDB}
is the scaling function of two polymers in independent environments, on its principal sheet ($0<\xi<1$, solid
blue) and beyond $\epsilon_c^{N=1}=0.737$ (dashed blue); dotted: the common asymptote $-\epsilon^3/3$; grey: the
large-negative-$\epsilon$ asymptotics \eqref{eq:mainLargeNegativeEpsilon}, \eqref{eq:mainLargeNegativeEpsilonDB};
the dots mark the two branch points. Middle: the ratio ${\cal E}/{\cal E}^{N=1}$, which interpolates between
$2^{2/3}$ for $\epsilon\to-\infty$, \eqref{eq:mainEComparisonNegative}, $2$ at $\epsilon=0$ (twice the mean) and
$1$ for $\epsilon\to+\infty$; the tangent at the origin has slope $0.392$, which encodes the variance ratio $11/4$
of \eqref{eq:mainc2ratioDB}. Right: the scaled rate functions $\Phi(Y)$ and $\Phi^{N=1}(Y)$ of
\eqref{eq:mainScaledRateFunction}, with minima at $Y=-4$ and $Y=-2$, the lower-tail asymptotics
\eqref{eq:mainScaledDeepLowerTail}, \eqref{eq:mainScaledDeepLowerTailDB}, and the common upper-tail behavior
$\frac23Y^{3/2}$. Details are given in Sec.~\ref{sec:largePositiveEpsilon} and Appendix~\ref{app:continuationAc}.}
\label{fig:scalingfunction}
\end{figure}

The scaling function ${\cal E}(\epsilon)$ and the scaled rate function $\Phi(Y)$ are shown in
Fig.~\ref{fig:scalingfunction}, together with the one-polymer functions of Brunet and Derrida, which in the
normalization of \eqref{eq:mainEnergyScalingDB} describe two polymers in independent environments: the comparison
isolates the effect of the non-crossing constraint, from the ratio $2^{2/3}$ of the lower tails to the coincidence
of the upper tails. The figure is discussed in Sec.~\ref{sec:largePositiveEpsilon}; the continuation of
${\cal E}(\epsilon)$ beyond the branch point $\epsilon_c$, which is needed for $\epsilon>\epsilon_c$ and for the
whole upper tail of $\Phi(Y)$, is constructed in Appendix~\ref{app:continuationAc}.

\paragraph{Remark on the scaling of the cumulants.}

The powers of $L$ appearing in the large-$L$ cumulants admit a simple
KPZ scaling interpretation. On the infinite line, the transverse scale
and the free-energy fluctuations behave as
\be
\ell_\perp(t)\sim\gamma^{1/3}t^{2/3},
\qquad
\delta\log Z(t)\sim(\gamma^2t)^{1/3}.
\ee
The time needed to explore a circle of circumference $L$ therefore scales as
\be
t^*(L)\sim\frac{L^{3/2}}{\sqrt{\gamma}},
\ee
and the free-energy fluctuations accumulated over such a time are of
order $\sqrt{\gamma L}$. Treating a long polymer as a succession of
approximately independent blocks of duration $t^*(L)$ then gives
\be
c_k(L)
\sim
\frac{(\gamma L)^{k/2}}{t^*(L)}
\sim
\gamma^{(k+1)/2}L^{(k-3)/2},
\ee
in agreement with \eqref{eq:mainCumulantsLargeLGamma}. This argument
only fixes the scaling exponents and is insensitive to the distinction
between one and two polymers; the non-trivial dependence on the
two-polymer constraint resides in the amplitudes computed above.

In the opposite regime $c\ll1$, the leading variance can be understood
from the spatial zero mode of the Gaussian disorder. This mode contributes
once to the free energy of a single polymer and twice to that of two
polymers, leading respectively to
$c_2^{N=1}(L)\sim\gamma/L$ and $c_2(L)\sim4\gamma/L$.
Since the zero mode is Gaussian, this argument does not account for the
higher cumulants, which involve the nonzero spatial modes.

\subsection{Direct numerical simulation}
\label{subsec:numerics}

We have tested the
finite-$L$ predictions \eqref{c1L} and \eqref{eq:c2exactIntro}, beyond the $O(c)$ terms, $-4c$ and $8c$, by a
direct simulation of two non-crossing polymers on a lattice. We use the high-temperature weak-disorder scaling of
Ref.~\cite{calabrese2010free}: directed polymers on a square lattice with i.i.d.\ Gaussian site energies of variance
$\sigma^2$, wrapped on a cylinder of $W_X$ sites, converge at large temperature $T$, with $x\propto X/T^2$,
$t\propto\hat t/T^4$ and $\gamma=\sigma^2$, to the continuum model of Sec.~\ref{subsec:model} with
$L\propto W_X/T^2$, so that $c=\gamma L/2=L\sigma^2/2$. For each disorder realization the non-crossing two-polymer
partition sum is obtained from the lattice version of the determinantal formula \eqref{eq:KarlMcGregCyc}: the two
one-polymer partition sums are propagated by transfer matrix in the same disorder, with the antiperiodic twist which
implements the sign of the winding sectors as in \eqref{eq:periodicity}, and $Z^{\rm NC}$ is their $2\times2$
determinant. Technical details are provided in Appendix~\ref{app:lattice}. The
cumulants $c_k(L)$ are then extracted from the linear growth in time of the sample cumulants of $\log Z^{\rm NC}$,
over $10^5$ to $5\cdot10^5$ disorder realizations per point, for $c=0.5,1,2,3$, where the $c^2$ terms of
\eqref{c1L} and \eqref{eq:c2exactIntro} are between $4\%$ and $23\%$ of the $O(c)$ ones; two further runs at
$c=1$, with a finer lattice and with a larger ring, control the residual lattice corrections.

\begin{figure}[t]
\centering
\includegraphics[width=0.98\linewidth]{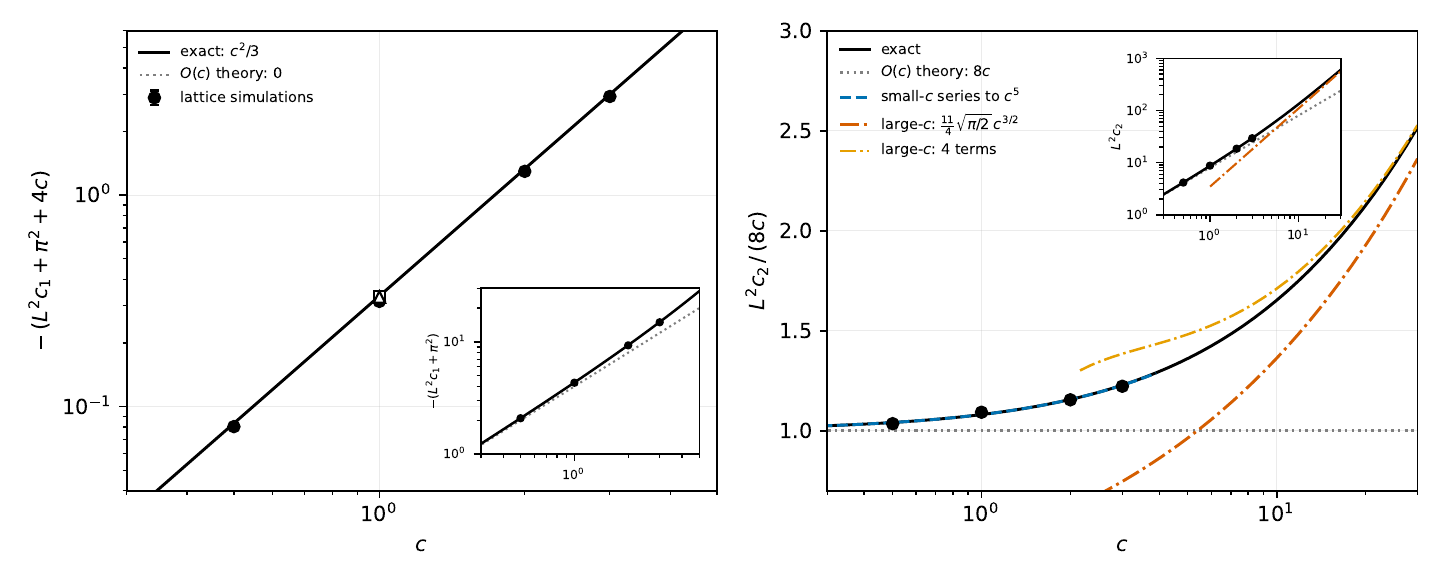}
\caption{Lattice simulations of two non-crossing polymers (Appendix~\ref{app:lattice}) compared with the
finite-$L$ predictions, as functions of $c=\gamma L/2$. Left: the $c^2$ term of $L^2c_1(L)$,
$-(L^2c_1+\pi^2+4c)$, versus $c^2/3$ from \eqref{c1L}; inset: $-(L^2c_1+\pi^2)$ with $4c$ (dotted) and
$4c+c^2/3$. The open symbols at $c=1$ are control runs with a finer lattice (squares) and a larger ring
(triangles), benchmarks of the convergence in the lattice resolution and in $L$. Right: $L^2c_2(L)/(8c)$ versus the
exact variance \eqref{eq:c2exactIntro} (solid), its small-$c$ series through $c^5$ (dashed) and its large-$c$
expansion \eqref{c2largeLintro} with one and four terms (dash-dotted); inset: $L^2c_2(L)$ with $8c$ (dotted) and
the leading large-$c$ term. Error bars are bootstrap estimates over $10^5$ to $5\cdot10^5$ disorder realizations
per point.}
\label{fig:latticeVsTheory}
\end{figure}

The results are shown in Fig.~\ref{fig:latticeVsTheory}. Since $L^2c_1(L)$ and $L^2c_2(L)$ are dominated by their
$O(c)$ terms, the figure displays the $c^2$ term of $L^2c_1$ and the ratio $L^2c_2/(8c)$. The exact variance
\eqref{eq:c2exactIntro} is confirmed at the level of $0.5$--$1\%$: the measured values agree with it within
$1$--$1.5$ standard deviations at all points, and resolve the terms beyond the linear order $8c$ with high
significance ($9$ standard deviations at $c=0.5$, more than $18$ at $c=2,3$), confirming the expansion up to the
quadratic term. The small-$c$ series \eqref{eq:smallcE} through $O(c^5)$ is
indistinguishable from the exact result up to $c=3$, while the large-$c$ expansion \eqref{c2largeLintro} approaches
it only for $c\gtrsim10$. The $c^2/3$ term of $L^2c_1$ in \eqref{c1L} is likewise confirmed, up to a residual
lattice shift of a few per mille of $L^2c_1$ which decreases with a finer lattice or a larger ring. The third and
fourth cumulants are not resolved with these statistics, their statistical errors being comparable to the predicted
values: a direct test of the non-Gaussian cumulants, and of the large-deviation function, requires
importance-sampling methods targeting the rare fluctuations, as discussed in the Conclusion.

\section{Determinant formula, fermions and Bethe ansatz }

\label{sec:determinant}
In this section we recall a determinantal formula for non-crossing polymers on the circle, and we formulate the problem of computing moments of partition sum in terms of a Lieb-Liniger Hamiltonian.

\subsection{Determinantal formula for
non-crossing polymers on the circle}

\begin{figure}[h]
    \centering
\includegraphics[width=0.75\linewidth]{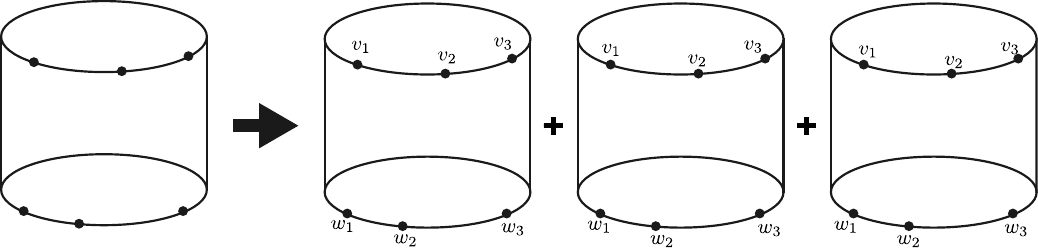}
    \caption{A graphical explanation of Eq.~\eqref{eq:ZtoWeylC} for $N=3$: on the circle, once the endpoint sets of the non-crossing paths have been fixed, there are still different ways to connect them, associated with cyclic permutations of the final points (or of the starting points). Lighter portions of the trajectories lie on the rear side of the cylinder.}
    \label{fig:circle-sketch}
\end{figure}

Let us go back to the partition function for the $N$ non-crossing polymers as defined in Eq.~\eqref{eq:ZNCcircle}. Similarly to what we did on the line, we want to make use of the fact that non-crossing paths preserve an appropriate notion of order.
Intuitively, given a set of points on the circle $W$, by following the circle counterclockwise, we can define an order among its points. However, since the starting point is arbitrary, it is clear that only the circular order is well-defined. Continuous paths on a circle that are constrained not to intersect must necessarily preserve the cyclic order~(see Fig.~\ref{fig:circle-sketch}).
To formalize this concept, following \cite{hobson1996non,grabiner2002random}, we introduce the generalization of the Weyl chamber as
\begin{equation}
     W^{(L)}_N = \{  \xx = (x_1,\ldots, x_N) \in \mathbb{R}^N  | x_1 < x_2 < \ldots < x_N < x_1 + L \} .
\end{equation}
Note the last inequality, absent if $L \to \infty$, ensures that the $N$ points on the real line can be identified with $N$ ordered points on the circle.

Thanks to the definition of the Weyl chamber, in analogy with Eq.~(\ref{eq:ZNCline},\ref{eq:ZtoWeyl}), we can define a partition sum for $N$ non-crossing polymers on the circle where
each path has fixed endpoints
\begin{equation}
\label{eq:ZfixedendpointC}
    \mathcal{Z}_t^{(L)}(\xx,  \yy) = \int_{\substack{w_i(0) = \mathsf{e}(y_i) \\ w_i(t) = \mathsf{e}(x_i) }}^{\rm non-crossing} \prod_i \mathcal{D} w_i e^{-\int_0^t d\tau \mathcal{L}[w_i, \dot w_i, \tau]} \;, \qquad \xx, \yy \in W_N^{(L)}
\end{equation}
Note that as this function only depends on the endpoints via the exponential map and $\eee(x + n L) = \eee(x)$ holds, this quantity satisfies
\begin{equation}
\label{eq:circle-periodicity}
\mathcal{Z}_t^{(L)}(\xx + n L, \yy + n' L) =  \mathcal{Z}_t^{(L)}(\xx, \yy) \;.
\end{equation}
Here and below, we use the shorthand
$\xx+a := (x_1+a,\ldots,x_N+a)$.
On the real line, we were able to write Eq.~\eqref{eq:ZtoWeyl} expressing the sum $Z_{t}(X,Y)$ directly in terms of the partition sum
$\mathcal{Z}_t(\xx,\yy)$ over non-crossing paths
with fixed end-points in the Weyl chamber. Since on the circle, it is rather the cyclic order that is preserved, it is useful to introduce the operator implementing the cyclic shift in the Weyl chamber
\begin{equation}
\label{eq:Tshiftdef}
    T: W_N^{(L)} \to W_N^{(L)} \;, \quad T[\xx] = (x_2, x_3, \ldots, x_N, x_1 + L)
\end{equation}
Consequently, the analogue of Eq.~\eqref{eq:ZtoWeyl} in the case of the circle becomes
\begin{equation}
\label{eq:ZtoWeylC}
    Z_t^{(L)}(W,V) = \sum_{i = 0}^{N-1} \mathcal{Z}_t^{(L)}(T^{i}[\xx], \yy)
\end{equation}
where $\xx, \yy \in W_N^{(L)}$ chosen such that $\{\eee(x_1), \ldots, \eee(x_N)\} \equiv W$ and $\{\eee(y_1), \ldots, \eee(y_N)\} \equiv V$.
Note that although there are multiple $\xx, \yy$ with this property, the right-hand side of Eq.~\eqref{eq:ZtoWeylC} is well-defined. Indeed, if there are $\xx$ and $\xx'$ both corresponding to the same set $W$, it must be true that $\xx' = T^k[\xx] + n L $, for $k \in 0,\ldots, N-1$ and $n \in \mathbb{Z}$, where $T^k$ means $k$ composition of $T$ in \eqref{eq:Tshiftdef}. But one has $\mathcal{Z}^{(L)}_t(\xx + n L, \yy) = \mathcal{Z}^{(L)}_t(\xx,  \yy) $ since in the definition \eqref{eq:ZfixedendpointC} only the image of the endpoints through the exponential map enters. Additionally, one can re-absorb any application of the shift operator $T$ in the sum, as $T^{N}[\xx] = \xx + L$. A similar argument can be made for a $\yy' = T^{\ell}[\yy] + n' L$, using that $ \mathcal Z_t^{(L)}(T[\xx],T[\yy]) = \mathcal Z_t^{(L)}(\xx,\yy)$. Thus, the right-hand side is actually a well-defined function of $W, V$.

Importantly, the partition function $\mathcal{Z}^{(L)}_t(\xx,\yy)$ with fixed end points in the Weyl chamber can be related to the solution of the multi-dimensional periodic SHE
\be
\partial_t \mathcal{Z}^{(L)} = \frac{1}{2} \sum_i \partial_{x_i}^2 \mathcal{Z}^{(L)}
+ \sqrt{\gamma} \sum_i \eta^{(L)}(x_i,t) \mathcal{Z}^{(L)} \label{SHEperiodic}
\ee
with initial condition $\mathcal{Z}^{(L)}_0(\xx,\yy) = \prod_{i=1}^N \delta^{(L)}(x_i-y_i)$, with $\delta^{(L)}(x)  = \sum_{j \in \mathbb{Z}} \delta(x + j L)$ the periodic $\delta$ comb
and absorbing boundary condition for all $1 \leq i \neq j \leq N$
\be \label{modulo}
\mathcal{Z}^{(L)}_t(\xx,\yy)|_{x_i=x_j \text{(modulo L)}} = 0
\ee
which amounts to impose that
$\mathcal{Z}^{(L)}_t(\xx,\yy)$ vanishes at the boundary of $ W_N^{(L)} $.
We recall that $\eta^{(L)}(x,t)$ is periodic, $\eta^{(L)}(x+L,t)=\eta^{(L)}(x,t)$
\\
At this point, it is worth mentioning that there is a generalization of the Karlin-McGregor formula, which allows us to express the partition function $Z^{(L)}(W,V)$ once again as a single determinant~\cite{liechty2016nonintersecting}.
The corresponding formula for non-intersecting Brownian motions on the circle is discussed in \cite{hobson1996non,liechty2016nonintersecting}; the same determinantal construction applies here, for each realization of the random potential.
Specifically,
\begin{equation}
\label{eq:KarlMcGregCyc}
    Z_t^{(L)}(W, V) =
     \sum_{i = 0}^{N-1} \mathcal{Z}_t^{(L)}(T^{i}[\xx], \yy) = \det_{1 \leq i,j \leq N} \left( \sum_{m=-\infty}^{\infty} z_t^{(L)}(x_i,y_j + L m) (-1)^{m(N-1)}\right) \;,
\end{equation}
We note that for odd $N$, the coefficient of the determinant reduces to the single-polymer partition function \eqref{eq:single-polymercyl}, and therefore the final formula is formally analogous to the Karlin-McGregor formula for the full line, Eq. \eqref{eq:KarlMcGregLine} \cite{karlin1959coincidence}. In the case of even $N$, however, it is necessary to assign a sign $(-1)$ to each winding $m$ of the polymer around the cylinder.  It is important to distinguish the fixed-endpoint partition sum
$\mathcal{Z}^{(L)}_t(\xx,\yy)$ from the partition sum
$Z_t^{(L)}(W,V)$ associated with unordered endpoint sets: the latter is given
by the single determinant in Eq.~\eqref{eq:KarlMcGregCyc}, while the former
corresponds to only one of the cyclic sectors entering the sum in
Eq.~\eqref{eq:ZtoWeylC} and does not, in general, admit such a simple
determinantal representation. For more details on the relationship between these two partition functions, see~\cite{hobson1996non,liechty2016nonintersecting}.

\subsection{Mapping onto free fermions}
The determinantal structure of Eq.~\eqref{eq:KarlMcGregCyc} suggests an interpretation of the right-hand side as the wave function of a set of $N$ non-interacting fermions. Indeed, Eq.~\eqref{eq:KarlMcGregCyc} can be naturally extended beyond the Weyl Chamber to $\xx, \yy \in \mathbb{R}^{N}$, introducing the wave function
\begin{equation}
\label{eq:psidef}
\psi_t(\xx, \yy) := \det_{1 \leq i,j \leq N} \left( \sum_{m=-\infty}^{\infty} z_t^{(L)}(x_i,y_j + L m) (-1)^{m(N-1)}\right) \;.
\end{equation}
Consistently with the fermionic interpretation, it satisfies the antisymmetry
\begin{equation}
\label{eq:antisymmetry}
    \psi_t(\sigma \xx, \yy) = (-1)^\sigma     \psi_t(\xx,\yy)
\end{equation}
where $\sigma \in S_N$ is a permutation acting on the indices. For later convenience our convention is that the vector ${\bf x}' = \sigma \xx$
has components $x'_i=x_{\sigma^{-1}_i}$.
This last equation also gives a more direct way to understand the need for the $(-1)$ weighting the winding number:
as the fermion travels around the circle it collects $-1$ exchange phases from each of the $N-1$ other particles.
It also satisfies the spatial boundary condition
\be
\label{eq:periodicity}
\psi_t(x_1,\dots,x_i+L, \dots,x_N , \yy) = (-1)^{N-1} \psi_t(x_1,\dots,x_N , \yy )
\ee
Note that these properties clearly imply
\bea \label{per}
&& \psi_t(x_1+L,x_2+L,\dots,x_N+L,\yy) = \psi_t(x_1,\dots,x_N,\yy)  \\
&& \psi_t(x_2,x_3,\dots,x_1+L, \yy)= \psi_t(x_1,\dots,x_N , \yy)
\eea
The first property is global periodicity,
and in the last equality we used that the parity of a $N$ cycle is precisely $(-1)^{N-1}$.
The second identity is the fermionic counterpart of the cyclic invariance underlying Eq.~\eqref{eq:ZtoWeylC}.

The wave function $\psi_t(\xx, \yy)$ can also be seen as the solution of the SHE \eqref{SHEperiodic}, satisfying (\ref{eq:antisymmetry}, \ref{eq:periodicity}) with the initial condition
\begin{equation}
\label{eq:inicondpsit}
\psi_{t=0}(\xx, \yy) := \det_{1 \leq i,j \leq N} \left( \sum_{m=-\infty}^{\infty} \delta(x_i-y_j - L m) (-1)^{m(N-1)}\right) \;.
\end{equation}
As a consequence, it also satisfies Eq.~\eqref{modulo}.

\subsection{Replicas and noise average}

We want to compute the symmetrized moments of the partition function, averaged over the noise, defined for $n \in \mathbb{N}$ as
\bea \label{productZ}
\sum_{\sigma' \in S_n} \overline{ \prod_{a=1}^n Z_t^{(L)}(W_a, V_{\sigma'_a}) }
\eea
Using \eqref{eq:KarlMcGregCyc} and \eqref{eq:psidef}, we can re-express this quantity in terms of the replica wave function
\begin{equation}
\label{eq:Psi}
    \Psi_t(\XX, \YY) := \sum_{\sigma' \in S_n} \overline{\prod_{a=1}^n \psi_t(\xx_a , \yy_{\sigma'_a})  }
\end{equation}
where $\XX, \YY$ are shorthands for
\begin{equation} \label{coordinates}
    \XX := \{\xx_a \}_{a=1}^n = \{x_{i,a}\}_{\substack{
1 \leq i \leq N \\
    1\leq a \leq n
    }} = \{ x_\ibig \}_{\mu=1}^{nN} \;, \qquad \YY := \{\yy_a\}_{a=1}^n = \{y_{i,a}\}_{\substack{
1 \leq i \leq N \\
    1\leq a \leq n
    }} = \{y_\ibig \}_{\mu=1}^{nN}
\end{equation}
and to simplify we introduced collective indices $\ibig = (i,a)$ so that $\ibig \in \{1,\ldots, n N\}$.

We will focus on the case of identical boundary sets, $V_a=V$ and $W_a=W$ for all $a$. Then \eqref{productZ}
reduces to
\begin{equation}
\Psi_t(\XX,\YY)=n!\,\overline{Z_t^{(L)}(W,V)^n},
\end{equation}
as in \eqref{eq:defE}. Note that the overall factor $n!$ does not affect the large-time
exponential rate defining $E^{(N)}(n,L)$. In this case,
anticipating the notation using Young tableaux, it is useful to arrange the
coordinates in an $N \times n$ rectangle, with the row index corresponding
to the polymer label $i$ and the column index to the replica label $a$.
For instance, for $n = 3$ and $N = 2$, one has the tableau
\begin{equation}
\label{eq:Trect}
\ytableausetup{mathmode, boxsize=2.0em}
\trect :=
\begin{ytableau}
{x_{1,1}} & {x_{1,2}} & {x_{1,3}} \\
{x_{2,1}} & {x_{2,2}} &  {x_{2,3}}
\end{ytableau}
\; .
\end{equation}
As each individual factor in Eq.~\eqref{eq:Psi} satisfies Eq.~\eqref{SHEperiodic},
the replica wave
function obeys the imaginary time Schrodinger equation
\begin{equation}
    \partial_t \Psi_t(\XX , \YY)
    = - H_{\rm LL} \Psi_t(\XX , \YY)
\end{equation}
where we have defined the Lieb--Liniger Hamiltonian with attractive coupling $\cb>0$ (note that $2H_{\rm LL}$ is the Lieb--Liniger Hamiltonian in the standard normalization $-\sum_\ibig\partial_{x_\ibig}^2+2c_{\rm LL}\sum_{\ibig<\jbig}\delta^{(L)}(x_\ibig-x_\jbig)$, with $c_{\rm LL}=-\cb$; the coupling $\cb$ should not be confused with the dimensionless parameter $c=\cb L/2$ used throughout)
\begin{equation}
\label{eq:HLL}
    H_{\rm LL} = -\frac 1 2\sum_{\ibig=1}^{N n} \partial_{x_\ibig}^2 - \cb\sum_{1\leq \ibig<\jbig\leq nN} \delta^{(L)}(x_\ibig - x_\jbig)
\end{equation}
where $\delta^{(L)}(x)=\sum_{m\in\mathbb{Z}}\delta(x+m L)$.
As a consequence of  \eqref{eq:periodicity}, one has the periodicity condition
\begin{equation}
\Psi_t\left( x_{1}, \ldots, x_\ibig + L, x_{\ibig+1}, \ldots, x_{nN} , \YY\right) = (-1)^{N-1}\Psi_t\left(\XX , \YY\right) \;.
    \label{eq:periodicitycond}
\end{equation}
Additionally, we have
certain symmetry properties for the replica wave function.
To fix the conventions, we treat the elements of the permutation group $\SSS_M$ as invertible functions $S: \ibig \to S(\ibig) = S_\ibig$, with the multiplication defined by  composition $SS'(\ibig) = S(S'(\ibig)) = S_{S'_\ibig}$. We also denote the standard action of the symmetric group $\SSS_{n N}$ on the coordinates as $(S \XX)_{\mu} = x_{S^{-1}_\mu}$. Then, there are two relevant subgroups $\AAA_{\trect}, \BBB_{\trect}$ of $\SSS_{n N}$ defined respectively as the permutations preserving the rows/columns of $\trect$ in Eq.~\eqref{eq:Trect}. Explicitly we can write
\begin{equation}
\label{subgroupA}
\AAA_{\trect} = \{ S = (\sigma_{(1)},  \sigma_{(2)} , \ldots, \sigma_{(N)}) \in \SSS_{n N}\;|\; \sigma_{(i)} \in \SSS_n \} \equiv \SSS_n \times \ldots \times \SSS_n = (\SSS_n)^{N}
\end{equation}
where each element $\sigma_{(i)}$ is a permutation acting non-trivially only on the $i$-th row of $\trect$. And similarly,
\begin{equation}
\label{subgroupB}
\BBB_{\trect} = \{ S = (\sigma^{(1)}, \ldots, \sigma^{(n)}) \in \SSS_{n N} \; |\; \sigma^{(i)} \in \SSS_N \} \equiv (\SSS_N)^n
\end{equation}
where each element $\sigma^{(j)}$ is now a permutation acting non-trivially only on the $j$-th column of $\trect$.
Then, from the definition \eqref{eq:Psi} and the antisymmetry property \eqref{eq:antisymmetry} one has
\begin{subequations}
\label{eq:PsiCond}
\begin{align}
    &\Psi_t( S \XX , \YY) = (-1)^S \Psi_t(\XX, \YY)  \;, \quad S \in \BBB_{\trect}
    \label{eq:antisymm} \\
    &\Psi_t( S \XX, \YY) = \Psi_t(\XX, \YY)  \;, \qquad S = (\sigma, \sigma, \ldots, \sigma) \in \AAA_{\trect}
    \label{eq:replicasymm}
\end{align}
\end{subequations}
where in the second line we consider the specific elements in $\AAA_{\trect}$ acting with the same $\sigma \in \SSS_n$ on each row of $\trect$. This simply permutes the replica vectors ${\bf x}_a \to {\bf x}_{\sigma^{-1}_a}$ as a whole in \eqref{eq:Psi}. The permutations appearing in \eqref{eq:PsiCond}, the exchanges within a column and the permutations of whole columns, generate the subgroup $\SSS_N\wr\SSS_n\subset\SSS_{nN}$ of the permutations of the boxes of $\trect$ which map columns onto columns, of order $(N!)^n\,n!$ (the hyperoctahedral group for $N=2$). Eqs.~\eqref{eq:PsiCond} state that under this subgroup $\Psi_t$ is multiplied by a factor $\pm1$, the product of the signs of the permutations performed within each column. For even $N$, in particular for $N=2$, since the exchange of two whole columns is a product of $N$ transpositions, this factor is simply the sign of the permutation of $\SSS_{nN}$. These symmetry properties will be used in Sec.~\ref{subsec:youngsymmetrizer} to determine which irreducible representations of $\SSS_{nN}$ can contribute to the replica wave function.
In the large-time limit, the Lieb--Liniger evolution projects onto the
lowest-energy state compatible with the symmetry conditions \eqref{eq:PsiCond}, specifically selecting, as we will see, a single irreducible representation—namely, the rectangular one.
 Denoting $E^{(N)}(n,L)$ the corresponding energy, we thus expect
\begin{equation}
\Psi_t(\XX, \YY) \sim C_n(\XX,\YY)\,e^{-E^{(N)}(n, L)\,t} \;.
\label{eq:En_sector_def}
\end{equation}
and we thus confirm, as anticipated, that the quantities $E^{(N)}(n,L)$ coincide with the ground state energies of the Lieb-Liniger Hamiltonian in the appropriate symmetry sector identified by \eqref{eq:PsiCond}
\begin{equation}
  E^{(N)}(n, L) = - \lim_{t\to\infty} \frac{1}{t} \log \overline{Z_t(W, V)^n} = -\lim_{t\to \infty} \frac 1 t \log \Psi_t(\XX, \YY)  \;.
\end{equation}
From this approach, it is also clear why the $E^{(N)}(n,L)$ are actually independent of the choice of the initial and final condition sets $\{V_a\}_a, \{W_a\}_a$.
Hence, for the determination of the exponential growth rate, we may
specialize to identical replica boundary configurations,
$\xx_a=\xx$ and $\yy_a=\yy$. We recall that to simplify the notation we simply denote $Z(t) = Z_t(W, V)$.
As a side note, we observe that the factor $C_n(\XX,\YY)$ depends on the value of the ground-state wave function at the initial and final conditions. However, this will not be the focus of this article, and we reserve it for future explorations.

\subsection{Nested Bethe Ansatz}
\label{subsec:cba_general}
The eigenvectors of the Lieb-Liniger Hamiltonian \eqref{eq:HLL} can be taken in the generalized Bethe Ansatz form in terms of $M$ momenta $\vec{\bf k} = (k_1, \ldots, k_M)$~\cite{yang1967some}
\begin{equation}
\Psi(\XX)=\sum_{P,Q\in \SSS_M}\vartheta_Q(\XX)
A_{Q}^{P} \exp\!\Big(i\sum_{\ibig=1}^{M} k_{P_\ibig}\,x_{Q_\ibig}\Big)
= \sum_{Q, P\in \SSS_M}\vartheta( Q^t \XX)\,
A_{Q}^P \exp\!\Big(i {\vec{\bf k}}^t P Q^t \XX \Big)
\label{eq:CBA_general}
\end{equation}
where $M = n N$ denotes the total number of variables,
$Q^t = Q^{-1}$ denotes the transposition of the matrix representation of $Q$ in $\mathbb{R}^{M}$ (and similarly for $P$), namely
$(Q \XX)_\mu = x_{Q^{-1}_\mu}$. For clarity, we specify that we use the same notation for a permutation and its matrix representation. Finally, $\vartheta_Q(\XX)$ is the indicator function
of the Weyl chamber labeled by the permutation $Q$
\begin{equation}
    W_Q = \{ (x_1,\ldots, x_M) \; | \; x_{Q_1} < x_{Q_2} < \ldots < x_{Q_M} <  x_{Q_1} + L \} \;.
\end{equation}
Note that one also has  $\vartheta_Q(\XX) = \vartheta_{\mathbb{I}}(Q^t \XX)$.
It is useful to see the amplitude data $A_{Q}^P$ as a tensor with one index labeled by the sector permutation $Q$ and one index labeled by the momentum permutation $P$.
Inside each sector, the action of the Hamiltonian \eqref{eq:HLL} reduces to the double derivative with respect to each variable $x_\mu$, thus leading to the energy eigenvalue
\begin{equation}
    \label{eq:HLLeigenBethe}
E(\vec{\bf k}) := \frac{1}{2} \sum_{\alpha = 1}^{N n} k_\alpha^2
\end{equation}
Other conserved quantities correspond to higher symmetric polynomials of the momenta $\vec{\bf k}$.
However, in order for this to be an eigenstate, we need to verify the conditions at the boundary between different sectors. Specifically, two adjacent sectors $W_Q$ and $W_{Q'}$ corresponds to configurations where two coordinates coincide. In particular, let $Q' = Q \tau_\mu$, where we denote by $\tau_{\mu,\nu}$ the transposition exchanging $\mu$ and $\nu$, and by $\tau_\mu := \tau_{\mu,\mu+1}$ the adjacent transposition. Then, the interface between $W_Q$ and $W_{Q\tau_\mu}$ is the hyperplane
\be
x_{Q_\mu} = x_{Q_{\mu+1}} \quad , \quad 1 \leq \mu \leq M-1
\ee
At this interface one must impose: i) the continuity of the wave function, and ii) the cusp condition induced by the $\delta$-interaction in \eqref{eq:HLL}. These conditions relate the behavior of the wave function $\Psi(\XX)$ in the two adjacent sectors $W_Q$ and $W_{Q\tau_\mu}$, and are obtained by considering the limits $x_{Q_\mu} \to x_{Q_{\mu+1}}^{\pm}$. They lead to
\begin{align}
&A_Q^{P \tau_\mu} + A_Q^P = A_{Q \tau_\mu}^P + A_{Q \tau_\mu}^{P \tau_\mu}, \\
&i \left(k_{P_\mu}-k_{P_{\mu+1}}\right)
\left(-A_Q^{P \tau_\mu} + A_{Q \tau_\mu}^P - A_{Q \tau_\mu}^{P \tau_\mu} + A_Q^P\right)
= 2 \cb \left(A_Q^{P \tau_\mu} + A_Q^P\right).
\end{align}
These two equations can be combined into
\begin{equation}
\label{eq:combinedcontcusp}
    A_{Q}^{P\tau_\mu} = -\frac{\cb}{\cb+i (k_{P_\mu}-k_{P_{\mu + 1}})} A_{Q}^{P} +  \frac{i(k_{P_\mu}-k_{P_{\mu+1}})}{\cb+i(k_{P_\mu}-k_{P_{\mu+1}})} A_{Q \tau_\mu}^{P} \;.
\end{equation}
It is thus useful to see $(A^P)_Q$ for fixed $P \in \SSS_M$ as a vector
$A^P$ in a vector space $\mathcal{A}$ of dimension $M!$. In this vector space, we can define the left/right actions of $\SSS_M$ on $A^P$ denoted respectively by $\mathsf{\ell}(\sigma)$ and $\mathsf{r}(\sigma)$ for $\sigma \in \SSS_M$, and defined in the permutation space as
\begin{equation}
\mathsf{\ell}(\sigma)_{Q,Q'}=\delta_{Q,\sigma Q'},
\qquad
\mathsf{r}(\sigma)_{Q,Q'}=\delta_{Q,Q'\sigma^{-1}} \;,
\label{eq:left_right_actions_matrix}
\end{equation}
so that
\begin{equation}
\label{eq:defLRrepr}
(\mathsf{\ell}(\sigma) A^P)_Q=\sum_{Q'}\mathsf{\ell}(\sigma)_{Q,Q'}A_{Q'}^P=A_{\sigma^{-1}Q}^P,
\qquad
(\mathsf{r}(\sigma) A^P)_Q=\sum_{Q'}\mathsf{r}(\sigma)_{Q,Q'} A^P_{Q'}=A^P_{Q\sigma}.
\end{equation}
In these conventions the two actions define bona fide commuting representations, i.e. for any $Q_1, Q_2 \in \SSS_M$, it holds
\begin{equation}
\mathsf{\ell}(Q_1)\mathsf{\ell}(Q_2)=\mathsf{\ell}(Q_1 Q_2),
\qquad
\mathsf{r}(Q_1)\mathsf{r}(Q_2)=\mathsf{r}(Q_1 Q
_2),
\qquad
\mathsf{\ell}(Q_1) \mathsf{r}(Q_2) =\mathsf{r}(Q_2)\mathsf{\ell}(Q_1) \;.
\label{eq:left_right_representation}
\end{equation}
We can reexpress Eq.~\eqref{eq:combinedcontcusp} as a linear equation between the vectors $A^{P\tau_\mu}$ and $A^P$
\begin{equation}
\label{eq:AwithY}
    A^{P \tau_\mu} = [Y^\mu_{P_{\mu} P_{\mu+1}}]
    A^P \;, \qquad Y^\mu_{\nu \delta} %
     = f(k_{\delta \nu}) \mathsf{r}(\tau_\ibig) + (f(k_{\delta \nu})-1) \mathsf{r}(I) \;, \quad f(u) = \frac{u}{u + i \cb}
\end{equation}
where $k_{\delta \nu} = k_\delta - k_\nu$ and $I$ denotes the identity permutation.
By decomposing any $P$ as a product of transpositions, one can then relate any vector $A^P$ to a reference one, say $A^I \equiv A$. One can verify the Yang Baxter relations \begin{align}
\label{consistency}
A_Q^{P \tau_\ibig^2 } &= A_Q^{P} &&\Longleftrightarrow [Y^{\ibig}_{\nu \delta}][Y^{\ibig}_{\delta \nu}]=1\\
A_Q^{P \tau_\ibig \tau_{\ibig+1} \tau_\ibig } &= A_Q^{P \tau_{\ibig+1} \tau_{\ibig} \tau_{\ibig+1}}  &&\Longleftrightarrow [Y^{\ibig}_{\nu \delta}][Y^{\ibig+1}_{\rho \delta}] [Y^{\ibig}_{\rho \nu}]=[Y^{\ibig+1}_{\rho \nu}][Y^{\ibig}_{\rho \delta}] [Y^{\ibig+1}_{\nu \delta}] \label{consistency2}
\end{align}
and therefore different decompositions are compatible \cite{yang1967some,gaudin1967un,gaudin2014bethe,sutherland1968further}.

\subsection{From periodicity to a commuting transfer matrix}
\label{subsec:periodicity_transfer}
We now consider the effect of imposing the condition \eqref{eq:periodicitycond}, which corresponds to
\be
\label{periodicWF}
\Psi(x_1, \ldots, x_\ibig + L, \ldots, x_M) = \Phi \Psi(x_1, \ldots, x_\mu, \ldots, x_M) \;, \quad , \quad \Phi = (-1)^{N-1}
\ee
To see the effect of this condition, as usual let us assume that $\XX \in W_Q$ for a given permutation $Q$. Imposing the condition \eqref{periodicWF} for $\ibig = Q_{1}$, we end up
in another sector $\tilde Q = Q R = \{ Q_2, \ldots, Q_M, Q_1\}$ characterized by $x_{Q_2} < \ldots < x_{Q_M} < x_{Q_1} + L$, with $R = \{2,3,\ldots, M,1\}$  the right shift permutation; therefore we get from \eqref{periodicWF}
\begin{equation}
 \vartheta_{\tilde Q}(\XX') \sum_P A_{\tilde Q}^P \exp(i \sum_{\mu=1}^M k_{P_{\mu}} x_{\tilde Q_{\mu}}')=
\Phi\vartheta_Q (\XX) \sum_P A_Q^P \exp(i \sum_{\mu} k_{P_\mu} x_{Q_\mu})
\end{equation}
where $\XX$' is obtained from $\XX$ replacing $x_{Q_1}$ with $x_{Q_1} + L$. Equating the terms with the same exponent we get
\be
\label{betheansatz0}
  A_{QR}^{PR} e^{i k_{P_1} L} = \Phi A_Q^P \;.
  \ee
which is thus true for any $P,Q$.
Thanks to the relations in \eqref{eq:AwithY}, one can
reduce this equation to one only acting on $A^{I}_Q$, which reads
\begin{align}
\label{XYangBethe}
&[\mathcal{T}_{\mu}]_{Q,Q'} := [X_{\ibig+1,\ibig} X_{\ibig+2,\ibig} .. X_{M,\ibig} X_{1\ibig} X_{2 \ibig} .. X_{\ibig-1,\ibig}]_{Q,Q'} \;, \\
&\sum_{Q'} [\mathcal{T}_{\mu}]_{Q,Q'} A_{Q'}^{I} =
 \Phi^{-1} e^{i k_\ibig L} A^I_Q \;, 
\end{align}
where we introduced
\be
\label{XYang}
X_{\ibig \jbig} \equiv f(k_{\jbig \ibig}) +  (f(k_{\jbig \ibig}) -1) \mathsf{r}(\tau_{\ibig \jbig})
\ee
which satisfies $X_{\mu \nu} X_{\nu \mu} = 1$.
In order to show \eqref{XYangBethe}, we decompose the cycle $R = \tau_1 \tau_2 \ldots \tau_{M-1}$ and we apply sequentially \eqref{eq:AwithY}. Setting $R_m = \tau_1 \tau_2 \ldots \tau_{m}$, we claim that
\begin{equation}
\label{eq:recASm}
    A_{Q R_m}^{R_m} = \sum_{Q'} [X_{1, m+1} X_{1,m} \ldots X_{1,2}]_{Q,Q'} A^I_{Q'}  = \sum_{Q'} [X_{1, m+1}]_{Q,Q'} A_{Q' R_{m-1}}^{R_{m-1}}
\end{equation}
Indeed, from Eq.~\eqref{eq:AwithY},
since $R_{m-1} = (2,3,\ldots, m,1, m+1, \ldots, M)$, we have
\begin{equation}
    A^{R_m} = Y_{1,m+1}^m A^{R_{m-1}} \;.
\end{equation}
Also, we note from Eq.~\eqref{eq:left_right_representation} that $A^{R_m}_{Q R_m} = [\mathsf{r}(R_m) A^{R_m}]_Q$
and then Eq.~\eqref{eq:recASm} follows verifying that
\begin{equation}
     \mathsf{r}(R_{m}) Y_{1, m+1}^m  \mathsf{r}^{-1}(R_{m-1}) = X_{1, m+1}
\end{equation}
Then, the case $\mu = 1$ in Eq.~\eqref{XYangBethe} is recovered using that $X_{\mu \nu} X_{\nu \mu} = 1$.
For the other values of $\mu$ one can cyclically permute the $k'$s and check that Eq.~\eqref{XYangBethe} holds generally.

Now, Eq. \eqref{XYangBethe} tells us that $A^I$ must be a common eigenvector for $M$ operators, corresponding to \eqref{XYangBethe} with $\mu=1,\dots,M$.
One can check that these $M$ operators commute because of the properties of the transpositions $\mathsf{r}(\tau_{\ibig \jbig})$
and the Yang Baxter relations \eqref{consistency}-\eqref{consistency2}.
The determination of this eigenvector together with the eigenvalues fixes a set of conditions
for the $k$'s, that can be seen as a generalization of the Bethe ansatz equations.
We notice that all the operators are written in terms of the $X_{\ibig \jbig}$, which in turn are expressed using the $\mathsf{r}(\tau_{\ibig \jbig})$.
The pair $(\mathcal{A}, \mathsf{r})$ is dubbed the regular (right) representation of $\SSS_M$, introduced in \eqref{eq:defLRrepr}.
It is clear that each eigenvector and therefore any possible solution of \eqref{XYangBethe} will lie in a subspace which is invariant by the action of $\mathsf{r}(Q')$ for any permutation $Q' \in \SSS_M$: it means that
we can reduce the dimensionality of the problem by looking at irreducible representation of $\SSS_M$ contained inside the regular one. For this reason, in the next section we briefly summarize the classification of the irreducible representation for the symmetric group.
\subsection{Irreducible representations of \texorpdfstring{$\SSS_M$}{SM} and the Young symmetrizer}
\label{subsec:youngsymmetrizer}
The irreducible representations $V_\lambda$ of $\SSS_M$ are in one-to-one correspondence with integer partitions $\lambda \vdash M = \{ \lambda_1, \lambda_2, \ldots \}$ of $M$, i.e. such that $\sum_i \lambda_i  = M$, with $\lambda_i \geq \lambda_{i+1} \geq 0$.
A natural way to represent a partition is obtained employing Young diagram; given the numbers $\lambda_i$, we represent them as a set of $k$ rows, aligned
on the left,
each containing $\lambda_1,\ldots,\lambda_k$ elements. As an example, we have for $M = 8$
$$ \Yvcentermath1
3+2+2+1 = 8 \quad \to \quad \yng(3,2,2,1) $$
In particular, within the regular representation, every irreducible representation appears with a multiplicity given by $\dim V_\lambda$. More explicitly, one has the isotypic decomposition
\begin{equation}
    \mathcal{A}  \equiv \oplus_{\lambda \vdash M} \oplus_{a = 1}^{\dim V_\lambda} V_\lambda^{(a)}
\end{equation}
where $V_\lambda^{(a)}$ indicates isomorphic copies of the same irreducible representation labeled by $\lambda$.
Consistently, one has the standard relation
\begin{equation}
    M! = \dim \mathcal{A} = \sum_{\lambda \vdash M} (\dim V_\lambda)^2 \;.
\end{equation}
For a given $\lambda$, we can construct projectors onto each $V_{\lambda}^{(a)}$ considering a standard tableau of shape $\lambda$.
Given a Young diagram,
one calls a standard Young tableau, a filling of each box with symbols chosen from a totally ordered set, such that the symbols increase on each row and each column. As an example,
using as symbols the integers from $1, \ldots, M$,
consider
$$\Yvcentermath1 \yng(3,2,2,1)  \quad \to \quad \young(123,45,67,8) \; .$$
Once this construction has been performed to obtain a tableau $\mathcal{T}$, two subgroups $\AAA_\mathcal{T}, \BBB_{\mathcal{T}}$ of $\SSS_M$ are naturally defined,
corresponding respectively to the subgroups sending
the rows/columns of $\mathcal{T}$ into themselves. We can define two elements of $\operatorname{End}(\mathcal{A})$
\begin{subequations}
    \label{subgroupsTableau}
\begin{align}
a_{\mathcal{T}} &= \sum_{Q \in \AAA_\mathcal{T}} \mathsf{\ell}(Q) \;,
\label{eq:asumdef}
\\
b_{\mathcal{T}} &= \sum_{Q\in \BBB_\mathcal{T}} (-1)^{Q} \mathsf{\ell}(Q) \;.
\label{eq:bsumdef}
\end{align}
\end{subequations}
The Young symmetrizer of the tableau $\mathcal{T}$ is then defined as $c_{\mathcal{T}} = b_{\mathcal{T}} a_{\mathcal{T}} / z_{\mathcal{T}} $, with $z_{\mathcal{T}}$ an appropriate constant, chosen to make it idempotent $c_{\mathcal{T}}^2 = c_{\mathcal{T}}$.
Because the left/right actions commute, clearly the $c_\mathcal{T} \mathcal{A}$ represents an invariant subspace under the right action $\mathsf{r}(\sigma)$ for any $\sigma \in \SSS_M$. A standard result ensures that we can identify $c_{\mathcal{T}}\mathcal{A} \equiv V_\lambda^{(a)}$, giving thus a bijection between standard tableaux $\mathcal{T}$ corresponding to the same diagram $\lambda$ and the ensemble $\{V_\lambda^{(a)}\}_a$~\cite{fulton2013representation}.
In our case, we are particularly interested in rectangular tableaux of shape $\lambda = (n, n, \ldots, n)$, with $M =N n$ and it is useful to consider the filling $\mathcal{T} = \trect$ corresponding to the definition of coordinates~\eqref{coordinates}
as introduced in Eq.~\eqref{eq:Trect}.
In this case, we observe that the groups $\AAA_{\mathcal{T}}, \BBB_{\mathcal{T}}$ in Eq.~\eqref{subgroupsTableau} reduce to the ones introduced in Eqs.~(\ref{subgroupA}, \ref{subgroupB}).

We now verify that the amplitude data $A_{Q}^P$ chosen inside the irreducible representation selected by the corresponding Young symmetrizer $c_{\trect}$, i.e. satisfying
\begin{equation}
\label{eq:APproj}
    c_{\trect} A^P = A^P
\end{equation}
enforce the symmetry properties \eqref{eq:PsiCond}. We recall it is enough to impose Eq.~\eqref{eq:APproj} for the identity $P = I$ and it is automatically valid any other $P$ thanks to Eq.~\eqref{eq:AwithY}. Recalling the standard action of $S \in \SSS_M$ on $\mathbb{R}^M$ as
$(S \xx)_{\mu} = x_{S^{-1}_\mu}$, we can rewrite
\begin{equation}
\label{eq:PsiSx}
    \Psi(S \XX)
    =\sum_{Q, P\in \SSS_M}\vartheta( Q^t S \XX)\,
A_{Q}^P \exp\!\Big(i \vec{{\bf k}}^t P Q^t S \XX \Big) =\\
    \left.\Psi(\XX)\right|_{A_Q^P \to A_{S Q}^P} \;,
\end{equation}
where in the second equality we relabeled in the sum $Q \to S Q$ and used that $(SQ)^t S =Q^t S^t S= Q^t$.
In other words, the permutation of the coordinates with a permutation $S$ amounts to acting on the vectors $A^P$ with the left action as $A^P_{S Q} = (\mathsf{\ell}(S^{-1}) A^P)_Q$.
To verify Eq.~\eqref{eq:antisymm},
consider $S \in \BBB_{\trect}$. Then, we have from \eqref{eq:bsumdef},
\begin{equation}
\label{eq:actionoverB}
    \ell(S^{-1}) b_{\trect} =
    \sum_{Q\in \BBB_\mathcal{T}} (-1)^{Q} \mathsf{\ell}(S^{-1})\mathsf{\ell}(Q) =
\sum_{Q\in \BBB_\mathcal{T}} (-1)^{Q} \mathsf{\ell}(S^{-1} Q) =
    (-1)^{S} b_{\trect}
\end{equation}
where in the first equality we used that the left action is a representation and in the last one we relabeled the sum $Q \to S Q$ and $(-1)^{S Q} = (-1)^S (-1)^{Q}$. And thus, we have
\begin{equation}
    \ell(S^{-1}) A^P =     \ell(S^{-1}) c_{\trect} A^P =  z_{\trect}^{-1} \ell(S^{-1}) b_{\trect} a_{\trect} A^P = (-1)^S A^P  \;, \quad \forall S \in \BBB_{\trect}
\end{equation}
which thus implies Eq.~\eqref{eq:antisymm}. Instead, to show Eq.~\eqref{eq:replicasymm}, we start observing that for $S = (\sigma, \ldots, \sigma) \in \AAA_{\trect}$, with $\sigma \in \SSS_n$, we have for any $(\sigma^{(1)}, \ldots, \sigma^{(n)}) \in \BBB_{\trect}$ that
\begin{equation}
    S (\sigma^{(1)}, \ldots, \sigma^{(n)}) S^{-1} =
        (\sigma^{(\sigma_1)}, \ldots, \sigma^{(\sigma_n)})
\end{equation}
in other words, this particular choice of the permutation $S$ simply acts as a relabeling of the columns of the diagram. As a consequence, any such $S$ acts as a permutation of the elements of $\BBB_{\trect}$.
As a conjugacy transformation preserves the parity of a permutation, we deduce, summing over all the elements of $\BBB_{\trect}$ in Eq.~\eqref{subgroupB}, that
\begin{equation}
    [\ell(S^{-1}), b_{\trect}] = 0 \;, \qquad S = (\sigma, \ldots, \sigma) \in \AAA_{\trect}
\end{equation}
Finally, as this $S \in \AAA_{\trect}$, it follows similarly to Eq.~\eqref{eq:actionoverB} that $\ell(S^{-1}) a_{\trect} = a_{\trect}$ and consequently
\begin{equation}
    \ell(S^{-1}) A^P = A^P \;, \quad \forall S = (\sigma, \ldots, \sigma) \in \AAA_{\trect}
\end{equation}
confirming the validity of Eq.~\eqref{eq:replicasymm}.
\\

We have thus shown that the rectangular diagram (with the filling Eq.~\eqref{eq:Trect}) has the appropriate set of symmetries, under the exchange of coordinates, selected by the replicated wave function. However, there could be other irreducible representations satisfying these properties. For instance, for even $N$, the completely antisymmetric representation (corresponding to a single-column Young diagram) also satisfies Eq.~\eqref{eq:PsiCond}, since exchanging the $N$ coordinates between two replicas leads to a factor $(-1)^N = 1$. This means that in principle the initial state in Eq.~(\ref{eq:inicondpsit}, \ref{eq:Psi}) could have a non-trivial overlap with the totally antisymmetric sector.
One can check for a generic initial condition of the SHE $\mathcal{Z}^{(L)}_{t=0}(\xx)$ this overlap does not a priori vanish. However, one can also verify that this overlap vanishes if the initial condition is a delta function, i.e. if the $y_{i,a} = y_{i}$ are taken independently from the replica index.

Which representations are compatible with Eq.~\eqref{eq:PsiCond} can be made precise. By Eq.~\eqref{eq:PsiSx},
the conditions \eqref{eq:PsiCond} require the vector $A^P$ to be multiplied, under the group $\SSS_N\wr\SSS_n$
introduced after Eq.~\eqref{eq:replicasymm}, by the product of the signs of the permutations within the columns.
The antisymmetry within the columns alone, Eq.~\eqref{eq:antisymm}, already implies by Pieri's rule that only the
Young diagrams with at most $n$ columns can appear, i.e.\ those obtained from the rectangle $(n^N)$ by moving
boxes to lower rows, the rectangle itself appearing exactly once.

For $N=2$ the list is complete and explicit. In this case
$\SSS_2\wr\SSS_n$ is the hyperoctahedral subgroup of $\SSS_{2n}$,
and the product of the signs of the permutations within the columns
is precisely the restriction of the sign character of $\SSS_{2n}$.
The classical multiplicity-free decomposition of this Gelfand pair
\cite[Chap.~VII, Sec.~2]{macdonald1995symmetric}, together with the
twist by the sign representation, then shows that the compatible
irreducible representations of $\SSS_{2n}$ are exactly those whose
Young diagram has all columns of even length, i.e.\ of the form
$(\mu_1,\mu_1,\mu_2,\mu_2,\ldots)$ with
$\mu_1\ge\mu_2\ge\ldots$ a partition of $n$. Each contains a
one-dimensional space of vectors transforming in the required way.
For $n=3$, for instance, these diagrams are $(3,3)$,
$(2,2,1,1)$ and the single-column diagram $(1^6)$ mentioned above.

For $N>2$ the complete list is more involved and will not be needed. In all cases the rectangle $(n^N)$ is
the only compatible diagram with $N$ rows, and it is the most symmetric of the compatible ones, in the sense that
every other one is obtained from it by moving boxes to lower rows. Since the attractive interaction in Eq.~\eqref{eq:HLL} (induced by the disorder average)
favours symmetric wave functions, one expects the rectangle to be the sector of lowest energy among the compatible
ones. This is indeed the case, for every value of $\cb$: it follows from an argument of Lieb and Mattis
\cite{lieb1962ordering}, adapted to the ring in Appendix~\ref{app:LiebMattis}, where it is also shown that the
ordering holds thanks to the boundary condition \eqref{eq:periodicitycond}. At large times the evolution therefore
projects onto the rectangular sector whenever the overlap of the initial condition with its ground state is
nonzero, and in the following we focus on the rectangular diagrams.
For the fixed-endpoint initial conditions considered here,
this overlap is nonzero by the positivity of the ground
state established in Appendix~\ref{app:LiebMattis}.

\subsection{Nested Bethe Ansatz equations}
\label{sec:nbaeq}
Having discussed the symmetries of the wave function identified by the corresponding Young’s diagram, we can now examine how the matrix equation \eqref{XYangBethe} can be converted into a set of equations that effectively determine the spectrum. As we have seen, the operator equation depends solely on the choice of representation—that is, on the Young diagram—and not on its filling, which instead determines the specific spatial symmetries of the wave function. Therefore, rather than using the construction explained in the previous section based on the Young symmetriser, it proves more useful to construct the representation of the symmetric group explicitly by introducing an auxiliary integrable \textit{spin} chain. To deal with a diagram with $N$ rows, the spins are chosen in the fundamental representations of $SU(N)$; the Schur-Weyl duality allows us to construct the required representation. In general, this produces a hierarchical family of $N$ levels of Bethe equations, known as the nested Bethe Ansatz. For clarity, in this section we consider the case relevant to this article, corresponding to $N=2$ non-crossing polymers, which thus corresponds to a standard $SU(2)$ spin chain with spin $1/2$.

We introduce the Hilbert space of the corresponding spin chain as $\mathbb{H} = \otimes_{i=1}^M \mathbb{C}^2$, i.e. $M$ copies of a spin $1/2$ $\mathbb{C}^2$ space. This space, as any tensor space, provides a natural representation $\mathcal{R}$ of $S_M$ by exchanging the vectors
among the spaces in the tensor products, i.e.
\be
\label{tensorprodrepr}
 \mathcal{R}(\tau_{\mu \nu}) (v_1 \otimes \ldots v_\mu \ldots v_\nu \otimes \ldots \otimes v_M) = v_1 \otimes \ldots v_\nu \ldots v_\mu \otimes \ldots \otimes v_M \; .
\ee
We introduce the standard $SU(2)$ $R$-matrix as an operator on the tensor product of two spin $1/2$ space labeled by indices $\mu,\nu$, $\mathbb{C}_\mu^2\otimes \mathbb{C}_\nu^2$, given by
\be
\label{Rmatrix}
R_{\mu \nu}(\lambda) = \frac{1}{\lambda - i \cb}\left(\begin{array}{llll} \lambda - i\cb & & &\\ & \lambda & -i\cb & \\ & -i\cb & \lambda & \\ & & & \lambda - i\cb  \end{array}\right) = \frac{\lambda}{\lambda - i\cb} \ide_{\mu \nu}  - \frac{i \cb}{\lambda - i\cb} \mathcal{R}(\tau_{\mu \nu}) \; .
\ee
where $\ide_{\mu \nu}$ is the identity and $\tau_{\mu \nu}$ is the exchange operator acting as in \eqref{tensorprodrepr}.
Comparing with \eqref{XYang}, it is easy to verify that, upon mapping the right representation $\mathsf{r}(.) \to \mathcal{R}(.)$, we can set
\be
\label{XfromR}
X_{\mu \nu} = S(k_{\nu\mu}) R_{\mu\nu}(k_{\nu\mu})  \;, \qquad S(k)  = \frac{k - i\cb}{k + i\cb}
\ee
where $S(k)$ is a pure phase, corresponding the standard Lieb-Liniger scattering phase. The $R$-matrix can be used to build an integrable spin model.
In order to do so, one first notices that the Yang-Baxter identity holds. To fix the notation, indexes of the form $\mu,\nu,\rho,\ldots$ will indicate sites in the real spin lattice, while indexes of the form $\alpha,\beta,\ldots$ will
indicate the auxiliary spaces, which are still $\mathbb{C}^2$. Then the Yang-Baxter identity can be written as an operator identity in $\mathbb{C}_\alpha^2\otimes \mathbb{C}_\beta^2\otimes \mathbb{C}_\mu^2$. Then we have
\be
\label{YB}
R_{\alpha,\beta} (\lambda - \lambda') R_{\mu,\alpha} (\lambda) R_{\mu,\beta} (\lambda') = R_{\mu,\beta}(\lambda')  R_{\mu,\alpha}(\lambda) R_{\alpha,\beta} (\lambda - \lambda')
\ee
and this property is satisfied just because of the algebra of the transpositions $\tau_{\mu \nu}$. This $R$-matrix is the same that enters
in the solution of the XXX spin chain. We will show that in the ``inhomogeneous'' case (that we will specify later), one can reobtain the LHS of \eqref{XYangBethe}. The first step requires the introduction of
the transfer matrix as
\be
\label{TM}
T_{M, \alpha} (\lambda, \{\xi\}) = R_{1,\alpha} (\lambda - \xi_1) R_{2,\alpha} (\lambda - \xi_{2}) \ldots R_{M,\alpha}(\lambda - \xi_M)
\ee
Here the $\xi$'s are an arbitrary set of numbers called the ``inhomogeneities''. The crucial fact is that employing \eqref{YB} many times, one can prove that an analogous relation holds also for the transfer matrix itself
\be
\label{YBTM}
R_{\alpha,\beta} (\lambda - \lambda') T_{M,\alpha} (\lambda, \{\xi\}) T_{M,\beta} (\lambda', \{\xi\}) = T_{M,\beta}(\lambda', \{\xi\})  T_{M,\alpha}(\lambda, \{\xi\}) R_{\alpha,\beta} (\lambda - \lambda')
\ee
From this relation it follows that if we set $F(\lambda, \{\xi\}) = \Tr_\alpha [T_{M,\alpha}(\lambda, \{\xi\})]$, i.e. the trace on the auxiliary space, we have that
\begin{equation}
\label{eq:FFcomm}
[F(\lambda, \{\xi\}), F(\lambda', \{\xi\})] = 0
\end{equation}
for any $\lambda, \lambda'$, telling us that
$F(\lambda, \{\xi\})$ can be diagonalized simultaneously for arbitrary $\lambda$.
Now, for the standard XXX Heisenberg chain, one has to set all the inhomogeneities to the same value $\xi_\mu \to \frac i 2$. Instead in our case we set $\xi_\mu = k_\mu$ and denote correspondingly $F(\lambda, \{ \xi \}) = F(\lambda, \vec{k})$. Now we claim that the transfer matrix introduced in Eq.~\eqref{XYangBethe} can be rewritten as
\be
\label{XYangABA}
\mathcal{T}_{\mu} =
X_{\mu+1,\mu} X_{\mu+2,\mu} \ldots X_{M,\mu} X_{1,\mu} \ldots X_{\mu-1,\mu} = \left[\prod_{\nu \neq \mu} S(k_\mu - k_\nu)\right]  F(k_\mu, \vec{k}) \;.
\ee
To prove this fact, we focus on $\mu=1$, for the other values it follows from the cyclic symmetry. Then from \eqref{TM} we have
\be
\label{TMpjk}
T_{M, \alpha} (k_1, \vec{k}) = \mathcal{R}(\tau_{1,\alpha}) R_{2,\alpha} (k_1 - k_2) \ldots R_{M,\alpha}(k_1 - k_M)
\ee
where we used that $R_{\mu \nu} (0) = \mathcal{R}(\tau_{\mu\nu})$ reduces to the swap of $\mu,\nu$. Then it is easy to see that
\be
\mathcal{R}(\tau_{1,\alpha}) R_{\mu,\alpha} (k_{1\mu}) = R_{\mu,1} (k_{1\mu}) \mathcal{R}(\tau_{1,\alpha})
\ee
which easily follows by $\tau_{1\alpha}\tau_{\mu\alpha} \tau_{1\alpha} = \tau_{\mu1}$. It means that we have
\be
\label{TMpjkSimplified}
T_{M, \alpha} (k_1, \vec{k}) = R_{2,1} (k_1 - k_2) \ldots R_{M,1}(k_1 - k_M) \mathcal{R}(\tau_{1,\alpha})
\ee
Now, we take the trace over the $\alpha$ Hilbert space; as the trace only affect the last term and $\Tr_\alpha \mathcal{R}(\tau_{1,\alpha})= \ide$. The product factor in the RHS of \eqref{XYangABA} then just reproduces the $X$ from the $R$ using
\eqref{XfromR}.
\\

The importance of this fact relies on the possibility of diagonalizing the family of operators $F(\lambda; \vec{k})$ that can be done with the standard methods of the algebraic Bethe ansatz~\cite{faddeev1996algebraicbetheansatzworks}. This is based on the following procedure. In the following we will omit
the dependence on $\vec{k}$ unless it is necessary to stress it. In the auxiliary space $\mathbb{C}^2_\alpha$ the transfer matrix $T_{M,\alpha}(\lambda)$ can be written as $2\times 2$ matrix of operators in the Hilbert space $\mathbb{H}$ of $M$ spin $1/2$. Explicitly
\be
\label{Tmatr}
T_{M,\alpha}(\lambda) = \left( \begin{array}{ll} A(\lambda) & B(\lambda) \\ C(\lambda) & D(\lambda)
\end{array}
 \right) \;.
\ee
With this notation it is clear that $F(\lambda) = A(\lambda) + D(\lambda)$ and moreover \eqref{YBTM} can be interpreted as a set of commutation relations for the operators $A,B,C,D$.
Those relevant for us are
\begin{subequations}
\label{FCR}
    \begin{align}
&[B(\lambda), B(\lambda')] = 0\, \\
&A(\lambda) B(\lambda') = f_+(\lambda - \lambda') B(\lambda') A(\lambda) + g_+(\lambda - \lambda') B(\lambda) A(\lambda') \;, \\
&D(\lambda) B(\lambda') = f_-(\lambda - \lambda') B(\lambda') D(\lambda) + g_-(\lambda - \lambda') B(\lambda) D(\lambda')\;,\\
&f_{\pm}(\lambda) = \frac{\lambda \pm i\cb}{\lambda}, \quad g_{\pm}(\lambda) = \mp \frac{i\cb}{\lambda}\;.
\end{align}
\end{subequations}
Let us introduce the reference state in $\mathbb{H}$ as $\ket{\Omega} = \ket{\uparrow \uparrow \ldots \uparrow}$. This is an eigenstate of $F(\lambda)$, since
\be
\label{OmegaEigen}
A(\lambda) \ket{\Omega} = \alpha(\lambda) \ket{\Omega}; \qquad D(\lambda) \ket{\Omega} = \delta(\lambda) \ket{\Omega}; \qquad C(\lambda) \ket{\Omega} = 0\;.
\ee
where we set
\begin{equation}
    \alpha(\lambda) = 1  \;, \qquad \delta(\lambda) =  \prod_\mu \frac{\lambda - k_\mu}{\lambda - k_\mu - i \cb} \;.
\end{equation}
Then, by the repeated action of a set of operators $B(\lambda)$
computed on a set of rapidities $\{\lambda_\alpha\}$, we
build the states
\begin{equation}
\label{eq:bethestates}
    \ket{\{\lambda_a\}} = \ket{\lambda_1, \ldots, \lambda_{\tilde n}} := B(\lambda_1) \ldots B(\lambda_{\tilde n}) \ket{\Omega} \;.
\end{equation}
Using the commutation relations \eqref{FCR}, one can verify that these states are eigenstates
\be
\label{eigenvaluesF}
F(\lambda) \ket{\{\lambda_a\}} = \Lambda(\lambda,\{\lambda_a\}) \ket{\{\lambda_a\}}
\ee
provided that the Bethe ansatz equations are satisfied
\be
\label{BAE}
\prod_{b \neq a} \frac{f_+(\lambda_a - \lambda_b)}{f_-(\lambda_a - \lambda_b)} = \frac{\delta(\lambda_a)}{\alpha(\lambda_a)}
\ee
where $a,b = 1,\ldots, \tilde n$. Each operator $B(\lambda)$ acts flipping precisely one spin, so that the state $\ket{\{\lambda_a\}_{a=1}^{\tilde n}}$ belongs to the sector with $M - \tilde n$ spins up and with total magnetization $S_z^{\rm tot} = \sum_{\alpha = 1}^M S_z^\alpha = M/2 - \tilde n$.

The eigenvalue $\Lambda(\lambda, \{\lambda_a\})$ of $F(\lambda)$ is given by
\begin{align}
\label{eigen}
\Lambda(\lambda, \{\lambda_a\}) = \alpha(\lambda) \prod_{a=1}^{\tilde n} f_+(\lambda - \lambda_a)  + \delta(\lambda) \prod_{a=1}^{\tilde n} f_-(\lambda-\lambda_a)
\end{align}

Now, we clarify the connection between this construction based on an integrable spin chain and the irreducible representation based on Young diagrams introduced in the previous section. We recall the statement of Schur-Weyl duality about the joint action of $S_M$ and of $SU(2)$ on $\mathbb{H}$, the first acting by permuting the copies as in \eqref{tensorprodrepr} and the second by the simultaneous action on each copy
\begin{equation} \rho(g) v_1 \otimes v_2 \ldots \otimes v_M = (g v_1) \otimes (g v_2) \ldots  \otimes (g v_M) \;.
\qquad g \in SU(2)
\end{equation}
The Schur-Weyl duality tells how
$\mathbb{H}$ decomposes onto the irreducible representations of $S_M$ and $SU(2)$
\begin{equation}
\label{schurweyl}
\mathbb{H} := \mathbb{C}^2 \otimes \mathbb{C}^2 \otimes \ldots \otimes \mathbb{C}^2 = \sum_D V_D \otimes \rho_D
\end{equation}
Here the sum runs over the Young diagram $D = (M - \tilde{n}, \tilde n)$ with $M$ boxes and at most $2$ rows of sizes $M - \tilde n, \tilde n$ (with $0\leq \tilde n \leq M/2$). $V_D$
is the corresponding representation of $S_M$, while $\rho_{D}$ is the irreducible representation of $SU(2)$ with total spin $S^{\rm tot} = M/2 - \tilde n$. As an example for $M=3$, this translates into
\begin{equation}
\mathbb{C}^2 \otimes \mathbb{C}^2 \otimes \mathbb{C}^2
= \frac 3 2 \oplus \frac 1 2 \oplus \frac 1 2 = \left(\frac 3 2 \otimes \Yvcentermath1 \yng(3)\right)
\oplus
\left(\frac 1 2 \otimes \Yvcentermath1 \yng(2,1) \right)\;.
\end{equation}
corresponding to the total spin $S^{\rm tot}$ being $3/2$ and $1/2$. Here, the $2-$dimensional representation of $S_{M=3}$ of shape $(2,1)$ acts exchanging the  two isomorphic copies of spin $1/2$. This is a general property valid for arbitrary $M$: the irreducible representations $V_D$ act on isomorphic copies of the same total spin $S^{\rm tot}$.

Crucially, the states constructed via Eq.~\eqref{eq:bethestates} are always highest-weights of the ${\rm SU}(2)$ representations, i.e. they correspond to $(S^{\rm tot} = S_z^{\rm tot}, S_z^{\rm tot})$. Thus, we have one state (the highest weight) for each copy of the same total spin. Thanks to the Schur--Weyl
duality \eqref{schurweyl}, the span of these highest-weight states provide with an explicit construction of the irreducible representation $V_D$, with $D = (M-\tilde n, \tilde n)$.

The solution of \eqref{XYangBethe} can be now written identifying $A_Q^I$ with an eigenvector of $F(\lambda)$ labeled by the set $\{\lambda_a\}$.
Using Eq.~\eqref{XYangABA}, this gives the equation for $\mu=1,\dots,M$
\begin{align}
\label{NBA}
&\left[\prod_{\nu \neq \mu} S(k_\mu - k_\nu)\right]  \Lambda(k_\mu, \{\lambda_a\}) = \Phi^{-1} e^{i k_\mu L}
\end{align}
Putting it together with Eq.~\eqref{BAE}
and
substituting the explicit expressions for $\alpha(\lambda), \delta(\lambda), f_{\pm}(\lambda)$,
we get, using that $\delta(k_\mu) = 0$, the combined equations in the form
\begin{subequations}
    \label{NBAexplicit0}
\begin{align}
&\prod_{b \neq a} \frac{\lambda_a - \lambda_b + i \cb}{\lambda_a - \lambda_b - i \cb} =
\prod_{\mu} \frac{\lambda_a - k_\mu}{\lambda_a - k_\mu - i \cb} \;, \quad a=1,\dots,\tilde{n} \\
&\prod_{\nu \neq \mu} \frac{k_\mu - k_\nu - i \cb}{k_\mu - k_\nu + i\cb}
\prod_{a} \frac{k_\mu - \lambda_a + i \cb}{k_\mu  - \lambda_a }
= \Phi^{-1} e^{i k_\mu L}
\quad , \quad \mu=1,\dots, M
\label{secondNBA}
\end{align}
\end{subequations}
After the shift $\lambda_a \to \lambda_a + i \cb/2$, they take the more symmetric form
\begin{subequations}
    \label{NBAexplicit}
\begin{align}
&\prod_{b\neq a} \frac{\lambda_a - \lambda_b + i \cb}{\lambda_a - \lambda_b - i \cb} =
\prod_{\mu} \frac{\lambda_a - k_\mu + i \cb/2}{\lambda_a - k_\mu - i \cb/2}
\quad , \quad a=1,\dots,\tilde{n}
\label{eq:auxbos}
\\
&\prod_{\nu \neq \mu} \frac{k_\mu - k_\nu - i \cb}{k_\mu - k_\nu + i\cb}
\prod_{a} \frac{k_\mu - \lambda_a + i \cb/2}{k_\mu  - \lambda_a - i \cb/2}
= \Phi^{-1} e^{i k_\mu L} \quad , \quad \mu=1,\dots, M
\label{eq:physbos}
\end{align}
\end{subequations}
These equations determine the spectrum of the Lieb-Liniger Hamiltonian in the symmetry sector identified by $D = (M - \tilde n, \tilde n)$. In our case, we are interested in the rectangular diagrams \eqref{eq:Trect}; thus we set $M = n N = 2n$ and $\tilde n = n$. The variables $k_\mu$ are the physical momenta and determine the conserved charges of the eigenstate
\be
Q_m  = \frac{1}{2} \sum_{\mu = 1}^{N n} (k_\mu)^m
\ee
where $m=2$ corresponds to the energy. By contrast, the $\lambda_a$ are auxiliary variables, with a priori no physical meaning, which need to be eliminated in order to determine the $k_\alpha$.
\\

More general diagrams (and in particular the rectangular ones for $N>2$) can also be treated with a similar construction still based on Schur-Weyl duality and
${\rm SU}(N)$ spin chains. This leads to a larger set of $N$ Bethe equations ~\cite{oelkers2006bethe, sutherland1968further}. We stress that here, as we are interested in non-intersecting paths on the circle, we have the non-trivial boundary conditions controlled by $\Phi = (-1)^{N-1}$.

Note that in their study of $N$ non-crossing directed polymers, Emig and Kardar use a different set of Bethe equations (of total number $n$) based on iteratively removing the columns of the diagram $\lambda \vdash M$. They arrive to these equations by considering fermions in a random potential.
We discuss in Appendix~\ref{app:row_column_nba} the relation between the two procedures. We show that although the equations are different the solutions for the physical momenta are generically in one to one correspondence.
Our approach however is much more convenient to study the analytical continuation in $n$, the number of replicas.

\section{Study of the replica Bethe equations for \texorpdfstring{$N=2$}{N=2}, analytic continuation in \texorpdfstring{$n$}{n}}
\label{sec:replicaN2}

\subsection{Bethe equations for \texorpdfstring{$N=2$}{N=2}}

Upon performing the change of variables in \eqref{NBAexplicit}
\be
i k_\alpha = \frac{2}{L} q_\alpha \quad , \quad
i \lambda_a = \frac{2}{L} p_a \quad , \quad c = \frac{\gamma L}{2}
\ee
and the choice $\Phi=-1$, we obtain
the Bethe equations for $N=2$ in the form
\bea
\label{BEA3}
&& \prod_{b \neq a, b=1}^n \frac{p_a - p_b - c}{p_a - p_b + c} =
\prod_{\alpha=1}^{2 n}  \frac{p_a - q_\alpha  - c/2}{p_a - q_\alpha + c/2} \quad , \quad a=1,\dots,n \\
\label{BEA4}
&&  \prod_{\substack{\beta=1\\ \beta\neq\alpha}}^{2n} \frac{q_\alpha  - q_\beta  +  c}{q_\alpha  - q_\beta - c}
\prod_{a=1}^{n}  \frac{q_\alpha - p_a  - c/2}{q_\alpha - p_a + c/2}
= -  e^{2 q_\alpha} \quad , \quad \alpha =1 , \cdots, 2 n
\eea
where the $\{ q_\alpha\}_{\alpha=1,\dots,2n}$ are the physical momenta and
$\{ p_a \}_{a=1,\dots,n}$ the auxiliary ones.
The aim is now to compute
the ground state energy
\be
E(n,L) :=  \frac{2}{L^2} \min_{\{q\}, \{p\}}\left[-\sum_{\alpha=1}^{2 n} q_\alpha^2
\right]\ee
where the minimum is taken over the solutions of Eq.~\eqref{BEA3},\eqref{BEA4}, corresponding to the lowest-energy
state in the $(n,n)$ symmetry sector.

Note that $E(n,L)$ is expressed only in terms of the physical momenta $q_{\alpha}$'s so a priori we need to eliminate the auxiliary momenta.
These are determined in terms of the $q_{\alpha}$'s by the polynomial equations \eqref{BEA3}. They have then to be substituted in the equations \eqref{BEA4} to obtain closed reduced BE for the $q_{\alpha}$'s, which then need to be solved.
This formidable looking program can be implemented for each fixed integer $n$, order by order in an expansion at small $c$. However, even this is non trivial since there are a priori many roots to the polynomial equations.
In fact, since we are interested in the analytic continuation in $n$
we will below introduce proper generating functions so that the dependence in $n$ is explicit, and extend the Brunet Derrida method of analytical continuation in $n$ to the present case. We will show that the elimination of auxiliary momenta is possible at the level of these generating functions, a quite remarkable and non-trivial fact.

{\bf Symmetries.}
We note that the Bethe equations are invariant under the two transformations
(i) simultaneous reflection of all the roots,
$\{q_\alpha,p_a\}\to\{-q_\alpha,-p_a\}$, and
(ii) complex conjugation,
$\{q_\alpha,p_a\}\to\{q_\alpha^*,p_a^*\}$.
Assuming that the ground state is non-degenerate, these transformations
must map the ground-state solution onto itself, up to a permutation of the
roots. We will therefore look for a ground-state solution satisfying
\be  \label{reflection}
\{q_\alpha\}=\{-q_\alpha\}=\{q_\alpha^*\},
\qquad
\{p_a\}=\{-p_a\}=\{p_a^*\}.
\ee
In particular, the ground-state energy is real and the total momentum
vanishes, $\sum_{\alpha=1}^{2n}q_\alpha=0$. This can also be obtained directly by multiplying all the
Bethe equations: all two-body scattering factors cancel pairwise, yielding
$e^{2\sum_{\alpha=1}^{2n}q_\alpha}=1$,
and the zero-momentum branch is selected by continuity from $c=0$.
\\

{\bf Root polynomials}. It is useful to introduce the two root polynomials
\begin{equation}
Q(X)=\prod_{\alpha=1}^{2n}(X-q_\alpha),
\qquad
P(X)=\prod_{a=1}^{n}(X-p_a).
\end{equation}
The reflection symmetry \eqref{reflection} implies
\begin{equation}
Q(-X)=Q(X),
\qquad
P(-X)=(-1)^n P(X),
\end{equation}
while complex-conjugation invariance implies that both polynomials can be
chosen with real coefficients. In particular, for odd $n$ the auxiliary
root set necessarily contains $p=0$.

Using that
\be
\prod_{\substack{b=1\\b\neq a}}^n
\frac{p_a-p_b-c}{p_a-p_b+c}
=
-\frac{P(p_a-c)}{P(p_a+c)} \quad , \quad
\prod_{\substack{\beta=1\\\beta\neq\alpha}}^{2n}
\frac{q_\alpha-q_\beta+c}{q_\alpha-q_\beta-c}
=
-\frac{Q(q_\alpha+c)}{Q(q_\alpha-c)} \;.
\ee
The Bethe equations \eqref{BEA3}, \eqref{BEA4} can be rewritten
as
\bea
\label{eq:polyP}
&& P(p_a-c)\,
Q\left(p_a+\frac{c}{2}\right)
+
P(p_a+c)\,
Q\left(p_a-\frac{c}{2}\right)
=0
\quad , \quad a=1,\ldots,n \;.
\\
&&
\label{eq:polyQ}
e^{-q_\alpha}
Q(q_\alpha+c)
P\left(q_\alpha-\frac{c}{2}\right)
-
e^{q_\alpha}
Q(q_\alpha-c)
P\left(q_\alpha+\frac{c}{2}\right)
=0
\quad , \quad \alpha=1,\ldots,2n \;.
\eea
These two relations are the nested analogue of the polynomial relation
used by Brunet and Derrida in the single-polymer problem. In that case,
denoting by $Q_{\rm DB}(X)$ the polynomial whose zeros are the Bethe roots,
one has
\be
e^{q_\alpha}Q_{\rm DB}(q_\alpha-c)
+
e^{-q_\alpha}Q_{\rm DB}(q_\alpha+c)
=0 \;.
\ee
The coupled relations \eqref{eq:polyP} and \eqref{eq:polyQ} suggest that,
rather than eliminating the auxiliary roots $p_a$ individually, it is more
natural to eliminate them directly at the level of the polynomials, or of
suitable generating functions constructed from them.

Finally, the case $n=1$ provides a useful exact check. Reflection symmetry
imposes $q_1=-q_2$ and $p_1=0$, and the two scattering factors in
Eq.~\eqref{BEA4} cancel identically. The ground-state roots are therefore
independent of $c$,
\begin{equation}
q_1=\frac{i\pi}{2},
\qquad
q_2=-\frac{i\pi}{2},
\qquad
p_1=0,
\qquad
E(1,L)=\frac{\pi^2}{L^2}.
\end{equation}
This is expected physically, since for $n=1$ the two-particle wave function
is fully antisymmetric and hence vanishes at coincidence, so that the
delta interaction has no effect.

For $c\to 0$, the physical Bethe equations reduce to
$1=-e^{2q_\alpha}$, so that the antiperiodic free-particle momenta are
$q_\alpha=i\pi(m+\frac12)$, with $m\in\mathbb Z$.
For $n=1$, antisymmetry of the two-particle wave function selects the
ground-state configuration
$\{q_1,q_2\}=\{-\frac{i\pi}{2},\frac{i\pi}{2}\}$.
More generally, in the $(n,n)$ Young sector the wave function is
antisymmetric under exchange of the two particles belonging to each
column of the tableau. Hence these two particles must occupy distinct
one-particle orbitals, and the ground-state branch is obtained by filling
the two lowest antiperiodic modes in each column. Thus
\begin{equation}
\{q_\alpha\}_{\alpha=1}^{2n}
\underset{c\to0}{\longrightarrow}
\left\{
\underbrace{\frac{i\pi}{2},\ldots,\frac{i\pi}{2}}_{n},
\underbrace{-\frac{i\pi}{2},\ldots,-\frac{i\pi}{2}}_{n}
\right\},
\qquad
E_0(n,L)\big|_{c=0}
=
\frac{n\pi^2}{L^2}.
\end{equation}
In the small-$c$ expansion, the ground-state solution is therefore selected
as the branch continuously connected to this free-particle configuration.
Excited-state branches can similarly be obtained by starting from higher
antiperiodic free-particle modes.

This branch selection will reappear below when the Bethe equations are
reformulated as functional equations. As in the Brunet--Derrida construction
\cite{brunet2000ground,brunet2000probability}, and in its extension to the
interval \cite{BarraquandLeDoussal2025KPZInterval}, the resulting functional
equations admit homogeneous ambiguities unless supplemented by additional
analyticity and growth conditions. We will use such a condition below to
select the continuation of the ground-state branch.

\subsection{Direct expansion of the Bethe equations at small \texorpdfstring{$c$}{c}}

For fixed integer $n$, the Bethe equations can be solved perturbatively
around the free-particle solution at $c=0$. This provides useful benchmarks
for the analytic continuation developed below. We illustrate the procedure
for $n=2$.

Using the reflection and complex-conjugation symmetries, we choose the roots
such that
\be
p_2=-p_1 \quad , \quad q_3=-q_1 \quad , \quad q_4=-q_2
\quad , \quad q_2=-q_1^* \;.
\ee
In the limit $c \to 0$, the auxiliary Bethe equation become simple at leading
order. Expanding it to first non-trivial order in $c$ gives
$3p_1^2-\frac{\pi^2}{4}=0$. Using this and inserting a series expansion in $\sqrt{c}$ in the Bethe equations, one finds for the ground-state branch
and $n=2$
\bea
p_1 & = &
\frac{\pi}{2\sqrt{3}}
+\frac{c}{2\sqrt{3}\pi}
+\frac{(15+8\pi^2)c^2}{12\sqrt{3}\pi^3}
+O(c^3)
\\
q_1 & = &
\frac{i\pi}{2}
-\frac{\sqrt{c}}{\sqrt{2}}
-\frac{ic}{2\pi}
-\frac{(3+\pi^2)c^{3/2}}{12\sqrt{2}\pi^2}
-\frac{(675-30\pi^2+11\pi^4)c^{5/2}}
{1440\sqrt{2}\pi^4}
+\frac{ic^3}{\pi^5}
+O(c^{7/2})
\\
q_2 & = &
\frac{i\pi}{2}
+\frac{\sqrt{c}}{\sqrt{2}}
-\frac{ic}{2\pi}
+\frac{(3+\pi^2)c^{3/2}}{12\sqrt{2}\pi^2}
+\frac{(675-30\pi^2+11\pi^4)c^{5/2}}
{1440\sqrt{2}\pi^4}
+\frac{ic^3}{\pi^5}
+O(c^{7/2}) \;.
\eea
We note in particular that $p_1$ is real and that $q_2=-q_1^*$,
consistently with the symmetries discussed above. We also note the absence
of a $c^2$ term in the expansion of the physical roots.

Using the symmetry of the physical roots, the small $c$ expansion for the ground-state energy for
$n=2$ reads
\bea \label{E2Lc3}
E(2,L)
&=&
-\frac{4}{L^2}(q_1^2+q_2^2)
\nn\\
&=&
-\frac{4}{L^2}
\left[
-\frac{\pi^2}{2}
+2c+\frac{c^2}{6}
+\left(\frac{1}{45}-\frac{1}{\pi^4}\right)c^3
+O(c^4)
\right].
\eea
In particular, the attractive interaction $c>0$ lowers the ground-state
energy, as expected. Determining the coefficient of the $O(c^4)$ term in
the energy requires pushing the perturbative solution of the Bethe
equations to higher order.

The same expansion is carried for $n=3$ in Appendix \ref{app:directexpansionc}.
It is also shown there that the $c \to 0$ limit of the auxiliary momenta for general $n$ take the form: $p_a
=
\frac{\pi}{2}
\cot\left(\frac{\pi a}{n+1}\right)$,
$a=1,\ldots,n$.

\subsection{Generating functions}

Let us define, in direct analogy with \cite{brunet2000probability}, the set of residues for the physical momenta
\be \label{def1}
\rho(q_\alpha) = \frac{Q(q_\alpha+c)}{cQ'(q_\alpha)} =
\prod_{\substack{\beta=1\\\beta\neq\alpha}}^{2n}   \frac{q_\alpha - q_\beta +c}{q_\alpha - q_\beta}
\quad , \quad
\frac{Q(X+c)}{Q(X)}= 1 + c \sum_{\alpha=1}^{2n} \frac{\rho(q_\alpha)}{X-q_\alpha}
\ee
as well as for the auxiliary momenta, a new feature here

\be \label{def2}
\hat \rho(p_a) = \frac{P(p_a+c)}{cP'(p_a)}
= \prod_{\substack{b=1\\b\neq a}}^{n}
\frac{p_a-p_b+c}{p_a-p_b}
\quad , \quad
\frac{P(X+c)}{P(X)}
=1+c\sum_{a=1}^{n}\frac{\hat\rho(p_a)}{X-p_a} \;.
\ee

Since we assume that the ground state obeys the reflection symmetry
\eqref{reflection}, the algebraic identities of Appendix A of
\cite{brunet2000probability} which follow only from the pole decomposition
and reflection symmetry remain valid. In particular, for the physical roots
one obtains, upon replacing $n\to 2n$,
\bea
\sum_{\alpha=1}^{2 n} \rho(q_\alpha) = 2 n \quad , \quad \sum_{\alpha=1}^{2 n} q_\alpha \rho(q_\alpha)
=c \binom{2 n}{2} \quad , \quad \sum_{\alpha=1}^{2 n} q_\alpha^2 \rho(q_\alpha)
= c^2 \binom{2 n}{3}- \frac{L^2}{2} E(n,L)
\eea
as well as
\be
\sum_{a=1}^{n}\hat\rho(p_a)=n
\quad , \quad
\sum_{a=1}^{n}p_a\hat\rho(p_a)
=
c\binom{n}{2} \;.
\ee

In addition, one can verify directly
(independently of their Appendix A) that
\be
\rho(-q_\alpha) =
\prod_{\substack{\beta=1\\q_\beta\neq -q_\alpha}}^{2n}
\frac{-q_\alpha-q_\beta+c}{-q_\alpha-q_\beta}
=
\prod_{\substack{\beta=1\\\beta\neq\alpha}}^{2n}
\frac{q_\alpha-q_\beta-c}{q_\alpha-q_\beta} \;.
\ee

Using the same identity for $\hat \rho(-p_a)$, we find that the
Bethe equations \eqref{BEA3},\eqref{BEA4} can be rewritten as
\bea
\label{BEA3new2}
&& \frac{\hat \rho(-p_a)}{\hat \rho(p_a)} = \frac{Q(p_a - c/2)}{Q(p_a + c/2)}
 \quad , \quad a=1,\dots,n \\
&& \label{BEA4new2}
\frac{\rho(q_\alpha)}{\rho(-q_\alpha)} \frac{P(q_\alpha - c/2)}{P(q_\alpha + c/2)}
=  - e^{2 q_\alpha} \quad , \quad \alpha =1 , \cdots, 2 n
\eea
Using the pole decompositions \eqref{def1}, \eqref{def2}, one has
\bea
&& \frac{Q(p_a + c/2)}{Q(p_a - c/2)} =
1 + c \sum_{\alpha =1}^{2 n} \frac{\rho(q_\alpha)}{p_a -c/2 -q_\alpha} \\
&& \frac{P(q_\alpha + c/2)}{P(q_\alpha - c/2)}
= 1 + c \sum_{a =1}^{n} \frac{\hat \rho(p_a)}{q_\alpha - c/2-p_a}
\eea
Thus the Bethe equations can be rewritten as
\bea  \label{BE10}
&& \frac{\rho(q_\alpha)}{\rho(-q_\alpha)}
= - e^{2 q_\alpha} \left(
1 + c \sum_{a =1}^{n} \frac{\hat \rho(p_a)}{q_\alpha - c/2-p_a} \right)
\quad , \quad \alpha =1 , \cdots, 2 n \\
&& \frac{\hat \rho(-p_a)}{\hat \rho(p_a)} \left( 1 + c
\sum_{\alpha =1}^{2 n} \frac{\rho(q_\alpha)}{p_a-c/2-q_\alpha}
\right) = 1 \quad , \quad a=1,\dots, n
\eea
\\

Next one introduces as in \cite{BarraquandLeDoussal2025KPZInterval}, the generating functions
\bea
&& L(z) = \sum_{\alpha=1}^{2 n} \rho(q_\alpha) e^{q_\alpha z} \quad , \quad M(z)= \frac{1}{2 n} e^{c z/2} L(z) = e^{- c z^2/4} B(1+z)  \\
&& \hat L(z) = \sum_{a=1}^{n} \hat \rho(p_a) e^{p_a z} \quad , \quad \hat M(z)= \frac{1}{n} e^{c z/2} \hat L(z)
\eea
where $B(u)$ is the analog of the generating function defined in \cite{brunet2000probability} (with $n \to 2 n$). They satisfy
the boundary conditions
\bea
&& M(0)=1 \quad , \quad M'(0)= n c \quad , \quad E(n,L)= - \frac{ 4 n}{L^2} (M''(0)- \frac{1}{12} c^2 (1 + 8 n^2)) \\
&& \hat M(0)=1 \quad , \quad \hat M'(0)= n c/2
\eea

\subsection{Integral equations}

By construction of these
functions, following \cite{brunet2000probability}, and for $M(z)$ following \cite{BarraquandLeDoussal2025KPZInterval},
these generating functions obey the following functional integral equations (containing $2n$ and $n$ as a parameter respectively)
\bea  \label{inteqM}
&& 2 n c \int_0^z M(-z') M(z-z') dz' = M(z) - M(-z) \\
&& n c \int_0^z \hat M(-z') \hat M(z-z') dz' = \hat M(z) - \hat M(-z)
\label{inteqMhat}
\eea

Once the reflection symmetry \eqref{reflection} is assumed, these two
equations follow solely from the pole decompositions \eqref{def1},
\eqref{def2} and do not use the detailed form of the nested Bethe equations.
The latter provide an additional relation coupling $M$ and $\hat M$, which
we now derive. If we further assume invariance of the ground state under complex conjugation in \eqref{reflection}, then the functions $L(z),\hat L(z),M(z), \hat M(z)$ are all real for real $z$.

In \cite{brunet2000probability}, the Bethe equations lead to the simple
parity relation $B(-u)=B(u)$. Here the nested Bethe equations instead
provide a non-trivial relation coupling the two generating functions.
To derive it, we first rewrite \eqref{BE10} as
\begin{subequations}
\label{eq:betherho}
\bea
-\rho(-q_\alpha)e^{2q_\alpha}
&=&
\rho(q_\alpha)
-
c\sum_{a=1}^n
\frac{\hat\rho(p_a)\rho(q_\alpha)}
{q_\alpha+p_a+c/2}
\label{bet}
\\
\hat\rho(-p_a)
&=&
\hat\rho(p_a)
-
c\sum_{\alpha=1}^{2n}
\frac{\rho(q_\alpha)\hat\rho(p_a)}
{q_\alpha+p_a+c/2} \;.
\label{eq:betaux}
\eea
\end{subequations}
Multiplying the first equation by $e^{q_\alpha z}$ and summing over
$\alpha$, and the second by $e^{p_a z}$ and summing over $a$, gives
\be
-L(-2-z) = L(z)-c\,{\cal A}(z)
\quad , \quad
\hat L(-z) = \hat L(z)-c\,\hat{\cal A}(z) \;,
\ee
where
\be
{\cal A}(z)
=
\sum_{a,\alpha}
\frac{\rho(q_\alpha)\hat\rho(p_a)}
{q_\alpha+p_a+c/2}
e^{q_\alpha z}
\quad , \quad
\hat{\cal A}(z)
=
\sum_{a,\alpha}
\frac{\rho(q_\alpha)\hat\rho(p_a)}
{q_\alpha+p_a+c/2}
e^{p_a z} \;.
\ee
The two double sums are related by
\be
\int_{-z}^0dw\,
e^{cw/2}\hat L(w+z)L(w)
=
\hat{\cal A}(z)
-
e^{-cz/2}{\cal A}(-z) \;.
\ee
Using the two preceding relations to eliminate ${\cal A}$ and
$\hat{\cal A}$, and then replacing $z\to-z$, one obtains the single
independent mixed integral equation
\be
\label{eq:mixedL}
\hat L(z)-\hat L(-z)
+
e^{cz/2}\left[L(z)+L(-z-2)\right]
=
c\int_0^z dz'\,
e^{cz'/2}L(z')\hat L(z'-z) \;.
\ee
The equation obtained from \eqref{eq:mixedL} by $z\to-z$ is equivalent
to it and does not provide an independent relation.

In terms of the normalized generating functions $M$ and $\hat M$,
Eq.~\eqref{eq:mixedL} becomes
\bea
\label{eq:mixedM}
2nc \int_0^z dw\, e^{-cw/2}\hat M(w)M(-z+w)
&=&
e^{-cz/2}\hat M(z)-e^{cz/2}\hat M(-z)
+\nn\\
&&
-2M(-z)-2e^{c(1-z)}M(-2+z) \;.
\eea
Together with Eqs.~\eqref{inteqM} and \eqref{inteqMhat},
Eq.~\eqref{eq:mixedM} provides a closed system of integral equations
for $M$ and $\hat M$, in which $n$ appears as a continuous parameter.

As a simple consequence, setting $z=0$ in Eq.~\eqref{eq:mixedM} gives
\be
M(-2)=-e^{-c} M(0) = -e^{-c}\;.
\ee
Since $M(-2)=e^{-c}B(-1)$, while $M(0)=B(1)=1$, this is equivalently
\be
B(-1)=-1
\quad , \quad
B(1)=1 \;.
\ee
This replaces the condition $B(-1)=B(1)=1$ of
\cite{brunet2000probability}.

\subsection{\texorpdfstring{$M, \hat M$}{M, M} system in Laplace}

We introduce the even and odd parts of the generating functions,
\bea
M^{\rm e}(z)&=&\frac{M(z)+M(-z)}{2}
\quad , \quad
M^{\rm o}(z)=\frac{M(z)-M(-z)}{2}
\nn\\
\hat M^{\rm e}(z)&=&\frac{\hat M(z)+\hat M(-z)}{2}
\quad , \quad
\hat M^{\rm o}(z)=\frac{\hat M(z)-\hat M(-z)}{2} \;,
\eea
and define their one-sided Laplace transforms by
\be
\mathcal L[f](u)=\int_0^\infty dz\,e^{-uz}f(z) \;,
\ee
and
\bea
\MM^{\rm e/o}(u)&=&\mathcal L[M^{\rm e/o}](u)
\quad , \quad
\hat\MM^{\rm e/o}(u)=\mathcal L[\hat M^{\rm e/o}](u) \;.
\eea
We also denote by
\be
\MM(u):=\mathcal L[M](u)
=\MM^{\rm e}(u)+\MM^{\rm o}(u)
\quad , \quad
\hat\MM(u):=\mathcal L[\hat M](u)
=\hat\MM^{\rm e}(u)+\hat\MM^{\rm o}(u) \;.
\ee

For integer $n$, the functions $M$ and $\hat M$ are finite sums of
exponentials. Their Laplace transforms therefore admit a meromorphic
continuation to the complex $u$ plane. In particular, if
${\bf f}(u)=\mathcal L[f](u)$, then
\be
\mathcal L[f(-z)](u)=-{\bf f}(-u) \;.
\ee
This identity is specific to this class of functions and does not hold
for a generic one-sided Laplace transform. More generally, it holds for suitable superpositions $f(z)=\int dp\,\mu(p)e^{-pz}$ for which the corresponding transforms
and their analytic continuations are well defined (compact support of
$\mu$ being a sufficient, but not necessary, condition).

This identity implies in particular
that $\MM^o(u)$ is even in $u$ and $\MM^e(u)$ is odd in $u$, i.e.
\be \label{parityMeo}
\MM^{\rm e}(-u)=-\MM^{\rm e}(u)
\quad , \quad
\MM^{\rm o}(-u)=\MM^{\rm o}(u) \;,
\ee
and similarly for $\hat\MM^{\rm e/o}$.

We can now rewrite Eqs.~\eqref{inteqM}, \eqref{inteqMhat} as
\begin{subequations}
\label{eq:oddfromeven}
\bea
&& \MM^o(u)  = n c (\MM^e(u)^2 - \MM^o(u)^2)
\label{eq:oddfromevenM}
\\
&& \hat \MM^o(u)  = \frac{n c}{2}   (\hat\MM^e(u)^2 - \hat\MM^o(u)^2)
\eea
\end{subequations}
These two equations determine the odd part in terms of the even part,
as
\be
\MM^o(u) = \frac{1}{2 c n} ( \sqrt{ 1 + 4 c^2 n^2 \MM^e(u)^2}-1) \quad , \quad
\hat \MM^o(u) = \frac{1}{c n} ( \sqrt{ 1 + c^2 n^2 \hat \MM^e(u)^2}-1)
\ee

The Laplace transform of the mixed equation is slightly more involved and reads
\begin{multline}
\label{eq:mixed}
2 n c \hat\MM(u + c/2) (\MM^{\rm e}(u) - \MM^{\rm o}(u)) =\\= \hat{\MM}^{\rm e}(\frac{c}{2}+u)-\hat{\MM}^{\rm e}(u-\frac{c}{2})+\hat{\MM}^{\rm o}(u+\frac{c}{2})+\hat{\MM}^{\rm o}(u-\frac{c}{2})
- 2 (\MM^{\rm e}(u)-\MM^{\rm o}(u)) +  2 e^c \mathsf{K}(u + c)
\end{multline}
where we set
\begin{equation}
\label{eq:Kdef}
    \mathsf{K}(u) := - \mathcal{L}[M(z - 2)](u) = - \int_0^\infty dz M(z-2) e^{-z u}
\end{equation}

Note that $\hat{\MM}^{\rm e}(\frac{c}{2}+u)-\hat{\MM}^{\rm e}(u-\frac{c}{2})$
and $\hat{\MM}^{\rm o}(u+\frac{c}{2})+\hat{\MM}^{\rm o}(u-\frac{c}{2})$ are even
in $u$ from \eqref{parityMeo}. One has the following asymptotics for $u \to +\infty$
\begin{subequations}
\label{eq:largeuconstraint}
\bea
&& M(0)=1 \quad , \quad M'(0)= n c  \quad \leftrightarrow \quad  \MM(u) \simeq \frac{1}{u} + \frac{n c}{u^2} +
\frac{M''(0)}{u^3} + O(u^{-4})
 \\
&& \hat M(0)=1 \quad , \quad \hat M'(0)= n c/2  \quad \leftrightarrow \quad  \hat \MM(u) \simeq \frac{1}{u} + \frac{n c}{2 u^2} +
\frac{\hat M''(0)}{u^3} + O(u^{-4})
\eea
\end{subequations}
By taking the odd part of Eq.~\eqref{eq:mixed}, we have
\begin{multline}
\label{eq:oddlapla}
n c [ \MM^{\rm e}(u) (\hat\MM(u + c/2) + \hat\MM(-u + c/2)) -
\MM^{\rm o}(u) (\hat\MM(u + c/2) - \hat\MM(-u + c/2)) ]
 =\\=
- 2 \MM^{\rm e}(u)+  e^c \mathsf{K}(u + c) -  e^c \mathsf{K}(-u + c)
\end{multline}

\subsection{Decoupling of the equations}

It is useful to introduce
\be
\MM_{\pm}(u)
=
\pm \MM(\pm u)
=
\MM^{\rm e}(u)\pm\MM^{\rm o}(u)
\quad , \quad
\hat\MM_{\pm}(u)
=
\pm \hat\MM(\pm u)
=
\hat\MM^{\rm e}(u)\pm\hat\MM^{\rm o}(u) \;.
\ee
In particular,
$\MM_+(u)=\MM(u)$ and $\hat\MM_+(u)=\hat\MM(u)$. Using
\be
\MM^{\rm o}
=
\frac{\MM_+-\MM_-}{2}
\quad , \quad
(\MM^{\rm e})^2-(\MM^{\rm o})^2
=
\MM_+\MM_- \;,
\ee
the Eqs.~\eqref{eq:oddfromeven} become
\be
\MM_+-\MM_-
=
2nc\,\MM_+\MM_- \quad , \quad
\hat\MM_+-\hat\MM_-
=
nc\,\hat\MM_+\hat\MM_- \;.
\ee
The negative components can therefore be expressed algebraically in
terms of the positive ones as
\be
\label{eq:minusfromplus}
\MM_-(u)
=
\frac{\MM_+(u)}
{1+2cn\,\MM_+(u)}
\quad , \quad
\hat\MM_-(u)
=
\frac{\hat\MM_+(u)}
{1+cn\,\hat\MM_+(u)} \;.
\ee
Equivalently, suppressing the subscript $+$,
\be
\label{eq:reflectionMM}
\MM(-u)
=
-\frac{\MM(u)}
{1+2cn\,\MM(u)}
\quad , \quad
\hat\MM(-u)
=
-\frac{\hat\MM(u)}
{1+cn\,\hat\MM(u)} \;.
\ee

We can now rewrite Eq.~\eqref{eq:mixed} in terms of the $\pm$ components as
\be
2nc\,\hat\MM_+\left(u+\frac{c}{2}\right)\MM_-(u)
=
\hat\MM_+\left(u+\frac{c}{2}\right)
-\hat\MM_-\left(u-\frac{c}{2}\right)
-2\MM_-(u)
+2e^c\mathsf K(u+c) \;.
\ee
Using Eq.~\eqref{eq:minusfromplus} to eliminate $\MM_-$ and
$\hat\MM_-$, and solving for
$\hat\MM_+(u+c/2)$, we obtain
\be
\label{eq:mmhatplus}
\hat\MM_+\left(u+\frac{c}{2}\right)
=
\frac{
\hat\MM_+\left(u-\frac{c}{2}\right)
\left(1+4cn\MM_+(u)\right)
+2\MM_+(u)
}{
1+cn\hat\MM_+\left(u-\frac{c}{2}\right)
}
-2e^c\mathsf K(c+u)
\left(1+2cn\MM_+(u)\right) \;.
\ee

Finally, Eq.~\eqref{eq:oddlapla} can be written in terms of the $\pm$
components as
\be
nc\left[
\MM_-(u)\hat\MM_+\left(u+\frac{c}{2}\right)
-
\MM_+(u)\hat\MM_-\left(u-\frac{c}{2}\right)
\right]
=
-\MM_+(u)-\MM_-(u)
+e^c\left[\mathsf K(c+u)-\mathsf K(c-u)\right] \;.
\ee
We now use Eq.~\eqref{eq:minusfromplus} together with
Eq.~\eqref{eq:mmhatplus}. Remarkably, all dependence on
$\hat\MM_+(u-c/2)$ cancels, and one obtains
\be
\frac{2cn\,\MM_+(u)^2}{1+2cn\,\MM_+(u)}
-
2cn\,e^c\MM_+(u)\mathsf K(c+u)
=
-\MM_+(u)
-\frac{\MM_+(u)}{1+2cn\,\MM_+(u)}
+e^c\left[\mathsf K(c+u)-\mathsf K(c-u)\right] \;.
\ee
Multiplying by $1+2cn\,\MM_+(u)$ and simplifying finally gives, upon suppressing
the subscript $+$,
\be
\label{eq:smalleq}
2\MM(u)
=
e^c\left[
\mathsf K(c+u)\left(1+2cn\,\MM(u)\right)
-\mathsf K(c-u)
\right] \;.
\ee
Hence the auxiliary function $\hat M$ has been eliminated.

Equation~\eqref{eq:smalleq} is a necessary consequence of the mixed
equation, but it is not equivalent to it, since it is obtained by taking
its odd part in $u$. The information contained in the even part of
Eq.~\eqref{eq:mixed} is therefore lost in deriving \eqref{eq:smalleq}.
Some of this additional information can be recovered from the large-$u$
behavior of Eq.~\eqref{eq:mixed}, or equivalently from the expansion of
the original mixed equation near $z=0$.
Expanding Eq.~\eqref{eq:mixed} at large $u$, using
Eq.~\eqref{eq:largeuconstraint}, gives
\be
\label{eq:Klargeu}
\mathsf K(u)
=
\frac{e^{-c}}{u}
-\frac{c(n-3)e^{-c}}{2u^2}
+O(u^{-3}) \;.
\ee
On the other hand, from the definition \eqref{eq:Kdef},
\be
\mathsf K(u)
=
-\frac{M(-2)}{u}
-\frac{M'(-2)}{u^2}
+O(u^{-3}) \;.
\ee
Hence
\be
M(-2)=-e^{-c}
\quad , \quad
M'(-2)=\frac{c}{2}(n-3)e^{-c} \;,
\ee
and therefore
\be
M(z)
=
-e^{-c}
+\frac{c}{2}(n-3)e^{-c}(z+2)
+O\left((z+2)^2\right) \;.
\ee
The same two conditions follow directly by evaluating the original mixed
equation at $z=0$ and differentiating it once with respect to $z$.

\subsection{Back to the reduced equations for \texorpdfstring{$M(z)$}{M(z)}}

We can now undo the Laplace transform and obtain the following closed set of equations for the generating function $M(z)$ alone
\begin{subequations}
\label{eq:fullset}
\begin{align}
   &  M(z) - M(-z) = 2 n c \int_0^z M(-z') M(z-z') dz'  \\
    & - 2 M(z) = e^{c(1-z)} M(z-2) +
    e^{c(1+z)} M(-z-2) + 2 c n \int_0^z dw e^{c (1-w)} M(w - 2) M(z - w)
    \label{reducedM}
\end{align}
\end{subequations}
The normalization is
\be
M(0)=1 \;.
\ee
The first equation then implies $M'(0)=nc$, while setting $z=0$ in the
second one gives
\be
M(-2)=-e^{-c} \;.
\ee
In addition, the complementary even sector of the original mixed equation
provides the condition
\be
M'(-2)=\frac{c}{2}(n-3)e^{-c} \;.
\ee
This formulation eliminates the auxiliary generating function $\hat M$
entirely. We stress, however, that the second equation in
\eqref{eq:fullset} was obtained from the odd part in Laplace space of the
original mixed equation. The condition on $M'(-2)$ represents only the
first piece of information inherited from the complementary even sector,
and we cannot yet claim that the reduced system above is fully equivalent
to the original coupled $(M,\hat M)$ system.
\\

{\bf Direct elimination of the auxiliary roots}. For integer $n$, the same reduction can also be carried
out directly at the level of the Bethe roots, without introducing the
Laplace-space formulation. Eliminating the auxiliary residues
$\hat\rho(p_a)$ from the nested Bethe equations gives the reduced Bethe
equation
\be
\label{reducedBE}
-2\rho(q_\alpha)
=
\rho(-q_\alpha)e^{2q_\alpha}
-
c\rho(q_\alpha)
\sum_{\beta=1}^{2n}
\frac{\rho(q_\beta)e^{-2q_\beta}}
{q_\beta-q_\alpha-c} \;.
\ee
Through the definition of $M(z)$, this equation is equivalent to the
second equation in \eqref{eq:fullset}, namely \eqref{reducedM}.
A direct derivation of the elimination procedure and of this equivalence is given in
Appendix~\ref{app:RBEresidue}, see also Appendix~\ref{app:RBEbaxter}.

\subsection{Expansion in \texorpdfstring{$c$}{c} for general \texorpdfstring{$n$}{n}}

We now solve the reduced equations perturbatively in $c$, working in
Laplace space, through order $c^5$. %
A complementary small-$c$ expansion, performed directly at the level
of the root polynomials $P$ and $Q$ and avoiding the generating-function
construction, is presented in Appendix~\ref{app:expansionpolynomial},
where it is worked out explicitly through order $c^2$.
We set $M(z)=\sum_{k \geq 0} M_k(z) c^k$, $\MM(u)=\sum_{k \geq 0} \MM_k(u) c^k$, ${\mathsf K}(u)= \sum_{k \geq 0} {\mathsf K}_k(u) c^k$,
with
\be
M_0(z)= \cos(\pi z/2) \quad , \quad
\MM_0(u)= {\mathsf K}_0(u)= \frac{u}{u^2 + \frac{\pi^2}{4}}
\ee
One finds
\bea
&& M_1(z) = \frac{(n (z+2)-z) \left(2 \sin \left(\frac{\pi  z}{2}\right)+\pi  z \cos
   \left(\frac{\pi  z}{2}\right)\right)}{4 \pi } \\
&& M_2(z) = \frac{z \left(\pi  z \left(2 n^2 (z+2)^2-n (z+2) (5 z-2)+3 z^2\right)
   \cos \left(\frac{\pi  z}{2}\right)+12 (z-n (z+2))^2 \sin
   \left(\frac{\pi  z}{2}\right)\right)}{96 \pi } \nn
\eea
Higher orders can be obtained recursively and were computed through order
$c^5$. %
The resulting functions obey the required boundary conditions at $z=0$
and $z=-2$.

Recalling the formula for the energy
\bea
E(n,L)= - \frac{4 n}{L^2} (M''(0)- \frac{1}{12} c^2 (1 + 8 n^2))
\eea
we have obtained its expansion to $O(c^5)$
\bea \label{Esmallc}
&& E(n,L) = \frac{n \pi^2}{L^2}
- \frac{4 n (n-1)}{L^2} \bigg(
c+\frac{c^2}{12}+\frac{\left(\pi ^4-45\right) c^3 n}{180 \pi
   ^4} \nn \\
   && +\frac{c^4 n \left(2835 (n-4)+\pi ^6 (5 n-6)-63 \pi ^4 (n-1)+945
   \pi ^2\right)}{7560 \pi ^6} \nn \\
   && -\frac{c^5}{226800 \pi
   ^8} n \big(56700 \pi ^2
   (n-4)+14175 (n (17 n-118)+180)
   \\
   && +\pi ^8 ((58-21 n) n-36)+60 \pi ^6 (n-1)
   (5 n-6)-1890 \pi ^4 (n (2 n-7)+2)\big)
   +O\left(c^6\right) \bigg) \nn
\eea
which agrees with the expansions of $E(2,L)$ given in \eqref{E2Lc3}
to $O(c^3)$, and of $E(3,L)$ given in Appendix \ref{app:directexpansionc}.
It is interesting to compare with the result of Brunet and Derrida \cite{brunet2000probability} for $N=1$, which has the simpler form
\bea
L^2 E^{(N=1)}(n,L)= - 2 n(n-1) \bigg( \frac{c}{2} + \frac{c^2}{12} + \frac{n c^3}{180}
+ ( \frac{n^2}{1512} - \frac{n}{1260} ) c^4 \bigg) + O(c^5)
\eea

{\bf Cumulants}. Using that $\overline{ e^{n \log Z(t)} }  \sim e^{- E(n,L)  t}$,
 the cumulants of $\log Z(t)$ are extracted as follows
\be
\overline{(\log Z(t))^p}^c \simeq c_p(L) t \quad , \quad c_p(L)=- \partial_n^p E(n,L)|_{n=0}
\ee
We then obtain the following expansions for the cumulants
\bea \label{cumsmallc0}
L^2 c_1(L) &=& -\pi ^2-4 c-\frac{c^2}{3}+O\left(c^6\right) \\
L^2 c_2(L) &=&
8 c+\frac{2
   c^2}{3}+\left(-\frac{2}{45}+\frac{2}{\pi ^4}\right)
   c^3+\left(\frac{2}{315}+\frac{12}{\pi ^6}-\frac{1}{\pi ^4}-\frac{1}{15
   \pi ^2}\right) c^4 \\
   &+& \left(-\frac{2}{1575}+\frac{90}{\pi
   ^8}-\frac{8}{\pi ^6}-\frac{2}{15 \pi ^4}+\frac{4}{315 \pi ^2}\right)
   c^5+O\left(c^6\right) \nn \\
  L^2 c_3(L) &=&
   \left(\frac{2}{15}-\frac{6}{\pi ^4}\right)
   c^3+\left(-\frac{11}{315}-\frac{45}{\pi ^6}+\frac{3}{\pi
   ^4}+\frac{2}{5 \pi ^2}\right) c^4
   \\
   &+& \left(\frac{47}{4725}-\frac{447}{\pi
   ^8}
   + \frac{30}{\pi ^6}+\frac{9}{5 \pi ^4}-\frac{34}{315 \pi ^2}\right)
   c^5+O\left(c^6\right) \nn \\
   L^2 c_4(L) &=&
   \left(\frac{4}{63}+\frac{36}{\pi ^6}-\frac{4}{5
   \pi ^2}\right) c^4+\left(-\frac{158}{4725}+\frac{810}{\pi
   ^8}-\frac{24}{\pi ^6}-\frac{36}{5 \pi ^4}+\frac{128}{315 \pi
   ^2}\right) c^5+O\left(c^6\right) \nn
\eea
The first line is consistent with the fact that the perturbative expansion of the first cumulant stops at order $c^2$.

\subsection{Equations using the function \texorpdfstring{$B$}{B}}

It is useful to perform the change of function $M(z) = e^{- c z^2/4} B(z+1)$ and rewrite the reduced system of equations in terms of the
function $B$, the analog to the function used by Brunet and Derrida.
The boundary conditions become
\be
B(1)=1 \quad B'(1)=n c \quad B(-1)=-1 \quad B'(-1) = \frac{(n-1) c}{2}
\ee
and the equations read
\begin{subequations}
\begin{align}
    &  -2 B(1 + z) = B(z-1) + B(-z-1) + 2 n c \int_0^z dv e^{\frac{c z v}{2}-\frac{c v^2}{2}} B(v-1)B(z - v + 1) \\
    & B(1 + z) - B(1-z) = 2 n c\int_0^z d v  e^{\frac{c z v}{2}-\frac{c v^2}{2}} B(1-v) B(z-v+1)
\end{align}
\end{subequations}
It is also convenient to introduce the even/odd part of the function
\be
B_{e/o}(z) = \frac{B(z) \pm B(-z)}{2}
\ee
Then, taking the sum of these two equations, we arrive at a symmetric and compact form which reads
\begin{subequations}
\label{eq:integraleqsymmetric2}
\begin{align}
    & \label{eq:newinteq2}
     - (B_e(1 + z) + B_e(1 - z))  = 2 n c \int_0^z dv e^{\frac{c z v}{2}-\frac{c v^2}{2}} B_e (v-1)B(z - v + 1) \\ \label{eq:usualinteq2}
    & B(1 + z) - B(1-z) = 2 n c\int_0^z d v  e^{\frac{c z v}{2}-\frac{c v^2}{2}} B(1-v) B(z-v+1)\\
    & B_e(z)  = \frac{(n+1) (z-1) c}{4} + O((z-1)^2) \;, \quad
    B_o(z) = 1 + \frac{(3n - 1)(z-1)c}{4} + O((z-1)^2)
\end{align}
\end{subequations}

\section{Expansion in \texorpdfstring{$n$}{n}}
\label{sec:nExpansion}

\subsection{Equations valid to all orders in \texorpdfstring{$n$}{n}}
\label{subsec:n_all_orders}
We rewrite the two reduced equations for $B$ in terms of their definite
parity components,
\bea
B_e(1+z)+B_e(1-z) &=& \Psi_e(z),
\label{eq:BePsi}
\\
B(1+z)-B(1-z) &=& \Phi_o(z),
\label{eq:BPhi}
\eea
where
\begin{subequations}
\begin{align}
\Psi_e(z)
&=
-2nc\int_0^z dv\,
e^{\frac{czv}{2}-\frac{cv^2}{2}}
B_e(v-1)B(z-v+1),
\label{eq:Psie}
\\
\Phi_o(z)
&=
2nc\int_0^z dv\,
e^{\frac{czv}{2}-\frac{cv^2}{2}}
B(1-v)B(z-v+1).
\label{eq:Phio}
\end{align}
\end{subequations}
By construction $\Psi_e$ is even and $\Phi_o$ is odd. We want to solve this system with the boundary
conditions
\be
B_e(1)=0,\qquad
B_e'(1)=\frac{(n+1)c}{4},\qquad
B_o(1)=1,\qquad
B_o'(1)=\frac{(3n-1)c}{4}.
\ee
Two of these conditions are already encoded in the integral equations:
at $z=0$,
\be
\label{eq:basicboundary}
2B_e(1)=\Psi_e(0)=0,
\qquad
2B'(1)=\Phi_o'(0)=2nc\,B(1)^2=2nc.
\ee
We may therefore take
\be
\label{eq:independentBCB}
B(1)=1,\qquad B'(-1)=\frac{(n-1)c}{2}
\ee
as the two independent boundary conditions.

\paragraph{Spectral decomposition.}

The system \eqref{eq:BePsi}-\eqref{eq:BPhi} has a convenient structure:
its left-hand sides are linear in $B$ and involve only the shifts
$z\to1\pm z$, whereas the nonlinearity is confined to the source terms
$\Psi_e$ and $\Phi_o$, which we treat initially as independent functions carrying the explicit  $n$-dependence. Since the shifts
act diagonally on the exponentials $e^{\lambda z}$, it would seem natural to
look for $B$ directly in the form of a bilateral Laplace transform. The
solution at $n=0$, where the sources vanish, shows that this is not
possible: anticipating Eq.~\eqref{eq:Bzero} of Sec.~\ref{subsec:zerothfirst},
$B_0(z)=\sin(\pi z/2)+\frac{c(z-1)}{2\pi}\cos(\pi z/2)$ is a combination of
the two lowest antiperiodic harmonics $e^{\pm i\pi z/2}$ with polynomial
coefficients.
Equivalently, it contains modes with complex Laplace exponents
$\lambda=\pm i\pi/2$,
and $B_0$ grows as $e^{\pi|u|/2}$ along the
imaginary axis $z=iu$, where it has no Fourier transform. Since the shifts
act on these harmonics by a simple multiplication,
$e^{\pm i\pi(z+1)/2}=\pm i\,e^{\pm i\pi z/2}$, we factor them out explicitly
at all orders in $n$ and write
\be
\label{eq:Bdecomposition}
B(z)=\BB_+(z)e^{i\pi z/2}+\BB_-(z)e^{-i\pi z/2}.
\ee
The choice of this decomposition is based on the fact that the amplitudes $\BB_\pm$ could now admit a Laplace transform, consistent with the case $n=0$, in which they reduce to first-degree polynomials. Specifically, for a complex-conjugation invariant ground state, for which $B(z)$ is real for real $z$, we choose the decomposition such
that, for real $z$, $\BB_-(z)=\BB_+(z)^*$, so that \eqref{eq:Bdecomposition}
reads $B(z)=2\,{\rm Re}[\BB_+(z)e^{i\pi z/2}]$ and involves a single complex
function. Even with this condition, the decomposition
\eqref{eq:Bdecomposition} has a residual freedom,
\be
\BB_+(z)\to\BB_+(z)+ig(z)e^{-i\pi z/2},
\qquad
\BB_-(z)\to\BB_-(z)-ig(z)e^{i\pi z/2},
\ee
with $g(z)$ real for real $z$. We fix this residual freedom by requiring
that, up to a polynomial of degree one in $z$, the functions $\BB_\pm(z)$
admit bilateral Laplace representations with sufficiently fast decay in
the spectral parameter, i.e.\ we write
\be
\BB_\pm(z)
=
\BB_\pm^{\rm(h)}(z)
+
\BB_\pm^{\rm(inh)}(z),
\qquad
\BB_\pm^{\rm(inh)}(z)=\int_{-\infty}^{\infty}d\lambda\,
e^{\lambda z}\tilde\BB_\pm(\lambda),
\label{integralrepresentation}
\ee
with $\BB_\pm^{\rm(h)}(z)$ a polynomial of degree one. Within this class, we dub as homogeneous parts $\BB_\pm^{\rm(h)}(z)$
the solutions of Eqs.~\eqref{eq:BePsi}-\eqref{eq:BPhi} with vanishing right-hand side, i.e.\ of the linear relations \eqref{BBPhi}-\eqref{BBPsi} below with vanishing sources, corresponding to polynomials of
degree at most one, Eq.~\eqref{eq:BBhomogeneous}, whose two coefficients are fixed
by the boundary conditions \eqref{eq:independentBCB}. The integral term, which we call the
non-homogeneous part, is instead uniquely determined by the sources $\Psi_e(z),\Phi_o(z)$.
This prescription also specifies the analytic class in which we solve the
functional equations. Indeed, the representation \eqref{integralrepresentation}
implies that on the imaginary
axis $z=iu$, $\BB_\pm^{\rm(inh)}(iu)$ admits a Fourier transform,
\be
\BB_\pm^{\rm(inh)}(iu)
=
\int_{-\infty}^{\infty}d\lambda\,
e^{i\lambda u}\tilde\BB_\pm(\lambda),
\qquad
\tilde\BB_\pm(\lambda)
=
\frac{1}{2\pi}\int_{-\infty}^{\infty}du\,
e^{-i\lambda u}\BB_\pm^{\rm(inh)}(iu).
\ee
and the spectral density $\tilde\BB_\pm(\lambda)$
is thus recovered by Fourier inversion from the values of $\BB_\pm^{\rm(inh)}$
on the imaginary axis. Consistently, $\BB_\pm^{\rm(inh)}(iu)$ can grow at most polynomially in $u$. Without this condition the
homogeneous relations admit additional solutions, involving functions of
period two in $z$, $e^{i\pi k z}$ with $k\neq0$, which behave as $e^{-\pi k u}$
on the imaginary axis and are compatible with the boundary conditions, as in
the periodic-function ambiguity encountered in the Brunet--Derrida construction
\cite{brunet2000ground,brunet2000probability} and in the corresponding
analysis on the interval \cite{BarraquandLeDoussal2025KPZInterval}. We
interpret the present minimal-growth prescription as selecting the analytic
continuation of the ground-state branch identified above at integer $n$.

Similarly, we decompose the functions of definite parity as
\be
\label{eq:PsiPhiDecomposition}
\Psi_e(z)
=
e^{i\pi z/2}\Psi(z)+e^{-i\pi z/2}\Psi(-z),
\qquad
\Phi_o(z)
=
e^{i\pi z/2}\Phi(z)-e^{-i\pi z/2}\Phi(-z),
\ee
and
assume that the function $\Psi(z)$ and $\Phi(z)$ can be written as bilateral Laplace transforms as
\be
\Psi(z)=\int_{-\infty}^{\infty}d\lambda\,
e^{\lambda z}\tilde\Psi(\lambda),
\qquad
\Phi(z)=\int_{-\infty}^{\infty}d\lambda\,
e^{\lambda z}\tilde\Phi(\lambda).
\ee
Equation \eqref{eq:basicboundary} implies
\be
\label{BCPhiz}
\Psi(0)=0,
\qquad
\Phi'(0)+\frac{i\pi}{2}\Phi(0)=nc.
\ee
If the ground state is invariant under complex conjugation, then for real
$z$ and real $\lambda$ one chooses again
\bea
\label{ccsym}
&&
\Psi(-z)=\Psi(z)^*,
\qquad
\Phi(-z)=-\Phi(z)^*,
\nn\\
&&
\tilde\Psi(-\lambda)=\tilde\Psi(\lambda)^*,
\qquad
\tilde\Phi(-\lambda)=-\tilde\Phi(\lambda)^*.
\eea
We will verify that the decompositions Eqs.~(\ref{eq:Bdecomposition},\ref{eq:PsiPhiDecomposition}) are indeed consistent with Eqs.~(\ref{eq:Psie},\ref{eq:Phio}).
The decompositions above turn \eqref{eq:BePsi}-\eqref{eq:BPhi} into the
linear relations
\begin{align}
\label{BBPhi}
&i\BB_-(1-z)+i\BB_+(1+z)=\Phi(z),
\\
\label{BBPsi}
&\frac{i}{2}\left[
\BB_-(-1-z)-\BB_-(1-z)
-\BB_+(z-1)+\BB_+(z+1)
\right]=\Psi(z).
\end{align}
As anticipated, the corresponding homogeneous solutions are
\be
\label{eq:BBhomogeneous}
\BB_\pm^{\rm(h)}(z)
=
a_\pm-\frac{a_++a_-}{2}\,z.
\ee
For the non-homogeneous parts, the bilateral transforms obey
\bea
&&
i\left(
e^{-\lambda}\tilde\BB_-(-\lambda)
+
e^\lambda\tilde\BB_+(\lambda)
\right)
=
\tilde\Phi(\lambda),
\nn\\
&&
i\sinh\lambda\left(
\tilde\BB_-(-\lambda)
+
\tilde\BB_+(\lambda)
\right)
=
\tilde\Psi(\lambda),
\eea
and hence
\be
\label{eq:soluBB}
\tilde\BB_\pm(\pm\lambda)
=
-\frac{i}{2\sinh\lambda}
\left[
\tilde\Psi(\lambda)(1\mp\coth\lambda)
\pm\tilde\Phi(\lambda)
\right].
\ee

After imposing the two boundary conditions \eqref{eq:independentBCB},
the result can be written directly in the form
\be
\label{eq:BBkernelrep}
\BB_\pm(z)
=
\int_{-\infty}^{\infty}d\lambda\,
\left[
\phi_\pm(\lambda,z)\tilde\Phi(\lambda)
+
\psi_\pm(\lambda,z)\tilde\Psi(\lambda)
\right]
-\frac12\left(
\pm i+\frac{z-1}{2\pi}(n-1)c
\right),
\ee
where
\bea
\label{phipsi}
&&
\phi_\pm(\lambda,z)
=
\frac{i}{2}\left[
\mp\frac{e^{\pm\lambda z}}{\sinh\lambda}
\pm\coth\lambda
+
\left(
\frac12-\frac{i\lambda}{\pi}
\right)(z-1)
\right],
\nn\\
&&
\psi_\pm(\lambda,z)
=
\frac{i}{2}\left[
\frac{(-1\pm\coth\lambda)e^{\pm\lambda z}}{\sinh\lambda}
\mp\frac{1}{\sinh^2\lambda}
+
\coth\lambda
\left(
-1+\frac{2i\lambda}{\pi}
\right)(z-1)
\right].
\eea
The kernels in \eqref{phipsi} invert the two linear relations
\eqref{BBPhi}. Explicitly,
\bea
&&
i\phi_-(\lambda,1-z)+i\phi_+(\lambda,1+z)
=e^{\lambda z},
\nn\\
&&
\frac{i}{2}\left[
\phi_-(\lambda,-1-z)-\phi_-(\lambda,1-z)
-\phi_+(\lambda,z-1)+\phi_+(\lambda,z+1)
\right]=0,
\nn\\
&&
i\psi_-(\lambda,1-z)+i\psi_+(\lambda,1+z)
=0,
\nn\\
&&
\frac{i}{2}\left[
\psi_-(\lambda,-1-z)-\psi_-(\lambda,1-z)
-\psi_+(\lambda,z-1)+\psi_+(\lambda,z+1)
\right]=e^{\lambda z}.
\eea

It is useful to record a few properties of these kernels. Defining
\be
{\cal P}_\chi(\lambda,z)
:=
\chi_+(\lambda,z)e^{i\pi z/2}
+
\chi_-(\lambda,z)e^{-i\pi z/2},
\qquad
\chi=\phi,\psi,
\ee
one finds
\bea
\label{phipsi1}
&&
{\cal P}_\phi(\lambda,1)
=
{\cal P}_\psi(\lambda,1)
=
{\cal P}_\phi(\lambda,-1)
=0,
\qquad
{\cal P}_\psi(\lambda,-1)=2,
\nn\\
&&
\partial_z{\cal P}_\phi(\lambda,-1)
=
\partial_z{\cal P}_\psi(\lambda,-1)
=
\partial_z{\cal P}_\psi(\lambda,1)
=0,
\nn\\
&&
\partial_z{\cal P}_\phi(\lambda,1)
=
\lambda+\frac{i\pi}{2}.
\eea
Together with \eqref{BCPhiz}, these identities ensure all four boundary
conditions for $B$. The kernels are also regular at $\lambda=0$:
\bea
&&
\phi_\pm(\lambda,z)
=
-\frac{i}{4}(1+z)+O(\lambda),
\nn\\
&&
\psi_\pm(\lambda,z)
=
\frac{i(z-1)\left[\pm\pi(z-1)+4i\right]}{4\pi}
+O(\lambda).
\eea
For real $z,\lambda$ they obey
\be
\phi_-(-\lambda,z)=-\phi_+(\lambda,z)^*,
\qquad
\psi_-(-\lambda,z)=\psi_+(\lambda,z)^*,
\ee
consistently with \eqref{ccsym}.

\paragraph{Nonlinear source terms and recursion in $n$.}
The source functions $\Phi$ and $\Psi$ can be reconstructed directly
from $B$ and $\BB_\pm$. We use the identity
\be
\label{eq:gshift}
\int_0^z dv\,g(v)
=
\int_0^\infty dv\,[g(v)-g(z+v)].
\ee
Applying this identity to Eq.~\eqref{eq:Phio}, we can identify $\Phi(z)$ consistent with the complex-conjugation symmetry \eqref{ccsym},
\be
\label{eq:phifromB3}
\Phi(z)
=
2nci\sum_{b=\pm1}
\int_0^\infty dv\,
e^{-cv^2/2}
B(1-v)\BB_b(1-v+bz)
e^{bcvz/2-i\pi bv/2}.
\ee
The corresponding on-shell representation of $\Psi$ is
\be
\label{eq:psifromB3}
\Psi(z)
=
-\frac{i}{2}nc
\sum_{a,b=\pm1}a
\int_0^\infty dv\,
e^{-cv^2/2}
B(b(v-1))
\BB_a(b(1-v)+az)
e^{abv(cz-i\pi)/2}.
\ee
The representation \eqref{eq:psifromB3} uses the evenness of $\Psi_e$,
and is therefore to be understood on shell, for a function $B$
satisfying the reduced integral equations.

Equations \eqref{eq:BBkernelrep}, \eqref{eq:phifromB3} and
\eqref{eq:psifromB3} form a closed recursive scheme in $n$. Indeed, write
\be
B(z)=\sum_{r\geq0}n^r B_r(z),
\qquad
\BB_\pm(z)=\sum_{r\geq0}n^r\BB_{\pm,r}(z),
\qquad
\Phi(z)=\sum_{r\geq1}n^r\Phi_r(z),
\qquad
\Psi(z)=\sum_{r\geq1}n^r\Psi_r(z).
\ee
Because the nonlinear source terms carry an explicit factor $n$, at
order $n^r$ one has
\bea
\label{eq:PhiRecursion}
\Phi_r(z)
&=&
2ci\sum_{b=\pm1}
\sum_{j+k=r-1}
\int_0^\infty dv\,
e^{-cv^2/2}
B_j(1-v)\BB_{b,k}(1-v+bz)
e^{bcvz/2-i\pi bv/2},
\\
\label{eq:PsiRecursion}
\Psi_r(z)
&=&
-\frac{i}{2}c
\sum_{a,b=\pm1}a
\sum_{j+k=r-1}
\int_0^\infty dv\,
e^{-cv^2/2}
B_j(b(v-1))
\BB_{a,k}(b(1-v)+az)
e^{abv(cz-i\pi)/2}.
\eea
Thus $\Phi_r$ and $\Psi_r$ depend only on the already known orders
$0,\ldots,r-1$, while \eqref{eq:BBkernelrep} gives the new
$\BB_{\pm,r}$ by the same linear kernel at every order, together
with the explicit homogeneous contribution displayed there. This is the
main advantage of the present formulation.

\paragraph{Energy.}
Finally, we recall that
\be  \label{energyGS2}
E(n,L)
=
-\frac{4n}{L^2}
\left[
M''(0)-\frac{c^2}{12}(1+8n^2)
\right],
\quad , \quad
M''(0)=B''(1)-\frac{c}{2}.
\ee
Using the kernel representation \eqref{eq:BBkernelrep}, one obtains
\bea
\label{eq:BppEnergyKernel}
B''(1)
&=&
\int_{-\infty}^{\infty}d\lambda\,
\Bigg[
\left(
-\lambda+\lambda(\lambda+i\pi)\coth\lambda-\frac{i\pi}{2}
\right)\tilde\Phi(\lambda)
\\
&&\qquad\qquad
+
\left(
(2\lambda+i\pi)\coth\lambda
-(\lambda+i\pi)\frac{\lambda}{\sinh^2\lambda}
\right)\tilde\Psi(\lambda)
\Bigg]
+\frac{c}{2}(n-1)-\frac{\pi^2}{4}. \nn
\eea
Hence the energy, and therefore the cumulants, can be extracted directly
from the successive coefficients of $\tilde\Phi$ and $\tilde\Psi$.

\subsection{Zeroth and first orders in \texorpdfstring{$n$}{n}}
\label{subsec:zerothfirst}

We now apply the recursive scheme of the previous section to the zeroth and first order
in $n$, which allows to obtain the first and second cumulants
$c_1(L)$ and $c_2(L)$.

\paragraph{Zeroth order and first cumulant.}
Since the source terms \eqref{eq:phifromB3} and \eqref{eq:psifromB3}
contain an explicit factor $n$, one has at zeroth order
\be
\Phi_0(z)=\Psi_0(z)=0.
\ee
Hence $\BB_{\pm,0}$ are entirely given by the homogeneous solution
\eqref{eq:BBhomogeneous},
\be
\BB_{\pm,0}(z)
=
a_{\pm,0}-\frac{a_{+,0}+a_{-,0}}{2}z.
\ee
The two boundary conditions \eqref{eq:independentBCB} at $n=0$ fix
\be
a_{\pm,0}
=
\pm\frac{1}{2i}-\frac{c}{4\pi},
\ee
and therefore
\be
\label{eq:BBzero}
\BB_{\pm,0}(z)
=
\frac{1}{2}
\left(
\frac{c(z-1)}{2\pi}\mp i
\right).
\ee
This gives
\be
\label{eq:Bzero}
B_0(z)
=
\sin\left(\frac{\pi z}{2}\right)
+
\frac{c(z-1)}{2\pi}
\cos\left(\frac{\pi z}{2}\right).
\ee
At this lowest order the role of the spectral prescription is particularly
transparent. The solution \eqref{eq:Bzero} is built from the lowest
antiperiodic harmonic. If the growth restriction imposed above is relaxed,
additional homogeneous solutions involving higher antiperiodic harmonics can
be added while preserving the functional equations and the boundary
conditions; these solutions have faster exponential growth along the
imaginary axis. Thus \eqref{eq:Bzero} is the minimal-growth solution, naturally
associated with the ground-state branch selected at $c=0$. This genuine
freedom in the solution for $B_0$ is distinct from the residual freedom in
the decomposition \eqref{eq:Bdecomposition}, which leaves $B_0$ itself
unchanged.

In particular,
\be
B_0''(1)
=
-\frac{\pi^2}{4}-\frac{c}{2}.
\ee
Using the energy formula together with $M''(0)=B''(1)-c/2$, this
immediately gives the exact first cumulant
\be
L^2 c_1(L)
=
-\pi^2-4c-\frac{c^2}{3}.
\ee

\paragraph{First order and second cumulant.}
We now write
\be
\Psi(z)=n\Psi_1(z)+O(n^2),
\qquad
\Phi(z)=n\Phi_1(z)+O(n^2).
\ee
The functions $\Phi_1$ and $\Psi_1$ are obtained from
\eqref{eq:phifromB3} and \eqref{eq:psifromB3} by replacing $B$ and
$\BB_\pm$ by their zeroth-order expressions \eqref{eq:Bzero} and
\eqref{eq:BBzero}.

It is convenient to invert directly the bilateral Laplace transforms
appearing in \eqref{eq:phifromB3} and \eqref{eq:psifromB3}. Since the
$z$ dependence occurs only through exponentials, this can be done using
the following inversion rules,
\bea
&&
\int_0^{+\infty}dv\,f(v)e^{\pm cvz/2}
\ \longrightarrow\
\frac{2}{c}
f\left(\pm\frac{2\lambda}{c}\right)
\theta(\pm\lambda),
\label{eq:FTrule1}
\\
&&
\int_0^{+\infty}dv\,f(v)z\,e^{\pm cvz/2}
\ \longrightarrow\
\mp\frac{2}{c}
\left[
\frac{2}{c}
f'\left(\pm\frac{2\lambda}{c}\right)
\theta(\pm\lambda)
+
f(0)\delta(\lambda)
\right].
\label{eq:FTrule2}
\eea
Here the arrow denotes the corresponding contribution to the bilateral
Laplace density. These rules can equivalently be justified by restricting
to $z=iu$ and using ordinary Fourier inversion. The second rule contains a Dirac
contribution which will be important below.

Using these rules one
finds
\bea
\tilde\Phi_1(\lambda)
&=&
\frac{
e^{-\frac{2\lambda(\lambda+i\pi)}{c}}
\left[
e^{\frac{2i\pi\lambda}{c}}
\left(c-2(\lambda-i\pi)^2\right)
-c+2\lambda^2+2i\pi\lambda
\right]
\text{sgn}(\lambda)
}{2\pi^2}
-\frac{2ic}{\pi}\delta(\lambda),
\label{firstorderPhi}
\\
\tilde\Psi_1(\lambda)
&=&
c\,
\frac{
e^{-\frac{2\lambda(\lambda+i\pi)}{c}}
\left[
2(\lambda-i\pi)e^{\frac{2i\pi\lambda}{c}}
-2\lambda-i\pi
\right]
\text{sgn}(\lambda)
}{4\pi^2}.
\label{firstorderPsi}
\eea
One can check explicitly the boundary and symmetry properties
\bea
&&
\int_{-\infty}^{\infty}d\lambda\,
\tilde\Psi_1(\lambda)=0,
\qquad
\int_{-\infty}^{\infty}d\lambda\,
\left(\lambda+\frac{i\pi}{2}\right)
\tilde\Phi_1(\lambda)=c,
\nn\\
&&
\tilde\Psi_1(-\lambda)
=
\tilde\Psi_1(\lambda)^*,
\qquad
\tilde\Phi_1(-\lambda)
=
-\tilde\Phi_1(\lambda)^*.
\eea
In the second relation of the first line, the contribution $c$ comes
entirely from the Dirac term in \eqref{firstorderPhi}.

The Dirac part of $\tilde\Phi_1$ also gives a simple contribution to
$\BB_{\pm,1}$. Using
\be
\phi_\pm(0,z)=-\frac{i}{4}(1+z),
\ee
one obtains
\be
\label{eq:BB1Dirac}
\left.\BB_{\pm,1}(z)\right|_{\rm Dirac}
=
-\frac{2ic}{\pi}\phi_\pm(0,z)
=
-\frac{c(z+1)}{2\pi}.
\ee

We now use the formula for the ground state energy.
Taking the
coefficient of $n$ in \eqref{eq:BppEnergyKernel} gives
\bea
\label{Bsecond1}
B_1''(1)
&=&
\int_{-\infty}^{\infty}d\lambda\,
\bigg[
\left(
-\lambda
+(\lambda+i\pi)\lambda\coth\lambda
-\frac{i\pi}{2}
\right)
\tilde\Phi_1(\lambda)
\nn\\
&&\qquad\qquad
+
\left(
(2\lambda+i\pi)\coth\lambda
-(\lambda+i\pi)
\frac{\lambda}{\sinh^2\lambda}
\right)
\tilde\Psi_1(\lambda)
\bigg]
+\frac{c}{2}.
\eea
We now insert \eqref{firstorderPhi} and \eqref{firstorderPsi} into this
formula. The resulting integrand is rather bulky. Factoring out
$e^{-2\lambda^2/c}$, we denote the remaining expression by $F(\lambda)$
and write
\be
\int_{-\infty}^{\infty}d\lambda\,
e^{-2\lambda^2/c}F(\lambda)
=
\int_0^{+\infty}d\lambda\,
e^{-2\lambda^2/c}
\left[F(\lambda)+F(-\lambda)\right].
\ee
The resulting integrand is real. The explicit $c/2$ term in
\eqref{Bsecond1}, together with the contribution $c$ of the Dirac term,
gives $3c/2$. Since $B(1)=1$ at all $n$, one has $B_1(1)=0$, and therefore the
coefficient of $n$ in $M''(0)$ is simply $B_1''(1)$. Hence one obtains
the following final formula for the ground state energy to order $n^2$ and for the second cumulant as a function of $L$ and $c=\gamma L/2$
\be \label{secondcum2}
\left.E(n,L)\right|_{n^2}
=
-\frac{4}{L^2}B_1''(1)
\quad , \quad
c_2(L)=
\frac{8}{L^2}B_1''(1),
\ee
where
\be
B_1''(1)=
\frac{3c}{2}
+
\int_0^{+\infty}
\frac{d\lambda}{2\pi^2}
e^{-\frac{2\lambda^2}{c}}
\bigg[
\tilde f_0(\lambda)
+
\tilde f_c(\lambda)
\cos\left(\frac{2\pi\lambda}{c}\right)
+
\tilde f_s(\lambda)
\sin\left(\frac{2\pi\lambda}{c}\right)
\bigg],
\label{eq:c2exact}
\ee
and the three auxiliary functions are given by
\bea
\tilde f_0(\lambda)
&=&
2\left[
c(3\lambda^2+\pi^2)
-2\lambda^2(\lambda^2+\pi^2)
\right]\coth\lambda
\nn\\
&&
-2\lambda
\left[
c(\lambda^2+\pi^2)\text{csch}^2\lambda
+c-2\lambda^2
\right],
\\
\tilde f_c(\lambda)
&=&
2\lambda(c-2\lambda^2+\pi^2)
+
\left[
c(\pi^2-6\lambda^2)
+4(\lambda-\pi)(\lambda+\pi)\lambda^2
\right]\coth\lambda
\nn\\
&&
+c\lambda(2\lambda^2-\pi^2)
\text{csch}^2\lambda,
\\
\tilde f_s(\lambda)
&=&
\pi\left[
(8\lambda^3-6c\lambda)\coth\lambda
+3c\lambda^2\text{csch}^2\lambda
+c-6\lambda^2
\right].
\eea

\paragraph{Small-$c$ expansion of the second cumulant.}

The small-$c$ expansion of \eqref{eq:c2exact} can be organized into
three contributions: the non-oscillating term, and the cosine and sine
oscillating terms.

For the non-oscillating part, which does not contain
$\cos(2\pi\lambda/c)$ or $\sin(2\pi\lambda/c)$, one changes variables
$\lambda\to\sqrt{c}\,\lambda$. One then expands the integrand in $c$ to
the desired order and integrates term by term the resulting expressions
of the form $\lambda^k e^{-2\lambda^2}$. For the cosine contribution, one instead changes variables
$\lambda\to c\lambda$, so that $
\cos\left(\frac{2\pi\lambda}{c}\right)
\longrightarrow
\cos(2\pi\lambda)$.
One then introduces an Abel regulator $e^{-\epsilon\lambda}$,
$\epsilon>0$, expands in $c$ to the desired order, and integrates each
term of the form $
\lambda^k\cos(2\pi\lambda)e^{-\epsilon\lambda}$.
The limit $\epsilon\to0^+$ is taken at the end. The sine contribution is
treated in exactly the same way, with
$\sin(2\pi\lambda)$ in place of $\cos(2\pi\lambda)$.
The above procedure gives
\bea
B_1''(1)
&=&
c+\frac{c^2}{12}
+
\left(
\frac{1}{4\pi^4}-\frac{1}{180}
\right)c^3
\nn\\
&&
+
\left(
\frac{1}{1260}
+\frac{3}{2\pi^6}
-\frac{1}{8\pi^4}
-\frac{1}{120\pi^2}
\right)c^4
\nn\\
&&
+
\left(
-\frac{1}{6300}
+\frac{45}{4\pi^8}
-\frac{1}{\pi^6}
-\frac{1}{60\pi^4}
+\frac{1}{630\pi^2}
\right)c^5
+O(c^6).
\eea
This leads to
\bea
L^2 c_2(L)
&=&
8c+\frac{2c^2}{3}
+
\left(
-\frac{2}{45}+\frac{2}{\pi^4}
\right)c^3
\nn\\
&&
+
\left(
\frac{2}{315}
+\frac{12}{\pi^6}
-\frac{1}{\pi^4}
-\frac{1}{15\pi^2}
\right)c^4
\nn\\
&&
+
\left(
-\frac{2}{1575}
+\frac{90}{\pi^8}
-\frac{8}{\pi^6}
-\frac{2}{15\pi^4}
+\frac{4}{315\pi^2}
\right)c^5
+O(c^6),
\eea
This is found to coincide with the small-$c$ expansion \eqref{cumsmallc0}
which was derived by a quite different method, providing a useful check.

\paragraph{Large-$c$ behavior.}

The large-$c$ behavior of \eqref{eq:c2exact} requires some care, since
two different regions of the integration variable contribute. In the
bulk region $\lambda\sim\sqrt{c}$, we set
$\lambda=\sqrt{c}\,x$. For fixed $x>0$ one can replace
$\coth(\sqrt{c}\,x)\to1$ and
$\operatorname{csch}^2(\sqrt{c}\,x)\to0$, which gives
\bea
B_1''(1)\big|_{\rm bulk}
&=&
\frac{11}{32}\sqrt{\frac{\pi}{2}}\,c^{3/2}
+\frac{3c}{2}
+\frac{5\pi^{5/2}}{128\sqrt{2}}\,c^{1/2}
\nn\\
&&
-\frac{7\pi^{9/2}}{256\sqrt{2}}\,c^{-1/2}
+O(c^{-3/2}).
\eea
This approximation is however not uniform near $\lambda=0$.
As shown in Appendix~\ref{app:largecvariance}, the boundary region
$\lambda=O(1)$ gives the additional contribution
\be
\Delta B_1''(1)
=
-\frac{3c}{2}
+\frac{5}{2}\zeta(3)
+O(c^{-1}).
\ee
In particular, the $O(c)$ terms cancel. Combining the two regions, one finds
\bea
\label{eq:B1largec}
B_1''(1)
&=&
\frac{11}{32}\sqrt{\frac{\pi}{2}}\,c^{3/2}
+\frac{5\pi^{5/2}}{128\sqrt{2}}\,c^{1/2}
+\frac{5}{2}\zeta(3)
\nn\\
&&
-\frac{7\pi^{9/2}}{256\sqrt{2}}\,c^{-1/2}
+O(c^{-1}).
\eea
For the leading large-$c$ behavior, this gives
\be
\label{c2largec}
L^2[E(n,L)]_{n^2}
\simeq
-\frac{11}{8}\sqrt{\frac{\pi}{2}}\,c^{3/2},
\qquad
L^2 c_2(L)
\simeq
\frac{11}{4}\sqrt{\frac{\pi}{2}}\,c^{3/2}.
\ee
For comparison, Eq.~(70) of Brunet and Derrida for a single polymer
($N=1$) gives
\be
L^2[E^{N=1}(n,L)]_{n^2}
\simeq
-\frac{\sqrt{\pi}}{2\sqrt{2}}\,c^{3/2},
\ee
so that the coefficient of the $n^2$ term is larger by a factor $11/4$
in the present two-polymer problem. Finally, using $c=\gamma L/2$,
\be
c_2(L)
\simeq
\frac{11\sqrt{\pi}}{16}
\frac{\gamma^{3/2}}{L^{1/2}}.
\ee

It is also useful to consider directly the ratio of the two-polymer and
one-polymer variance amplitudes,
\be
R(c)=\frac{c_2^{N=2}(L)}{c_2^{N=1}(L)}.
\label{eq:varianceRatio}
\ee
Combining the small-$c$ expansions given above, one finds
\be
R(c)
=
4-\frac{c}{3}
+\left(\frac{7}{90}+\frac{1}{\pi^4}\right)c^2
+O(c^3),
\qquad c\to0,
\label{eq:varianceRatioSmallC}
\ee
while the large-$c$ expansions give
\be
R(c)
=
\frac{11}{4}
+\frac{5\pi^2}{16c}
+\frac{9\sqrt{2}\,\zeta(3)}
{\sqrt{\pi}\,c^{3/2}}
+O(c^{-2}),
\qquad c\to+\infty.
\label{eq:varianceRatioLargeC}
\ee
Thus the relative enhancement of the variance crosses over from $4$
at weak coupling to $11/4$ in the large-$L$ KPZ regime.

\section{Derivation of the large-\texorpdfstring{$L$}{L} scaling equations}
\label{sec:scalingEquations}

We now consider the scaling limit
\be
c\to+\infty,
\qquad
n=\frac{\epsilon}{\sqrt{c}},
\qquad
\epsilon=n\sqrt{c}\quad\text{fixed}.
\label{eq:scalingeps}
\ee
As in the Brunet--Derrida analysis, the relevant variations of $B(z)$
occur on a scale $1/\sqrt{c}$ around $z=\pm1$.

It is useful to distinguish two steps. We first derive identities for the
rescaled functions which are exact at finite $c$. We then use the spectral
representation derived in the previous section to take the strict algebraic
scaling limit $c\to+\infty$ at fixed $\epsilon$ and obtain a closed system
of equations for the leading scaling functions. Their solution by
Wiener--Hopf methods will be developed in Sec.~\ref{sec:WHscaling}.

\subsection{Exact rescaled identities}

Let us first rescale directly the exact equations
\eqref{eq:BPhi} and \eqref{eq:BePsi}. Setting
\be
z=\frac{y}{\sqrt{c}},
\qquad
v=\frac{w}{\sqrt{c}},
\qquad
n=\frac{\epsilon}{\sqrt{c}},
\ee
in \eqref{eq:BPhi}, one obtains the exact relation
\bea
&&
B\left(1+\frac{y}{\sqrt{c}}\right)
-
B\left(1-\frac{y}{\sqrt{c}}\right)
\nn\\
&&\qquad =
2\epsilon
\int_0^y dw\,
e^{\frac12(yw-w^2)}
B\left(1-\frac{w}{\sqrt{c}}\right)
B\left(1+\frac{y-w}{\sqrt{c}}\right).
\label{eq:exactscaledplus}
\eea

A second exact identity follows by combining \eqref{eq:BePsi} with
\eqref{eq:BPhi}. Indeed, one first obtains
\bea
&&
2B(1+z)+B(-1+z)+B(-1-z)
\nn\\
&&\qquad =
-2nc\int_0^z dv\,
e^{\frac{c}{2}(zv-v^2)}
B(v-1)B(1+z-v).
\eea
Hence
\bea
&&
2B\left(1+\frac{y}{\sqrt{c}}\right)
+
B\left(-1+\frac{y}{\sqrt{c}}\right)
+
B\left(-1-\frac{y}{\sqrt{c}}\right)
\nn\\
&&\qquad =
-2\epsilon
\int_0^y dw\,
e^{\frac12(yw-w^2)}
B\left(-1+\frac{w}{\sqrt{c}}\right)
B\left(1+\frac{y-w}{\sqrt{c}}\right).
\label{eq:exactscaledminus}
\eea
Equations \eqref{eq:exactscaledplus} and
\eqref{eq:exactscaledminus} are exact at finite $c$. They will not be
used to derive the leading scaling equations below, which requires the
additional information contained in the spectral representation. They
are nevertheless useful as consistency checks, and become important if
one wishes to study subleading algebraic corrections in $1/\sqrt{c}$.
For the derivation of the leading scaling equations, we now return to
the representation in terms of $\BB_\pm$, $\Phi$ and $\Psi$.

\subsection{Scaled spectral representation}

We define the scaled bilateral Laplace densities
\be
\label{defhat}
\hat\Phi(\lambda)
=
\sqrt{c}\,\tilde\Phi(\sqrt{c}\lambda),
\qquad
\hat\Psi(\lambda)
=
\sqrt{c}\,\tilde\Psi(\sqrt{c}\lambda),
\ee
so that
\be
\Phi\left(\frac{y}{\sqrt{c}}\right)
=
\int_{\mathbb R}d\lambda\,
e^{\lambda y}\hat\Phi(\lambda),
\qquad
\Psi\left(\frac{y}{\sqrt{c}}\right)
=
\int_{\mathbb R}d\lambda\,
e^{\lambda y}\hat\Psi(\lambda).
\ee

\paragraph{Behavior near $z=1$.}

We first set
\be
z=1+\frac{y}{\sqrt{c}}.
\ee
For fixed $\lambda\neq0$, the kernels defined in \eqref{phipsi} behave at large $c$,
up to corrections exponentially small in $\sqrt{c}|\lambda|$, as
\bea
\label{limitphipsia}
\phi_a\left(\sqrt{c}\lambda,1+\frac{y}{\sqrt{c}}\right)
&=&
-i\theta(a\lambda)e^{a\lambda y}
+\frac{\lambda y}{2\pi}
+\frac{ia\,\sgn(\lambda)}{2}
+\frac{iy}{4\sqrt{c}},
\\
\psi_a\left(\sqrt{c}\lambda,1+\frac{y}{\sqrt{c}}\right)
&=&
-\frac{\lambda y\,\sgn(\lambda)}{\pi}
-\frac{iy\,\sgn(\lambda)}{2\sqrt{c}}. \nn
\eea
with $a=\pm 1$.
Here $\theta$ denotes the Heaviside function.

Inserting the asymptotic expressions \eqref{limitphipsia} into the
kernel representation \eqref{eq:BBkernelrep}, and using the rescaled
spectral densities defined in \eqref{defhat}, one obtains
\bea
\label{Bexpandplus}
\BB_a\left(1+\frac{y}{\sqrt{c}}\right)
&=&
-i\int_0^{+\infty}d\lambda\,
e^{\lambda y}\hat\Phi(a\lambda)
+iaA+iCy,
\eea
where we have introduced the two constants
\bea
A
&=&
\frac12\int_{\mathbb R}d\lambda\,
\sgn(\lambda)\hat\Phi(\lambda)-\frac12,
\label{eq:Adef}
\\
C
&=&
\frac14\int_{\mathbb R}d\lambda\,
\left(\frac{2i\lambda}{\pi}-\frac{1}{\sqrt{c}}\right)
\left[
-\hat\Phi(\lambda)
+2\sgn(\lambda)\hat\Psi(\lambda)
\right]
\nn\\
&&
-\frac{i\sqrt{c}}{4\pi}
\left(1-\frac{\epsilon}{\sqrt{c}}\right).
\label{eq:Cdef}
\eea
The last terms in \eqref{eq:Adef}--\eqref{eq:Cdef} originate from the
homogeneous contribution
\be
\BB_{h,a}(z)
=
-\frac{i}{2}a
+\frac{c(1-n)}{4\pi}(z-1).
\ee
We denote the second constant by $C$, rather than $B$, in order not to
confuse it with the function $B(z)$.

\paragraph{Behavior near $z=-1$.}
We now write
\be
z=-1-\frac{y}{\sqrt{c}}.
\ee
For fixed $\lambda\neq0$ the corresponding kernel large $c$ asymptotics obtained from
\eqref{phipsi} are
\bea
\label{limitphipsiminusa}
\phi_a\left(\sqrt{c}\lambda,-1-\frac{y}{\sqrt{c}}\right)
&=&
i\theta(-a\lambda)e^{-a\lambda y}
+\frac{ia\,\sgn(\lambda)}{2}
-\frac{\sqrt{c}\lambda}{\pi}
\nn\\
&&
+y\left(
-\frac{\lambda}{2\pi}
-\frac{i}{4\sqrt{c}}
\right)
-\frac{i}{2},
\\
\psi_a\left(\sqrt{c}\lambda,-1-\frac{y}{\sqrt{c}}\right)
&=&
2ia\theta(-a\lambda)e^{-a\lambda y}
\nn\\
&&
+\sgn(\lambda)
\left[
\frac{2\sqrt{c}\lambda}{\pi}
+y\left(
\frac{\lambda}{\pi}
+\frac{i}{2\sqrt{c}}
\right)
+i
\right].
\eea
Inserting \eqref{limitphipsiminusa} into
\eqref{eq:BBkernelrep} and using \eqref{defhat}, this gives
\bea
\label{Bexpandminus}
\BB_a\left(-1-\frac{y}{\sqrt{c}}\right)
&=&
i\int_0^{+\infty}d\lambda\,e^{\lambda y}
\left[
\hat\Phi(-a\lambda)
+2a\hat\Psi(-a\lambda)
\right]
\nn\\
&&
+iaA-2i\sqrt{c}\,C-iCy.
\eea
In particular, combining \eqref{Bexpandplus} and
\eqref{Bexpandminus} eliminates $\hat\Phi$ and gives
\be
\label{eq:PsiCombination}
\BB_{-a}\left(-1-\frac{y}{\sqrt{c}}\right)
+
\BB_a\left(1+\frac{y}{\sqrt{c}}\right)
=
-2ai
\int_0^\infty d\lambda\,
e^{\lambda y}\hat\Psi(a\lambda)
-2i\sqrt{c}\,C.
\ee
\\

{\bf Remark}.
We emphasize that the kernel limits
\eqref{limitphipsia} and \eqref{limitphipsiminusa} are pointwise
large-$c$ limits for fixed nonzero $\lambda$. They are not uniform in a
neighborhood of $\lambda=0$. This deserves some care, since the spectral
density contains a point mass precisely there: the term
$-\frac{2ic}{\pi}\delta(\lambda)$ in \eqref{firstorderPhi} becomes, after the
rescaling \eqref{defhat} with $n=\epsilon/\sqrt{c}$, a contribution
$-\frac{2i\epsilon\sqrt{c}}{\pi}\delta(\lambda)$ to $\hat\Phi(\lambda)$, of
order $\sqrt{c}$. This is nevertheless harmless at leading order. Indeed, near
$z=1$ the pointwise limits $\lambda\to0^\pm$ of
$\phi_a\left(\sqrt{c}\lambda,1+\frac{y}{\sqrt{c}}\right)$ give
$-\frac{i}{2}+\frac{iy}{4\sqrt{c}}$ for both $a=\pm1$, whose leading part
coincides with the exact value
$\phi_\pm(0,z)=-\frac{i}{4}(1+z)=-\frac{i}{2}-\frac{iy}{4\sqrt{c}}$. The mismatch
at order $1/\sqrt{c}$ produces an $a$-independent correction to $\BB_a$ which,
after summation over $a$, is multiplied by
$\sin\left(\frac{\pi y}{2\sqrt{c}}\right)$ and is therefore subleading in $B$;
the same holds near $z=-1$. When the half-line integrals in
\eqref{Bexpandplus} and the constant $A$ in \eqref{eq:Adef} are written
separately, the point mass at $\lambda=0$ is to be counted with weight
$\frac12$ in each half-line integral, with ${\rm sgn}(0)=0$. 
Consequently, once the spectral
integrals are performed, these formulas are sufficient for the strict leading
algebraic scaling limit derived below, but should not be interpreted as
providing a complete expansion to all orders in $1/\sqrt{c}$. For subleading
algebraic corrections one must return to the exact kernels, or equivalently
keep track separately of the small-$\lambda$ region.

\subsection{Equation around \texorpdfstring{$z=1$}{z=1}}

We now use the nonlinear representation for $\Phi(z)$ given in
\eqref{eq:phifromB3}.
Setting $z=y/\sqrt{c}$, rescaling $v\to v/\sqrt{c}$, and using
$n=\epsilon/\sqrt{c}$ gives
\be
\Phi\left(\frac{y}{\sqrt{c}}\right)
=
2i\epsilon
\sum_{b=\pm1}
\int_0^\infty dv\,
e^{-v^2/2}
B\left(1-\frac{v}{\sqrt{c}}\right)
\BB_b\left(
1+\frac{by-v}{\sqrt{c}}
\right)
e^{bvy/2-i\pi bv/(2\sqrt{c})}.
\label{eq:PhiScaled}
\ee

We now insert the representation \eqref{Bexpandplus}, with
$y$ replaced by $by-v$,
\be
\BB_b\left(1+\frac{by-v}{\sqrt{c}}\right)
=
-i\int_0^{+\infty}d\lambda\,
e^{\lambda(by-v)}
\hat\Phi(b\lambda)
+ibA+iC(by-v).
\ee
Substitution into \eqref{eq:PhiScaled} gives
\bea
\Phi\left(\frac{y}{\sqrt{c}}\right)
&=&
2\epsilon
\sum_{b=\pm1}
\int_0^\infty dv\,
e^{-v^2/2}
B\left(1-\frac{v}{\sqrt{c}}\right)
e^{bvy/2-i\pi bv/(2\sqrt{c})}
\nn\\
&&\times
\left[
\int_0^{+\infty}d\lambda\,
e^{\lambda(by-v)}
\hat\Phi(b\lambda)
-bA-C(by-v)
\right].
\label{eq:PhiScaledInserted}
\eea

Now we recall from \eqref{defhat} that
\be
\Phi\left(\frac{y}{\sqrt{c}}\right)
=
\int_{\mathbb R}d\mu\,e^{\mu y}\hat\Phi(\mu).
\ee
Hence we can identify the bilateral Laplace density in
\eqref{eq:PhiScaledInserted}. Using
\be
e^{\mu_0y}
=
\int_{\mathbb R}d\mu\,
e^{\mu y}\delta(\mu-\mu_0),
\qquad
y e^{\mu_0y}
=
-\int_{\mathbb R}d\mu\,
e^{\mu y}\delta'(\mu-\mu_0),
\label{eq:deltalaplace}
\ee
we obtain
\bea
\label{eq:PhiHatScaled}
\hat\Phi(\mu)
&=&
2\epsilon
\sum_{b=\pm1}
\int_0^\infty dv\,
e^{-v^2/2}
B\left(1-\frac{v}{\sqrt{c}}\right)
e^{-i\pi bv/(2\sqrt{c})}
\nn\\
&&\times
\bigg[
\int_0^{+\infty}d\lambda\,
e^{-\lambda v}
\delta\left(
\mu-b\left(\lambda+\frac{v}{2}\right)
\right)
\hat\Phi(b\lambda)
\nn\\
&&\hspace{13mm}
-(bA-vC)
\delta\left(\mu-\frac{bv}{2}\right)
+bC\,
\delta'\left(\mu-\frac{bv}{2}\right)
\bigg].
\eea

Inserting this expression in \eqref{Bexpandplus}, because of the presenco of the delta functions, only $b=a$ contributes
to the integral over the positive spectral half-axis. Performing the
$\mu$ integration then gives
\bea
\BB_a\left(1+\frac{y}{\sqrt{c}}\right)
&=&
iaA+iCy
\nn\\
&&
-2i\epsilon
\int_0^\infty dv\,
e^{-v^2/2}
B\left(1-\frac{v}{\sqrt{c}}\right)
e^{-i\pi av/(2\sqrt{c})}
\nn\\
&&\times
\bigg[
\int_0^\infty d\lambda\,
e^{-\lambda v}
e^{(\lambda+v/2)y}\hat\Phi(a\lambda)
\nn\\
&&\hspace{14mm}
-(aA-vC)e^{vy/2}
-Cy\,e^{vy/2}
\bigg].
\eea
The expression in square brackets is
\be
i\,e^{vy/2}
\BB_a\left(
1+\frac{y-v}{\sqrt{c}}
\right),
\ee
as follows immediately from \eqref{Bexpandplus}. Hence
\bea
\BB_a\left(1+\frac{y}{\sqrt{c}}\right)
&=&
iaA+iCy
\nn\\
&&
+2\epsilon
\int_0^\infty dv\,
e^{-v^2/2}
B\left(1-\frac{v}{\sqrt{c}}\right)
e^{-i\pi av/(2\sqrt{c})}
e^{vy/2}
\nn\\
&&\hspace{25mm}\times
\BB_a\left(
1+\frac{y-v}{\sqrt{c}}
\right).
\label{eq:BBplusRecursive}
\eea

Multiplying \eqref{eq:BBplusRecursive} by
$e^{i\pi a(1+y/\sqrt{c})/2}$ and summing over $a=\pm1$, we obtain
\bea
\label{Bsummaryplus}
B\left(1+\frac{y}{\sqrt{c}}\right)
&=&
2\epsilon
\int_0^\infty dv\,
e^{-v^2/2+vy/2}
B\left(1-\frac{v}{\sqrt{c}}\right)
B\left(1+\frac{y-v}{\sqrt{c}}\right)
\nn\\
&&
-2A\cos\left(\frac{\pi y}{2\sqrt{c}}\right)
-2iCy\sin\left(\frac{\pi y}{2\sqrt{c}}\right).
\eea

\subsection{Equation around \texorpdfstring{$z=-1$}{z=-1}}

We now use the non linear representation \eqref{eq:psifromB3} for $\Psi(z)$ which we recall here
\bea
\Psi(z)
=
-\frac{i}{2}nc
\sum_{a,b=\pm1}a
\int_0^\infty dv\,
e^{-cv^2/2}
B(b(v-1))
\BB_a(b(1-v)+az)
e^{abv(cz-i\pi)/2}.
\eea
Setting again $z=y/\sqrt{c}$, rescaling $v\to v/\sqrt{c}$, and using
$n=\epsilon/\sqrt{c}$, one obtains
\bea
\Psi\left(\frac{y}{\sqrt{c}}\right)
&=&
-\frac{i}{2}\epsilon
\sum_{a,b=\pm1}a
\int_0^\infty dv\,
e^{-v^2/2}
B\left[b\left(\frac{v}{\sqrt{c}}-1\right)\right]
\nn\\
&&\times
\BB_a\left[
b\left(1-\frac{v}{\sqrt{c}}\right)
+\frac{ay}{\sqrt{c}}
\right]
e^{abvy/2-i\pi abv/(2\sqrt{c})}.
\label{eq:PsiScaled}
\eea

It is useful to treat separately the two values $b=\pm1$ in
\eqref{eq:PsiScaled}.

For $b=+1$, one has
\bea
\Psi^{(b=+1)}\left(\frac{y}{\sqrt{c}}\right)
&=&
-\frac{i}{2}\epsilon
\sum_{a=\pm1}a
\int_0^\infty dv\,
e^{-v^2/2}
B\left(-1+\frac{v}{\sqrt{c}}\right)
\nn\\
&&\times
\BB_a\left(
1+\frac{ay-v}{\sqrt{c}}
\right)
e^{avy/2-i\pi av/(2\sqrt{c})}.
\eea
Inserting \eqref{Bexpandplus}, with $y$ replaced by $ay-v$, gives
\bea
\Psi^{(b=+1)}\left(\frac{y}{\sqrt{c}}\right)
&=&
\frac{\epsilon}{2}
\sum_{a=\pm1}a
\int_0^\infty dv\,
e^{-v^2/2}
B\left(-1+\frac{v}{\sqrt{c}}\right)
e^{avy/2-i\pi av/(2\sqrt{c})}
\nn\\
&&\times
\left[
-\int_0^\infty d\lambda\,
e^{\lambda(ay-v)}
\hat\Phi(a\lambda)
+aA+C(ay-v)
\right].
\label{eq:PsiPlusPart}
\eea
Using the bilateral Laplace representation
\be
\Psi^{(b=+1)}\left(\frac{y}{\sqrt{c}}\right)
=
\int_{\mathbb R}d\mu\,
e^{\mu y}\hat\Psi^{(b=1)}(\mu),
\ee
together with \eqref{eq:deltalaplace}, we obtain
\bea
\hat\Psi^{(b=+1)}(\mu)
&=&
\frac{\epsilon}{2}
\sum_{a=\pm1}a
\int_0^\infty dv\,
e^{-v^2/2}
B\left(-1+\frac{v}{\sqrt{c}}\right)
e^{-i\pi av/(2\sqrt{c})}
\nn\\
&&\times
\bigg[
-\int_0^\infty d\lambda\,
e^{-\lambda v}
\hat\Phi(a\lambda)
\delta\left(
\mu-a\left(\lambda+\frac{v}{2}\right)
\right)
\nn\\
&&\hspace{12mm}
+(aA-Cv)
\delta\left(\mu-\frac{av}{2}\right)
-aC\,
\delta'\left(\mu-\frac{av}{2}\right)
\bigg].
\label{eq:PsiHatPlus}
\eea

For $b=-1$, \eqref{eq:PsiScaled} gives instead
\bea
\Psi^{(b=-1)}\left(\frac{y}{\sqrt{c}}\right)
&=&
-\frac{i}{2}\epsilon
\sum_{a=\pm1}a
\int_0^\infty dv\,
e^{-v^2/2}
B\left(1-\frac{v}{\sqrt{c}}\right)
\nn\\
&&\times
\BB_a\left(
-1+\frac{v+ay}{\sqrt{c}}
\right)
e^{-avy/2+i\pi av/(2\sqrt{c})}.
\eea
We now use \eqref{Bexpandminus} with
$y$ replaced by $-(v+ay)$,
\bea
\BB_a\left(
-1+\frac{v+ay}{\sqrt{c}}
\right)
&=&
i\int_0^\infty d\lambda\,
e^{-\lambda(v+ay)}
\left[
\hat\Phi(-a\lambda)
+2a\hat\Psi(-a\lambda)
\right]
\nn\\
&&
+iaA-2i\sqrt{c}\,C+iC(v+ay).
\eea
It follows that
\bea
\Psi^{(b=-1)}\left(\frac{y}{\sqrt{c}}\right)
&=&
\frac{\epsilon}{2}
\sum_{a=\pm1}a
\int_0^\infty dv\,
e^{-v^2/2}
B\left(1-\frac{v}{\sqrt{c}}\right)
e^{-avy/2+i\pi av/(2\sqrt{c})}
\nn\\
&&\times
\bigg[
\int_0^\infty d\lambda\,
e^{-\lambda(v+ay)}
\left(
\hat\Phi(-a\lambda)
+2a\hat\Psi(-a\lambda)
\right)
\nn\\
&&\hspace{12mm}
+aA-2\sqrt{c}\,C+C(v+ay)
\bigg].
\label{eq:PsiMinusPart}
\eea
Hence its bilateral Laplace density is
\bea
\hat\Psi^{(b=-1)}(\mu)
&=&
\frac{\epsilon}{2}
\sum_{a=\pm1}a
\int_0^\infty dv\,
e^{-v^2/2}
B\left(1-\frac{v}{\sqrt{c}}\right)
e^{i\pi av/(2\sqrt{c})}
\nn\\
&&\times
\bigg[
\int_0^\infty d\lambda\,
e^{-\lambda v}
\left(
\hat\Phi(-a\lambda)
+2a\hat\Psi(-a\lambda)
\right)
\delta\left(
\mu+a\left(\lambda+\frac{v}{2}\right)
\right)
\nn\\
&&\hspace{9mm}
+
\left(
aA-2\sqrt{c}\,C+Cv
\right)
\delta\left(\mu+\frac{av}{2}\right)
-aC\,
\delta'\left(\mu+\frac{av}{2}\right)
\bigg].
\label{eq:PsiHatMinus}
\eea

We now fix a sign $\sigma=\pm1$ and consider the spectral component
of $\Psi$ on the half-axis of sign $\sigma$, namely
\be
\int_0^\infty d\mu\,
e^{\mu y}\hat\Psi(\sigma\mu),
\qquad
\hat\Psi(\mu)
=
\hat\Psi^{(b=1)}(\mu)
+
\hat\Psi^{(b=-1)}(\mu).
\label{eq:PsiHalfAxis}
\ee
Since $\mu,\lambda,v\geq0$, the support of the delta functions in
\eqref{eq:PsiHatPlus} implies
\be
\sigma\mu
=
a\left(\lambda+\frac{v}{2}\right),
\ee
and therefore selects
\be
a=\sigma.
\ee
Similarly, in \eqref{eq:PsiHatMinus} the support condition is
\be
\sigma\mu
=
-a\left(\lambda+\frac{v}{2}\right),
\ee
which selects
\be
a=-\sigma.
\ee
The same choices of $a$ apply to the $\delta$ and $\delta'$ terms,
whose support is obtained by setting $\lambda=0$ in the corresponding
continuous contribution.

Using these two support conditions, we find
\bea
&&
-2\sigma i
\int_0^\infty d\mu\,
e^{\mu y}\hat\Psi(\sigma\mu)
\nn\\
&&=
-i\epsilon
\int_0^\infty d\mu\,e^{\mu y}
\int_0^\infty dv\,
e^{-v^2/2}
B\left(-1+\frac{v}{\sqrt{c}}\right)
e^{-i\pi\sigma v/(2\sqrt{c})}
\nn\\
&&\quad\times
\bigg[
-\int_0^\infty d\lambda\,
e^{-\lambda v}\hat\Phi(\sigma\lambda)
\delta\left(
\mu-\lambda-\frac{v}{2}
\right)
+
(\sigma A-Cv)
\delta\left(\mu-\frac{v}{2}\right)
-C\,
\delta'\left(\mu-\frac{v}{2}\right)
\bigg]
\nn\\
&&\quad
+i\epsilon
\int_0^\infty d\mu\,e^{\mu y}
\int_0^\infty dv\,
e^{-v^2/2}
B\left(1-\frac{v}{\sqrt{c}}\right)
e^{-i\pi\sigma v/(2\sqrt{c})}
\nn\\
&&\quad\times
\bigg[
\int_0^\infty d\lambda\,
e^{-\lambda v}
\left(
\hat\Phi(\sigma\lambda)
-2\sigma\hat\Psi(\sigma\lambda)
\right)
\delta\left(
\mu-\lambda-\frac{v}{2}
\right)
\nn\\
&&\hspace{17mm}
+
\left(
-\sigma A-2\sqrt{c}\,C+Cv
\right)
\delta\left(\mu-\frac{v}{2}\right)
+C\,
\delta'\left(\mu-\frac{v}{2}\right)
\bigg].
\label{eq:PsiProjectedDelta}
\eea

We can now perform the $\mu$ integral explicitly. Using
\be
\int_0^\infty d\mu\,
e^{\mu y}
\delta\left(\mu-\mu_0\right)
=
e^{\mu_0y},
\qquad
\int_0^\infty d\mu\,
e^{\mu y}
\delta'\left(\mu-\mu_0\right)
=
-y e^{\mu_0y},
\ee

one obtains
\bea
&&
-2\sigma i
\int_0^\infty d\mu\,
e^{\mu y}\hat\Psi(\sigma\mu)
\nn\\
&&=
-i\epsilon
\int_0^\infty dv\,
e^{-v^2/2+vy/2}
e^{-i\pi\sigma v/(2\sqrt{c})}
\bigg\{
B\left(-1+\frac{v}{\sqrt{c}}\right)
\nn\\
&&\qquad\times
\left[
-\int_0^\infty d\lambda\,
e^{\lambda(y-v)}
\hat\Phi(\sigma\lambda)
+\sigma A-Cv+Cy
\right]
\nn\\
&&\qquad
-
B\left(1-\frac{v}{\sqrt{c}}\right)
\Big[
\int_0^\infty d\lambda\,
e^{\lambda(y-v)}
\left(
\hat\Phi(\sigma\lambda)
-2\sigma\hat\Psi(\sigma\lambda)
\right)
\nn\\
&&\hspace{48mm}
-\sigma A-2\sqrt{c}\,C+Cv-Cy
\Big]
\bigg\}.
\label{eq:PsiProjectedBeforeRecognition}
\eea

The two expressions in square brackets can now be recognized directly
from \eqref{Bexpandplus} and \eqref{Bexpandminus}. Namely,
\bea
&&
-\int_0^\infty d\lambda\,
e^{\lambda(y-v)}
\hat\Phi(\sigma\lambda)
+\sigma A-Cv+Cy
\nn\\
&&\hspace{30mm}
=
-i\BB_\sigma\left(
1+\frac{y-v}{\sqrt{c}}
\right),
\eea
whereas
\bea
&&
\int_0^\infty d\lambda\,
e^{\lambda(y-v)}
\left(
\hat\Phi(\sigma\lambda)
-2\sigma\hat\Psi(\sigma\lambda)
\right)
-\sigma A-2\sqrt{c}\,C+Cv-Cy
\nn\\
&&\hspace{30mm}
=
-i\BB_{-\sigma}\left(
-1-\frac{y-v}{\sqrt{c}}
\right).
\eea
Therefore
\bea
-2\sigma i
\int_0^\infty d\mu\,
e^{\mu y}\hat\Psi(\sigma\mu)
&=&
-\epsilon
\int_0^\infty dv\,
e^{-v^2/2+vy/2}
e^{-i\pi\sigma v/(2\sqrt{c})}
\nn\\
&&\times
\bigg[
B\left(-1+\frac{v}{\sqrt{c}}\right)
\BB_\sigma\left(
1+\frac{y-v}{\sqrt{c}}
\right)
\nn\\
&&\hspace{15mm}
-
B\left(1-\frac{v}{\sqrt{c}}\right)
\BB_{-\sigma}\left(
-1-\frac{y-v}{\sqrt{c}}
\right)
\bigg].
\label{eq:PsiProjected}
\eea

Using \eqref{eq:PsiCombination}, the left hand side is
\bea
&&
\BB_{-\sigma}\left(-1-\frac{y}{\sqrt{c}}\right)
+
\BB_\sigma\left(1+\frac{y}{\sqrt{c}}\right)
+2i\sqrt{c}\,C.
\eea
Multiplying by
$e^{i\pi\sigma(1+y/\sqrt{c})/2}$ and summing over
$\sigma=\pm1$ finally gives
\bea
\label{Bsummaryminus}
&&
B\left(-1-\frac{y}{\sqrt{c}}\right)
+
B\left(1+\frac{y}{\sqrt{c}}\right)
-4i\sqrt{c}\,C
\sin\left(\frac{\pi y}{2\sqrt{c}}\right)
\nn\\
&&\qquad =
-\epsilon
\int_0^\infty dv\,
e^{-v^2/2+vy/2}
\bigg[
B\left(-1+\frac{v}{\sqrt{c}}\right)
B\left(1+\frac{y-v}{\sqrt{c}}\right)
\nn\\
&&\hspace{48mm}
-
B\left(1-\frac{v}{\sqrt{c}}\right)
B\left(-1-\frac{y-v}{\sqrt{c}}\right)
\bigg].
\eea

The constants $A$ and $C$ depend on $\epsilon$ and $c$ through the
spectral densities, and are therefore not independent parameters. They
have to be determined self-consistently together with $B(z)$. In the
strict scaling limit below, their leading contributions will be
parametrized by two constants $a$ and $b$, which will be fixed by the
boundary conditions.

Equations \eqref{Bsummaryplus} and \eqref{Bsummaryminus} are the
spectral large-$c$ equations from which we shall extract the strict
scaling limit. The explicit trigonometric factors have been retained
because they make the origin of the polynomial terms transparent.
However, the fixed-$\lambda$ kernel limits used in their derivation are
not uniform in a neighborhood of $\lambda=0$. Consequently,
\eqref{Bsummaryplus} and \eqref{Bsummaryminus} are sufficient for the
strict leading algebraic scaling limit, but should not be interpreted
as providing a complete expansion to all orders in $1/\sqrt{c}$.
Subleading algebraic corrections require a separate treatment of the
small-$\lambda$ region, or equivalently a return to the exact kernels.
The exact rescaled identities
\eqref{eq:exactscaledplus} and \eqref{eq:exactscaledminus} remain valid
at finite $c$ and provide useful consistency conditions for such an
analysis.

\paragraph{Check at $\epsilon=0$.}
For $\epsilon=0$, the spectral densities vanish and the preceding
equations reduce to the homogeneous solution,
\bea
B_0\left(1+\frac{y}{\sqrt{c}}\right)
&=&
-2A_0\cos\left(\frac{\pi y}{2\sqrt{c}}\right)
-2iC_0y\sin\left(\frac{\pi y}{2\sqrt{c}}\right),
\\
B_0\left(-1-\frac{y}{\sqrt{c}}\right)
&=&
2A_0\cos\left(\frac{\pi y}{2\sqrt{c}}\right)
\nn\\
&&
+2i(y+2\sqrt{c})C_0
\sin\left(\frac{\pi y}{2\sqrt{c}}\right).
\eea
They coincide with the exact zeroth-order result
\bea
\label{Bn0}
B_0(z)
&=&
\sin\left(\frac{\pi z}{2}\right)
+
\frac{c(z-1)}{2\pi}
\cos\left(\frac{\pi z}{2}\right),
\\
B_0\left(1+\frac{y}{\sqrt{c}}\right)
&=&
\cos\left(\frac{\pi y}{2\sqrt{c}}\right)
-\frac{\sqrt{c}\,y}{2\pi}
\sin\left(\frac{\pi y}{2\sqrt{c}}\right),
\\
B_0\left(-1-\frac{y}{\sqrt{c}}\right)
&=&
\frac{\sqrt{c}\,y+2c}{2\pi}
\sin\left(\frac{\pi y}{2\sqrt{c}}\right)
-\cos\left(\frac{\pi y}{2\sqrt{c}}\right),
\eea
with
\be
2A_0=-1,
\qquad
2iC_0=\frac{\sqrt{c}}{2\pi}.
\ee

\subsection{Leading algebraic scaling equations}

We now take the strict limit $c\to+\infty$ at fixed $\epsilon$. Motivated
also by the exact zeroth-order solution \eqref{Bn0}, we write
\begin{subequations}
\label{eq:scalingexpansions}
\bea
B\left(1+\frac{y}{\sqrt{c}}\right)
&=&
\beta(y)
+\frac{\alpha(y)}{\sqrt{c}}
+O(c^{-1}),
\\
B\left(-1-\frac{y}{\sqrt{c}}\right)
&=&
\sqrt{c}\,\tilde\beta(y)
+\tilde\alpha(y)
+O(c^{-1/2}),
\\
-2A
&=&
a+\frac{a'}{\sqrt{c}}+O(c^{-1}),
\\
-i\pi C
&=&
b\sqrt{c}+b'+O(c^{-1/2}) \;,
\eea
\end{subequations}
and $\beta,\tilde\beta,a,b$ depend on $\epsilon$ but not on $c$.

The exact boundary conditions
\be
B(1)=1,
\qquad
B'(1)=nc=\epsilon\sqrt{c},
\qquad
B(-1)=-1,
\qquad
B'(-1)=\frac{c}{2}
\left(
\frac{\epsilon}{\sqrt{c}}-1
\right)
\ee
give
\bea
&&
\beta(0)=1,
\qquad
\beta'(0)=\epsilon,
\qquad
\tilde\beta(0)=0,
\qquad
\tilde\beta'(0)=\frac12,
\label{eq:betabcs}
\\
&&
\alpha(0)=0,
\qquad
\alpha'(0)=0,
\qquad
\tilde\alpha(0)=-1,
\qquad
\tilde\alpha'(0)=-\frac{\epsilon}{2}.
\eea

Taking the leading order of
\eqref{Bsummaryplus}--\eqref{Bsummaryminus} gives the closed system
\bea
\label{eq:compactbetas}
\beta(y)
&=&
2\epsilon
\int_0^\infty dv\,
e^{-v^2/2+vy/2}
\beta(-v)\beta(y-v)
+a+by^2,
\\
\tilde\beta(y)
&=&
\epsilon
\int_0^\infty dv\,
e^{-v^2/2+vy/2}
\left[
\beta(-v)\tilde\beta(y-v)
-\tilde\beta(-v)\beta(y-v)
\right]
-2by. \nn
\eea
These are the two equations which characterize the leading algebraic
large-$c$ scaling problem.

They are consistent with the exact finite-interval equations derived
above. For instance, subtracting the first equation in
\eqref{eq:compactbetas} evaluated at $y$ and at $-y$, and shifting the
integration variable in the second integral, gives
\be
\beta(y)-\beta(-y)
=
2\epsilon
\int_0^y dv\,
e^{(yv-v^2)/2}
\beta(-v)\beta(y-v),
\ee
which is precisely the $c\to+\infty$ limit of
\eqref{eq:exactscaledplus}.

Two of the four leading boundary conditions in \eqref{eq:betabcs} are
automatic. First, the second equation of \eqref{eq:compactbetas}
immediately gives $\tilde\beta(0)=0$. Second, differentiating the first equation of \eqref{eq:compactbetas}
\bea
&&
\beta'(0) = 2\epsilon
\left.
\partial_y
\int_0^\infty dv\,
e^{-v^2/2+vy/2}
\beta(-v)\beta(y-v)
\right|_{y=0}
\nn\\
&&\qquad =
-\epsilon
\left[
e^{-v^2/2}\beta(-v)^2
\right]_{v=0}^{v=+\infty}
=
\epsilon\beta(0)^2.
\label{totalder}
\eea
Hence, since $\beta(0)=1$, one obtains automatically
$\beta'(0)=\epsilon$.

The two remaining boundary conditions determine $a$ and $b$. From
$\beta(0)=1$ one obtains
\bea
\label{aexact}
a
=
1
-
2\epsilon
\int_0^\infty dv\,
e^{-v^2/2}\beta(-v)^2,
\eea
while $\tilde\beta'(0)=1/2$ gives
\bea
\label{bexact}
b
=
-\frac14
-\frac{\epsilon}{2}
\int_0^\infty dv\,
e^{-v^2/2}
\left[
\tilde\beta(-v)\beta'(-v)
-\beta(-v)\tilde\beta'(-v)
\right].
\eea

\subsection{Energy in the scaling regime}

We recall that
\bea
E(n,L)
&=&
-\frac{4n}{L^2}
\left[
B''(1)
-\frac{c}{2}
-\frac{c^2}{12}(1+8n^2)
\right],
\\
B''(1)
&=&
c\,\beta''(0)
+\sqrt{c}\,\alpha''(0)
+\dots.
\eea
with $n=\epsilon/\sqrt{c}$.
It is therefore natural to define the leading nontrivial scaling part of
the energy by
\be
{\cal E}(\epsilon)
:=
\lim_{c\to+\infty}
\frac{
L^2E(\epsilon/\sqrt{c},L)
-\frac{\epsilon}{3}c^{3/2}
}{\sqrt{c}}.
\ee
We shall refer to ${\cal E}(\epsilon)$ as the large-$L$ scaling
function of the SCGF.
One then finds
\be
\label{eq:scalingenergybeta}
{\cal E}(\epsilon)
=
-4\epsilon
\left[
\beta''(0)
-\frac12
-\frac{2}{3}\epsilon^2
\right].
\ee
Equivalently this reads
\be
L^2 E\left(\frac{\epsilon}{\sqrt{c}},L\right)
=
\frac{\epsilon}{3}c^{3/2}
-4\epsilon\sqrt{c}
\left[
\beta''(0)
-\frac12
-\frac{2}{3}\epsilon^2
\right]
+o(\sqrt{c}).
\label{eq:scalingenergyasympt}
\ee

We recall that the cumulants of the free energy are obtained from
$
c_k(L)
=
-\left.\partial_n^k E(n,L)\right|_{n=0}$.

To extract them from \eqref{eq:scalingenergyasympt}, we expand
\be
\beta(y)=\sum_{j\geq0}\epsilon^j\beta_j(y),
\qquad
\tilde\beta(y)=\sum_{j\geq0}\epsilon^j\tilde\beta_j(y),
\qquad
a=\sum_{j\geq0}\epsilon^j a_j,
\qquad
b=\sum_{j\geq0}\epsilon^j b_j.
\label{eq:betaepsexpansion}
\ee
In particular,
\be
\beta''(0)
=
\sum_{j\geq0}\epsilon^j\beta_j''(0),
\qquad
\bar\beta_j''
=
\beta_j''(0)-\frac23\delta_{j,2},
\qquad
\epsilon=n\sqrt{c}.
\label{eq:betaepssecond}
\ee

Using $\beta_0''(0)=-1/2$, which follows directly from
\eqref{Bn0} and will also be recovered below in
\eqref{eq:beta0scaling}, one obtains
\be
L^2 c_1(L) =
-\frac{c^2}{3}-4c+O(1),
\label{eq:c1largescale}
\ee
and for $k \geq 2$
\be
L^2c_k(L)
=
4k!\,\bar\beta_{k-1}''
\,c^{(k+1)/2}
+o\left(c^{(k+1)/2}\right).
\label{eq:cklargescale}
\ee

For later use it is useful to express $\beta''(0)$ directly in terms of
the scaling functions. Differentiating the first equation in
\eqref{eq:compactbetas} twice at $y=0$ and using \eqref{bexact} gives
\bea
\label{betasecond02}
\beta''(0)
&=&
-\frac12
+
\epsilon
\int_0^{+\infty}dv\,
e^{-v^2/2}
\bigg\{
\frac12\beta(-v)
\left[
v^2\beta(-v)
+4\left(
v\beta'(-v)+\beta''(-v)
\right)
\right]
\nn\\
&&\hspace{34mm}
-
\left[
\tilde\beta(-v)\beta'(-v)
-\beta(-v)\tilde\beta'(-v)
\right]
\bigg\}.
\eea

\subsection{Checks from the expansion in \texorpdfstring{$\epsilon$}{epsilon}}

We now check the reduced equations \eqref{eq:compactbetas}
perturbatively in $\epsilon$.

For $\epsilon=0$, expansion of the exact result \eqref{Bn0} gives
\bea
B_0\left(1+\frac{y}{\sqrt{c}}\right)
&=&
1-\frac{y^2}{4}
+
\frac{\pi^2y^2(y^2-12)}{96c}
+O(c^{-2}),
\\
B_0\left(-1-\frac{y}{\sqrt{c}}\right)
&=&
\frac{\sqrt{c}\,y}{2}
+\frac{y^2-4}{4}
-\frac{\pi^2y^3}{48\sqrt{c}}
+O(c^{-1}).
\eea
Together with \eqref{aexact} and \eqref{bexact} evaluated at
$\epsilon=0$, this gives
\be
\beta_0(y)=1-\frac{y^2}{4},
\qquad
\tilde\beta_0(y)=\frac{y}{2},
\qquad
a_0=1,
\qquad
b_0=-\frac14.
\label{eq:beta0scaling}
\ee

At first order in $\epsilon$, inserting $\beta_0$ and
$\tilde \beta_0$ in \eqref{aexact}--\eqref{bexact} and performing the integrals, one obtains
\be \label{eq:a1b1direct}
a_1
=
-\frac{11}{8}\sqrt{\frac{\pi}{2}},
\qquad
b_1
=
\frac{5}{16}\sqrt{\frac{\pi}{2}}.
\ee

Inserting $\beta_0$ in the first equation of
\eqref{eq:compactbetas}, together with the above values of $a_1$ and
$b_1$, and performing the integral gives
\bea
\beta_1(y)
=
\frac{1}{256}
\bigg[
&\sqrt{2\pi}\,
e^{y^2/8}
\left(y^4-40y^2+176\right)
\left(
\erf\left(\frac{y}{2\sqrt{2}}\right)+1
\right)
\nn\\
&+40\sqrt{2\pi}\,y^2
+4(y^2+20)y
-176\sqrt{2\pi}
\bigg].
\eea
In particular,
\be   \label{eq:beta1secondDirect}
\beta_1''(0)
=
\frac{11}{32}\sqrt{\frac{\pi}{2}}.
\ee
This gives the leading large-$c$ behavior of the second cumulant,
from \eqref{eq:cklargescale},
\be
L^2c_2(L)
\simeq
8\beta_1''(0)c^{3/2} =  \frac{11}{4}\sqrt{\frac{\pi}{2}}\,c^{3/2},
\label{kappa2largec}
\ee
in agreement with the direct large-$c$ expansion of the exact formula
for $c_2(L)$, see \eqref{c2largec}.

At the same order, expanding the second equation of
\eqref{eq:compactbetas} to first order in $\epsilon$ gives
\bea
\tilde\beta_1(y)
&=&
\int_0^\infty dv\,
e^{-v^2/2+vy/2}
\left[
\beta_0(-v)\tilde\beta_0(y-v)
-\tilde\beta_0(-v)\beta_0(y-v)
\right]
-2b_1y.
\label{eq:tildebeta1integral}
\eea
Using \eqref{eq:beta0scaling} and the value of $b_1$ obtained above,
the integral can be performed explicitly and yields
\be
\tilde\beta_1(y)
=
\frac{y}{64}
\bigg[
-\sqrt{2\pi}\,
e^{y^2/8}
(y^2-20)
\left(
\erf\left(\frac{y}{2\sqrt{2}}\right)+1
\right)
-4\left(y+5\sqrt{2\pi}\right)
\bigg].
\ee

At second order in $\epsilon$, one first expands the self-consistency
relations \eqref{aexact} and \eqref{bexact} using the previously
determined functions $\beta_0,\tilde\beta_0,\beta_1,\tilde\beta_1$,
which gives $a_2$ and $b_2$ in terms of Gaussian integrals involving
polynomials and error functions. Independently, $\beta_2''(0)$ is
obtained most directly by expanding \eqref{betasecond02} to first order
in $\epsilon$, which again leads to integrals of the same type. One finds
\bea \label{eq:a2b2direct}
a_2
&=&
\left(
\frac{61}{32}
-\frac{64}{27\sqrt{3}}
\right)\pi,
\\
b_2
&=&
\frac{1}{24}
+
\left(
\frac{14}{27\sqrt{3}}
-\frac{13}{32}
\right)\pi,
\eea
and
\be   \label{eq:beta2secondDirect}
\beta_2''(0)
=
\frac34
+
\left(
\frac{64}{81\sqrt{3}}
-\frac{61}{128}
\right)\pi.
\ee

Consequently, using \eqref{eq:cklargescale}
\be
L^2c_3(L)
\simeq
24
\left(
\beta_2''(0)-\frac23
\right)c^2 = \left[
2+
\left(
\frac{512}{27\sqrt{3}}
-\frac{183}{16}
\right)\pi
\right]c^2   \label{eq:c3largec}
\ee

The perturbative calculation above provides a direct check of the
coupled scaling equations derived in this section. We have thus reduced
the original replica Bethe-ansatz problem, in the limit
$c\to\infty$ with $\epsilon=n\sqrt{c}$ fixed, to the closed system
\eqref{eq:compactbetas}, together with the boundary conditions
\eqref{eq:betabcs} and the expression
\eqref{eq:scalingenergybeta} for the large-$L$ scaling function.

We now turn to the solution of this system. As in the one-polymer
analysis of Brunet--Derrida, the appropriate tool is a Wiener--Hopf
factorization. In the present two-polymer problem a second coupled
equation is present, but remarkably it can also be reduced to a scalar
Wiener--Hopf decomposition.

\section{Wiener--Hopf solution and large-\texorpdfstring{$L$}{L} scaling function}
\label{sec:WHscaling}

The coupled equations obtained in the previous section still involve
the two real-space functions $\beta$ and $\tilde\beta$ introduced in Eqs.~\eqref{eq:scalingexpansions}. We now solve
this system by adapting the Wiener--Hopf construction of
Brunet--Derrida. This will reduce the problem to two scalar
self-consistency conditions and will lead to a convenient parametric
representation of the large-$L$ scaling function ${\cal E}(\epsilon)$.

\subsection{Reduction to Wiener--Hopf form}

Following the same idea as in Brunet--Derrida, we absorb the Gaussian
factors by defining
\be
\label{eq:defomega}
\beta(y)=e^{y^2/4}\omega(y),
\qquad
\tilde\beta(y)=-e^{y^2/4}\tilde\omega(y).
\ee
Substitution in \eqref{eq:compactbetas} gives
\begin{subequations}
\label{eq:coupledeq}
\bea
\label{eq:omega}
\omega(y)
&=&
2\epsilon
\int_0^\infty dv\,
\omega(-v)\omega(y-v)
+
(a+by^2)e^{-y^2/4},
\\
\label{eq:omegatilde}
\tilde\omega(y)
&=&
\epsilon
\int_0^\infty dv\,
\left[
\omega(-v)\tilde\omega(y-v)
-\tilde\omega(-v)\omega(y-v)
\right]
+
2by\,e^{-y^2/4}.
\eea
\end{subequations}
The two independent boundary conditions become simply
\be
\omega(0)=1,
\qquad
\tilde\omega'(0)=-\frac12.
\ee
Equations \eqref{eq:omega}--\eqref{eq:omegatilde}, together with the
conditions determining $a$ and $b$, will be the starting point for the
Wiener--Hopf analysis.

\subsection{Wiener--Hopf solution}

We now solve the coupled equations \eqref{eq:omega}--\eqref{eq:omegatilde}
by adapting the Wiener--Hopf construction of Brunet--Derrida
\cite{brunet2000probability}. Their scalar equation, Eq.~(66) of Sec.~IV,
and its analysis in Appendix~B will apply directly to the equation for
$\omega$, while the equation for $\tilde\omega$ requires one additional
Wiener--Hopf decomposition.

We introduce
\be
H(y)=(a+by^2)e^{-y^2/4},
\qquad
J(y)=2by\,e^{-y^2/4},
\label{eq:HJdef}
\ee
whose Fourier transforms, with the convention
$\hat f(q)=\int_{\mathbb R}dy\,e^{iqy}f(y)$, are
\bea
\hat H(q)
&=&
2\sqrt{\pi}\,
\left(a+2b-4bq^2\right)e^{-q^2},
\\
\hat J(q)
&=&
8ib\sqrt{\pi}\,q\,e^{-q^2}.
\label{eq:HJFourier}
\eea
We also define the half-Fourier transforms
\be
\hat f_\pm(q)
=
\int_{-\infty}^{+\infty}dy\,
e^{iqy}\theta(\pm y)f(y),
\qquad
\hat f=\hat f_++\hat f_-,
\label{eq:halfFourierWH}
\ee
which are analytic in the upper and lower half-planes, respectively.

{\it Wiener--Hopf projections}. The decomposition
$\hat f=\hat f_++\hat f_-$ can be written directly in Fourier space.
For a sufficiently regular $\hat f(q)$ with fast enough decay at
infinity, the decomposition into parts analytic in the upper and lower
half-planes is unique (or unique up to constants or polynomials,
depending on the decay conditions), and it is given by
\be
\hat f_\pm(q)
=
\pm\int_{\mathbb{R}}\frac{dk}{2i\pi}\,
\frac{\hat f(k)}{k-q\mp i0^+}
=
\frac12 \hat f(q)
\pm\frac{i}{2\pi}
\dashint_{\mathbb{R}}\frac{\hat f(k)\,dk}{q-k}
=
\frac12\left(\hat f\pm i{\cal H}\hat f\right)(q),
\label{eq:WHprojections}
\ee
whose analytic continuations to the upper and lower half-planes are
obtained by replacing $q\pm i0^+$ with the complex variable $z$,
$\Im z\gtrless0$, in the first expression. The second equality in \eqref{eq:WHprojections} follows from the
Sokhotski--Plemelj formula
\be
\frac{1}{x\mp i0^+}
=
{\rm PV}\frac{1}{x}
\pm i\pi\delta(x),
\ee
and ${\cal H}$ denotes the Hilbert transform,
\be
({\cal H}\hat f)(q)
=
\frac{1}{\pi}
\dashint_{\mathbb{R}}
\frac{\hat f(k)\,dk}{q-k}.
\label{eq:defHilbert}
\ee
By uniqueness, \eqref{eq:WHprojections} reproduces the half-Fourier
transforms \eqref{eq:halfFourierWH}, which justifies the common
notation. The two representations are thus equivalent: the
half-Fourier transforms are natural in real space, while
\eqref{eq:WHprojections} is convenient when a function is known only
in Fourier space.
We write $\hat f_\pm$ when $\hat f$ is the Fourier transform of a
given $f(y)$, and $[\,\cdot\,]_\pm$ for functions known only in
Fourier space, e.g.\ $[\log{\cal K}]_\pm$ and $[{\cal R}]_\pm$ below.

\paragraph{Solution of the equation for $\omega$.}

The first equation in \eqref{eq:coupledeq} reads
\be
\omega(y)
=
H(y)
+
2\epsilon\int_0^\infty dv\,
\omega(-v)\omega(y-v).
\label{eq:omegaNonlinearWH}
\ee
This is exactly of the form considered by Brunet--Derrida in their
Eq.~(B.1), with their parameter $\epsilon$ replaced here by
$2\epsilon$. As shown in Appendix~B.1 of
Ref.~\cite{brunet2000probability}, Eqs.~(B.1)--(B.8), the nonlinear equation
is perturbatively equivalent to the linear half-line equation
\be
\omega(y)
=
H(y)
+
2\epsilon\int_0^\infty dv\,
H(y-v)\omega(v).
\label{eq:omegaLinearWH}
\ee
This equivalence has a simple origin in Fourier space, which
we make explicit because the same mechanism will be used below for
$\tilde\omega$. The Fourier transform of \eqref{eq:omegaNonlinearWH}
reads
\be
\hat\omega(q)\left[1-2\epsilon\hat\omega_-(-q)\right]=\hat H(q).
\label{eq:omegaFourierNL}
\ee
The nonlinearity enters only through the \emph{reflected} factor
$\hat\omega_-(-q)$, which is analytic in the upper half-plane although
it encodes $\omega$ on the negative half-line. Setting
$P_+(q):=\hat\omega_+(q)\left[1-2\epsilon\hat\omega_-(-q)\right]$,
analytic in the upper half-plane, and $P(q):=P_+(q)+\hat\omega_-(q)$
(so that $P_+=[P]_+$), Eq.~\eqref{eq:omegaFourierNL} states that
$P(q)=\hat H(q)+2\epsilon\hat\omega_-(-q)\hat\omega_-(q)$,
which is even in $q$ since $H(y)$ is even. Hence
$P_+(q)-\hat\omega_-(-q)=P_+(-q)-\hat\omega_-(q)$: the left-hand side is
analytic in the upper half-plane, the right-hand side in the lower one,
and both vanish at infinity, so that both vanish identically. This
gives the reflection identity
\be
\left[1+2\epsilon\hat\omega_+(q)\right]
\left[1-2\epsilon\hat\omega_-(-q)\right]
=1 \;.
\label{eq:omegaReflection}
\ee
Using \eqref{eq:omegaReflection} to eliminate
$\hat\omega_-(-q)$ from \eqref{eq:omegaFourierNL} gives
\be
\hat\omega(q) = \hat\omega_+(q)+\hat\omega_-(q)
=
\hat H(q)
\left[
1+2\epsilon\hat\omega_+(q)
\right],
\ee
which is the Fourier transform of \eqref{eq:omegaLinearWH}.
Equivalently,
\be
\frac{
1-2\epsilon\hat\omega_-(q)
}{
1+2\epsilon\hat\omega_+(q)
}
=
{\cal K}(q),
\qquad
{\cal K}(q)
:=
1-2\epsilon\hat H(q).
\label{eq:WHfactor}
\ee
Explicitly,
\be
{\cal K}(q)
=
1
-
4\sqrt{\pi}\,\epsilon\,
\left(a+2b-4bq^2\right)e^{-q^2}.
\label{eq:Kexplicit}
\ee
We look for solutions such that ${\cal K}(q)>0$ for
$q\in\mathbb{R}$, so that $\log{\cal K}(q)$ is real, even and
decaying. Since ${\cal K}$ depends on the parameters $a$ and $b$, which
are determined only at the end by the conditions \eqref{eq:fWH} and
\eqref{eq:gWH}, this is an assumption to be verified a posteriori. As we will see, it
holds for small $\epsilon$, where ${\cal K}(q)\to1$, and along the
branch of solutions continuously connected to $\epsilon=0$ up to the
point where ${\cal K}$ develops a pair of real double zeros (the
continuation beyond this point is discussed in
Appendix~\ref{app:continuationAc}). The Wiener--Hopf method
consists in factorizing multiplicatively
${\cal K}(q)={\cal K}_-(q)/{\cal K}_+(q)$, with ${\cal K}_+$
(${\cal K}_-$) analytic and nonvanishing in the upper (lower)
half-plane and ${\cal K}_\pm\to1$ at infinity. Such a factorization is
obtained from the additive decomposition \eqref{eq:WHprojections} of
the logarithm,
\be
{\cal K}_+(q)=e^{-[\log{\cal K}]_+(q)},
\qquad
{\cal K}_-(q)=e^{[\log{\cal K}]_-(q)}.
\label{eq:Kpmexplicit}
\ee
Writing \eqref{eq:WHfactor} as
$\left[1-2\epsilon\hat\omega_-(q)\right]/{\cal K}_-(q)
=\left[1+2\epsilon\hat\omega_+(q)\right]/{\cal K}_+(q)$,
the two sides define a single entire function tending to $1$ at
infinity, which by Liouville's theorem is identically equal to $1$.
Hence
\begin{equation}
\label{eq:WHKappa}
1-2\epsilon\hat\omega_-(q)
= \mathcal{K}_-(q) \;, \qquad
1+2\epsilon\hat\omega_+(q)
= \mathcal{K}_+(q).
\end{equation}
This is the complete solution of the equation for $\omega$ in Fourier
space. To translate it into real space, we follow Eqs.~(B.11)--(B.13)
of Ref.~\cite{brunet2000probability} and introduce the even auxiliary
function
\be
\omega^\ast(y)
=
-\frac{1}{2\epsilon}
\int_{-\infty}^{+\infty}
\frac{dq}{2\pi}\,
e^{-iqy}\log{\cal K}(q),
\label{eq:omegastar}
\ee
i.e.\ $\hat\omega^\ast(q)=-\log{\cal K}(q)/(2\epsilon)$.
From the definition \eqref{eq:omegastar} and the Wiener--Hopf
projections introduced above, one has
\be
-2\epsilon\hat\omega^\ast_\pm(q)
=
\left[\log{\cal K}\right]_\pm(q).
\label{eq:omegastarprojections}
\ee
On the other hand, taking the logarithm of
\eqref{eq:WHfactor} gives
\be
\log{\cal K}(q)
=
\log\left[
1-2\epsilon\hat\omega_-(q)
\right]
-
\log\left[
1+2\epsilon\hat\omega_+(q)
\right].
\ee
The first term on the right-hand side is a lower-half-plane function,
whereas the second is an upper-half-plane function. By uniqueness of
the Wiener--Hopf decomposition,
\bea
\left[\log{\cal K}\right]_+(q)
&=&
-\log\left[
1+2\epsilon\hat\omega_+(q)
\right],
\\
\left[\log{\cal K}\right]_-(q)
&=&
\log\left[
1-2\epsilon\hat\omega_-(q)
\right].
\label{eq:logKpm}
\eea
Combining \eqref{eq:omegastarprojections} and
\eqref{eq:logKpm}, we obtain
\bea
1+2\epsilon\hat\omega_+(q)
&=&
e^{2\epsilon\hat\omega^\ast_+(q)}, \nn
\\
1-2\epsilon\hat\omega_-(q)
&=&
e^{-2\epsilon\hat\omega^\ast_-(q)}.
\label{eq:omegaFactors}
\eea
Furthermore, since $\log{\cal K}(q)$ is even, the symmetry of its
Wiener--Hopf projections gives
\be
\left[\log{\cal K}\right]_-(-q)
=
\left[\log{\cal K}\right]_+(q),
\ee
and therefore
\be
\hat\omega^\ast_-(-q)
=
\hat\omega^\ast_+(q).
\label{eq:omegastarSymmetry}
\ee
which, together with Eq.~\eqref{eq:omegaFactors}, is equivalent to \eqref{eq:omegaReflection}.

It is also convenient to express these factors directly in terms
of ${\cal K}$ and the odd function
\be
\Omega(q)
=
-\frac{1}{2\epsilon}
{\cal H}\log{\cal K}(q) \;.
\label{eq:OmegaWH}
\ee
Using \eqref{eq:WHprojections} and
\eqref{eq:OmegaWH}, one obtains from \eqref{eq:omegastarprojections}

\bea
-2\epsilon\hat\omega^\ast_+(q)
&=&
\frac12\log{\cal K}(q)
-i\epsilon\Omega(q),
\\
-2\epsilon\hat\omega^\ast_-(q)
&=&
\frac12\log{\cal K}(q)
+i\epsilon\Omega(q).
\eea
Hence
\bea
 \mathcal{K}_+^{-1}(q)
= e^{-2\epsilon\hat\omega^\ast_+(q)}
&=&
\sqrt{{\cal K}(q)}\,
e^{-i\epsilon\Omega(q)},
\\
\mathcal{K}_-(q)
= e^{-2\epsilon\hat\omega^\ast_-(q)}
&=&
\sqrt{{\cal K}(q)}\,
e^{+i\epsilon\Omega(q)}.
\label{eq:omegastarHilbert}
\eea

The exponential relation
\eqref{eq:omegaFactors} is the convenient one to
extract the boundary values of $\omega$ at $y=0^+$, which fix $a$ and
$b$. Evaluating its first line at $q=iX$, $X>0$, we obtain, writing explicitly the Fourier transforms (compare
with Eqs.~(B.16)--(B.23) of Ref.~\cite{brunet2000probability})

\be
2\epsilon
\int_0^\infty dy\,e^{-Xy}\omega^\ast(y)
=
\log\left[
1+
2\epsilon
\int_0^\infty dy\,e^{-Xy}\omega(y)
\right].
\label{eq:DBidentityomega}
\ee
Expanding at large $X$, as in Eqs.~(B.14)--(B.15) of
Ref.~\cite{brunet2000probability}, yields
\bea
\omega(0)
&=&
\omega^\ast(0),
\\
\omega'(0)
&=&
\epsilon\,\omega(0)^2,
\\
\omega''(0)
&=&
\omega^{\ast\prime\prime}(0)
+
\frac{2}{3}\epsilon^2\omega(0)^3.
\label{eq:omegaDerWH}
\eea
The second relation is consistent with the boundary condition
$\beta'(0)=\epsilon$. Since $\omega(0)=1$, the first relation gives
the first equation determining $a$ and $b$,
\be
f(a,b,\epsilon)
:=
-\frac{1}{2\epsilon}
\int_{-\infty}^{+\infty}
\frac{dq}{2\pi}\,
\log{\cal K}(q)
=
1.
\label{eq:fWH}
\ee

\paragraph{Solution of the equation for $\tilde\omega$.}

We now turn to the second equation in \eqref{eq:coupledeq},
\be
\tilde\omega(y)
=
J(y)
+
\epsilon\int_0^\infty dv\,
\left[
\omega(-v)\tilde\omega(y-v)
-
\tilde\omega(-v)\omega(y-v)
\right].
\label{eq:tomegaWHstart}
\ee
Taking the Fourier transform gives
\be
\hat{\tilde\omega}(q)
=
\hat J(q)
+
\epsilon
\left[
\hat\omega_-(-q)\hat{\tilde\omega}(q)
-
\hat{\tilde\omega}_-(-q)\hat\omega(q)
\right].
\label{eq:tomegaFourier}
\ee
Equation \eqref{eq:tomegaFourier} is linear in
$\tilde\omega$, with coefficients fixed by the solution for $\omega$
obtained above.
However, the form of the right-hand side involves a reflected term $\tilde \omega_-(-q)$ —which is analytic in the positive half-plane but depends on the function in the negative half-plane— inside mixed terms $\hat{\tilde\omega}_-(-q)\hat\omega_-(q)$, in which the
reflected unknown multiplies a function analytic in the opposite
half-plane. These aspects complicate the direct application of the WH method.
As for
$\omega$, the term $\hat{\tilde\omega}_-(-q)$ is removed by a parity argument. Similarly to what was done below Eq.~\eqref{eq:omegaFourierNL}, we
introduce the auxiliary functions $G_+$ and $G$ (compare with $P_+$
and $P$ there), so that $G_+=[G]_+$:
\bea
G_+(q)
&:=&
\hat{\tilde\omega}_+(q)
\left[
1-\epsilon\hat\omega_-(-q)
\right]
+
\epsilon
\hat{\tilde\omega}_-(-q)\hat\omega_+(q),
\\
G(q)
&:=&
G_+(q)+\hat{\tilde\omega}_-(q).
\eea
Equation \eqref{eq:tomegaFourier} implies
\be
G(q)
=
\hat J(q)
+
\epsilon
\left[
\hat\omega_-(-q)\hat{\tilde\omega}_-(q)
-
\hat{\tilde\omega}_-(-q)\hat\omega_-(q)
\right].
\ee
The right-hand side is odd under $q\to-q$, since $\hat J(q)$ is odd (in contrast with the even function $P(q)$ below \eqref{eq:omegaFourierNL})
and the two terms in the bracket are exchanged. Hence
\be
G(q)+G(-q)=0.
\ee
Using $G(q)=G_+(q)+\hat{\tilde\omega}_-(q)$, this can be written as
\bea
0
&=&
\left[
G_+(q)+\hat{\tilde\omega}_-(-q)
\right]
+
\left[
G_+(-q)+\hat{\tilde\omega}_-(q)
\right].
\label{eq:Gpmdecomposition}
\eea
The first bracket is analytic in the upper half-plane, since both
$G_+(q)$ and $\hat{\tilde\omega}_-(-q)$ are $+$ functions, while the
second bracket is analytic in the lower half-plane. Thus
\eqref{eq:Gpmdecomposition} is a Wiener--Hopf decomposition of zero.
Under the assumed decay at infinity, uniqueness of this decomposition
(or equivalently Liouville's theorem) implies that both brackets vanish.
Therefore
\be
G_+(q)+\hat{\tilde\omega}_-(-q)=0.
\ee
Using the definition of $G_+$, we obtain
\be
\hat{\tilde\omega}_+(q)
\left[
1-\epsilon\hat\omega_-(-q)
\right]
+
\hat{\tilde\omega}_-(-q)
\left[
1+\epsilon\hat\omega_+(q)
\right]
=
0.
\label{eq:tomegaIdentity}
\ee
This is the analogue for $\tilde\omega$ of the reflection
identity \eqref{eq:omegaReflection}.

Similarly to what we did below \eqref{eq:Kexplicit}, we assume in the following that the factors
$1+\epsilon\hat\omega_+(q)$ and
$1-\epsilon\hat\omega_-(q)$ are nonvanishing
in their respective half-planes, as holds
perturbatively for sufficiently small $\epsilon$.

We now make use of \eqref{eq:tomegaIdentity} to eliminate
$\hat{\tilde\omega}_-(-q)$ and it is useful to proceed in two steps.
First, \eqref{eq:tomegaFourier} can be rewritten as
\be
\hat{\tilde\omega}(q)
\left[
1-\epsilon\hat\omega_-(-q)
\right]
=
\hat J(q)
-
\epsilon\hat{\tilde\omega}_-(-q)\hat\omega(q).
\label{eq:tomegaIntermediate1}
\ee
From \eqref{eq:tomegaIdentity},
\be
\hat{\tilde\omega}_-(-q)
=
-\hat{\tilde\omega}_+(q)
\frac{
1-\epsilon\hat\omega_-(-q)
}{
1+\epsilon\hat\omega_+(q)
}.
\label{eq:tomegaEliminate}
\ee
Substituting this into \eqref{eq:tomegaIntermediate1} and dividing by
$1-\epsilon\hat\omega_-(-q)$ gives
\be
\hat{\tilde\omega}(q)
=
\frac{\hat J(q)}
{1-\epsilon\hat\omega_-(-q)}
+
\epsilon\hat\omega(q)
\frac{\hat{\tilde\omega}_+(q)}
{1+\epsilon\hat\omega_+(q)}.
\label{eq:tomegaIntermediate2}
\ee
Using
$\hat{\tilde\omega}=\hat{\tilde\omega}_+
+\hat{\tilde\omega}_-$ and
$\hat\omega=\hat\omega_++\hat\omega_-$, this becomes
\be
\hat{\tilde\omega}_+(q)
\frac{
1-\epsilon\hat\omega_-(q)
}{
1+\epsilon\hat\omega_+(q)
}
+
\hat{\tilde\omega}_-(q)
=
\frac{\hat J(q)}
{1-\epsilon\hat\omega_-(-q)}.
\label{eq:tomegaIntermediate3}
\ee

Dividing finally by $1-\epsilon\hat\omega_-(q)$, one obtains the
scalar additive Wiener--Hopf equation
\be
\frac{\hat{\tilde\omega}_+(q)}
{1+\epsilon\hat\omega_+(q)}
+
\frac{\hat{\tilde\omega}_-(q)}
{1-\epsilon\hat\omega_-(q)}
=
{\cal R}(q),
\label{eq:tomegaAdditiveWH}
\ee
where the left-hand side is now explicitly split into functions analytic in opposite halves and we introduced the odd function
\be
{\cal R}(q)
:=
\frac{\hat J(q)}
{
\left[1-\epsilon\hat\omega_-(-q)\right]
\left[1-\epsilon\hat\omega_-(q)\right]
}.
\label{eq:RWH}
\ee
Using
the Wiener--Hopf projections \eqref{eq:WHprojections} onto functions analytic in the
upper and lower half-planes, respectively, the solution is
\bea
\hat{\tilde\omega}_+(q)
&=&
\left[
1+\epsilon\hat\omega_+(q)
\right]
[{\cal R}(q)]_+, \nn
\\
\hat{\tilde\omega}_-(q)
&=&
\left[
1-\epsilon\hat\omega_-(q)
\right]
[{\cal R}(q)]_-.
\label{eq:tomegaWHsolution}
\eea
Thus the second equation, namely
\eqref{eq:tomegaWHstart}, also reduces to a scalar Wiener--Hopf
decomposition; no matrix factorization is required.

Using \eqref{eq:omegaFactors}, \eqref{eq:omegaReflection} and
\eqref{eq:omegastarHilbert}, we then obtain for the denominator of $\mathcal{R}(q)$
\bea
&&
\left[
1-\epsilon\hat\omega_-(-q)
\right]
\left[
1-\epsilon\hat\omega_-(q)
\right]
\nn\\
&&\qquad =
\frac14
\left[
1+e^{-2\epsilon\hat\omega^\ast_+(q)}
\right]
\left[
1+e^{-2\epsilon\hat\omega^\ast_-(q)}
\right]
\nn\\
&&\qquad =
\frac12
\left[
1-\epsilon\hat H(q)
+
\sqrt{{\cal K}(q)}
\cos\left(\epsilon\Omega(q)\right)
\right],
\label{eq:denominatorWH}
\eea
where in the last equality we used
\be
{\cal K}(q)=1-2\epsilon\hat H(q).
\ee

Therefore the function ${\cal R}(q)$ defined in \eqref{eq:RWH}
can be written entirely in terms of ${\cal K}$ and its Hilbert transform \eqref{eq:OmegaWH} as
\be
{\cal R}(q)
=
\frac{
2\hat J(q)
}{
1-\epsilon\hat H(q)
+
\sqrt{{\cal K}(q)}
\cos\left(\epsilon\Omega(q)\right)
}.
\label{eq:RWHexplicit}
\ee
Using \eqref{eq:HJFourier}, this becomes explicitly
\be
{\cal R}(q)
=
\frac{
16ib\sqrt{\pi}\,q\,e^{-q^2}
}{
1-\epsilon\hat H(q)
+
\sqrt{{\cal K}(q)}
\cos\left(\epsilon\Omega(q)\right)
}.
\label{eq:RWHexplicit2}
\ee
Since $\hat H(q)$ and ${\cal K}(q)$ are even, while
$\Omega(q)$ and $\hat J(q)$ are odd, the denominator in
\eqref{eq:RWHexplicit} is even and ${\cal R}(q)$ is indeed odd.

It remains to impose the second independent boundary condition
$\tilde\omega'(0)=-1/2$. We evaluate the first equation in
\eqref{eq:tomegaWHsolution} at $q=iX$, with $X\to+\infty$. Since
$\tilde\omega(0)=0$,
\be
\hat{\tilde\omega}_+(iX)
=
\int_0^\infty dy\,e^{-Xy}\tilde\omega(y)
=
\frac{\tilde\omega'(0)}{X^2}
+O(X^{-3}).
\label{eq:tomegaLargeX}
\ee
Moreover,
$1+\epsilon\hat\omega_+(iX)=1+O(X^{-1})$. From the definition of the
$+$ Wiener--Hopf projection,
\bea
[{\cal R}]_+(iX)
&=&
\int_{\mathbb{R}}\frac{dq}{2i\pi}\,
\frac{{\cal R}(q)}{q-iX}
\nn\\
&=&
\frac{1}{X^2}
\int_{\mathbb{R}}\frac{dq}{2i\pi}\,
q\,{\cal R}(q)
+O(X^{-3}),
\eea
where the term of order $X^{-1}$ vanishes since ${\cal R}(q)$ is odd.
Comparing with \eqref{eq:tomegaLargeX} gives
\be
\tilde\omega'(0)
=
\int_{\mathbb{R}}\frac{dq}{2i\pi}\,
q\,{\cal R}(q).
\label{eq:tomegaDerivativeR}
\ee

Using \eqref{eq:RWH},
\be
\hat J(q)
=
8ib\sqrt{\pi}\,q\,e^{-q^2},
\ee
together with \eqref{eq:denominatorWH}, the boundary condition
$\tilde\omega'(0)=-1/2$ becomes
\be
g(a,b,\epsilon)=1,
\label{eq:gcondition}
\ee
where
\bea
g(a,b,\epsilon)
&:=&
-32b\sqrt{\pi}
\int_{-\infty}^{+\infty}
\frac{dq}{2\pi}\,
\frac{
q^2e^{-q^2}
}{
1-\epsilon\hat H(q)
+
\sqrt{{\cal K}(q)}
\cos\left(\epsilon\Omega(q)\right)
}.
\label{eq:gWH}
\eea
Together with \eqref{eq:fWH}, this gives the two scalar equations
which determine $a$ and $b$ for fixed $\epsilon$.

\paragraph{Formal solution for the scaling energy.}

For a fixed value of $\epsilon$, the two equations
\eqref{eq:fWH} and \eqref{eq:gWH} determine
\be
a=a_\epsilon,
\qquad
b=b_\epsilon.
\ee
The derivatives entering the energy follow directly from
\eqref{eq:omegaDerWH}. Since
$\beta(y)=e^{y^2/4}\omega(y)$ and $\omega(0)=1$,
\be
\beta''(0)
=
\frac12
+
\frac{2}{3}\epsilon^2
+
h(a_\epsilon,b_\epsilon,\epsilon),
\label{eq:betaSecondWH}
\ee
where
\be
h(a,b,\epsilon)
:=
\frac{1}{2\epsilon}
\int_{-\infty}^{+\infty}
\frac{dq}{2\pi}\,
q^2\log{\cal K}(q).
\label{eq:hWH}
\ee
Using \eqref{eq:scalingenergybeta}, the scaling energy therefore takes
the particularly simple form
\be
{\cal E}(\epsilon)
=
-2
\int_{-\infty}^{+\infty}
\frac{dq}{2\pi}\,
q^2\log{\cal K}(q)
\label{eq:energyWH}
\ee
where ${\cal K}(q)$ is given by \eqref{eq:Kexplicit}, evaluated at
$a=a_\epsilon$, $b=b_\epsilon$ determined by
\eqref{eq:fWH}--\eqref{eq:gWH}.

Equations \eqref{eq:fWH}, \eqref{eq:gWH}, and
\eqref{eq:energyWH} thus provide a closed formal parametric solution of
the leading large-$c$ scaling problem.

All logarithms, square roots and Wiener--Hopf factors above are taken
on the branch continuously connected to $\epsilon=0$, for which
${\cal K}(q)\to1$. As in the Brunet--Derrida construction in
Appendix~B, the derivation is directly justified perturbatively for
sufficiently small $\epsilon$. The analytic continuation relevant for
the large-deviation regime is discussed in Appendix.~\ref{app:continuationAc}.

\subsection{Summary of the computation of \texorpdfstring{$\beta''(0)$}{''(0)}}

We summarize the practical procedure for computing $\beta''(0)$ at a
given value of $\epsilon$. For given $a$ and $b$, define
\be
\hat H(q)
=
2\sqrt{\pi}\,
\left(a+2b-4bq^2\right)e^{-q^2}.
\label{eq:summaryH}
\ee
We also define
\be
\Omega(q)
=
-\frac{1}{2\pi\epsilon}
\dashint_{\mathbb R} dk\,
\frac{
\log\left[1-2\epsilon\hat H(k)\right]
}{q-k}.
\label{eq:summaryOmega}
\ee
Then introduce the three scalar functions
\bea
f(a,b,\epsilon)
&=&
-\frac{1}{2\epsilon}
\int_{-\infty}^{+\infty}
\frac{dq}{2\pi}\,
\log\left[1-2\epsilon\hat H(q)\right],
\label{eq:summaryf}
\\
g(a,b,\epsilon)
&=&
-32b\sqrt{\pi}
\int_{-\infty}^{+\infty}
\frac{dq}{2\pi}\,
\frac{
q^2e^{-q^2}
}{
1-\epsilon\hat H(q)
+
\sqrt{1-2\epsilon\hat H(q)}
\cos\left(\epsilon\Omega(q)\right)
},
\label{eq:summaryg}
\\
h(a,b,\epsilon)
&=&
\frac{1}{2\epsilon}
\int_{-\infty}^{+\infty}
\frac{dq}{2\pi}\,
q^2
\log\left[1-2\epsilon\hat H(q)\right].
\label{eq:summaryh}
\eea

For fixed $\epsilon$, the two parameters $a$ and $b$ are determined by
the two equations
\be
f(a,b,\epsilon)=1,
\qquad
g(a,b,\epsilon)=1,
\label{eq:summaryfg}
\ee
choosing the branch continuously connected to
\be
a=1,
\qquad
b=-\frac14
\qquad
\text{at }\epsilon=0.
\label{eq:summaryab0}
\ee
Denoting this solution by $a_\epsilon,b_\epsilon$, the desired
quantity is then simply
\be
\beta''(0)
=
\frac12
+
\frac{2}{3}\epsilon^2
+
h(a_\epsilon,b_\epsilon,\epsilon)
.
\label{eq:summaryBetaSecond}
\ee

Thus the computation of $\beta''(0)$ requires only the following
steps: for a trial pair $(a,b)$ evaluate $\hat H$ and $\Omega$,
solve $f=1$ and $g=1$ for $(a,b)$, and finally evaluate $h$ at this
solution. In particular, it is not necessary to reconstruct either
$\omega(y)$ or $\tilde\omega(y)$.

In Appendix \ref{app:perturbativeWH} we perform the perturbative expansion of the equations \eqref{eq:summaryf}--\eqref{eq:summaryBetaSecond} at small $\epsilon$ up to $O(\epsilon^2)$. We check explicitly that
the results \eqref{eq:a1b1direct} and the formula \eqref{eq:a2b2direct} for $a_1,b_1,a_2,b_2$, and \eqref{eq:beta1secondDirect} and
\eqref{eq:beta2secondDirect} for $\beta_1''(0)$ and $\beta_2''(0)$, obtained
previously by solving the coupled
integral equations directly, are recovered.

\subsection{Parametric representation of the scaling function}
\label{sec:parametricSCGF}

The Wiener--Hopf formulation above is particularly convenient for
deriving the closed equations and for checking their perturbative
solution at small $\epsilon$. To obtain higher orders more efficiently,
and to put the large-$L$ scaling function of the SCGF in a parametric
form closer to the one found by Brunet--Derrida for a single polymer,
it is useful to introduce a new pair of variables which absorbs the
explicit dependence on $\epsilon$.

We define
\be
A=(a+2b)\epsilon,
\qquad
B=\frac{-4b}{a+2b}.
\label{eq:ABdef}
\ee
Equivalently,
\be
a=\frac{A}{\epsilon}\left(1+\frac{B}{2}\right),
\qquad
b=-\frac{AB}{4\epsilon}.
\label{eq:abFromAB}
\ee
With this parametrization the combination $\epsilon\hat H$ which enters
the Wiener--Hopf equations takes the simple form
\be
H_{A,B}(q)
:=
\epsilon\hat H(q)
=
2\sqrt{\pi}\,A(1+Bq^2)e^{-q^2}.
\label{eq:HABdef}
\ee
It is then natural to introduce the rescaled phase
\be
\Omega_{A,B}(q)
:=
-\epsilon\Omega(q)
=
\frac12{\cal H}
\log\left[1-2H_{A,B}\right](q),
\label{eq:OmegaABdef}
\ee
or explicitly
\be
\Omega_{A,B}(q)
=
\frac{1}{2\pi}\dashint_{\mathbb R}
\frac{\log\left[1-2H_{A,B}(k)\right]}{q-k}\,dk.
\ee
The three functions entering the solution are correspondingly rescaled
as
\be
\hat f(A,B)=\epsilon f(a,b,\epsilon),
\qquad
\hat g(A,B)=\epsilon g(a,b,\epsilon),
\qquad
\hat h(A,B)=\epsilon h(a,b,\epsilon).
\label{eq:fghHatDef}
\ee
Using \eqref{eq:HABdef}--\eqref{eq:fghHatDef}, their explicit expressions
become
\be
\hat f(A,B)
=
-\frac12
\int_{\mathbb R}\frac{dq}{2\pi}
\log\left[1-2H_{A,B}(q)\right],
\label{eq:fhatAB}
\ee
\be
\hat h(A,B)
=
\frac12
\int_{\mathbb R}\frac{dq}{2\pi}\,
q^2\log\left[1-2H_{A,B}(q)\right],
\label{eq:hhatAB}
\ee
and
\be
\hat g(A,B)
=
AB\int_{\mathbb R}\frac{dq}{2\pi}
\frac{8\sqrt{\pi}\,q^2e^{-q^2}}
{1-H_{A,B}(q)
+\sqrt{1-2H_{A,B}(q)}
\cos\left(\Omega_{A,B}(q)\right)}.
\label{eq:ghatAB}
\ee

The advantage of this change of variables is that the explicit factors
of $\epsilon$ have disappeared from the integral equations. The physical
solution is now obtained parametrically. Indeed, the two conditions
$f=1$ and $g=1$ become
\be
\hat f(A,B)=\hat g(A,B)=\epsilon.
\label{eq:fgHatPhysical}
\ee
For a fixed value of $A$, we first determine $B=B_A$ from the single
equation
\be
\hat f(A,B_A)=\hat g(A,B_A),
\label{eq:BAequation}
\ee
The branch relevant near the origin satisfies
\be
A\simeq\frac{\epsilon}{2},
\qquad
B_A\simeq2,
\qquad
\epsilon\to0. \label{eq:BAorigin}
\ee
The replica scaling variable is then obtained as
\be
\epsilon=\hat f(A,B_A),
\label{eq:epsilonParametric}
\ee
while, using \eqref{eq:summaryBetaSecond}, the scaling function becomes
\be
{\cal E}(\epsilon)
=
-4\epsilon\,h(a_\epsilon,b_\epsilon,\epsilon)
=
-4\hat h(A,B_A).
\label{eq:EParametric}
\ee
Equations \eqref{eq:BAequation}--\eqref{eq:EParametric} therefore give
the scaling function in parametric form,
\be
\epsilon=\hat f(A,B_A),
\qquad
{\cal E}=-4\hat h(A,B_A),
\qquad
\hat f(A,B_A)=\hat g(A,B_A).
\label{eq:SCGFParametricSummary}
\ee
This representation is the natural two-polymer analogue of the
Brunet--Derrida parametrization: the auxiliary variable $A$
parametrizes the scaling function, while the additional function
$B_A$, absent in the one-polymer problem, encodes the non-trivial
two-layer structure.

Returning to the original ground-state energy, the scaling function
${\cal E}(\epsilon)$ enters as
\be
E(n,L)
=
\frac{4}{L^2}
\left[
\frac{nc^2}{12}
+
\frac{\sqrt{c}}{4}\,
{\cal E}(n\sqrt{c})
\right]
+
o\left(\frac{\sqrt{c}}{L^2}\right),
\qquad
c\to\infty,
\quad
n\sqrt{c}=\epsilon \, \, \text{fixed}.
\label{eq:EnergyScalingFinal}
\ee
Thus the parametric equations
\eqref{eq:SCGFParametricSummary} determine directly the universal
large-$L$ contribution to $E(n,L)$, and hence all the scaled cumulants
of the two-polymer free energy.

We now use this representation to obtain the systematic small-$A$
expansion, and hence the expansion of ${\cal E}(\epsilon)$ to higher
orders.

\subsection{Series expansion of \texorpdfstring{$\hat f$}{f} and \texorpdfstring{$\hat h$}{h}}

The functions $\hat f(A,B)$ and $\hat h(A,B)$ admit simple power-series
expansions in the parameter $A$. Using
\be
H_{A,B}(q)
=
2\sqrt{\pi}\,A(1+Bq^2)e^{-q^2},
\ee
and expanding
\be
-\log(1-x)
=
\sum_{k=1}^\infty\frac{x^k}{k},
\ee
one obtains
\be
\hat f(A,B)
=
\sum_{k=1}^\infty A^k F_k(B),
\label{eq:fhatSeries}
\ee
with
\be
F_k(B)
=
\frac{
2^{2k-2}\pi^{(k-2)/2}
}{
k^{3/2}
}
\sum_{m=0}^k
\binom{k}{m}
\Gamma\left(m+\frac12\right)
\left(\frac{B}{k}\right)^m.
\label{eq:FkB}
\ee

Similarly,
\be
\hat h(A,B)
=
\sum_{k=1}^\infty A^k H_k(B),
\label{eq:hhatSeries}
\ee
with
\be
H_k(B)
=
-\frac{
2^{2k-2}\pi^{(k-2)/2}
}{
k^{5/2}
}
\sum_{m=0}^k
\binom{k}{m}
\Gamma\left(m+\frac32\right)
\left(\frac{B}{k}\right)^m.
\label{eq:HkB}
\ee
These expressions follow directly from the Gaussian moments
\be
\int_{-\infty}^{+\infty}
\frac{dq}{2\pi}\,
q^{2m}e^{-kq^2}
=
\frac{
\Gamma\left(m+\frac12\right)
}{
2\pi\,k^{m+1/2}
}.
\ee

The first two orders in $A$ are
\bea
\hat f(A,B)
&=&
A\left(1+\frac{B}{2}\right)
+
\sqrt{2\pi}\,A^2
\left(
1+\frac{B}{2}+\frac{3B^2}{16}
\right)
+O(A^3),
\label{eq:fhatFirstOrders}
\\
\hat h(A,B)
&=&
-A\left(
\frac12+\frac{3B}{4}
\right)
-
\sqrt{2\pi}\,A^2
\left(
\frac14+\frac{3B}{8}+\frac{15B^2}{64}
\right)
+O(A^3).
\label{eq:hhatFirstOrders}
\eea

\subsection{Expansion of the parametric function \texorpdfstring{$\hat g$}{g}}

We now study the small-$A$ expansion of the function
$\hat g(A,B)$ defined in \eqref{eq:ghatAB}. It is useful to introduce
\be
P_B(q)=2\sqrt{\pi}\,(1+Bq^2)e^{-q^2},
\qquad
H_{A,B}(q)=A P_B(q),
\label{eq:PBdef}
\ee
and
\be
\Phi_r(q)={\cal H}\left[P_B^r\right](q).
\label{eq:PhiBdef}
\ee
The function entering the cosine can then be expanded as
\bea
\Omega_{A,B}(q)
&=&
\frac12{\cal H}\log\left[1-2A P_B\right](q)
\nn\\
&=&
-\sum_{r\geq1}\frac{2^{r-1}}{r}A^r\Phi_r(q).
\label{eq:OmegaABseries}
\eea

The Hilbert transforms which enter this expansion remain rather
explicit. In particular,
\be
{\cal H}\left(e^{-q^2}\right)
=
e^{-q^2}{\rm erfi}(q),
\ee
and hence
\be
\Phi_1(q)
=
2\sqrt{\pi}
\left[
(1+Bq^2)e^{-q^2}{\rm erfi}(q)
-\frac{Bq}{\sqrt{\pi}}
\right].
\label{eq:Phi1Explicit}
\ee

Let us denote
\be
D_{A,B}(q)
=
1-A P_B(q)
+
\sqrt{1-2A P_B(q)}
\cos\left(\Omega_{A,B}(q)\right).
\label{eq:DABdef}
\ee
A direct expansion gives
\bea
\frac{2}{D_{A,B}}
&=&
1+A P_B
+\frac{A^2}{4}\left(5P_B^2+\Phi_1^2\right)
\nn\\
&&
+\frac{A^3}{4}
\left(
7P_B^3+P_B\Phi_1^2+2\Phi_1\Phi_2
\right)
\nn\\
&&
+\frac{A^4}{24}
\left(
63P_B^4+6P_B^2\Phi_1^2+12P_B\Phi_1\Phi_2
+16\Phi_1\Phi_3+6\Phi_2^2+\Phi_1^4
\right)
\nn\\
&&
+O(A^5).
\label{eq:DABinverseSeries}
\eea
From \eqref{eq:ghatAB}, the definition of $\hat g$ becomes
\be
\hat g(A,B)
=
AB
\int_{-\infty}^{+\infty}\frac{dq}{2\pi}\,
4\sqrt{\pi}\,q^2e^{-q^2}
\frac{2}{D_{A,B}(q)}.
\label{eq:ghatExpansionStart}
\ee

Equation \eqref{eq:DABinverseSeries} shows the general structure of
the expansion. At any fixed order in $A$, only a finite number of
integrals involving products of Hilbert transforms of powers of
$P_B$ occurs. Since $P_B$ is a polynomial times a Gaussian, these
can be reduced to Gaussian integrals and integrals involving powers
of ${\rm erfi}$. For example, $\Phi_1^2$ generates ${\rm erfi}^2$,
while the term $\Phi_1^4$, which contributes for the first time to
$G_5$, generates integrals containing ${\rm erfi}^4$. The details
of this reduction are given in Appendix~\ref{app:ghatIntegrals}.

Writing
\be
\hat g(A,B)=\sum_{r\geq1}A^r G_r(B),
\label{eq:ghatSeries}
\ee
and performing the integrals, one obtains
\bea
G_1(B)&=&B,
\\
G_2(B)&=&
\sqrt{\frac{\pi}{2}}\,
B\left(1+\frac{3B}{4}\right),
\\
G_3(B)&=&
B\left[
\frac23
+\frac{16\sqrt{3}\pi}{27}(1+B)
+\frac{20\sqrt{3}\pi}{81}B^2
\right],
\\
G_4(B)&=&
\frac{\sqrt{\pi}\,B}{256}
\bigg[
512(\pi+\sqrt{2})
+(576\pi+128\sqrt{2})B
\nn\\
&&\hspace{32mm}
+360\pi B^2+105\pi B^3
\bigg].
\label{eq:G1G4}
\eea
At the next order one finds
\bea
G_5(B)
&=&
B\bigg[
4\pi\left(
1+\frac{8\sqrt{3}}{9}
+\frac{64\sqrt{5}\pi}{125}
\right)
\nn\\
&&
+B\left(
-\frac{4}{15}
+2\pi
+\frac{32\sqrt{3}\pi}{27}
+\frac{1536\sqrt{5}\pi^2}{625}
\right)
\nn\\
&&
+B^2\left(
\frac{\pi}{4}
+\frac{8\sqrt{3}\pi}{27}
+\frac{1152\sqrt{5}\pi^2}{625}
\right)
\nn\\
&&
+\frac{2688\sqrt{5}\pi^2}{3125}B^3
+\frac{3024\sqrt{5}\pi^2}{15625}B^4
\bigg].
\label{eq:G5}
\eea
The calculation of $G_5$ involves at intermediate stages integrals
containing ${\rm erfi}(q)^4$, originating from the term
$\Phi_1(q)^4=[{\cal H}P_B(q)]^4$ in
\eqref{eq:DABinverseSeries}. Their reduction is discussed in
Appendix~\ref{app:ghatIntegrals}; the final result
\eqref{eq:G5} is nevertheless elementary.

Combining \eqref{eq:ghatSeries} with the expansion of $\hat f(A,B)$,
the equation $\hat f(A,B_A)=\hat g(A,B_A)$ can be solved recursively
on the branch $B_A\to2$ as $A\to0$. One finds
\bea
B_A
&=&
2+\sqrt{\frac{\pi}{2}}\,A
+
\left[
\left(
\frac{128}{27\sqrt{3}}-\frac32
\right)\pi-\frac83
\right]A^2
\nn\\
&&
+\frac{\sqrt{\pi}}{432}
\left[
\left(
4293+891\sqrt{2}-1792\sqrt{6}
\right)\pi
-3744\sqrt{2}
\right]A^3
\nn\\
&&
+\bigg[
\frac{256}{45}
-\pi\left(
14+\frac{5120}{81\sqrt{3}}
\right)
\nn\\
&&\hspace{12mm}
+\pi^2\left(
-\frac{85219}{3888}
-\frac{803}{16\sqrt{2}}
+\frac{424}{9\sqrt{3}}
+\frac{253952}{3125\sqrt{5}}
\right)
\bigg]A^4
+O(A^5).
\label{eq:BASeries}
\eea
Numerically,
\be
B_A
=
2+1.25331\,A+1.21970\,A^2
-6.72599\,A^3-92.3848\,A^4+O(A^5).
\ee

\subsection{Perturbative expansion of the large \texorpdfstring{$L$}{L} scaling function}

We can now combine the expansions of $\hat f$ and $\hat h$ obtained
in \eqref{eq:fhatSeries} and \eqref{eq:hhatSeries} with the solution
$B=B_A$ of $\hat f=\hat g$, whose expansion is given in
\eqref{eq:BASeries}. Using the parametric relations
\eqref{eq:epsilonParametric} and \eqref{eq:EParametric}, we first obtain
\bea
\epsilon
&=&
n\sqrt c
=
2A
+
3\sqrt{2\pi}\,A^2
+
\left[
-\frac43+\frac{\pi}{2}
+\frac{64\sqrt{3}\pi}{9}
\right]A^3
\nn\\
&&
+
\frac{\sqrt{\pi}}{324}
\left[
-2484\sqrt{2}
+
\pi\left(
13203-243\sqrt{2}+640\sqrt{6}
\right)
\right]A^4
\nn\\
&&
+
\bigg[
\frac{128}{45}
-\pi\left(
\frac{89}{3}+\frac{5248\sqrt{3}}{243}
\right)
\nn\\
&&\hspace{10mm}
+\pi^2\left(
\frac{155129}{11664}
+\frac{489\sqrt{2}}{32}
-\frac{68\sqrt{3}}{9}
+\frac{202752\sqrt{5}}{3125}
\right)
\bigg]A^5
+
O(A^6).
\label{eq:epsilonSeriesA}
\eea
Similarly,
\bea
{\cal E}(\epsilon(A))
&=&
-4\hat h(A,B_A)
\nn\\
&=&
8A
+
\frac{37}{4}\sqrt{2\pi}\,A^2
+
\left[
-8+\frac{3\pi}{4}
+\frac{4864\sqrt{3}\pi}{243}
\right]A^3
\nn\\
&&
+
\frac{\sqrt{\pi}}{5184}
\left[
-207360\sqrt{2}
+
\pi\left(
566757-6318\sqrt{2}+16384\sqrt{6}
\right)
\right]A^4
\nn\\
&&
+
\bigg[
\frac{256}{15}
-\pi\left(
138+\frac{24832\sqrt{3}}{243}
\right)
\nn\\
&&\hspace{10mm}
+\pi^2\left(
\frac{1566599}{69984}
+\frac{819\sqrt{2}}{32}
-\frac{1048\sqrt{3}}{81}
+\frac{13180928\sqrt{5}}{78125}
\right)
\bigg]A^5
+
O(A^6).
\label{eq:ESeriesA}
\eea

Eliminating $A$ between \eqref{eq:epsilonSeriesA} and
\eqref{eq:ESeriesA} gives the scaling SCGF directly as a series in
$\epsilon=n\sqrt c$:
\bea
{\cal E}(\epsilon)
&=&
4\epsilon
-
\frac{11}{8}\sqrt{\frac{\pi}{2}}\,\epsilon^2
+
\left[
-\frac13
+
\left(
\frac{61}{32}
-\frac{256}{81\sqrt{3}}
\right)\pi
\right]\epsilon^3
\nn\\
&&
-
\frac{\sqrt{\pi}}{27648}
\left[
1728\sqrt{2}
+
\pi\left(
92745+91746\sqrt{2}-91136\sqrt{6}
\right)
\right]\epsilon^4
\nn\\
&&
+
\bigg[
\frac{1}{90}
+
\pi\left(
\frac14-\frac{32}{81\sqrt{3}}
\right)
\nn\\
&&\hspace{10mm}
+
\pi^2\left(
\frac{46098029}{2239488}
-\frac{1490}{27\sqrt{3}}
-\frac{221696}{15625\sqrt{5}}
+\frac{6377}{256\sqrt{2}}
\right)
\bigg]\epsilon^5
+
O(\epsilon^6).
\label{eq:ESeriesEpsilon}
\eea
Numerically,
\be
{\cal E}(\epsilon)
=
4\epsilon
-1.72330694\,\epsilon^2
-0.07717403\,\epsilon^3
-0.00697120\,\epsilon^4
-0.000675336\,\epsilon^5
+O(\epsilon^6).
\ee

Using ${\cal E}(\epsilon)=-4\hat h(A,B_A)=-4\epsilon h(a_\epsilon,b_\epsilon,\epsilon)$,
the first terms of \eqref{eq:ESeriesEpsilon} agree with the direct
perturbative expansion \eqref{eq:hWHpertSolution}; since the latter was
obtained through $O(\epsilon^2)$ in $h$, this checks
\eqref{eq:ESeriesEpsilon} through $O(\epsilon^3)$.

Finally, let us recall that the large-$c$ ground-state
energy $E(n,L)$ takes the form \eqref{eq:EnergyScalingFinal}.
If we denote
${\cal E}(\epsilon)=\sum_{k\geq1}e_k\epsilon^k$, then
the leading large-$c$ behavior of the cumulants for $k\geq2$ follows
directly as
\be
L^2c_k(L)
=
-k!\,e_k\,c^{(k+1)/2}
+
o\left(c^{(k+1)/2}\right).
\label{eq:ckFromScalingSCGF}
\ee

Inserting \eqref{eq:ESeriesEpsilon} into
\eqref{eq:ckFromScalingSCGF} reproduces for $c_2$ and $c_3$ the
previous results \eqref{kappa2largec} and \eqref{eq:c3largec}.
The next two cumulants are new and read
\bea
L^2c_4(L)
&\simeq&
\frac{\sqrt{\pi}}{1152}
\left[
1728\sqrt{2}
+
\pi\left(
92745+91746\sqrt{2}-91136\sqrt{6}
\right)
\right]c^{5/2}
\nn\\
&\simeq&
0.167308885\,c^{5/2},
\label{eq:c4largec}
\\
L^2c_5(L)
&\simeq&
-120\bigg[
\frac{1}{90}
+
\pi\left(
\frac14-\frac{32}{81\sqrt{3}}
\right)
\nn\\
&&\hspace{12mm}
+
\pi^2\left(
\frac{46098029}{2239488}
-\frac{1490}{27\sqrt{3}}
-\frac{221696}{15625\sqrt{5}}
+\frac{6377}{256\sqrt{2}}
\right)
\bigg]c^3
\nn\\
&\simeq&
0.0810403524\,c^3.
\label{eq:c5largec}
\eea

\subsection{Large negative \texorpdfstring{$\epsilon$}{epsilon} and left tail \texorpdfstring{$Y \to - \infty$}{Y to - infty}}
\label{sec:largeNegativeEpsilon}

We now consider the limit $\epsilon\to-\infty$. We shall see
that the scaling SCGF has the asymptotic behavior
\be
{\cal E}(\epsilon)
\simeq
-\frac{4}{5}(3\pi)^{2/3}|\epsilon|^{5/3},
\qquad
\epsilon\to-\infty.
\label{eq:ElargeNegative}
\ee
We first derive this result by a simple argument, postponing the
justification of the behavior of $B_A$ to
Appendix~\ref{app:largeNegativeB}.

It is convenient to parametrize $A<0$ as
\be
4\sqrt{\pi}A=-e^\eta,
\qquad
\eta\to+\infty.
\label{eq:Aeta}
\ee
As shown in Appendix~\ref{app:largeNegativeB}, on the physical
solution $B=B_A$ one has
\be
B_A\simeq\frac{9\pi^2z_*^2}{\eta^3},
\qquad
\tanh z_*=\frac{z_*}{2},
\qquad
z_*\simeq1.91500805,
\label{eq:BlargeNegativeMain}
\ee
and hence $B_A\eta\to0$. Thus, in the region
$q=O(\sqrt{\eta})$ which controls $\hat f$ and $\hat h$,
\be
1-2H_{A,B}(q)
=
1+(1+Bq^2)e^{\eta-q^2}
\simeq
1+e^{\eta-q^2}.
\label{eq:HlargeNegative}
\ee
Using
\be
\log(1+e^{\eta-q^2})
\simeq
(\eta-q^2)\vartheta(\eta-q^2),
\label{eq:FermiApprox}
\ee
the parametric equation $\epsilon=\hat f(A,B_A)$ gives
\be
\epsilon
\simeq
-\frac12
\int_{-\sqrt{\eta}}^{\sqrt{\eta}}
\frac{dq}{2\pi}\,
(\eta-q^2)
=
-\frac{\eta^{3/2}}{3\pi}.
\label{eq:epsilonLargeNegative}
\ee
Thus
\be
\eta
\simeq
(3\pi|\epsilon|)^{2/3}.
\label{eq:etaEpsilonLargeNegative}
\ee
Similarly,
\be
\hat h(A,B_A)
\simeq
\frac12
\int_{-\sqrt{\eta}}^{\sqrt{\eta}}
\frac{dq}{2\pi}\,
q^2(\eta-q^2)
=
\frac{\eta^{5/2}}{15\pi}.
\label{eq:hLargeNegative}
\ee
Since ${\cal E}=-4\hat h$, eliminating $\eta$ between
\eqref{eq:etaEpsilonLargeNegative} and \eqref{eq:hLargeNegative}
gives \eqref{eq:ElargeNegative}.

The origin of the powers appearing here is particularly transparent.
The interval which contributes to the integrals has width
$q_*\sim\sqrt{\eta}$, so that
\be
|\epsilon|\sim q_*^3,
\qquad
-{\cal E}\sim q_*^5.
\label{eq:phaseSpace53}
\ee
Consequently ${\cal E}\sim-|\epsilon|^{5/3}$.
Upon Legendre transformation this corresponds to a $5/2$ power for
the deep lower tail $Y\to-\infty$ of the scaled rate function
$\Phi(Y)$.

It is instructive to compare \eqref{eq:ElargeNegative} with the
$N=1$ result of Brunet and Derrida. Writing their auxiliary
polylogarithm parameter as $-e^{\eta_1}$, their parametric solution
gives at large $\eta_1$
\be
\epsilon
\simeq
-\frac{2}{3\pi}\eta_1^{3/2},
\qquad
{\cal E}_1(\epsilon)
\simeq
-\frac{8}{15\pi}\eta_1^{5/2},
\ee
and hence
\be
{\cal E}_1(\epsilon)
\simeq
-2^{-2/3}\frac45(3\pi)^{2/3}|\epsilon|^{5/3}.
\label{eq:ElargeNegativeN1}
\ee
Thus the normalized scaling functions satisfy
\be
{\cal E}_2(\epsilon)
\simeq
2^{2/3}{\cal E}_1(\epsilon).
\label{eq:EratioN2N1}
\ee

Comparing the physical energy scalings
\eqref{eq:mainEnergyScaling} and
\eqref{eq:mainEnergyScalingDB}, the $N=2$ scaling contribution carries
an additional factor $2$ in its overall prefactor,
$\sqrt{c}/L^2$ instead of $\sqrt{c}/(2L^2)$ for $N=1$.
Combining this with the asymptotic relation between
${\cal E}$ and ${\cal E}^{N=1}$ obtained above reproduces the
finite-$L$ rate-function relation
\eqref{eq:mainPhiComparison},
\be
\Phi_L(H)
\simeq
\Phi_L^{N=1}\left(\frac{H}{2}\right).
\label{eq:PhiN2N1}
\ee

This relation has a simple physical interpretation. A very large
negative fluctuation of $\log Z(t)$ is produced by a
collective rare fluctuation of the common random environment: in the
polymer language the random potential is anomalously positive, or
equivalently the KPZ noise is anomalously negative. Since both
polymers experience the same disorder, it is more favorable to create
a single extended unfavorable region which depresses both free
energies simultaneously than two independent rare fluctuations. The
cost of the disorder fluctuation is therefore paid only once, while
the total shift in $\log Z$ is approximately doubled, leading
precisely to the relation \eqref{eq:PhiN2N1}. The
non-crossing constraint determines how the two polymers arrange
themselves inside this unfavorable region, but does not modify the
leading $|H|^{5/2}$ cost.

We finally note, as a comparison, that the same $5/2$
exponent also occurs in the deep lower tail of the one-layer KPZ
height on the infinite line
\cite{sasorov2017large,corwin2018coulomb,krajenbrink2018simple,
tsai2022exact}, beyond the cubic left tail associated with the
typical Tracy--Widom fluctuation regime.

\subsection{Large positive \texorpdfstring{$\epsilon$}{epsilon} and right tail \texorpdfstring{$Y \to +\infty$}{Y to +infty}}
\label{sec:largePositiveEpsilon}

We now consider the continuation of the SCGF to $\epsilon>\epsilon_c$,
and in particular the limit $\epsilon\to+\infty$, which determines the
right tail $Y\to+\infty$ of the scaled rate function $\Phi(Y)$.

The parametric solution derived in Sec.~\ref{sec:parametricSCGF},
see in particular Eqs.~\eqref{eq:fhatAB}--\eqref{eq:ghatAB},
describes the principal sheet connected continuously to $A=0$.
It reaches a branch point at
\be
\begin{aligned}
&A=A_c\simeq0.118672,
\qquad
B=B_c\simeq1.918303,\\
&\epsilon=\epsilon_c\simeq0.700325,
\qquad
{\cal E}={\cal E}_c\simeq1.927791,
\end{aligned}
\label{eq:mainCriticalPoint}
\ee
where a pair of symmetric zeros of
$K_{A,B}(q)=1-2H_{A,B}(q)$ collides with the real axis.
The SCGF itself
remains regular at this branch point and can be continued to a second sheet. The
construction of this continuation, based on the corresponding change of
Wiener--Hopf factorization, is given in detail in
App.~\ref{app:continuationAc}.
On the second sheet the parametric equations
take the form
\be
\epsilon
=
\hat f_{\rm II}(A,B)
=
\hat g_{\rm II}(A,B),
\qquad
{\cal E}(\epsilon)
=
-4\hat h_{\rm II}(A,B),
\label{eq:mainSecondSheet}
\ee
where, denoting by $z=p+i\kappa$, $p,\kappa>0$, the complex zero
$K_{A,B}(z)=0$ in the first quadrant, given explicitly in
\eqref{eq:contExplicitzNew},
\be
\hat f_{\rm II}(A,B)
=
\hat f(A,B)+2\kappa,
\qquad
\hat h_{\rm II}(A,B)
=
\hat h(A,B)
-\frac{2}{3}{\rm Im}\,z^3,
\label{eq:mainSecondSheetFH}
\ee
while $\hat g_{\rm II}$ is obtained by the continued Wiener--Hopf
factorization, see \eqref{eq:contgIINew}. The two determinations coincide at
$\kappa=0$.

Following this second sheet to large positive $\epsilon$, one finds
$A\to0$, $\kappa\to+\infty$, and
\be
{\cal E}(\epsilon)
=
-\frac{\epsilon^3}{3}
+
\frac{\pi^2}{\epsilon}
-\frac{4\pi^2}{\epsilon^3}
+
O(\epsilon^{\,-5}),
\qquad
\epsilon\to+\infty,
\label{eq:mainRightTailE}
\ee
see App.~\ref{app:continuationAc} for the asymptotic analysis.
Consequently, using the Legendre transform
\eqref{eq:mainScaledRateFunction} relating ${\cal E}(\epsilon)$ to the
scaled rate function $\Phi(Y)$, one obtains the right-tail behavior
\be
\Phi(Y)
=
\frac{2}{3}Y^{3/2}
+
\frac{\pi^2}{\sqrt{Y}}
+
O(Y^{-3/2}),
\qquad
Y\to+\infty.
\label{eq:mainRightTailPhi}
\ee

It is useful to compare these results with the one-polymer periodic
problem \cite{appert1999universal,brunet2000probability}. Recall that
the auxiliary parameter $\xi$ in
\eqref{eq:mainParametricScalingFunctionDB} is denoted $\epsilon$ in
Ref.~\cite{brunet2000probability}. The endpoint $\xi=1$ of the
principal branch corresponds, in our scaling variable, to
\be
\epsilon_c^{N=1}
=
\frac{\zeta(3/2)}{2\sqrt{\pi}}
\simeq 0.737.
\label{eq:mainN1CriticalEpsilon}
\ee
Beyond this point the $N=1$ solution is continued to a second branch.
In the normalization used here, this branch can be parametrized by
$-C=e^{-u^2}$, $u>0$, as
\bea
\epsilon_1^{\rm II}(u)
&=&
2u+\frac{1}{2\sqrt{\pi}}\,
{\rm Li}_{3/2}(e^{-u^2}),
\nn\\
{\cal E}_1^{\rm II}(u)
&=&
-\frac{8}{3}u^3
+\frac{1}{\sqrt{\pi}}\,
{\rm Li}_{5/2}(e^{-u^2}).
\label{eq:mainN1SecondSheet}
\eea
Thus, as $\epsilon\to+\infty$,
\be
{\cal E}_1(\epsilon)
=
-\frac{\epsilon^3}{3}
+
\frac{1+\epsilon^2/2}{\sqrt{\pi}}\,
e^{-\epsilon^2/4}
+\cdots .
\label{eq:mainN1RightTail}
\ee

The leading cubic asymptotics is therefore the same for $N=1$ and
$N=2$, and in both cases the corresponding scaled rate function has
the leading right tail $\Phi(Y) \simeq \frac{2}{3}Y^{3/2}$ as $Y\to+\infty$.
The first correction is, however, qualitatively different: it is
exponentially small for $N=1$, whereas for $N=2$ it is algebraic, as
shown in Eqs.~\eqref{eq:mainRightTailE} and
\eqref{eq:mainRightTailPhi}.

Although the scaled $N=1$ and $N=2$ rate functions have the same
leading right-tail behavior,
$\Phi^{N=1}(Y)\simeq\Phi^{N=2}(Y)\simeq \frac23 Y^{3/2}$,
the corresponding physical finite-$L$ rate functions differ because
of the different overall scalings in \eqref{eq:mainPhiScaling} and
\eqref{eq:mainPhiScalingDB}. One finds, in the corresponding
large-$L$ right-tail regime,
\be
\Phi_L(H)
\simeq
2\,\Phi_L^{N=1}\left(\frac{H}{2}\right).
\label{eq:mainRightTailPhysicalComparisonSec6}
\ee
For a single polymer, the $3/2$ right tail has the standard
optimal-fluctuation interpretation in terms of a favorable localized
corridor, whose transverse scale results from a balance between
confinement and disorder costs. The corresponding optimal fluctuation
has a solitonic structure, as found within optimal-fluctuation and
weak-noise approaches, inverse scattering, and rigorous weak-noise
large-deviation theory
\cite{kolokolov2007optimal,kolokolov2009explicit,meerson2016large,
krajenbrink2021inverse,gaudreaulamarre2023kpz,tsai2023integrability}.
For two non-crossing polymers, it is instead favorable to create two
spatially separated such corridors, each supporting one polymer and
producing half of the total excess in $\log Z$. The two disorder costs
then add: the factor $H/2$ in
\eqref{eq:mainRightTailPhysicalComparison} reflects the equal sharing
of the free-energy fluctuation between the two polymers, while the
overall factor $2$ reflects the cost of creating the two favorable
corridors. The non-crossing constraint is crucial, since it leads to
two orthogonal fermionic one-particle states and hence favors this
splitting. An optimal-fluctuation derivation of
\eqref{eq:mainRightTailPhysicalComparison} is given in
App.~\ref{app:rightTailOF}.

\paragraph{Global comparison with the one-polymer scaling function.}
Figure~\ref{fig:scalingfunction} in Subsec.~\ref{susubsec:largeL} compares ${\cal E}(\epsilon)$ on both sheets with ${\cal E}^{N=1}(\epsilon)$,
continued beyond $\xi=1$ as above. With the normalization of \eqref{eq:mainEnergyScalingDB}, ${\cal E}^{N=1}$ is
exactly the scaling function of two polymers in \emph{independent} environments,
$2E^{N=1}(n,L)=\frac{nc^2}{3L^2}+\frac{\sqrt c}{L^2}{\cal E}^{N=1}(n\sqrt c)$ having the same extensive part as
\eqref{eq:mainEnergyScaling}, so that the comparison isolates the effect of the non-crossing constraint on the
fluctuating part; the other natural reference, two unconstrained polymers in the \emph{same} environment, has
${\cal E}_{\rm b}(\epsilon)=\frac12{\cal E}^{N=1}(2\epsilon)$ and $\Phi_{\rm b}(Y)=\frac12\Phi^{N=1}(Y)$. The deep
lower tail of the non-crossing pair coincides with the latter, \eqref{eq:mainScaledPhiComparison}, and the upper
tail with the former, \eqref{eq:mainRightTailPhysicalComparison}. The ratio ${\cal E}/{\cal E}^{N=1}$ (middle
panel) increases monotonically from $2^{2/3}$ at $\epsilon\to-\infty$, \eqref{eq:mainEComparisonNegative} (it
equals $1.65$ at $\epsilon=-3.1$, the end of the first-sheet data), to $2$ at $\epsilon\to0$ (the mean free
energy, $e_1=4$ against $2$), with slope $(e_2e_1^{N=1}-e_1e_2^{N=1})/(e_1^{N=1})^2=0.392$ at the origin, which
encodes the enhancement $11/8$ of the variance with respect to two polymers in independent environments ($11/4$
with respect to one, \eqref{eq:mainc2ratioDB}). For $\epsilon>0$ the two functions behave differently at their
branch points: ${\cal E}^{N=1}$ has its maximum exactly at $\epsilon_c^{N=1}$, where
$d{\cal E}^{N=1}/d\epsilon=2{\rm Li}_{3/2}(\xi)/{\rm Li}_{1/2}(\xi)$ vanishes, while ${\cal E}$ crosses
$\epsilon_c$ with finite slope and reaches its maximum $2.2023$ at $\epsilon=1.072$ on the second sheet; both tend
to $-\epsilon^3/3$, so that the ratio tends to $1$. Correspondingly (right panel) the scaled rate functions have
their minima at $Y=-4$ and $Y=-2$, $\Phi(Y)\simeq\frac12\Phi^{N=1}(Y)$ for $Y\to-\infty$, and both behave as
$\frac23Y^{3/2}$ for $Y\to+\infty$, the two-polymer function being larger by $\pi^2Y^{-1/2}$.

\section{Conclusion}

We have developed a finite-volume replica approach to the calculation of the free energy of two non-crossing continuum directed polymers on a cylinder.
It extends the seminal work of Brunet and Derrida \cite{brunet2000ground}
for a single directed polymer on the cylinder. The problem is
mapped to the attractive Lieb--Liniger model for $2 n$ particles, in the symmetry sector associated
with the Young diagram $(n,n)$.
It is treated using the two-component bosonic
nested Bethe ansatz involving
$2n$ physical rapidities and $n$ auxiliary rapidities.
By reformulating the nested Bethe equations in terms of
functional relations adapted to analytic continuation in the number of replica, we obtained a systematic expansion around $n=0$.
A remarkable simplifying feature of the calculation
is that the auxiliary rapidities
can be completely eliminated.
This yields explicit
expressions for the mean and variance of the two-polymer free energy
at long time and arbitrary circumference $L$, together with a scheme for computing the higher cumulants.

We also studied the joint limit $L\to\infty$, $n\to0$, with
$n L^{1/2}$ fixed. Extending the Brunet--Derrida analysis and its
Wiener--Hopf formulation to the two-layer problem, we derived a universal
scaling function governing the long-time large-deviation statistics of the
two-polymer free energy. This provides a nontrivial multilayer generalization
of the single-polymer periodic KPZ problem and should characterize the
corresponding universal scaling regime of the one-dimensional two-layer
periodic SHE/KPZ class.

This work calls for several natural extensions. The first one would be to adapt the present calculation to the interval, as was done in \cite{BarraquandLeDoussal2025KPZInterval} for a single polymer. A challenging next step is to extend the construction to an arbitrary number
$N$ of non-crossing polymers. The relevant replica sector is then associated
with the rectangular Young diagram $(n^N)$ and involves a higher-rank nested
Bethe ansatz. It would be interesting to determine whether the functional
relations and analytic-continuation strategy developed here can be generalized
to that setting, and to clarify their relation to the thermodynamic large-$N$,
large-$L$ regime considered in earlier fermionic approaches.

Another interesting open direction would be to obtain an independent microscopic
derivation of the results presented here from an integrable discrete model.
At a first level, one would like to recover the finite-$L$ two-layer cumulants
by a scaling limit from a discrete or semi-discrete multilayer polymer, in the
same spirit in which exact results for the open ASEP provide an independent
derivation of the one-layer KPZ results on an interval. There are several
natural starting points. Non-intersecting discrete polymers are known to
converge in the intermediate-disorder regime to the multilayer SHE
\cite{corwin2017intermediate}, with analogous results for the multilayer
semi-discrete O'Connell--Yor polymer \cite{nica2021intermediate}, while the
Hall--Littlewood RSK construction of \cite{bufetov2018hall} provides an
integrable multilayer ASEP with exact joint observables and a conjectured
scaling relation to the multilayer SHE. What seems to be missing at present
is a finite-volume tilted-generator or deformed matrix-product formulation
giving direct access to the long-time cumulant generating function of the
two-layer free energy.

A stronger test of universality would be to derive directly from a microscopic
periodic model the large-$L$ scaling function obtained in this work, in
analogy with the agreement between the Brunet--Derrida scaling function for a
single polymer and the universal ASEP/TASEP current large-deviation function.
For the two-layer problem such a derivation appears to remain open. Recent
developments on periodic O'Connell--Yor and inverse-gamma polymers, periodic
Pitman transforms, and multipath partition functions
\cite{corwin2026periodicpitman,engel2026periodic}
provide promising structures from which such a program might eventually be
developed. Establishing a discrete realization of this type would provide a
particularly stringent test of the universality of the two-layer scaling
function derived here.
In the meantime, it would be interesting to perform lattice simulations
of two non-crossing polymers in order to compute the full
large-deviation rate function, providing a further test of our results
and of their universality. The lattice simulations performed here in
the high-temperature weak-disorder scaling regime, following the
approach of Ref.~\cite{calabrese2010free}, already test our predictions
for $c_1(L)$ and $c_2(L)$ at the percent level. Extending such
simulations to rare fluctuations would provide a direct numerical test
of the large-deviation scaling function derived in this work.

\subsection*{Acknowledgments}  We thank G. Barraquand for discussions and ongoing collaborations on related topics, and A. Borodin for pointing out useful references. 
P. Le Doussal was supported by ANR grant ANR-23-CE30-0020-01 (EDIPS).
We both thank the Harvard University Center of Mathematical Sciences and
Applications, for hospitality and partial support during the 2025 program Classical, Quantum, and
Probabilistic Integrable Systems.

AI statement: all the important results of this work were obtained before use of AI. AI was used to help with final writing, bibliography, efficient code for numerics and plots.

\printbibliography

\newpage
\appendix
\begin{center}
    \Large\bfseries Appendix
\end{center}

\section{Lattice simulations}
\label{app:lattice}

In this appendix we describe the lattice model used to test numerically the finite-$L$ predictions
\eqref{c1L} and \eqref{eq:c2exactIntro}, and in particular a parity effect which is specific to two polymers
on a bipartite lattice and must be taken into account for the continuum limit to reproduce the model of
Sec.~\ref{subsec:model}.

\paragraph{Model and continuum limit.}
We consider directed polymers on the square lattice with i.i.d.\ Gaussian site energies $V_{X,\hat t}$ of
variance $\sigma^2$, in the doubled transverse coordinate $X=i-j\in\{0,\dots,W_X-1\}$, which changes by
$\pm1$ at each time step $\hat t\to\hat t+1$. The one-polymer partition sum obeys
\be
\hat z_{\hat t+1}(X)=e^{-V_{X,\hat t+1}/T}\big[\hat z_{\hat t}(X-1)+\hat z_{\hat t}(X+1)\big],
\qquad
\hat z_{\hat t}(X+W_X)=\Phi\,\hat z_{\hat t}(X),
\label{eq:latticerecursion}
\ee
with $\Phi=(-1)^{N-1}=-1$ for $N=2$: a polymer which winds once around the cylinder picks up the sign of
Eq.~\eqref{eq:periodicity}. The circumference $W_X$ must be even, so that the periodic identification is
compatible with the bipartite structure of the lattice. In the high-temperature scaling
\be
x=\frac{2}{T^2}X,\qquad t=\frac{4}{T^4}\hat t,\qquad L=\frac{2}{T^2}W_X,
\label{eq:latticescaling}
\ee
the normalized partition sum
$\tilde z_{\hat t}=[2e^{\sigma^2/(2T^2)}]^{-\hat t}\hat z_{\hat t}$ converges, as $T\to\infty$ at fixed $(x,t,L)$,
to the solution of the SHE \eqref{SHEsingle} with periodic noise and
\be
\gamma=\sigma^2,\qquad c=\frac{\gamma L}{2}=\frac{L\sigma^2}{2}.
\label{eq:latticedictionary}
\ee
(The additional factor $2$ in $t=4\hat t/T^4$ comes from the doubled coordinate; it is fixed unambiguously by the exact
free benchmark discussed below.) For given target values of $L$ and $c$ one therefore chooses
$W_X=2\lceil LT^2/4\rceil$ and $\sigma=\sqrt{2c/L}$, with $L$ the value actually realized by $W_X$.

\paragraph{Two non-crossing polymers.}
For one disorder realization, the two-polymer partition sum with the non-crossing constraint and the cyclic
summation of Sec.~\ref{sec:determinant} is obtained from the lattice Karlin--McGregor determinant
\be
\hat Z^{\rm NC}_{\hat t}(X_1,X_2|Y_1,Y_2)=
\det\begin{pmatrix}
\hat z_{\hat t}(X_1|Y_1) & \hat z_{\hat t}(X_1|Y_2)\\
\hat z_{\hat t}(X_2|Y_1) & \hat z_{\hat t}(X_2|Y_2)
\end{pmatrix},
\label{eq:latticeKM}
\ee
where $\hat z_{\hat t}(X|Y)$ solves \eqref{eq:latticerecursion} with the twist $\Phi=-1$ and initial
condition $\delta_{X,Y}$: this is the lattice version of \eqref{eq:KarlMcGregCyc}. The determinant is
antisymmetric in $X_1\leftrightarrow X_2$, hence vanishes at coincidence, and the twist implements the sign of the
winding sectors. In practice the two one-polymer sums are propagated in the \emph{same} disorder with a cost
$O(W_X)$ per step, and the determinant is evaluated at the recording times. Numerically one works with
$\log|\hat Z^{\rm NC}|$, keeping track of the overall scale of the vectors to avoid overflow, and with the
deterministic normalization $[2e^{\sigma^2/(2T^2)}]^{-2\hat t}$. Disorder averages are performed over
$S$ independent realizations, which we propagate simultaneously as arrays of shape $(S,2,W_X)$.

One numerical subtlety deserves to be mentioned. The two one-polymer vectors
$\hat z_{\hat t}(\cdot|Y_1)$ and $\hat z_{\hat t}(\cdot|Y_2)$ are propagated by the same random transfer
matrices and for both of them, the leading contribution is the dominant Lyapunov vector of their product. The determinant
\eqref{eq:latticeKM} is therefore the difference of two nearly equal products and, with finite machine precision, is lost in rounding noise once $t$ is sufficiently large. Since the determinant is invariant under
$\hat z(\cdot|Y_2)\to\hat z(\cdot|Y_2)-\alpha\,\hat z(\cdot|Y_1)$, and this invariance is preserved by the
linear evolution, we remove at every rescaling step the component of the second vector along the first
(Gram--Schmidt, as in the computation of Lyapunov exponents), with $\alpha$ chosen to minimize its norm. The
two vectors are then rescaled separately and the determinant is computed without cancellation at arbitrarily
long times. Without this precaution the data at $ct/L^2\gtrsim 6$ are unreliable.

\paragraph{Parity of the endpoints.}
One important observation is that the lattice is bipartite and the discretization conserves the parity of
$X+\hat t$ along each path: a polymer starting at $Y$ lives forever on the sublattice of parity $Y$. With an odd
initial separation, such as adjacent starting sites, the two polymers (in every term of the determinant and in
every replica) live on the two complementary sublattices, which carry independent site energies: they never occupy
the same site, the contact interaction generated by the disorder average is never realized between them, and one
simulates, at any lattice spacing, two polymers in \emph{independent} environments (two decoupled Lieb--Liniger
gases at the replica level), with a variance $4c$ instead of $8c$ at small $c$. The initial separation must therefore be even: all runs
use $(Y_1,Y_2,X_1,X_2)=(0,2,0,2)$, with the observables recorded at even $\hat t$.

\paragraph{Free benchmark.}
For $n=1$ the disorder average of the normalized $\hat Z^{\rm NC}$ is deterministic: it is the determinant
\eqref{eq:latticeKM} built with the flat antiperiodic walk $\hat z_{\hat t+1}(X)=\frac12[\hat z_{\hat t}(X-1)+
\hat z_{\hat t}(X+1)]$. Its large-time decay rate converges, with the scaling \eqref{eq:latticescaling}, to the
free non-crossing value $E(1,L)=\pi^2/L^2$ of Sec.~\ref{sec:replicaN2} (deviation $1.6\cdot10^{-3}$ at $T=5$,
$L=2.56$, $t=125$), which fixes the dictionary \eqref{eq:latticescaling}. The single-polymer problem on the same
lattice reproduces the Brunet--Derrida finite-volume energies, with $c=L\sigma^2/2$, which fixes
\eqref{eq:latticedictionary}.

\paragraph{Extraction of the cumulants.}
At each recording time $t_j$ we compute the sample cumulants $\kappa_k(t_j)$, $k=1,\dots,4$, of
$\log\hat Z^{\rm NC}_{\rm norm}(t_j)$ over the $S$ realizations (unbiased $k$-statistics). After a transient of
order $L^{3/2}/\sqrt\gamma$ they grow linearly in $t$, and $c_k(L)$ is obtained as the slope of a least-squares fit
over the second half of the time window, $t\in[t_{\max}/2,t_{\max}]$ with $t_{\max}\simeq10\,L^{3/2}/\sqrt\gamma$;
errors are estimated by bootstrap: the whole procedure (cumulant time series and slope fit) is repeated on $200$
resampled sets of $S$ realizations drawn with replacement from the $S$ available ones, and the error is the standard
deviation of the resulting slopes; this takes into account the correlation between the cumulants at different
times, which are computed from the same samples. The simulated values of $c$ (and the corresponding $L$ and $T$), chosen so that the $O(c^2)$ terms of
\eqref{c1L} and \eqref{eq:c2exactIntro} are at least a few
percent of the $O(c)$ ones, are listed in Table~\ref{tab:lattice}; the two values of $L$ at $c=1$ (i.e.\ two values
of $\sigma$) and the two values of $T$ control the residual lattice corrections.

Three conclusions can be drawn. (i) The exact variance \eqref{eq:c2exactIntro} is confirmed, at the level of
$0.5$--$1\%$, for $0.5\le c\le3$: the measured $L^2c_2$ agree with it within $1$--$1.5$ standard deviations at all
points. The data resolve the terms beyond the linear order $8c$ with high significance (the deviation from $8c$
amounts to $9\sigma$ at $c=0.5$, where the $c^2$ term is $4\%$ of the total, to $10\sigma$ at $c=1$ and to more than
$18\sigma$ at $c=2,3$, where it is $16\%$ and $23\%$), i.e.\ they confirm the expansion beyond the linear order up to
and including the quadratic term $2c^2/3$. The terms of higher order, which amount to $-0.5$ at $c=3$ ($-1.8\%$ of the
total), are at the limit of resolution: there the quadratic truncation $8c+2c^2/3$ is disfavoured by $3.2\sigma$,
while the full result is within $0.7\sigma$. (ii) The
$c^2/3$ term of $L^2c_1$ (equal to $0.33$ at $c=1$ and $3$ at $c=3$) is likewise confirmed. Note that, for the first cumulant $c_1$, at too small values of $c$, (e.g. $c = 0.1$), the discretization of the kinetic term becomes relevant, the
free two-polymer energy on the lattice being $-2\log\cos(\pi/W_X)$ per step, i.e.\ $\pi^2/L^2+(2\pi^4/3)/(L^4T^4)+\ldots$ in continuum units. This is a deterministic shift which affects the average but is irrelevant for the higher cumulants.

The third and fourth cumulants are
not resolved by this campaign. Since the errors decrease only as
$S^{-1/2}$, a $3\sigma$ determination of $c_3$ at $c=3$ would require $3\cdot10^6$--$10^7$ realizations and $c_4$
is out of reach; the higher cumulants, and the rate function itself, are better studied by importance-sampling
methods that target the large deviations directly, which we leave for future work.

\begin{table}[t]
\centering
\begin{tabular}{ccccc|cc|ccc}
$c$ & $L$ & $T$ & $\sigma$ & $S$ & $-L^2c_1$ meas. & \eqref{c1L} & $L^2c_2$ meas. & \eqref{eq:c2exactIntro} & $8c$ \\
\hline
0.1 & 2.56 & 5 & 0.280 & $1.1\cdot10^5$ & $10.2877(9)$ & $10.2729$ & $0.795(6)$ & $0.807$ & $0.80$ \\
0.5 & 5 & 6 & 0.447 & $5.2\cdot10^5$ & $11.9500(12)$ & $11.9529$ & $4.140(15)$ & $4.164$ & $4.00$ \\
1 & 10 & 6 & 0.447 & $1.2\cdot10^5$ & $14.1842(45)$ & $14.2029$ & $8.738(68)$ & $8.644$ & $8.00$ \\
1 & 10 & 8 & 0.447 & $1.2\cdot10^5$ & $14.2000(44)$ & $14.2029$ & $8.684(65)$ & $8.644$ & $8.00$ \\
1 & 20 & 6 & 0.316 & $1.2\cdot10^5$ & $14.1996(53)$ & $14.2029$ & $8.662(70)$ & $8.644$ & $8.00$ \\
2 & 20 & 6 & 0.447 & $1.2\cdot10^5$ & $19.1666(82)$ & $19.2029$ & $18.47(14)$ & $18.500$ & $16.00$ \\
3 & 24 & 6 & 0.500 & $1.2\cdot10^5$ & $24.802(12)$ & $24.8696$ & $29.33(21)$ & $29.468$ & $24.00$ \\
\end{tabular}
\caption{Lattice results for the first two cumulants (slope of $\kappa_k(t)$ over $[t_{\max}/2,t_{\max}]$,
$t_{\max}=125,300,800,800,2500,2000,2400$ from top to bottom, bootstrap errors on the last digits) compared with
the exact finite-$L$ formulas \eqref{c1L} and \eqref{eq:c2exactIntro}; the last column is the $O(c)$ part of
$L^2c_2$. $S$ is the number of disorder realizations, $\sigma=\sqrt{2c/L}$ the lattice disorder strength.}
\label{tab:lattice}
\end{table}

\section{Row and column versions of the nested Bethe ansatz}
\label{app:row_column_nba}

Given the symmetry properties we required of the wave function, we considered a rectangular representation \eqref{eq:Trect} of $\SSS_M$ with $N$ rows and $n$ columns and $M = n N$. To study the spectrum of the Lieb-Liniger Hamiltonian in this symmetry class, it was convenient to perform a ‘row-by-row’ iteration of the Nested Bethe ansatz: each nesting corresponds to removing the top row of the diagram. For the case of $N=2$ polymers, this leads to two levels and the two equations \eqref{NBAexplicit}. This approach is useful in that the number of equations $N$ remains fixed, and we were able to implement the Derrida-Brunet analytical approach on a fixed number of equations, subsequently performing the analytical continuation in the variable $n$. However, in the literature on the Nested-Bethe ansatz, a ‘column-by-column’ iteration is often used instead~\cite{yang1967some}, even in the context of polymers in random media~\cite{emig2001probability}. In this appendix, we clarify the relationship between the two approaches.

Let $\lambda$ be a Young diagram with $M$ boxes, and let $V_\lambda$ denote the corresponding irreducible representation of $\SSS_M$. In the notation of Sec.~\ref{subsec:periodicity_transfer}, restricting the amplitude vector $A^I$ to $V_\lambda$ means replacing the right regular action $\mathsf r$ in Eq.~\eqref{XYang} by its action on $V_\lambda$. We will denote this restricted action by $\mathsf r_\lambda$. Thus Eq.~\eqref{XYang} restricted to the irreducible representation $\lambda$ is denoted as
\be
\label{app:Xlambda}
X_{\mu\nu}^{(\lambda)}
= f(k_{\nu\mu})
  +\bigl(f(k_{\nu\mu})-1\bigr)\mathsf r_\lambda(\tau_{\mu\nu}) .
\ee
Correspondingly, the commuting transfer matrices appearing in Eq.~\eqref{XYangBethe} are
\be
\label{app:Tlambda}
\mathcal T_\mu^{(\lambda)}
=X_{\mu+1,\mu}^{(\lambda)}X_{\mu+2,\mu}^{(\lambda)}\cdots
X_{M,\mu}^{(\lambda)}
X_{1,\mu}^{(\lambda)}\cdots X_{\mu-1,\mu}^{(\lambda)} ,
\ee
and the Bethe equations require
\be
\label{app:Tlambdaeig}
\mathcal T_\mu^{(\lambda)} A^I
= \Phi^{-1} e^{i k_\mu L} A^I,
\qquad \mu=1,\ldots,M .
\ee

Let us now recall what transposition means for a Young diagram in the
case most relevant for the column formulation. Consider for instance a
diagram with two columns,
\(\lambda=(2^{\tilde n},1^{M - 2\tilde n})\). The transposed diagram \(\lambda^t\) is obtained
by exchanging rows and columns, hence it has two rows,
\(\lambda^t=(M - \tilde n, \tilde n)\). For $\tilde n = 3$ and $M = 7$, we have
\be
\label{app:young_transpose_example}
\Yvcentermath1
\lambda=\yng(2,2,2,1)
\qquad \longrightarrow \qquad
\lambda^t=\yng(4,3).
\ee
Equivalently, the height of the $j$-th column of \(\lambda\) becomes the
length of the $j$-th row of \(\lambda^t\).

The key identity is that the representation associated with the transposed Young diagram $\lambda^t$ is obtained by tensoring with the sign representation,
\be
\label{app:conjugate}
V_{\lambda^t}\simeq V_\lambda\otimes {\rm sgn}.
\ee
Therefore, after identifying the two vector spaces through this tensoring, every transposition changes sign:
\be
\label{app:signtwist}
\mathsf r_{\lambda^t}(\tau_{\mu\nu})
=-\mathsf r_\lambda(\tau_{\mu\nu}) .
\ee
Inserting Eq.~\eqref{app:signtwist} in Eq.~\eqref{app:Xlambda} gives
\be
\label{app:Xconjugate}
X_{\mu\nu}^{(\lambda^t)}
= f(k_{\nu\mu})
  -\bigl(f(k_{\nu\mu})-1\bigr)\mathsf r_\lambda(\tau_{\mu\nu}) .
\ee
The advantage of this expression is that it allows expressing the two-body operator for the transpose representation $\lambda^t$ in terms of the representation $\lambda$.
Additionally, one can explicitly relate
\begin{equation}
X_{\mu\nu}^{(\lambda)}(\cb)
=
\frac{k_\nu-k_\mu-i\cb}{k_\nu-k_\mu+i\cb}\,
X_{\mu\nu}^{(\lambda^t)}(-\cb) = S(k_{\nu\mu}; \gamma) X_{\mu\nu}^{(\lambda^t)}(-\cb)
\;,
\end{equation}
where we wrote explicitly the dependence on the interaction $\gamma$ and $S(k; \gamma)$ was already introduced in Eq.~\eqref{XfromR}.
One obtains, for the transfer-matrices
\be
\label{app:YangRelationCurrentNotation}
\mathcal{T}_{\mu}^{(\lambda)}(\gamma) =
\left[\prod_{\nu\neq\mu}
S(k_{\mu\nu}; \gamma)\right]
\,
\mathcal{T}_{\mu}^{(\lambda^t)}(-\gamma)  \;.
\ee
As a consequence, we can relate the eigenvalues of the transfer matrices $\mathcal{T}_\mu^{(\lambda)}$ through those of $\mathcal{T}_\mu^{(\lambda^t)}$ by changing the sign of the interaction $\gamma \to -\gamma$ and including the scattering phase. For instance, for the example of Eq.~\eqref{app:young_transpose_example}, $\lambda^t$ is once again a two-row Young diagram that can be treated directly with the procedure of Sec.~\ref{sec:nbaeq}. Specifically, using Eq.~\eqref{XYangABA} to express $\mathcal{T}_{\mu}^{(\lambda^t)}(-\gamma)$, we have
\begin{equation}
    \mathcal{T}_{\mu}^{(\lambda)}(\gamma)  = F^{(\lambda^t)}(k_\mu, \vec{k}; -\gamma)
\end{equation}
where $F^{(\lambda^t)}(k_\mu, \vec{k}; -\gamma)$ denotes the commuting transfer matrices introduced in Sec.~\ref{sec:nbaeq} (above \eqref{eq:FFcomm}) specified for the representation $\lambda^t$. Note that the product of factors $S(k; \gamma)$ cancels between Eq.~\eqref{app:YangRelationCurrentNotation} and \eqref{XYangABA}, using that $S(k; \gamma) S(k; -\gamma) = 1$.
Specifically, for $\lambda^t = (M - \tilde n, \tilde n)$, such a representation can be obtained by looking at Bethe states with $\tilde n$ auxiliary rapidities. Denoting as $\Lambda(\lambda, \{\lambda_a\}_{a=1}^{\tilde n} ; -\gamma)$ the eigenvalue of $F(\lambda; -\gamma)$ on the corresponding Bethe state \eqref{eq:bethestates}, we can rewrite Eqs.~\eqref{app:Tlambdaeig} as
\begin{equation}
    \Lambda(k_\mu, \{\lambda_a\}_{a=1}^{\tilde n} ; -\gamma) = \Phi^{-1} e^{i k_\mu L}  \;.
\end{equation}
Again, note that, in contrast with \eqref{NBA}, the  scattering phases are removed. From this procedure, in analogy with Sec.~\ref{sec:nbaeq} and Eqs.~\eqref{NBAexplicit0}, we can derive the Nested Bethe equations for $\lambda = (2^{\tilde n} 1^{M - 2 \tilde n})$ as (in this case we shift $\lambda \to \lambda - i \gamma /2$)
\begin{subequations}
    \label{NBAexplicitFerm}
\begin{align}
&\prod_{b\neq a} \frac{\lambda_a - \lambda_b - i \cb}{\lambda_a - \lambda_b + i \cb} =
\prod_{\mu} \frac{\lambda_a - k_\mu - i \cb/2}{\lambda_a - k_\mu + i \cb/2}
\quad , \quad a=1,\dots,\tilde{n}
\label{eq:auxferm}
\\
& \prod_{a} \frac{k_\mu - \lambda_a - i \cb/2}{k_\mu  - \lambda_a + i \cb/2}
= \Phi^{-1} e^{i k_\mu L} \quad , \quad \mu=1,\dots, M
\label{eq:physferm}
\end{align}
\end{subequations}
Note that when compared with the equations \eqref{NBAexplicit} for $\lambda = (M - \tilde n, \tilde n)$, the auxiliary equations (\ref{eq:auxbos},\ref{eq:auxferm}) are equivalent, since the exchange $\gamma \to - \gamma$ is equivalent to taking the reciprocal of both sides. Conversely, for Eqs.~(\ref{eq:physbos},\ref{eq:physferm}
), the difference between the two cases lies in the elimination of the scattering phase between the physical particles alongside the swap $\gamma \to -\gamma$. This change is natural if one considers the relation linking $\lambda \leftrightarrow \lambda^t$ as a mapping between bosons and fermions. For fermions, the antisymmetry of the wave function means that the delta interaction is inessential. This procedure, which can link the diagonalisation of a symmetry sector $\lambda$ to that of its conjugate $\lambda^t$, can be generalised to diagrams with multiple rows/columns. In general, it remains true that all the auxiliary equations remain unchanged, whilst the equation determining the physical roots differs substantially due to the elimination of the scattering phase and the sign change $\gamma \to -\gamma$. See \cite{oelkers2006bethe} for the complete form of the Bethe equations in both cases.
It is interesting to note that with the full nested procedure, one has thus the possibility to diagonalize the problem with two equivalent procedure: either the 'row-by-row' nesting applied directly to $\lambda$, or by applying the 'column-by-column' nesting by using the correspondence explained here $\lambda \to \lambda^t$. The two procedures diagonalize the same Hamiltonian \eqref{eq:HLL} in the same symmetry class and thus produce the same set of physical momenta $\{k_\mu\}_\mu$ for each eigenstate. However, the relation of the auxiliary variables is highly non-trivial. Nonetheless, we verified in a few cases for small diagrams that perfect agreement emerges as expected.

\section{Ordering of the symmetry sectors}
\label{app:LiebMattis}

In this appendix we show that, for every value of $\cb$, the ground state of $H_{\rm LL}$ in the
rectangular sector $(n^N)$ lies strictly below the ground states of all the other sectors $\lambda$ with
$\lambda_1\le n$. These include all the sectors compatible with Eq.~\eqref{eq:PsiCond}. We work on the
ring with the boundary condition \eqref{eq:periodicitycond}. The argument is a variant of the one of Lieb
and Mattis \cite{lieb1962ordering}. Instead of their spin (here: colour) degrees of freedom, it uses
only the permutation symmetry of the replicated wave function, in line with the rest of the paper. For
concreteness we illustrate each step on the case $n=N=2$.

\subsection{Symmetry sectors and their ground-state energies}

Let $\mathcal H$ be the space of square-integrable functions $\Psi(\XX)$ of the $M=nN$ coordinates
$x_{i,a}$ ($i=1,\ldots,N$ labels the polymer, $a=1,\ldots,n$ the replica, as in the filling
\eqref{eq:Trect}) obeying Eq.~\eqref{eq:periodicitycond}. The group $\SSS_M$ acts on $\mathcal H$ by
permuting the coordinates, $(R(S)\Psi)(\XX)=\Psi(S^{-1}\XX)$, and $R(S)$ commutes with $H_{\rm LL}$.

For each partition $\lambda\vdash M$, let $V_\lambda$ be the corresponding irreducible representation, of
dimension $d_\lambda$, with matrices $D_\lambda(S)$ in a basis $e_1,\ldots,e_{d_\lambda}$. A
\emph{multiplet of type $\lambda$} is a $d_\lambda$-tuple of functions $\Psi_k=\phi(e_k)$ transforming as
this basis, $R(S)\phi(v)=\phi(D_\lambda(S)v)$. In other words, it is a linear map
$\phi:V_\lambda\to\mathcal H$ intertwining the two actions. Call $\mathcal M_\lambda$ the space of such
maps. The \emph{sector} $\lambda$ is the span $\mathcal H_\lambda$ of all the components of all the
multiplets of type $\lambda$. Equivalently, $\mathcal H_\lambda=P_\lambda\mathcal H$ with the central
projector $P_\lambda=\frac{d_\lambda}{M!}\sum_{S}\chi_\lambda(S)R(S)$, and
$\mathcal H=\bigoplus_\lambda\mathcal H_\lambda$. The map $v\otimes\phi\mapsto\phi(v)$ identifies
\be
\label{eq:LMisotypic}
\mathcal H_\lambda\cong V_\lambda\otimes\mathcal M_\lambda\;,\qquad
H_{\rm LL}\big|_{\mathcal H_\lambda}=\ide\otimes h_\lambda\;,\qquad h_\lambda\phi:=H_{\rm LL}\circ\phi\;.
\ee
The permutations act on the first factor only, and the dynamics on the second one. An eigenvector of
$h_\lambda$ is a whole multiplet of eigenfunctions with the same energy. The Bethe states of the diagram
$\lambda$ (one for each choice of the Bethe quantum numbers) are precisely these multiplets. The wave
function constructed in Sec.~\ref{subsec:youngsymmetrizer} with a given filling is one particular
component $\phi(v)$, $v\in V_\lambda$, of such a multiplet. We define
\be
E^\lambda:=\min{\rm spec}\,h_\lambda=\min{\rm spec}\,\big(H_{\rm LL}\big|_{\mathcal H_\lambda}\big)\;.
\ee
The energy $E^{(N)}(n,L)$ of Eq.~\eqref{eq:En_sector_def} is the minimum of $E^\lambda$ over the sectors
compatible with \eqref{eq:PsiCond}.

\subsection{The enlarged space \texorpdfstring{$\mathcal H_Y$}{HY}}

We keep only the antisymmetry within each replica, i.e.\ Eq.~\eqref{eq:antisymm}, and we drop
the symmetry between replicas, Eq.~\eqref{eq:replicasymm}:
\be
\label{eq:LMHY}
\mathcal H_Y=\big\{\Psi\in\mathcal H:\ \Psi(S\XX)=(-1)^S\,\Psi(\XX)\quad\forall S\in\BBB_{\trect}\big\}\;,
\ee
with $\BBB_{\trect}\cong(\SSS_N)^n$ the column group \eqref{subgroupB}. A function of $\mathcal H_Y$ has no
prescribed symmetry under the exchange of particles belonging to different replicas. The physical
replicated wave function lies in $\mathcal H_Y$, and in fact in a smaller subspace. The projector on
$\mathcal H_Y$ commutes with the central projectors $P_\lambda$ and acts on the factor $V_\lambda$ only,
so that
\be
\label{eq:LMHYdec}
\mathcal H_Y=\bigoplus_\lambda V_\lambda^{Y}\otimes\mathcal M_\lambda\;,\qquad
V_\lambda^{Y}=\{v\in V_\lambda:\ D_\lambda(S)v=(-1)^Sv\ \ \forall S\in\BBB_{\trect}\}\;.
\ee
As a trivial but relevant remark, since $H_{\rm LL}$ acts as $\ide\otimes h_\lambda$, each nonzero block $V^Y_\lambda\otimes\mathcal M_\lambda$
carries the full spectrum of $h_\lambda$; in particular, it contains states of energy $E^\lambda$.

\medskip\noindent\textbf{Lemma.} \emph{$m_\lambda:=\dim V_\lambda^{Y}=K_{\lambda',(N^n)}$, the Kostka
number of the conjugate diagram $\lambda'$ and of the weight $(N^n)$. In particular, $m_\lambda>0$ if and
only if $\lambda_1\le n$.}

\medskip\noindent\emph{Proof.} By Frobenius reciprocity, $m_\lambda$ is the multiplicity of $V_\lambda$ in
the representation induced from the sign of $\BBB_{\trect}$. Since $V_\lambda\otimes{\rm sgn}=V_{\lambda'}$,
this is the multiplicity of $V_{\lambda'}$ in the representation induced from the trivial representation of
the Young subgroup $(\SSS_N)^n$, which by Young's rule is $K_{\lambda',(N^n)}$
\cite{macdonald1995symmetric}. Finally, $K_{\nu\mu}>0$ if and only if $\nu$ dominates $\mu$, and a
partition $\nu$ of $nN$ dominates $(N^n)$ if and only if it has at most $n$ rows. For $\nu=\lambda'$ this
means $\lambda_1\le n$. $\square$

\medskip
This makes precise the statement of Sec.~\ref{subsec:youngsymmetrizer} based on Pieri's rule, and adds
that \emph{all} the diagrams with at most $n$ columns occur. Combining the Lemma with the remark after Eq.~\eqref{eq:LMHYdec},
\be
\label{eq:LMminHY}
\min{\rm spec}\big(H_{\rm LL}\big|_{\mathcal H_Y}\big)=\min_{\lambda_1\le n}E^\lambda\;.
\ee
For $n=N=2$ the diagrams with $\lambda_1\le2$ are $(2,2)$, $(2,1,1)$ and $(1^4)$, each with $m_\lambda=1$.
Of the three diagrams, $(2,2)$ and $(1^4)$ are compatible
with the full set of conditions \eqref{eq:PsiCond}. In $(2,1,1)$, instead, the vector antisymmetric
within the replicas changes sign under the exchange of the two replicas. Proving the ordering in the
larger space $\mathcal H_Y$ is a stronger statement. The sectors $(4)$ and $(3,1)$ are not contained in
$\mathcal H_Y$: the ordering established here holds only among the diagrams with $\lambda_1\le n$.

\subsection{Ordering each replica: the domain \texorpdfstring{$\mathcal D_Y$}{DY}}

Consider first the infinite line. A function $\Psi\in\mathcal H_Y$ is determined by its values on
\be
\label{eq:LMDY}
\mathcal D_Y=\{\XX:\ x_{1,a}<x_{2,a}<\dots<x_{N,a}\ \ \text{for every } a\}\;,
\ee
since any configuration with distinct coordinates within each replica is brought into $\mathcal D_Y$ by a
unique element of $\BBB_{\trect}$. The hyperplanes $x_{i,a}=x_{j,b}$ fall into two classes.
\begin{itemize}
\item \emph{Same replica} ($a=b$). On the other side of the hyperplane the values of $\Psi$ are fixed
by antisymmetry, and $\Psi=0$ on it. These hyperplanes form the boundary $\partial\mathcal D_Y$, with
Dirichlet condition. The delta interaction between particles of the same replica is therefore
ineffective.
\item \emph{Different replicas} ($a\ne b$). No symmetry relates the values of $\Psi$ on the two sides,
which are independent points of $\mathcal D_Y$. These hyperplanes lie in the interior of $\mathcal D_Y$.
There $\Psi$ is continuous and its normal derivative jumps because of the delta. In the relative
coordinate $r=x_{j,b}-x_{i,a}$,
\be
\label{eq:LMjump}
\partial_r\Psi\big|_{r=0^+}-\partial_r\Psi\big|_{r=0^-}=-\cb\,\Psi\big|_{r=0}\;.
\ee
\end{itemize}
The problem in $\mathcal H_Y$ is therefore a Schr\"odinger problem for an ordinary scalar function on
$\mathcal D_Y$: Dirichlet conditions on $\partial\mathcal D_Y$ and attractive deltas on interior
hyperplanes, with no sign relating values at different points. $\mathcal D_Y$ is an intersection of
half-spaces, hence convex and in particular connected.

\paragraph{Example $n=N=2$.} The coordinates are $x_{1,1}<x_{2,1}$ (replica 1) and $x_{1,2}<x_{2,2}$
(replica 2). $\mathcal D_Y$ is the union of $4!/(2!)^2=6$ of the $24$ Weyl chambers of $\R^4$, labelled by
the order in which the two replicas appear from left to right: $1122$, $1212$, $1221$, $2112$, $2121$,
$2211$. Two chambers are adjacent across an interior hyperplane $x_{i,1}=x_{j,2}$ when they differ by the
exchange of two neighbouring particles of different replicas:
\begin{center}
\begin{tikzpicture}[x=1.15cm,y=1cm,every node/.style={draw,rounded corners,inner sep=3pt,font=\small}]
\node (a) at (0,0) {$1122$};
\node (b) at (1.8,0) {$1212$};
\node (c) at (3.6,0.7) {$1221$};
\node (d) at (3.6,-0.7) {$2112$};
\node (e) at (5.4,0) {$2121$};
\node (f) at (7.2,0) {$2211$};
\draw (a)--(b) (b)--(c) (b)--(d) (c)--(e) (d)--(e) (e)--(f);
\end{tikzpicture}
\end{center}
The faces where two particles of the same replica are adjacent (for instance the face $x_{1,1}=x_{2,1}$
of the chamber $1122$) are instead Dirichlet walls. The graph of the chambers is connected: particles of
different replicas can pass each other, while particles of the same replica cannot.

\subsection{The ring}

On the ring, $\Psi$ also obeys Eq.~\eqref{eq:periodicitycond}. Since $\BBB_{\trect}$ and the translations
act on each replica separately, a fundamental domain is the product $\Omega=\prod_a\Omega_a$, with
\be
\Omega_a=\{0\le x_{1,a}<x_{2,a}<\dots<x_{N,a}\le L\}\;.
\ee
$\Omega_a$ describes each configuration of $N$ distinct points of replica $a$ on the circle exactly once.
Its faces $x_{i,a}=x_{i+1,a}$ are Dirichlet walls, as before. The faces $x_{1,a}=0$ and $x_{N,a}=L$ are not
physical boundaries: they are identified with each other by moving the particle at $0$ once around the
ring, so that it becomes the last one. Writing only the coordinates of replica $a$,
\be
\label{eq:LMcyclic}
\begin{aligned}
\Psi(x_{2,a},\ldots,x_{N,a},L)&=(-1)^{N-1}\,\Psi(x_{2,a},\ldots,x_{N,a},0)\\
&=(-1)^{N-1}(-1)^{N-1}\,\Psi(0,x_{2,a},\ldots,x_{N,a})=\Psi(0,x_{2,a},\ldots,x_{N,a})\;.
\end{aligned}
\ee
The first sign is the twist \eqref{eq:periodicitycond}, the second one is the sign of the cyclic
permutation of the $N$ particles of the replica, which belongs to $\BBB_{\trect}$. They cancel for every
$N$. The identification thus carries no sign: $\Psi$ is an ordinary function on the space obtained from
$\Omega_a$ by gluing the two faces $x_{1,a} = 0$ and $x_{N,a} = L$, which is the configuration space of $N$ unlabelled distinct points on
the circle. The space $\Omega$ is connected, and the deltas between different replicas,
$x_{i,a}\equiv x_{j,b}$ mod $L$, lie in its interior.

\begin{figure}[ht]
\centering
\begin{tikzpicture}[scale=3.2,>=stealth]
  \fill[blue!8] (0,0) -- (0,1) -- (1,1) -- cycle;
  \draw[->] (-0.05,0) -- (1.18,0) node[right] {$x_{1,a}$};
  \draw[->] (0,-0.05) -- (0,1.18) node[above] {$x_{2,a}$};
  \node[below left] at (0,0) {$0$};
  \draw (1,0.02) -- (1,-0.02) node[below] {$L$};
  \draw[gray,densely dotted] (1,0) -- (1,1);
  \draw[red!75!black,line width=1.4pt] (0,0) -- (1,1);
  \node[red!75!black,rotate=45,below] at (0.52,0.52) {\small $x_{1,a}=x_{2,a}$: $\Psi=0$};
  \draw[blue!70!black,line width=1.1pt] (0,0) -- (0,1) -- (1,1);
  \draw[blue!70!black,line width=1.1pt,->>] (0,0) -- (0,0.62);
  \draw[blue!70!black,line width=1.1pt,->>] (0,1) -- (0.62,1);
  \node[left,blue!70!black] at (0,0.78) {\small $x_{1,a}=0$};
  \node[above,blue!70!black] at (0.8,1) {\small $x_{2,a}=L$};
  \fill (0,0.3) circle (0.018) node[left] {\small $P=(0,y)$};
  \fill (0.3,1) circle (0.018) node[above] {\small $P'=(y,L)$};
  \draw[->,dashed,thick,gray!70!black] (0.03,0.31) .. controls (0.12,0.6) and (0.2,0.85) .. (0.29,0.97);
  \node at (0.27,0.70) {\small $\Omega_a$};
\end{tikzpicture}
\caption{Fundamental domain $\Omega_a=\{0\le x_{1,a}<x_{2,a}\le L\}$ of one replica for $N=2$. The
hypotenuse $x_{1,a}=x_{2,a}$ (coincidence of the two particles of the replica) is a Dirichlet wall. The
vertical side $x_{1,a}=0$ and the horizontal side $x_{2,a}=L$ are glued, with the orientation
indicated by the double arrows: the point $P=(0,y)$ is identified with $P'=(y,L)$, obtained by moving the
particle at $0$ once around the ring and relabelling the two particles. By Eq.~\eqref{eq:LMcyclic}
$\Psi(P')=\Psi(P)$, without sign. After the gluing, $\Omega_a$ is the configuration space of two
distinct unlabelled points on the circle; the free ground state $\sin\big(\pi(x_{2,a}-x_{1,a})/L\big)$ is
positive on it.}
\label{fig:LMtriangle}
\end{figure}
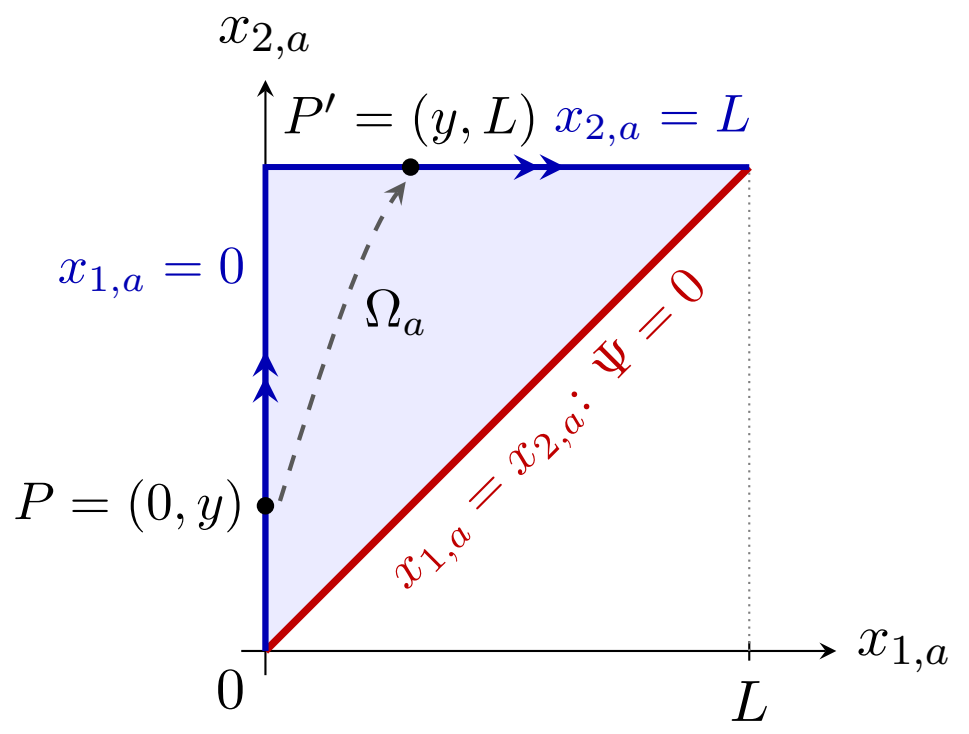

For $N=2$, $\Omega_a=\{0\le x_{1,a}<x_{2,a}\le L\}$ is a triangle (Fig.~\ref{fig:LMtriangle}). Its hypotenuse $x_{1,a}=x_{2,a}$ is a
Dirichlet wall, and its two other sides are glued by $(0,y)\leftrightarrow(y,L)$. The latter is the
exchange of the two particles (sign $-1$) combined with the antiperiodic twist (sign $-1$). With periodic
boundary conditions the gluing would carry a sign $-1$ and the argument below would fail. This is the
same mechanism that makes the ground state of two free fermions on a periodic ring degenerate
($k=0$ together with $k=\pm2\pi/L$). The twist $(-1)^{N-1}$, which is inherited from the polymer problem,
is exactly the one that compensates the sign of the cyclic permutation.

\subsection{Uniqueness of the ground state}

We now show that the ground state of $H_{\rm LL}$ in $\mathcal H_Y$ is non-degenerate. Since
$H_{\rm LL}$ is real, the real and imaginary parts of a ground state are separately ground states, and we
can restrict to real functions. The ground states are the minimizers of $\mathcal E[\Psi]/\|\Psi\|^2$,
with
\be
\label{eq:Ecaldef}
\mathcal E[\Psi]=\frac12\int_\Omega|\nabla\Psi|^2-\cb\sum_{(i,a),(j,b),\ a<b}\ \int_\Omega
\delta^{(L)}(x_{i,a}-x_{j,b})\,|\Psi|^2\;,
\ee
over the admissible functions: those vanishing on the Dirichlet walls and continuous, without sign,
across the interior hyperplanes and across the glued faces. Let $\Psi$ be a generic ground state, of
energy $E_0$.

\paragraph{Splitting into positive and negative parts.} If $\Psi$ is admissible, so is $|\Psi|$, with
$\mathcal E[|\Psi|]=\mathcal E[\Psi]$ and $\||\Psi|\|=\|\Psi\|$, since $|\nabla|\Psi||=|\nabla\Psi|$ almost
everywhere and the potential depends on $|\Psi|^2$ only. Hence $|\Psi|$ is also a ground state. This is
where the absence of signs in the gluings is essential: a gluing with a sign,
$\Psi(\XX')=-\Psi(\XX)$ for two identified points $\XX$ and $\XX'$ of the glued faces, would not be satisfied
by $|\Psi|$. The ground states form a linear space,
so also the positive and negative parts
\be
\Psi_\pm=\tfrac12\big(|\Psi|\pm\Psi\big)\ \ge0
\ee
are ground states, unless they vanish identically.

We report for completeness a standard argument
(based on positivity improving semigroups, see \cite[Sec.~XIII.12]{reed1978methods} and
\cite[Ch.~11]{lieb2001analysis}) ensuring that a non-negative ground state $\Phi\ge0$, not identically zero, must actually satisfy
$\Phi>0$ in the interior of a connected domain. A ground state satisfies $\Phi=e^{tE_0}e^{-tH_{\rm LL}}\Phi$
for every $t>0$, i.e.\ $\Phi(\XX')=e^{tE_0}\int_\Omega K_t(\XX',\XX)\,\Phi(\XX)\,d\XX$. By the
Feynman--Kac formula the heat kernel $K_t(\XX',\XX)$ is strictly positive whenever $\XX$ and $\XX'$ are
joined by a path in the interior, and on a \emph{connected} domain such paths always exist. Since
$\Phi\ge0$ is not identically zero, $\Phi(\XX')>0$ at every interior point.

\paragraph{Non-degeneracy.} Since $\Psi_+\Psi_-=0$ pointwise, $\Psi_+$ and $\Psi_-$ cannot both be
strictly positive, so one of them vanishes identically and without loss of generality we can assume that $\Psi(\XX) = \Psi_+(\XX)$: \emph{every} ground state has a definite
sign and does not vanish at any interior point. This is enough to ensure non-degeneracy, because from two linearly independent ground states, $\Psi_1$ and $\Psi_2$
one could always construct a nonzero linear combination vanishing at an interior point $\XX_0$.
In summary, \emph{the ground state $\Psi_0$ of $H_{\rm LL}$ in $\mathcal H_Y$ is non-degenerate, and can be chosen
strictly positive on $\Omega$} (Perron--Frobenius).
For clarity, we stress that the uniqueness holds in $\mathcal H_Y$, not in
$\mathcal H$. For $S\in\SSS_M$, $R(S)\Psi_0$ is again an eigenfunction of $H_{\rm LL}$ with energy $E_0$,
but in general it does not belong to $\mathcal H_Y$: it is antisymmetric under the conjugate group
$S\BBB_{\trect}S^{-1}$ instead of $\BBB_{\trect}$. Since the projectors $P_\lambda$ commute with
$H_{\rm LL}$ and preserve $\mathcal H_Y$, each $P_\lambda\Psi_0$ is either zero or proportional to
$\Psi_0$, so that $\Psi_0$ lies in a single sector $\lambda^*$. 
The functions $R(S)\Psi_0$ then span a whole multiplet of type $\lambda^*$: in $\mathcal H$, $E_0$ is a degenerate eigenvalue, with multiplicity at least $d_{\lambda^*}$. What is unique is the multiplet, which meets $\mathcal H_Y$ in a single direction.

By \eqref{eq:LMminHY}, $E^{\lambda^*}=\min_{\lambda_1\le n}E^\lambda$,
and in fact $E^{\lambda^*}<E^\lambda$ for every other $\lambda$ with $\lambda_1\le n$: if $E^\lambda=E^{\lambda^*}$ for some other $\lambda$ with $\lambda_1\le n$,
the Lemma and the remark after Eq.~\eqref{eq:LMHYdec} would provide a second ground state in $\mathcal H_Y$.

Each $E^\lambda(\cb)$ is the infimum of the ratios $\mathcal E[\Psi]/\|\Psi\|^2$, Eq.~\eqref{eq:Ecaldef},
over the nonzero $\Psi$ in the block $V^Y_\lambda\otimes\mathcal M_\lambda$. This set does not depend
on $\cb$, and each ratio is an affine function of $\cb$. Hence $E^\lambda(\cb)$ is concave, and
therefore continuous in $\cb$.
If $\lambda^*$ changed between $0$ and a given value of $\cb$, then at some intermediate value two
sectors with $\lambda_1\le n$ would share the lowest energy, contradicting the non-degeneracy just proved
at that value. The absence
of level crossings for the lowest level is therefore not an assumption: it follows from Perron--Frobenius
holding at every $\cb$. Thus $\lambda^*$ is independent of $\cb$, and can be determined at $\cb=0$.

\paragraph{Identification at $\cb=0$.}
At $\cb=0$ the replicas are independent in $\mathcal H_Y$, each consisting of $N$ free fermions with the
twist $(-1)^{N-1}$. The single-particle levels are periodic ($k=0,\pm2\pi/L,\ldots$) for odd $N$ and
antiperiodic ($k=\pm\pi/L,\pm3\pi/L,\ldots$) for even $N$. In both cases the $N$ fermions of each replica
exactly fill the lowest shells: the ground state is a closed shell, hence unique, as it must be.

For $N=2$ the two orbitals are $u_\pm(x)=e^{\pm i\pi x/L}$, and
\be
\label{eq:LMpsi0}
\Psi_0(\XX)=\prod_{a=1}^n s\big(x_{2,a}-x_{1,a}\big)\;,\qquad s(u)=\sin\frac{\pi u}{L}\;,\qquad
E=n\,\frac{\pi^2}{L^2}=nE(1,L)\;,
\ee
up to a constant, since $u_+(x)u_-(y)-u_-(x)u_+(y)=-2i\,s(y-x)$. Note that $\Psi_0>0$ on $\Omega$, where
$0<x_{2,a}-x_{1,a}<L$, in agreement with the previous subsection.

To identify the sector of $\Psi_0$, observe that every function obtained from $\Psi_0$ by permuting the
coordinates is a combination of products $u_{\sigma_1}(x_1)\cdots u_{\sigma_M}(x_M)$ of two orbitals
only. As an $\SSS_M$-module, the span of these products is $(\C^2)^{\otimes M}$, and it is annihilated by
any antisymmetrization over three or more coordinates. Now, if $V_\lambda$ occurs in a module, the Young
symmetrizer $c_{\mathcal T}=z_{\mathcal T}^{-1}b_{\mathcal T}a_{\mathcal T}$ of any tableau of shape
$\lambda$ acts on it non-trivially. But $b_{\mathcal T}$ antisymmetrizes over the columns of
$\mathcal T$, and it annihilates $(\C^2)^{\otimes M}$ as soon as $\lambda$ has a column of length $\ge3$.
Hence $\lambda^*$ has at most two rows. Together with $\lambda^*_1\le n$ and $|\lambda^*|=2n$, this leaves
only $\lambda^*=(n,n)$.

For general $N$ the same argument, with $N$ orbitals, gives at most $N$ rows and therefore
$\lambda^*=(n^N)$ and one has
\be
E^{(n^N)}<E^\lambda\qquad\text{for all }\lambda\ne(n^N)\text{ with }\lambda_1\le n\text{ and all }\cb\;.
\ee
To conclude, we observe that the constructed ground state in $\mathcal{H}_Y$ also has the symmetries \eqref{eq:PsiCond}. Indeed, the exchange $\sigma$ of two whole replicas commutes
with $H_{\rm LL}$ and preserves $\mathcal H_Y$ and $\Omega$; by uniqueness $R(\sigma)\Psi_0=\pm\Psi_0$, and
the positivity of $\Psi_0$ on $\Omega$ fixes the sign to $+1$. So $\Psi_0$ is indeed the ground state with
the required symmetries and in particular $E^{(N)}(n,L)=E^{(n^N)}$.

As a final remark, we observe that this argument did not use the sign of the interaction $\cb$. For $\cb>0$, the case of interest here, the attraction favours the amplitude at the contacts
between replicas, and one expects the gap to grow. However, even for repulsive interaction $\cb<0$, the argument holds and the gap
remains open for every finite repulsion. It closes only in the hard-core limit $\cb\to-\infty$, where the
hyperplanes between different replicas become impenetrable walls, $\Omega$ disconnects into separate
chambers, and the uniqueness argument fails.

\paragraph{Explicit check for $n=N=2$.} Label the coordinates $x_1=x_{1,1}$, $x_2=x_{2,1}$, $x_3=x_{1,2}$,
$x_4=x_{2,2}$, and set $s_{\mu\nu}=s(x_\nu-x_\mu)$. Then $\Psi_0=s_{12}s_{34}$.
\begin{itemize}
\item The permutations of the coordinates produce, up to signs, only the three pairings $s_{12}s_{34}$,
$s_{13}s_{24}$, $s_{14}s_{23}$, one for each coset of $\SSS_4$ modulo the group
$\SSS_2\wr\SSS_2$ of order $8$ generated by $(12)$, $(34)$ and $(13)(24)$, under which
$\Psi_0\to(-1)^S\Psi_0$.
\item The three pairings obey the Pl\"ucker relation
\be
s_{12}s_{34}-s_{13}s_{24}+s_{14}s_{23}=0\;,
\ee
so they span a two-dimensional space carrying a representation of $\SSS_4$.
\item This representation is not a sum of one-dimensional ones, because the three-cycle $(123)$ maps
$s_{12}s_{34}$ to $s_{23}s_{14}$, which is not proportional to it, whereas three-cycles act trivially in
both one-dimensional representations. Hence it is the unique two-dimensional irreducible representation
of $\SSS_4$, $V_{(2,2)}$.
\end{itemize}
This two-dimensional space is the multiplet of the ground state. Within it, $\Psi_0=s_{12}s_{34}$ is, up
to normalization, the only vector satisfying Eq.~\eqref{eq:PsiCond}, i.e.\ the one selected by the
filling \eqref{eq:Trect}. At strong coupling the ordering is also transparent: in the sector $\lambda$ the
ground state consists of bound clusters of sizes $\lambda_1,\lambda_2,\ldots$, with energy
$-\frac{\cb^2}{24}\sum_i\lambda_i(\lambda_i^2-1)$, which, for $\lambda_i\le n$, is minimal for the
rectangle.

\section{Direct expansion of the Bethe equations in \texorpdfstring{$c$}{c} at fixed \texorpdfstring{$n$}{n}}
\label{app:directexpansionc}

For $n=3$, reflection symmetry allows us to choose
$p_2=0$ and $p_3=-p_1$. Expanding the cross-multiplied auxiliary Bethe
equation to the first non-trivial order as $c\to0$ gives
\be
p_1\left(p_1^2-\frac{\pi^2}{4}\right)=0 \;.
\ee
The solution $p_1=0$ corresponds to a coalescence of the three auxiliary
roots and does not belong to the ground-state branch considered here.
Choosing $p_1>0$, we therefore have $p_1\to\pi/2$ as $c\to0$.
Inserting a series expansion in $\sqrt{c}$ into the Bethe equations, one
finds
\bea
p_1 & = &
\frac{\pi}{2}
+\frac{3(4+\pi^2)c^2}{4\pi^3}
+O(c^4)
\\
q_1 & = &
\frac{i\pi}{2}
-\sqrt{\frac{3}{2}}\sqrt{c}
-\frac{ic}{\pi}
-\frac{(6+\pi^2)c^{3/2}}{4\sqrt{6}\pi^2}
+\frac{ic^2}{6\pi^3}
-\frac{(900-60\pi^2+19\pi^4)c^{5/2}}
{480\sqrt{6}\pi^4}
+\frac{i(129-\pi^2)c^3}{36\pi^5}
+O(c^{7/2})
\nn\\
q_2 & = &
\frac{i\pi}{2}
-\frac{ic}{\pi}
-\frac{ic^2}{3\pi^3}
+\frac{i(33+\pi^2)c^3}{18\pi^5}
+O(c^4) \nn
\\
q_3 & = &
\frac{i\pi}{2}
+\sqrt{\frac{3}{2}}\sqrt{c}
-\frac{ic}{\pi}
+\frac{(6+\pi^2)c^{3/2}}{4\sqrt{6}\pi^2}
+\frac{ic^2}{6\pi^3}
+\frac{(900-60\pi^2+19\pi^4)c^{5/2}}
{480\sqrt{6}\pi^4}
+\frac{i(129-\pi^2)c^3}{36\pi^5}
+O(c^{7/2}) \nn  \;.
\eea
We note that $q_3=-q_1^*$, $q_2=-q_2^*$ and $p_1=p_1^*$, consistently
with the symmetries of the ground-state solution. Interestingly, the
first correction to $p_1$ occurs only at order $c^2$. Determining this
coefficient requires the expansion of the physical roots up to order
$c^3$.

Using the reflection symmetry of the remaining physical roots, the
ground-state energy is
\bea
E(3,L)
&=&
-\frac{4}{L^2}\left(q_1^2+q_2^2+q_3^2\right)
\nn\\
&=&
-\frac{4}{L^2}
\left[
-\frac{3\pi^2}{4}
+6c+\frac{c^2}{2}
+\left(\frac{1}{10}-\frac{9}{2\pi^4}\right)c^3
+O(c^4)
\right] \;.
\eea

{\it General $n$}. The limiting auxiliary roots can in fact be obtained explicitly for
arbitrary $n$. Recall the polynomial form
\eqref{eq:polyP}
of the auxiliary Bethe equations,
\be
P(p_a-c)Q\left(p_a+\frac{c}{2}\right)
+
P(p_a+c)Q\left(p_a-\frac{c}{2}\right)
=0 \;.
\ee
Since $P(p_a)=0$, expanding the shifted polynomials around $p_a$
the terms of order $c$ cancel identically, and the first non-trivial
equation is
\be
Q(p_a)P''(p_a)-Q'(p_a)P'(p_a)=0 \;.
\ee
On the ground-state branch, as $c\to0$, the physical roots consist of
$n$ copies of $i\pi/2$ and $n$ copies of $-i\pi/2$, so that
\be
Q(X)\underset{c\to0}{\longrightarrow}
Q_0(X)
=
\left(X^2+\frac{\pi^2}{4}\right)^n \;.
\ee
Denoting by $P_n(X)$ the monic polynomial whose zeros are the limiting
auxiliary roots, we thus obtain, at every zero $p_a$ of $P_n$,
\be
\left[
\left(X^2+\frac{\pi^2}{4}\right)P_n''(X)
-
2nX P_n'(X)
\right]_{X=p_a}
=0 \;.
\ee
The polynomial in brackets has degree $n$ and vanishes at all the zeros
of $P_n$. It is therefore proportional to $P_n$. Comparing the leading
coefficients gives
\be \label{polynPn}
\left(X^2+\frac{\pi^2}{4}\right)P_n''(X)
-
2nX P_n'(X)
+
n(n+1)P_n(X)
=0 \;.
\ee
The monic polynomial solution is
\be \label{P_n}
P_n(X)
=
\frac{
\left(X+\frac{i\pi}{2}\right)^{n+1}
-
\left(X-\frac{i\pi}{2}\right)^{n+1}
}
{i\pi(n+1)} \;.
\ee
Consequently, the limiting auxiliary roots are
\be
p_a
=
\frac{\pi}{2}
\cot\left(\frac{\pi a}{n+1}\right)
\quad , \quad
a=1,\ldots,n \;.
\ee

\section{Expansion in powers of \texorpdfstring{$c$}{c} from the polynomial Bethe equations}
\label{app:expansionpolynomial}

We present here a complementary approach to the small-$c$ expansion,
working directly with the root polynomials $P$ and $Q$, in the spirit of
Appendix D of \cite{brunet2000probability}. For arbitrary integer $n$,
this method determines the coefficients of the small-$c$ expansion
directly as functions of $n$, without introducing the generating
functions $M,\hat M$ and their Laplace transforms used in the main text.
At each order in $c$ the new polynomial corrections enter linearly, while
polynomiality and regularity provide the corresponding solvability
conditions. The construction can therefore in principle be iterated to
higher orders, although the algebra rapidly becomes more involved and we
do not attempt here a general all-orders proof of uniqueness. We illustrate
the method through the first non-trivial correction, which reproduces the
first two terms of the small-$c$ expansion of $E(n,L)$ obtained in the
main text.

We suppress below the dependence on $c$ of all polynomials and functions.
The two Bethe equations \eqref{eq:polyP}, \eqref{eq:polyQ}
can be written as
\be
\label{eq:Pfunctional}
P(X-c)Q\left(X+\frac{c}{2}\right)
+
P(X+c)Q\left(X-\frac{c}{2}\right)
=
T(X)P(X) \;,
\ee
and
\be
\label{eq:Qfunctional}
e^{-X}Q(X+c)P\left(X-\frac{c}{2}\right)
-
e^XQ(X-c)P\left(X+\frac{c}{2}\right)
=
R(X)Q(X) \;.
\ee
Here $T(X)$ is a polynomial of degree $2n$, while $R(X)$ is an entire
function. These properties follow from the fact that the left-hand sides
of \eqref{eq:Pfunctional} and \eqref{eq:Qfunctional} vanish respectively
at all the zeros of $P$ and $Q$, and that $Q$ has degree $2n$.

We look for an expansion
\be
P=\sum_{k\geq0}c^kP_k
\quad , \quad
Q=\sum_{k\geq0}c^kQ_k
\quad , \quad
T=\sum_{k\geq0}c^kT_k
\quad , \quad
R=\sum_{k\geq0}c^kR_k \;.
\ee
The polynomials $P$ and $Q$ are chosen monic, so that
$\deg P_k<n$ and $\deg Q_k<2n$ for $k\geq1$.

At each order in $c$, Eqs.~\eqref{eq:Pfunctional}-\eqref{eq:Qfunctional}
give linear equations for the new corrections. The regular parts determine
$T_k$ and $R_k$, while polynomiality of $T_k$ and regularity of $R_k$
provide the solvability conditions for $P_k$ and $Q_k$. In this way the
expansion can be constructed without expanding the individual Bethe roots.

At $c=0$ the physical roots give
\be
Q_0(X)=\left(X^2+\frac{\pi^2}{4}\right)^n .
\ee
The polynomial $P_0$, on the other hand, is not determined by the
zeroth-order equation alone. Indeed, expanding \eqref{eq:Pfunctional},
the orders $c^0$ and $c^1$ only give $T_0=2Q_0$ and $T_1=2Q_1$,
and impose no condition on $P_0$. The first non-trivial solvability
condition appears at order $c^2$, where
\be
T_2
=
2Q_2+\frac14 Q_0''
+
\frac{Q_0P_0''-Q_0'P_0'}{P_0}.
\label{eq:T2smallc}
\ee

Since $T_2$ must be a polynomial, the numerator of the last term must
vanish at every zero of $P_0$. Using
$Q_0'/Q_0=2nX/(X^2+\pi^2/4)$, the resulting degree-$n$ polynomial
must therefore be proportional to $P_0$. Matching its leading
coefficient gives
\be
\left(X^2+\frac{\pi^2}{4}\right)P_0''
-2nX P_0'
+n(n+1)P_0=0 \;.
\ee
This is precisely the equation \eqref{polynPn} obtained independently
in the previous appendix from the limiting auxiliary Bethe equations.
Thus $P_0=P_n$, with $P_n$ given explicitly in \eqref{P_n}.

We now determine the first correction to $Q$. At order $c^0$,
Eq.~\eqref{eq:Qfunctional} gives
\be
R_0=-2\sinh(X)P_0 \;,
\ee
while at order $c$ one finds
\be
R_1=
-2\sinh(X)P_1-\cosh(X)P_0'
+
2\cosh(X)P_0\frac{Q_0'}{Q_0} \;.
\ee
Regularity of $R_1$ imposes no new condition: the term containing
$P_1$ is entire, while the possible poles of $Q_0'/Q_0$ at
$X=\pm i\pi/2$ are cancelled by the zeros of $\cosh X$.
Thus neither $P_1$ nor $Q_1$ is fixed at this order.
To determine $Q_1$, one must therefore consider the regularity of
$R_2$. Dividing \eqref{eq:Qfunctional} by $Q(X)$, we use
\be
\frac{Q(X\pm c)}{Q(X)}
=
1\pm c\frac{Q_0'}{Q_0}
+c^2\left[
\frac12\frac{Q_0''}{Q_0}
\pm\left(\frac{Q_1}{Q_0}\right)'
\right]
+O(c^3).
\ee
The terms in $R_2$ which may be singular at the zeros
$X=\pm i\pi/2$ of $Q_0$ are then
\be
\label{eq:R2sing}
2\cosh(X)P_0
\left(\frac{Q_1}{Q_0}\right)'
-
\sinh(X)
\left[
P_0\frac{Q_0''}{Q_0}
-
P_0'\frac{Q_0'}{Q_0}
\right] \;.
\ee

Requiring this expression to be regular at $X=\pm i\pi/2$
determines $Q_1$. Since $Q_0$ has zeros of order $n$ at these two
points, cancellation of poles of order higher than two requires
$Q_1$ to contain the factor $(X^2+\pi^2/4)^{n-2}$. Since $Q$ is
even and monic, one can therefore write
\be
Q_1(X)=
\left(X^2+\frac{\pi^2}{4}\right)^{n-2}
\left(\alpha X^2+\beta\right) \;.
\ee

Expanding \eqref{eq:R2sing} around $X=i\pi/2$, cancellation of the double
and simple poles gives respectively
\be
\alpha-\frac{4\beta}{\pi^2}=-n(n-1)
\quad , \quad
\alpha+\frac{4\beta}{\pi^2}=-n(n-1) \;.
\ee
Thus $\alpha=-n(n-1)$, $\beta=0$, and we obtain
\be
\label{eq:Qsmallc1}
Q(X)=
\left(X^2+\frac{\pi^2}{4}\right)^n
-
c\,n(n-1)X^2
\left(X^2+\frac{\pi^2}{4}\right)^{n-2}
+O(c^2) \;.
\ee

Finally, writing
\be
Q(X)=X^{2n}+A_2X^{2n-2}+\ldots \;,
\ee
one has $A_2=-\frac12\sum_{\alpha=1}^{2n}q_\alpha^2$, and hence
\be
E(n,L)=-\frac{2}{L^2}\sum_{\alpha=1}^{2n}q_\alpha^2
=\frac{4}{L^2}A_2 \;.
\ee
Equation~\eqref{eq:Qsmallc1} therefore yields
the first two terms in the small $c$ expansion,
\be
E(n,L)=
\frac{n\pi^2}{L^2}
-\frac{4n(n-1)}{L^2}c
+O(c^2) \;.
\ee
in agreement with \eqref{Esmallc}.

\section{Direct elimination of the auxiliary roots}
\label{app:RBEresidue}

In this Appendix we derive the reduced Bethe equation \eqref{reducedBE}
directly from the Bethe equations by eliminating the auxiliary residues
$\hat\rho(p_a)$. This provides an alternative derivation of the reduced
equation \eqref{reducedM} for $M(z)$, without using Laplace transforms.

We introduce
\be
\label{eq:Salpha}
S_\alpha
:=
\sum_{a=1}^{n}
\frac{\hat\rho(p_a)}
{q_\alpha+p_a+c/2} \;.
\ee
The physical Bethe equation \eqref{bet} can then be written as
\be
\label{eq:physicalS}
-\rho(-q_\alpha)e^{2q_\alpha}
=
\rho(q_\alpha)(1-cS_\alpha) \;.
\ee
We shall also use the identity
\be
\label{eq:rhoidentityApp}
\sum_{\beta=1}^{2n}
\frac{\rho(q_\beta)}
{q_\beta-q_\alpha+c}
=
\frac1c \;,
\ee
which follows directly from the pole decomposition \eqref{def1},
evaluated at $X=q_\alpha-c$.

To eliminate the auxiliary residues, we use
\be
\frac{1}{
(q_\alpha+p_a-c/2)(q_\beta+p_a+c/2)}
=
\frac{1}{q_\alpha-q_\beta-c}
\left[
\frac{1}{q_\beta+p_a+c/2}
-
\frac{1}{q_\alpha+p_a-c/2}
\right] \;.
\ee
Multiplying by $-c\,\rho(q_\beta)$, summing over $\beta$, and using
Eq.~\eqref{eq:rhoidentityApp}, one obtains
\bea
&&
\frac{1}{q_\alpha+p_a-c/2}
\left[
1-c\sum_{\beta=1}^{2n}
\frac{\rho(q_\beta)}
{q_\beta+p_a+c/2}
\right]
\nn\\
&&\hspace{20mm}
=
-c\sum_{\beta=1}^{2n}
\frac{\rho(q_\beta)}
{(q_\alpha-q_\beta-c)(q_\beta+p_a+c/2)} \;.
\eea

Multiplying by $\hat\rho(p_a)$ and using the auxiliary Bethe equation
\eqref{eq:betaux} gives
\be
\frac{\hat\rho(-p_a)}
{q_\alpha+p_a-c/2}
=
-c\hat\rho(p_a)
\sum_{\beta=1}^{2n}
\frac{\rho(q_\beta)}
{(q_\alpha-q_\beta-c)(q_\beta+p_a+c/2)} \;.
\ee
We now replace $q_\alpha\to -q_\alpha$, which is allowed since the
physical root set is invariant under reflection. This gives
\be
\frac{\hat\rho(-p_a)}
{p_a-q_\alpha-c/2}
=
c\hat\rho(p_a)
\sum_{\beta=1}^{2n}
\frac{\rho(q_\beta)}
{(q_\alpha+q_\beta+c)(q_\beta+p_a+c/2)} \;.
\ee
Summing over $a$, we use the reflection symmetry
$\{p_a\}=\{-p_a\}$ to relabel $p_a\to -p_a$ on the left-hand side.
Hence
\be
\sum_{a=1}^n
\frac{\hat\rho(-p_a)}
{p_a-q_\alpha-c/2}
=
-\sum_{a=1}^n
\frac{\hat\rho(p_a)}
{q_\alpha+p_a+c/2}
=
-S_\alpha \;.
\ee
The sum over $a$ on the right-hand side is simply $S_\beta$, and one
therefore obtains
\be
\label{eq:Sclosed}
S_\alpha
=
-c\sum_{\beta=1}^{2n}
\frac{\rho(q_\beta)S_\beta}
{q_\alpha+q_\beta+c} \;.
\ee

On the other hand, Eq.~\eqref{eq:physicalS} gives
\be
cS_\beta
=
1+
\frac{\rho(-q_\beta)}{\rho(q_\beta)}
e^{2q_\beta} \;.
\ee

Inserting this expression into Eq.~\eqref{eq:Sclosed}, using
Eq.~\eqref{eq:rhoidentityApp} with the root $-q_\alpha$, and changing
$q_\beta\to-q_\beta$ in the remaining sum, yields
\be
\label{eq:Sphysical}
S_\alpha
=
-\frac1c
-
\sum_{\beta=1}^{2n}
\frac{\rho(q_\beta)e^{-2q_\beta}}
{q_\alpha-q_\beta+c} \;.
\ee
It follows that
\be
1-cS_\alpha
=
2
+
c\sum_{\beta=1}^{2n}
\frac{\rho(q_\beta)e^{-2q_\beta}}
{q_\alpha-q_\beta+c}
=
2
-
c\sum_{\beta=1}^{2n}
\frac{\rho(q_\beta)e^{-2q_\beta}}
{q_\beta-q_\alpha-c} \;.
\ee
Using this in Eq.~\eqref{eq:physicalS} gives
\be
\label{eq:RBEapp}
-2\rho(q_\alpha)
=
\rho(-q_\alpha)e^{2q_\alpha}
-
c\rho(q_\alpha)
\sum_{\beta=1}^{2n}
\frac{\rho(q_\beta)e^{-2q_\beta}}
{q_\beta-q_\alpha-c} \;,
\ee
which is the reduced Bethe equation \eqref{reducedBE}.

We now show that the reduced Bethe equation \eqref{eq:RBEapp} is
equivalent to the reduced integral equation \eqref{reducedM}. Introduce
\be
L(z)=
\sum_{\alpha=1}^{2n}
\rho(q_\alpha)e^{q_\alpha z}
\quad , \quad
M(z)=\frac{1}{2n}e^{cz/2}L(z) \;.
\ee
In terms of $L$, Eq.~\eqref{reducedM} is equivalent to
\bea
-2L(z)
&=&
e^{-cz}L(z-2)+L(-z-2)
\nn\\
&&
+c\int_0^z dw\,e^{-cw}L(w-2)L(z-w) \;.
\label{eq:reducedLapp}
\eea

We first assume the reduced Bethe equation \eqref{eq:RBEapp}.
Multiplying it by $e^{q_\alpha z}$ and summing over $\alpha$ gives
\be
-2L(z)
=
L(-z-2)
-c\sum_{\alpha,\beta}
\frac{
\rho(q_\alpha)\rho(q_\beta)e^{-2q_\beta}e^{q_\alpha z}
}{
q_\beta-q_\alpha-c
} \;, \label{summedBE}
\ee
where in the first term we used the reflection symmetry of the
physical root set. Exchanging $\alpha$ and $\beta$ in the double sum,
the last term becomes
\be
-c\sum_{\alpha,\beta}
\frac{
\rho(q_\alpha)\rho(q_\beta)e^{-2q_\alpha}e^{q_\beta z}
}{
q_\alpha-q_\beta-c
} \;.
\ee
On the other hand,
\bea
&&
c\int_0^z dw\,e^{-cw}L(w-2)L(z-w)
\nn\\
&&=
c\sum_{\alpha,\beta}
\rho(q_\alpha)\rho(q_\beta)e^{-2q_\alpha}e^{q_\beta z}
\frac{
e^{(q_\alpha-q_\beta-c)z}-1
}{
q_\alpha-q_\beta-c} \;.
\eea
Using Eq.~\eqref{eq:rhoidentityApp}, the contribution of the first
term in the numerator is
\be
-e^{-cz}L(z-2) \;.
\ee
Therefore
\be
-c\sum_{\alpha,\beta}
\frac{
\rho(q_\alpha)\rho(q_\beta)e^{-2q_\alpha}e^{q_\beta z}
}{
q_\alpha-q_\beta-c
}
=
e^{-cz}L(z-2)
+
c\int_0^z dw\,e^{-cw}L(w-2)L(z-w) \;.
\ee
Substituting this into the summed reduced Bethe equation \eqref{summedBE} gives exactly
\eqref{eq:reducedLapp}, and hence \eqref{reducedM}.

Conversely, assume Eq.~\eqref{eq:reducedLapp}. Expanding its integral
term as above and again using Eq.~\eqref{eq:rhoidentityApp}, the term
$-e^{-cz}L(z-2)$ generated by the numerator cancels the explicit
$e^{-cz}L(z-2)$ term. One is left with
\bea
0
=
\sum_{\alpha=1}^{2n} e^{q_\alpha z}
\bigg[
2\rho(q_\alpha)
+\rho(-q_\alpha)e^{2q_\alpha}
-c\rho(q_\alpha)
\sum_{\beta=1}^{2n}
\frac{\rho(q_\beta)e^{-2q_\beta}}
{q_\beta-q_\alpha-c}
\bigg] \;.
\eea
Since the $q_\alpha$ are distinct, the exponentials
$e^{q_\alpha z}$ are linearly independent. Each coefficient must
therefore vanish, which is precisely Eq.~\eqref{eq:RBEapp}.

This proves the equivalence
between the reduced Bethe equation and the reduced equation for $M$.

\section{Baxter-polynomial derivation of the reduced Bethe equation}
\label{app:RBEbaxter}

Here we give a second, more algebraic way, to eliminate the auxiliary roots.
Recall the Bethe polynomials
\be
Q(X)=\prod_{\alpha=1}^{2n}(X-q_\alpha)
\quad , \quad
P(X)=\prod_{a=1}^{n}(X-p_a) \;.
\ee
The auxiliary Bethe equations \eqref{eq:polyP} imply, see \eqref{eq:Pfunctional}
\be
\label{eq:BaxterApp}
Q\left(X+\frac c2\right)P(X-c)
+
Q\left(X-\frac c2\right)P(X+c)
=
T(X)P(X) \;,
\ee
where $T(X)$ is a polynomial of degree $2n$ with leading coefficient
$2$. Reflection symmetry implies that $Q$ and $T$ are even, while
$P(-X)=(-1)^nP(X)$.

Evaluating Eq.~\eqref{eq:BaxterApp} at
$X=q_\alpha-c/2$ gives
\be
T\left(q_\alpha-\frac c2\right)
P\left(q_\alpha-\frac c2\right)
=
Q(q_\alpha-c)
P\left(q_\alpha+\frac c2\right) \;.
\ee
Using the physical Bethe equation \eqref{eq:polyQ},
\be
\frac{P(q_\alpha+c/2)}
{P(q_\alpha-c/2)}
=
e^{-2q_\alpha}
\frac{Q(q_\alpha+c)}
{Q(q_\alpha-c)} \;,
\ee
we obtain
\be
\label{eq:TminusApp}
T\left(q_\alpha-\frac c2\right)
=
e^{-2q_\alpha}Q(q_\alpha+c) \;.
\ee
Similarly, evaluating Eq.~\eqref{eq:BaxterApp} at
$X=q_\alpha+c/2$ gives
\be
\label{eq:TplusApp}
T\left(q_\alpha+\frac c2\right)
=
e^{2q_\alpha}Q(q_\alpha-c) \;.
\ee
Thus the auxiliary polynomial $P$ has disappeared from these
interpolation relations.

Consider now the rational function $T(X-c/2)/Q(X)$. Since numerator and
denominator have the same degree and the leading coefficient of $T$ is
$2$, its partial-fraction decomposition reads
\be
\frac{T(X-c/2)}{Q(X)}
=
2+
\sum_{\beta=1}^{2n}
\frac{
T(q_\beta-c/2)
}{
Q'(q_\beta)(X-q_\beta)
} \;.
\ee
Using Eq.~\eqref{eq:TminusApp} and the definition \eqref{def1} of the
physical residues, this becomes
\be
\label{eq:Tpartialfraction}
\frac{T(X-c/2)}{Q(X)}
=
2+
c\sum_{\beta=1}^{2n}
\frac{
\rho(q_\beta)e^{-2q_\beta}
}{
X-q_\beta
} \;.
\ee

We now evaluate Eq.~\eqref{eq:Tpartialfraction} at
$X=q_\alpha+c$. Using Eq.~\eqref{eq:TplusApp}, its left-hand side becomes
\be
\frac{T(q_\alpha+c/2)}{Q(q_\alpha+c)}
=
e^{2q_\alpha}
\frac{Q(q_\alpha-c)}{Q(q_\alpha+c)} \;.
\ee
To express the last ratio in terms of the residues, we use
\eqref{def1}. Since $Q$ is even, $Q'$ is odd, and therefore
\be
\rho(-q_\alpha)
=
\frac{Q(-q_\alpha+c)}{cQ'(-q_\alpha)}
=
-\frac{Q(q_\alpha-c)}{cQ'(q_\alpha)} \;.
\ee
Together with
$\rho(q_\alpha)=Q(q_\alpha+c)/(cQ'(q_\alpha))$, this gives
\be
\frac{Q(q_\alpha-c)}{Q(q_\alpha+c)}
=
-\frac{\rho(-q_\alpha)}{\rho(q_\alpha)} \;.
\ee
Hence
\be
\frac{T(q_\alpha+c/2)}{Q(q_\alpha+c)}
=
-e^{2q_\alpha}
\frac{\rho(-q_\alpha)}{\rho(q_\alpha)} \;.
\ee
Comparing with \eqref{eq:Tpartialfraction} at $X=q_\alpha+c$ we obtain
\be
-e^{2q_\alpha}
\frac{\rho(-q_\alpha)}{\rho(q_\alpha)}
=
2-
c\sum_{\beta=1}^{2n}
\frac{
\rho(q_\beta)e^{-2q_\beta}
}{
q_\beta-q_\alpha-c
} \;.
\ee
Multiplying by $\rho(q_\alpha)$ immediately yields
\be
-2\rho(q_\alpha)
=
\rho(-q_\alpha)e^{2q_\alpha}
-
c\rho(q_\alpha)
\sum_{\beta=1}^{2n}
\frac{\rho(q_\beta)e^{-2q_\beta}}
{q_\beta-q_\alpha-c} \;,
\ee
namely the reduced Bethe equation \eqref{reducedBE}.

This derivation gives a direct algebraic interpretation of the reduced
Bethe equation. The auxiliary polynomial $P$ is eliminated through the
Baxter relation, leaving the two relations
\eqref{eq:TminusApp}-\eqref{eq:TplusApp} for the transfer polynomial
$T$. The reduced Bethe equation \eqref{reducedBE} then follows from the
partial-fraction decomposition of $T(X-c/2)/Q(X)$.
\\

\section{Large-\texorpdfstring{$c$}{c} expansion of the exact variance}
\label{app:largecvariance}

In this appendix we derive the large-$c$ expansion of the exact
expression \eqref{eq:c2exact}. We shall determine $B_1''(1)$ through
order $c^{-1/2}$. As we will see, the bulk contribution produces
half-integer powers of $c$, whereas a boundary region
$\lambda=O(1)$ produces integer powers. At the order considered here
the final error is $O(c^{-1})$.

We start from
\be
B_1''(1)
=
\frac{3c}{2}
+
\int_0^\infty d\lambda\,
{\cal I}_c(\lambda),
\label{eq:appB1start}
\ee
where
\be
{\cal I}_c(\lambda)
=
\frac{e^{-2\lambda^2/c}}{2\pi^2}
\left[
\tilde f_0(\lambda)
+
\tilde f_c(\lambda)
\cos\left(\frac{2\pi\lambda}{c}\right)
+
\tilde f_s(\lambda)
\sin\left(\frac{2\pi\lambda}{c}\right)
\right].
\label{eq:appIc}
\ee

The natural bulk region is $\lambda\sim\sqrt{c}$. In this region,
for fixed $\lambda/\sqrt{c}>0$,
\be
\coth\lambda\to1,
\qquad
\operatorname{csch}^2\lambda\to0.
\ee
We therefore define $\tilde f_j^{\rm bulk}$ by making these
replacements in the exact functions $\tilde f_j$. Explicitly,
\bea
\tilde f_0^{\rm bulk}(\lambda)
&=&
6c\lambda^2-2c\lambda+2\pi^2c
-4\lambda^4+4\lambda^3-4\pi^2\lambda^2,
\nn\\
\tilde f_c^{\rm bulk}(\lambda)
&=&
-6c\lambda^2+2c\lambda+\pi^2c
+4\lambda^4-4\lambda^3
-4\pi^2\lambda^2+2\pi^2\lambda,
\nn\\
\tilde f_s^{\rm bulk}(\lambda)
&=&
\pi\left(
-6c\lambda+c+8\lambda^3-6\lambda^2
\right).
\label{eq:appfbulk}
\eea
We then introduce
\be
{\cal I}^{\rm bulk}_c(\lambda)
=
\frac{e^{-2\lambda^2/c}}{2\pi^2}
\left[
\tilde f_0^{\rm bulk}
+
\tilde f_c^{\rm bulk}
\cos\left(\frac{2\pi\lambda}{c}\right)
+
\tilde f_s^{\rm bulk}
\sin\left(\frac{2\pi\lambda}{c}\right)
\right],
\ee
and write the exact identity
\be
B_1''(1)
=
B_{1,{\rm bulk}}''(1)
+
\Delta B_1''(1),
\label{eq:appExactSplit}
\ee
with
\bea
B_{1,{\rm bulk}}''(1)
&=&
\frac{3c}{2}
+
\int_0^\infty d\lambda\,
{\cal I}^{\rm bulk}_c(\lambda),
\label{eq:appBulkDef}
\\
\Delta B_1''(1)
&=&
\int_0^\infty d\lambda\,
\left[
{\cal I}_c(\lambda)
-
{\cal I}^{\rm bulk}_c(\lambda)
\right].
\label{eq:appEdgeDef}
\eea
This decomposition is exact for any $c$. In particular, no region of
$\lambda$ has been discarded or counted twice.

\paragraph{Bulk contribution.}

In \eqref{eq:appBulkDef} we set
$\lambda=\sqrt{c}\,x$. The Gaussian factor becomes $e^{-2x^2}$,
while
\be
\cos\left(\frac{2\pi x}{\sqrt{c}}\right),
\qquad
\sin\left(\frac{2\pi x}{\sqrt{c}}\right)
\ee
can be expanded in powers of $c^{-1/2}$. Since the functions in
\eqref{eq:appfbulk} are polynomials, the resulting terms reduce to
Gaussian moments
\be
\int_0^\infty dx\,x^m e^{-2x^2}
=
\frac{1}{2}\,2^{-(m+1)/2}
\Gamma\left(\frac{m+1}{2}\right).
\ee
Keeping all terms down to order $c^{-1/2}$ gives
\bea
B_{1,{\rm bulk}}''(1)
&=&
\frac{11}{32}\sqrt{\frac{\pi}{2}}\,c^{3/2}
+\frac{3c}{2}
+\frac{5\pi^{5/2}}{128\sqrt{2}}\,c^{1/2}
\nn\\
&&
-\frac{7\pi^{9/2}}{256\sqrt{2}}\,c^{-1/2}
+O(c^{-3/2}).
\label{eq:appBulkResult}
\eea
Notice in particular that the bulk integral itself has no $O(c)$
contribution; the term $3c/2$ in \eqref{eq:appBulkResult} is the
explicit term already present in \eqref{eq:appB1start}.

\paragraph{Boundary contribution.}

The bulk approximation is not uniform near $\lambda=0$. To isolate
what is missed there, define
\be
\delta\tilde f_j
=
\tilde f_j-\tilde f_j^{\rm bulk}.
\ee
From the exact expressions one obtains
\bea
\delta\tilde f_0
&=&
2\left[
c(3\lambda^2+\pi^2)
-2\lambda^2(\lambda^2+\pi^2)
\right](\coth\lambda-1)
\nn\\
&&
-2c\lambda(\lambda^2+\pi^2)
\operatorname{csch}^2\lambda,
\label{eq:appdf0}
\\
\delta\tilde f_c
&=&
\left[
c(\pi^2-6\lambda^2)
+4\lambda^2(\lambda^2-\pi^2)
\right](\coth\lambda-1)
\nn\\
&&
+c\lambda(2\lambda^2-\pi^2)
\operatorname{csch}^2\lambda,
\label{eq:appdfc}
\\
\delta\tilde f_s
&=&
\pi(8\lambda^3-6c\lambda)
(\coth\lambda-1)
+
3\pi c\lambda^2
\operatorname{csch}^2\lambda.
\label{eq:appdfs}
\eea
Hence
\bea
\Delta B_1''(1)
=
\int_0^\infty
\frac{d\lambda}{2\pi^2}
e^{-2\lambda^2/c}
\bigg[
&&
\delta\tilde f_0
+
\delta\tilde f_c
\cos\left(\frac{2\pi\lambda}{c}\right)
\nn\\
&&
+
\delta\tilde f_s
\sin\left(\frac{2\pi\lambda}{c}\right)
\bigg].
\label{eq:appEdgeExact}
\eea
Since
\be
\coth\lambda-1=O(e^{-2\lambda}),
\qquad
\operatorname{csch}^2\lambda=O(e^{-2\lambda}),
\qquad
\lambda\to+\infty,
\ee
the integral \eqref{eq:appEdgeExact} is exponentially localized in
the region $\lambda=O(1)$. We may therefore expand at fixed
$\lambda$,
\bea
e^{-2\lambda^2/c}
&=&
1-\frac{2\lambda^2}{c}+O(c^{-2}),
\nn\\
\cos\left(\frac{2\pi\lambda}{c}\right)
&=&
1+O(c^{-2}),
\nn\\
\sin\left(\frac{2\pi\lambda}{c}\right)
&=&
\frac{2\pi\lambda}{c}+O(c^{-3}).
\eea
Substitution into \eqref{eq:appEdgeExact} gives
\bea
\Delta B_1''(1)
&=&
-\frac{3c}{2}
\int_0^\infty d\lambda\,
\left[
\lambda\operatorname{csch}^2\lambda
+1-\coth\lambda
\right]
\nn\\
&&
+
\int_0^\infty d\lambda\,\lambda^2
\left[
6\lambda\operatorname{csch}^2\lambda
+13(1-\coth\lambda)
\right]
+O(c^{-1}).
\label{eq:appEdgeExpanded}
\eea
The combinations appearing here are regular enough at the origin and
decay exponentially at infinity, so that the expansion can be
integrated term by term.

The first integral is elementary since
\be
\lambda\operatorname{csch}^2\lambda
+1-\coth\lambda
=
\frac{d}{d\lambda}
\left[
\lambda(1-\coth\lambda)
\right],
\ee
which gives
\be
\int_0^\infty d\lambda\,
\left[
\lambda\operatorname{csch}^2\lambda
+1-\coth\lambda
\right]
=1.
\label{eq:appEdgeInt1}
\ee
For the second one we use
\be
\int_0^\infty d\lambda\,
\lambda^3\operatorname{csch}^2\lambda
=
\frac{3}{2}\zeta(3),
\qquad
\int_0^\infty d\lambda\,
\lambda^2(1-\coth\lambda)
=
-\frac{1}{2}\zeta(3),
\ee
and obtain
\be
\Delta B_1''(1)
=
-\frac{3c}{2}
+
\frac{5}{2}\zeta(3)
+
O(c^{-1}).
\label{eq:appEdgeResult}
\ee

Combining \eqref{eq:appBulkResult} and
\eqref{eq:appEdgeResult}, the $O(c)$ terms cancel and we finally find
\bea
B_1''(1)
&=&
\frac{11}{32}\sqrt{\frac{\pi}{2}}\,c^{3/2}
+
\frac{5\pi^{5/2}}{128\sqrt{2}}\,c^{1/2}
+
\frac{5}{2}\zeta(3)
\nn\\
&&
-
\frac{7\pi^{9/2}}{256\sqrt{2}}\,c^{-1/2}
+
O(c^{-1}).
\label{eq:appB1Final}
\eea
Using $c_2(L)=8B_1''(1)/L^2$, this yields
\bea
L^2c_2(L)
&=&
\frac{11}{4}\sqrt{\frac{\pi}{2}}\,c^{3/2}
+
\frac{5\pi^{5/2}}{16\sqrt{2}}\,c^{1/2}
+
20\zeta(3)
\nn\\
&&
-
\frac{7\pi^{9/2}}{32\sqrt{2}}\,c^{-1/2}
+
O(c^{-1}).
\label{eq:appc2Final}
\eea

Let us finally stress why the decomposition
\eqref{eq:appExactSplit} does not miss any contribution. It is an
exact add--subtract decomposition of the original integral. The bulk
part keeps the complete contribution obtained after replacing the
hyperbolic functions by their large-$\lambda$ limits, while the
difference is exponentially localized at $\lambda=O(1)$. Thus there
is no separate intermediate region to be added. The bulk expansion is
an expansion at fixed $\lambda/\sqrt{c}$ under a Gaussian weight,
whereas the remainder admits an ordinary fixed-$\lambda$ expansion
because of its exponential decay. Together the two expansions exhaust
the exact integral to the stated order.

\paragraph{Single-polymer case.}

It is instructive to apply the same decomposition to the exact
Brunet--Derrida result \eqref{eq:c2DBexact},
\be
L^2c_2^{N=1}(L)
=
8\int_0^\infty d\lambda\,
\lambda^2e^{-2\lambda^2/c}\coth\lambda.
\ee
Writing $\coth\lambda=1+(\coth\lambda-1)$ gives an exact separation
into a bulk contribution
\be
8\int_0^\infty d\lambda\,
\lambda^2e^{-2\lambda^2/c}
=
\sqrt{\frac{\pi}{2}}\,c^{3/2},
\ee
and a boundary contribution
\be
\Delta_{\rm DB}
=
8\int_0^\infty d\lambda\,
\lambda^2e^{-2\lambda^2/c}
(\coth\lambda-1).
\ee
Since $\coth\lambda-1$ decays exponentially, the latter is localized
at $\lambda=O(1)$ and the Gaussian can be expanded at fixed $\lambda$:
\bea
\Delta_{\rm DB}
&=&
8\int_0^\infty d\lambda\,\lambda^2(\coth\lambda-1)
-
\frac{16}{c}
\int_0^\infty d\lambda\,\lambda^4(\coth\lambda-1)
+O(c^{-2})
\nn\\
&=&
4\zeta(3)-\frac{24\zeta(5)}{c}+O(c^{-2}),
\eea
where we used
\be
\int_0^\infty d\lambda\,
\lambda^m(\coth\lambda-1)
=
\frac{m!}{2^m}\zeta(m+1).
\ee
Thus
\be
L^2c_2^{N=1}(L)
=
\sqrt{\frac{\pi}{2}}\,c^{3/2}
+
4\zeta(3)
-
\frac{24\zeta(5)}{c}
+
O(c^{-2}).
\label{eq:appc2DBlargec}
\ee
Hence also in the single-polymer problem the naive bulk expansion
misses a boundary contribution, whose leading term is proportional to
$\zeta(3)$. In the common integration variable used here, this
contribution comes from $\lambda=O(1)$.

\section{Perturbative check of the Wiener--Hopf formulas}
\label{app:perturbativeWH}

We now check explicitly that the scalar formulation
\eqref{eq:summaryf}--\eqref{eq:summaryBetaSecond} reproduces the
perturbative results obtained previously by solving the coupled
integral equations directly.

We first expand the functions $f$ and $h$. From
\be
-\log\left[1-2\epsilon\hat H(q)\right]
=
\sum_{k=1}^{\infty}
\frac{(2\epsilon)^k}{k}\,
\hat H(q)^k,
\ee
one obtains
\be
f(a,b,\epsilon)
=
\sum_{k=1}^{\infty}
\frac{(2\epsilon)^{k-1}}{k}
\int_{-\infty}^{+\infty}
\frac{dq}{2\pi}\,
\hat H(q)^k,
\label{eq:fPowerH}
\ee
and
\be
h(a,b,\epsilon)
=
-\sum_{k=1}^{\infty}
\frac{(2\epsilon)^{k-1}}{k}
\int_{-\infty}^{+\infty}
\frac{dq}{2\pi}\,
q^2\hat H(q)^k.
\label{eq:hPowerH}
\ee
Thus, to the order needed below,
\bea
f
&=&
\int\frac{dq}{2\pi}\,\hat H
+
\epsilon
\int\frac{dq}{2\pi}\,\hat H^2
+
\frac{4}{3}\epsilon^2
\int\frac{dq}{2\pi}\,\hat H^3
+
O(\epsilon^3),
\label{eq:fPowerH2}
\\
h
&=&
-\int\frac{dq}{2\pi}\,q^2\hat H
-
\epsilon
\int\frac{dq}{2\pi}\,q^2\hat H^2
-
\frac{4}{3}\epsilon^2
\int\frac{dq}{2\pi}\,q^2\hat H^3
+
O(\epsilon^3).
\label{eq:hPowerH2}
\eea
All these integrals are elementary Gaussian integrals.

For their explicit evaluation, it is convenient to introduce temporarily
\be
s=a+2b,
\qquad
t=-4b,
\ee
so that
\be
\hat H(q)
=
2\sqrt{\pi}\,
\left(s+tq^2\right)e^{-q^2}.
\label{eq:Hst}
\ee
One then finds
\bea
\int\frac{dq}{2\pi}\,\hat H
&=&
s+\frac{t}{2},
\\
\int\frac{dq}{2\pi}\,\hat H^2
&=&
\sqrt{2\pi}
\left(
s^2+\frac12st+\frac{3}{16}t^2
\right),
\\
\int\frac{dq}{2\pi}\,\hat H^3
&=&
\sqrt{3}\pi
\left(
\frac43s^3
+\frac23s^2t
+\frac13st^2
+\frac{5}{54}t^3
\right),
\label{eq:Hmoments}
\eea
and similarly
\bea
\int\frac{dq}{2\pi}\,q^2\hat H
&=&
\frac{s}{2}+\frac{3t}{4},
\\
\int\frac{dq}{2\pi}\,q^2\hat H^2
&=&
\sqrt{2\pi}
\left(
\frac14s^2
+\frac38st
+\frac{15}{64}t^2
\right),
\\
\int\frac{dq}{2\pi}\,q^2\hat H^3
&=&
\sqrt{3}\pi
\left(
\frac29s^3
+\frac13s^2t
+\frac{5}{18}st^2
+\frac{35}{324}t^3
\right).
\label{eq:q2Hmoments}
\eea
Substitution into \eqref{eq:fPowerH2} and
\eqref{eq:hPowerH2} gives
\bea
f
&=&
s+\frac{t}{2}
+
\sqrt{2\pi}\,\epsilon
\left(
s^2+\frac12st+\frac{3}{16}t^2
\right)
\nn\\
&&
+
\sqrt{3}\pi\,\epsilon^2
\left(
\frac{16}{9}s^3
+\frac89s^2t
+\frac49st^2
+\frac{10}{81}t^3
\right)
+
O(\epsilon^3),
\label{eq:fWHpert}
\\
h
&=&
-\frac{s}{2}
-\frac{3t}{4}
-
\sqrt{2\pi}\,\epsilon
\left(
\frac14s^2
+\frac38st
+\frac{15}{64}t^2
\right)
\nn\\
&&
-
\sqrt{3}\pi\,\epsilon^2
\left(
\frac{8}{27}s^3
+\frac49s^2t
+\frac{10}{27}st^2
+\frac{35}{243}t^3
\right)
+
O(\epsilon^3).
\label{eq:hWHpert}
\eea

We next expand $g$. It is useful first to do this formally, before
performing any integrals. From \eqref{eq:summaryOmega},
\be
\Omega
=
{\cal H}\hat H
+
\epsilon\,{\cal H}\left(\hat H^2\right)
+
\frac{4}{3}\epsilon^2
{\cal H}\left(\hat H^3\right)
+
O(\epsilon^3).
\label{eq:OmegaPert}
\ee
Since $\Omega=O(1)$ as $\epsilon\to0$, the denominator appearing in
\eqref{eq:summaryg} expands as
\bea
&&
1-\epsilon\hat H
+
\sqrt{1-2\epsilon\hat H}
\cos(\epsilon\Omega)
\nn\\
&&\qquad =
2
-
2\epsilon\hat H
-
\frac{\epsilon^2}{2}
\left[
\hat H^2+
\left({\cal H}\hat H\right)^2
\right]
+
O(\epsilon^3).
\label{eq:gDenPert}
\eea
Its inverse is therefore
\be
\frac{1}{
1-\epsilon\hat H
+
\sqrt{1-2\epsilon\hat H}
\cos(\epsilon\Omega)
}
=
\frac12
\left\{
1
+
\epsilon\hat H
+
\frac{\epsilon^2}{4}
\left[
5\hat H^2+
\left({\cal H}\hat H\right)^2
\right]
\right\}
+
O(\epsilon^3).
\label{eq:gDenInvPert}
\ee
Consequently,
\bea
g(a,b,\epsilon)
&=&
-16b\sqrt{\pi}
\int\frac{dq}{2\pi}\,
q^2e^{-q^2}
\bigg\{
1
+
\epsilon\hat H(q)
\nn\\
&&\qquad\qquad
+
\frac{\epsilon^2}{4}
\left[
5\hat H(q)^2
+
\left({\cal H}\hat H(q)\right)^2
\right]
\bigg\}
+
O(\epsilon^3).
\label{eq:gFormalPert}
\eea
Thus the Hilbert transform does not contribute at first order in
$\epsilon$; its first contribution appears at order $\epsilon^2$.

The Gaussian terms in \eqref{eq:gFormalPert} are again elementary.
The only additional integral needed at this order is
\be
\sqrt{\pi}
\int\frac{dq}{2\pi}\,
q^2e^{-q^2}
\left({\cal H}\hat H(q)\right)^2
=
\frac23s^2
+
\frac{\sqrt{3}\pi}{27}
\left(s^2+st\right)
+
\frac{5\sqrt{3}\pi}{324}t^2.
\label{eq:HilbertIntegralPert}
\ee
Together with the Gaussian contribution, this gives
\bea
g
&=&
t
+
\sqrt{\frac{\pi}{2}}\,\epsilon
\left(
st+\frac34t^2
\right)
\nn\\
&&
+
\epsilon^2
\bigg[
\frac23s^2t
+
\frac{16\sqrt{3}\pi}{27}
\left(
s^2t+st^2
\right)
+
\frac{20\sqrt{3}\pi}{81}t^3
\bigg]
+
O(\epsilon^3).
\label{eq:gWHpert}
\eea

We can now recover the perturbative solution. Write
\be
a
=
1+a_1\epsilon+a_2\epsilon^2+O(\epsilon^3),
\qquad
b
=
-\frac14+b_1\epsilon+b_2\epsilon^2+O(\epsilon^3),
\label{eq:abWHexpansion}
\ee
so that
\bea
s
&=&
\frac12
+
(a_1+2b_1)\epsilon
+
(a_2+2b_2)\epsilon^2
+
O(\epsilon^3),
\\
t
&=&
1
-
4b_1\epsilon
-
4b_2\epsilon^2
+
O(\epsilon^3).
\label{eq:stWHexpansion}
\eea
Substituting these expressions into \eqref{eq:fWHpert} and
\eqref{eq:gWHpert} gives
\bea
f
&=&
1
+
\epsilon
\left[
a_1
+
\frac{11}{8}\sqrt{\frac{\pi}{2}}
\right]
\nn\\
&&
+
\epsilon^2
\left[
a_2
+
3\sqrt{\frac{\pi}{2}}\,a_1
+
\sqrt{\frac{\pi}{2}}\,b_1
+
\frac{64\pi}{27\sqrt{3}}
\right]
+
O(\epsilon^3),
\label{eq:fWHpertab}
\\
g
&=&
1
+
\epsilon
\left[
-4b_1
+
\frac54\sqrt{\frac{\pi}{2}}
\right]
\nn\\
&&
+
\epsilon^2
\left[
-4b_2
+
\sqrt{\frac{\pi}{2}}\,a_1
-
6\sqrt{\frac{\pi}{2}}\,b_1
+
\frac16
+
\frac{56\pi}{27\sqrt{3}}
\right]
+
O(\epsilon^3).
\label{eq:gWHpertab}
\eea

Imposing $f=g=1$ at first order gives
\be
a_1
=
-\frac{11}{8}\sqrt{\frac{\pi}{2}},
\qquad
b_1
=
\frac{5}{16}\sqrt{\frac{\pi}{2}},
\label{eq:a1b1WHcheck}
\ee
in agreement with the direct result \eqref{eq:a1b1direct}.

At second order one obtains
\be
a_2
=
\left(
\frac{61}{32}
-
\frac{64}{27\sqrt{3}}
\right)\pi,
\qquad
b_2
=
\frac{1}{24}
+
\left(
\frac{14}{27\sqrt{3}}
-
\frac{13}{32}
\right)\pi,
\label{eq:a2b2WHcheck}
\ee
in agreement with \eqref{eq:a2b2direct}.

Finally, inserting these coefficients into \eqref{eq:hWHpert} gives
\be
h(a_\epsilon,b_\epsilon,\epsilon)
=
-1
+
\frac{11}{32}\sqrt{\frac{\pi}{2}}\,\epsilon
+
\left[
\frac{1}{12}
+
\left(
\frac{64}{81\sqrt{3}}
-
\frac{61}{128}
\right)\pi
\right]\epsilon^2
+
O(\epsilon^3).
\label{eq:hWHpertSolution}
\ee
Using \eqref{eq:summaryBetaSecond}, we finally obtain
\be
\beta''(0)
=
-\frac12
+
\frac{11}{32}\sqrt{\frac{\pi}{2}}\,\epsilon
+
\left[
\frac34
+
\left(
\frac{64}{81\sqrt{3}}
-
\frac{61}{128}
\right)\pi
\right]\epsilon^2
+
O(\epsilon^3).
\label{eq:betaSecondWHpert}
\ee
Thus
\bea
\beta_1''(0)
&=&
\frac{11}{32}\sqrt{\frac{\pi}{2}},
\\
\beta_2''(0)
&=&
\frac34
+
\left(
\frac{64}{81\sqrt{3}}
-
\frac{61}{128}
\right)\pi,
\eea
in agreement with the direct results
\eqref{eq:beta1secondDirect} and
\eqref{eq:beta2secondDirect}.

\section{Hilbert-transform integrals entering the expansion of \texorpdfstring{$\hat g$}{g}}
\label{app:ghatIntegrals}

We collect here the integrals needed to evaluate the coefficients
$G_r(B)$ in the expansion of $\hat g$. Recall from
\eqref{eq:DABinverseSeries} that the first non-trivial
Hilbert-transform contribution is $\Phi_1^2$ in $G_3$. At the next
orders one encounters $\Phi_1\Phi_2$, and then
$\Phi_1\Phi_3$, $\Phi_2^2$ and $\Phi_1^4$ in $G_5$, where
$\Phi_r={\cal H}[P_B^r]$. We first reduce a single $\Phi_r$, and then
discuss successively the resulting integrals containing one, two,
three and four ${\rm erfi}$ factors.

\paragraph{Reduction of a single Hilbert transform.}

Using the definition \eqref{eq:defHilbert} of the Hilbert transform,
one has
\be
{\cal H}\left(e^{-tq^2}\right)
=
e^{-tq^2}{\rm erfi}(q\sqrt{t}).
\label{eq:HilbertGaussianApp}
\ee
Since
\be
P_B(q)^r
=
(2\sqrt{\pi})^r(1+Bq^2)^r e^{-rq^2},
\ee
it follows that
\be
\Phi_r(q)
=
(2\sqrt{\pi})^r
\left.
(1-B\partial_t)^r
\left[
e^{-tq^2}{\rm erfi}(q\sqrt{t})
\right]
\right|_{t=r}.
\label{eq:PhiDerivativeApp}
\ee
The derivatives in \eqref{eq:PhiDerivativeApp} can be decomposed as
\bea
\partial_t^k
\left[
e^{-tq^2}{\rm erfi}(q\sqrt{t})
\right]
&=&
t^{-k}Q_k(q\sqrt{t})
\nn\\
&&
+
(-q^2)^k e^{-tq^2}{\rm erfi}(q\sqrt{t}),
\label{eq:QkDerivativeApp}
\eea
where
\be
Q_k(z)
=
\frac{(-1)^{k-1}}{\sqrt{\pi}}
\sum_{p=0}^{k-1}
\frac{\Gamma(k-p-\frac12)}{\Gamma(\frac12)}
z^{2p+1}.
\label{eq:QkExplicitApp}
\ee
Thus
\be
\Phi_r(q)
=
(2\sqrt{\pi})^r
\left[
(1+Bq^2)^r e^{-rq^2}{\rm erfi}(q\sqrt r)
+
R_{r,B}(q)
\right],
\label{eq:PhiErfiPolynomial}
\ee
where the odd polynomial $R_{r,B}$ is
\be
R_{r,B}(q)
=
\sum_{k=1}^r
\binom{r}{k}
(-B)^k r^{-k}Q_k(q\sqrt r).
\label{eq:RrB}
\ee
For $r=1$ this reproduces
\be
\Phi_1(q)
=
2\sqrt{\pi}
\left[
(1+Bq^2)e^{-q^2}{\rm erfi}(q)
-\frac{Bq}{\sqrt{\pi}}
\right],
\ee
as given in \eqref{eq:Phi1Explicit}.

Equation \eqref{eq:PhiErfiPolynomial} shows that products of the
$\Phi_r$ reduce to ordinary Gaussian integrals containing a certain
number of ${\rm erfi}$ factors. The argument associated with
$\Phi_r$ is $\sqrt r\,q$, so different arguments occur when different
$\Phi_r$'s are multiplied.

\paragraph{One ${\rm erfi}$ factor.}

Terms containing a single ${\rm erfi}$ reduce to integrals of the form
\be
S_n(s)
=
\int_{\mathbb R}dq\,
q\,e^{-(s+n)q^2}{\rm erfi}(\sqrt n\,q).
\label{eq:SnDef}
\ee
Integrating by parts gives
\be
S_n(s)
=
\frac{\sqrt n}{(s+n)\sqrt s}.
\label{eq:SnResult}
\ee
All the higher polynomial moments follow by differentiation,
\be
\int_{\mathbb R}dq\,
q^{2m+1}e^{-(s+n)q^2}{\rm erfi}(\sqrt n\,q)
=
(-\partial_s)^m S_n(s).
\label{eq:singleErfiMoments}
\ee
Hence all terms containing a single ${\rm erfi}$ are elementary.

\paragraph{Two ${\rm erfi}$ factors.}

Products such as $\Phi_1^2$, $\Phi_1\Phi_2$,
$\Phi_1\Phi_3$ and $\Phi_2^2$ require integrals with two
${\rm erfi}$ factors. Their arguments are in general different.
Introduce
\bea
T_{n_1,n_2}(s)
&:=&
\int_{\mathbb R}dq\,
e^{-(s+n_1+n_2)q^2}
{\rm erfi}(\sqrt{n_1}q)
{\rm erfi}(\sqrt{n_2}q).
\label{eq:Tn1n2Def}
\eea
The basic principal-value integral which enters its evaluation is
\be
I_2(c)
=
\frac{1}{\pi^2}
\dashint_{\mathbb R^2}
\frac{dk_1dk_2}{k_1k_2}\,
e^{-k_1^2-k_2^2+c k_1k_2}
=
\frac{2}{\pi}\arcsin\left(\frac c2\right).
\label{eq:I2Def}
\ee
Using the principal-value representation associated with
\eqref{eq:HilbertGaussianApp} and performing the Gaussian $q$
integral, one obtains
\bea
T_{n_1,n_2}(s)
&=&
\sqrt{\frac{\pi}{s+n_1+n_2}}\,
I_2\left(
2\sqrt{
\frac{n_1n_2}{(s+n_1)(s+n_2)}
}
\right)
\nn\\
&=&
\frac{2}{\sqrt{\pi(s+n_1+n_2)}}
\arcsin
\sqrt{
\frac{n_1n_2}{(s+n_1)(s+n_2)}
}.
\label{eq:twoErfiMasterApp}
\eea
This includes both the equal-argument and unequal-argument cases.
For instance, $\Phi_1^2$ involves $(n_1,n_2)=(1,1)$,
$\Phi_1\Phi_2$ involves $(1,2)$, $\Phi_1\Phi_3$ involves
$(1,3)$, while $\Phi_2^2$ involves $(2,2)$.

Again, all additional even powers of $q$ follow by differentiation,
\be
\int_{\mathbb R}dq\,
q^{2m}e^{-(s+n_1+n_2)q^2}
{\rm erfi}(\sqrt{n_1}q)
{\rm erfi}(\sqrt{n_2}q)
=
(-\partial_s)^m T_{n_1,n_2}(s).
\label{eq:twoErfiMoments}
\ee
Polynomial factors such as $(1+Bq^2)^r$ can similarly be generated
by powers of $(1-B\partial_s)$.

\paragraph{Three ${\rm erfi}$ factors.}

Up to the order $G_5$, three ${\rm erfi}$ factors arise only upon
expanding the term $\Phi_1^4$. Their arguments are therefore all
equal. The corresponding basic integral is
\be
V_3(s)
=
\int_{\mathbb R}dq\,
q\,e^{-(s+3)q^2}{\rm erfi}(q)^3.
\label{eq:V3Def}
\ee
No new integral is needed in this case. Indeed, using
\be
\frac{d}{dq}{\rm erfi}(q)^3
=
\frac{6}{\sqrt{\pi}}e^{q^2}{\rm erfi}(q)^2,
\ee
an integration by parts gives
\be
V_3(s)
=
\frac{3}{(s+3)\sqrt{\pi}}\,U_2(s),
\label{eq:V3U2}
\ee
where the two-${\rm erfi}$ integral with equal arguments is
\be
U_2(s)
:=
T_{1,1}(s)
=
\frac{2}{\sqrt{\pi(s+2)}}
\arcsin\left(\frac{1}{s+1}\right).
\label{eq:U2Result}
\ee
Thus
\be
V_3(s)
=
\frac{6}{\pi(s+3)\sqrt{s+2}}
\arcsin\left(\frac{1}{s+1}\right).
\label{eq:V3Result}
\ee
Higher odd moments are again obtained from
\be
\int_{\mathbb R}dq\,
q^{2m+1}e^{-(s+3)q^2}{\rm erfi}(q)^3
=
(-\partial_s)^m V_3(s).
\label{eq:threeErfiMoments}
\ee

For example, the part of the $\Phi_1^4$ contribution containing
three ${\rm erfi}$ factors is proportional to
\be
\left.
(-\partial_s)(1-B\partial_s)^3V_3(s)
\right|_{s=1}.
\label{eq:threeErfiG5}
\ee
Thus the three-${\rm erfi}$ sector of $G_5$ is entirely reduced to
the already known two-${\rm erfi}$ integral.

\paragraph{Four ${\rm erfi}$ factors.}

The only genuinely new type of integral at order $G_5$ comes from
the part of $\Phi_1^4$ containing four ${\rm erfi}$ factors. From
\eqref{eq:Phi1Explicit}, this contribution contains
\be
\int_{\mathbb R}dq\,
q^2(1+Bq^2)^4e^{-5q^2}{\rm erfi}(q)^4.
\label{eq:fourErfiIntegral}
\ee
Introduce
\be
U_4(s)
=
\int_{\mathbb R}dq\,
e^{-(s+4)q^2}{\rm erfi}(q)^4.
\label{eq:U4Def}
\ee
Then
\be
\int_{\mathbb R}dq\,
q^2(1+Bq^2)^4e^{-5q^2}{\rm erfi}(q)^4
=
\left.
(-\partial_s)(1-B\partial_s)^4U_4(s)
\right|_{s=1}.
\label{eq:fourErfiFromU4}
\ee
More explicitly, the corresponding four-${\rm erfi}$ contribution
to $G_5$ is
\be
G_5^{(4{\rm erfi})}(B)
=
\frac{4}{3}\pi^{3/2}B
\left.
(-\partial_s)(1-B\partial_s)^4U_4(s)
\right|_{s=1}.
\label{eq:G5fourErfi}
\ee

The analogue of \eqref{eq:I2Def} for four variables is
\be
I_4(c)
=
\frac{1}{\pi^4}
\dashint_{\mathbb R^4}
\frac{dk_1dk_2dk_3dk_4}{k_1k_2k_3k_4}
\exp\left[
-\sum_{j=1}^4k_j^2
+
c\sum_{1\leq i<j\leq4}k_i k_j
\right].
\label{eq:I4Def}
\ee
Performing the Gaussian $q$ integral gives the direct relation
\be
U_4(s)
=
\sqrt{\frac{\pi}{s+4}}\,
I_4\left(\frac{2}{s+3}\right).
\label{eq:U4I4}
\ee
The integral entering $G_5$ corresponds to $s=1$, hence to
$c=1/2$. In particular,
\be
U_4(1)
=
\sqrt{\frac{\pi}{5}}\,
I_4\left(\frac12\right).
\label{eq:U4I4half}
\ee

The derivative of $I_4$ can be reduced to the two-variable function
$I_2$. One finds
\be
I_4'(c)
=
\frac{12}{\pi\sqrt{4-c^2}}\,
I_2\left(\frac{c}{1-c}\right).
\label{eq:I4Recursion}
\ee
Using \eqref{eq:I2Def} and $I_4(0)=0$ therefore gives
\be
I_4(c)
=
\frac{24}{\pi^2}
\int_0^c
\frac{dx}{\sqrt{4-x^2}}\,
\arcsin\left(\frac{x}{2(1-x)}\right).
\label{eq:I4Integral}
\ee
At the value needed in \eqref{eq:U4I4half},
\be
I_4\left(\frac12\right)
=
\frac{4}{\pi}\arcsin\left(\frac14\right)-\frac15.
\label{eq:I4Half}
\ee
For example, integrating \eqref{eq:I4Integral} by parts gives
\bea
I_4\left(\frac12\right)
&=&
\frac{4}{\pi}\arcsin\left(\frac14\right)
\nn\\
&&
-
\frac{24}{\pi^2}
\int_0^{1/2}
\arcsin\left(\frac{x}{2}\right)
\,d\arcsin\left(\frac{x}{2(1-x)}\right),
\eea
where the remaining integral is $\pi^2/120$.

All derivatives of $I_4$ at $c=1/2$ generated by
\eqref{eq:fourErfiFromU4} reduce through \eqref{eq:I4Recursion} to
$I_2$ and hence to elementary inverse trigonometric functions.
The only new constant in the four-${\rm erfi}$ sector is therefore
$I_4(1/2)$ itself.

Combining the different contributions to $G_5$, the inverse
trigonometric terms simplify further. In particular one may use
\bea
&&
4\arcsin\left(\frac14\right)
+3\arcsin\left(\frac23\right)
+6\arcsin\left(\frac{1}{\sqrt6}\right)
\nn\\
&&\qquad
+8\arcsin\left(\frac{\sqrt6}{4}\right)
=
\frac{7\pi}{2},
\label{eq:TrigIdentityG5}
\eea
which, together with \eqref{eq:I4Half}, leads to the elementary
expression \eqref{eq:G5} quoted in the main text.

The same construction can be extended to higher even powers of
${\rm erfi}$ with equal arguments. However, $G_5$ is the last order
for which the four-${\rm erfi}$ sector involves only equal arguments.
At the next order, terms such as $\Phi_1^3\Phi_2$ appear, leading to
integrals containing
${\rm erfi}(q)^3{\rm erfi}(\sqrt{2}\,q)$. Their evaluation requires
a multivariable generalization of $I_4$, since the permutation
symmetry underlying the one-parameter recursion
\eqref{eq:I4Recursion} is then lost. We do not pursue this extension
here.

\section{Large negative \texorpdfstring{$\epsilon$}{epsilon}: asymptotics of \texorpdfstring{$B_A$}{BA}}
\label{app:largeNegativeB}

We now justify the assumption $B\eta\to0$ used in
Sec.~\ref{sec:largeNegativeEpsilon}. From
\eqref{eq:Aeta}, together with
$A=(a+2b)\epsilon$ and \eqref{eq:epsilonLargeNegative}, one has
\be
a+2b
=
\frac{A}{\epsilon}
\simeq
\frac{3\sqrt{\pi}}{4\eta^{3/2}}e^\eta.
\label{eq:aplus2bLargeNegative}
\ee
A naive estimate of $g$ based on the region
$|q|\lesssim\sqrt{\eta}$ does not satisfy the condition $g=1$.
Indeed, neglecting the oscillatory term in the denominator in this
region gives
\be
g_{\rm bulk}
\simeq
-\frac{64}{\sqrt{\pi}}\,
b\,e^{-\eta}\eta^{3/2}C(B\eta),
\qquad
C(x)=\int_0^1du\,\frac{u^2}{1+xu^2}.
\label{eq:gbulk}
\ee
Using
\be
b=-\frac{B}{4}(a+2b)
\ee
and \eqref{eq:aplus2bLargeNegative}, this becomes
\be
g_{\rm bulk}
\simeq
12B\,C(B\eta).
\label{eq:gbulk2}
\ee
Since $C(0)=1/3$ while $C(x)\simeq1/x$ for $x\to+\infty$,
this contribution cannot remain of order unity for any of the
algebraic scalings of $B$ relevant below. The condition $g=1$ is
therefore controlled by another region of momentum space.

To identify it, consider the large-$\eta$ behavior of the phase
$\Omega_{A,B}$. Since its Hilbert transform is dominated by
$q=O(\sqrt{\eta})$, where $Bq^2$ is negligible to leading order,
\eqref{eq:FermiApprox} gives, setting $q=\sqrt{\eta}\,u$,
\be
\Omega_{A,B}(\sqrt{\eta}\,u)
\simeq
\frac{\eta}{2\pi}
\left[
2u+(1-u^2)
\log\left|\frac{u+1}{u-1}\right|
\right].
\label{eq:OmegaLargeNegative}
\ee
For $u\gg1$ this reduces to
\be
\Omega_{A,B}(q)
\simeq
\frac{2\eta^{3/2}}{3\pi q}.
\label{eq:OmegaLargeqNegative}
\ee
The denominator entering $\hat g$ therefore develops narrow
quasipoles whenever
\be
\Omega_{A,B}(q_m)
\simeq
(2m+1)\pi,
\qquad m=0,1,2,\ldots,
\ee
namely
\be
q_m
\simeq
\frac{2\eta^{3/2}}
{3\pi^2(2m+1)}
=
\frac{q_c}{2m+1},
\qquad
q_c=\frac{2\eta^{3/2}}{3\pi^2}.
\label{eq:qmLargeNegative}
\ee
Notice the strong separation of scales
\be
q_{\hat f,\hat h}=O(\eta^{1/2}),
\qquad
q_g=O(\eta^{3/2}).
\label{eq:scaleSeparation}
\ee

To evaluate the quasipoles, let
\be
s(q)
:=
-2H_{A,B}(q)
=
(1+Bq^2)e^{\eta-q^2}.
\label{eq:sLargeNegative}
\ee
For $q$ close to $q_m$, $s(q_m)$ is exponentially small. Writing the
denominator in \eqref{eq:ghatAB} as
\be
D(q)
=
1+\frac{s(q)}{2}
+\sqrt{1+s(q)}\cos\Omega_{A,B}(q),
\ee
and expanding about $q=q_m$, one finds
\be
D(q)
\simeq
\frac{s(q_m)^2}{8}
+
\frac12
\left[\Omega'_{A,B}(q_m)\right]^2
(q-q_m)^2.
\label{eq:Dquasipole}
\ee
Although each quasipole is exponentially narrow, its height is
exponentially large, and the two effects compensate in the integral
for $g$.

Introduce
\be
x=Bq_c^2.
\label{eq:xBdef}
\ee
The contribution of the pair of quasipoles at $q=\pm q_m$ is then
\be
g_m
\simeq
\frac{16}{\pi^2}
\frac{x}
{(2m+1)^2\left[(2m+1)^2+x\right]}.
\label{eq:gmQuasipole}
\ee
Summing over all quasipoles gives
\bea
g
&\simeq&
\frac{16x}{\pi^2}
\sum_{m=0}^\infty
\frac{1}
{(2m+1)^2\left[(2m+1)^2+x\right]}
\nn\\
&=&
2-\frac{4}{\pi\sqrt{x}}
\tanh\left(\frac{\pi\sqrt{x}}{2}\right).
\label{eq:gLargeNegative}
\eea
The self-consistency condition $g=1$ therefore determines a finite
constant $x=x_*$ satisfying
\be
\tanh z_*=\frac{z_*}{2},
\qquad
z_*=\frac{\pi\sqrt{x_*}}{2}.
\label{eq:zstar}
\ee
The non-zero solution is
\be
z_*\simeq1.91500805,
\qquad
x_*=\frac{4z_*^2}{\pi^2}\simeq1.48628281.
\label{eq:xstarNumeric}
\ee
No numerical integration is involved here: these numbers are obtained
only by solving the scalar transcendental equation
\eqref{eq:zstar}. We finally obtain
\be
B_A
\simeq
\frac{x_*}{q_c^2}
=
\frac{9\pi^4x_*}{4\eta^3}
=
\frac{9\pi^2z_*^2}{\eta^3}
\simeq
\frac{325.7493}{\eta^3}.
\label{eq:BlargeNegative}
\ee

The approach to this asymptotic regime is rather slow. Indeed, the
quasipole approximation for the lowest mode requires
$q_c^2\gg\eta$, which, using
$q_c^2=4\eta^3/(9\pi^4)$, amounts to
$\eta\gg3\pi^2/2\simeq15$. Since $\eta$ grows only logarithmically
with $|A|$, sizeable pre-asymptotic corrections may persist up to very
large values of $|A|$. In particular,
\be
B_A\eta=O(\eta^{-2})\to0.
\label{eq:BetaSmall}
\ee
Thus in the region $q=O(\sqrt{\eta})$ controlling $\hat f$ and
$\hat h$ one indeed has
\be
B_Aq^2=O(\eta^{-2}),
\ee
which proves self-consistently that $B_A$ does not modify the leading
asymptotics \eqref{eq:epsilonLargeNegative}--\eqref{eq:ElargeNegative}.
The leading SCGF is therefore controlled by the broad central region
$q=O(\sqrt{\eta})$, while the second self-consistency condition is
enforced by the remote quasipoles at $q=O(\eta^{3/2})$.
Numerically, the quasipoles at $q_m\simeq q_c/(2m+1)$, of width $\propto e^{\eta-q_m^2}$, must be resolved by the
$q$ grid in the evaluation of $\hat g$, which limits the direct evaluation of the parametric representation to
$\eta\lesssim9$ ($A\gtrsim-10^3$, $\epsilon\gtrsim-3.1$). The data therefore only show the onset of the
$|\epsilon|^{5/3}$ behavior: at $A=-10^3$ one finds $\epsilon=-3.11$ and ${\cal E}=-27.3$, to be compared with
$-2.80$ and $-23.7$ from \eqref{eq:etaEpsilonLargeNegative}--\eqref{eq:ElargeNegative}.

\section{Analytic continuation beyond \texorpdfstring{$A_c$}{Ac} and the positive large-deviation tail}
\label{app:continuationAc}

In Sec.~\ref{sec:parametricSCGF} we obtained the large-$L$ scaling
function in the parametric form
\begin{align}
&\hat f(A,B)=\hat g(A,B),
\label{eq:fgeqABNew}
\\
&\epsilon=\hat f(A,B),
\qquad
{\cal E}(\epsilon)=-4\hat h(A,B),
\label{eq:contParametricRecallNew}
\end{align}
with $\hat f,\hat h,\hat g$ given in \eqref{eq:fhatAB},
\eqref{eq:hhatAB} and \eqref{eq:ghatAB}. These formulas rely on the
Wiener--Hopf factorization of Sec.~\ref{sec:parametricSCGF}, which assumes
that the kernel \eqref{eq:Kexplicit}, written in the variables
$(A,B)$,
\be
{\cal K}(q)
=
1-2H_{A,B}(q)
=
1-4\sqrt{\pi}\,A(1+Bq^2)e^{-q^2},
\label{eq:contKdefNew}
\ee
is positive on the real axis. (In Sec.~\ref{sec:largePositiveEpsilon}
the same function is denoted $K_{A,B}$.) Following the branch which
starts at $A=0$, $B=2$, this assumption fails at
$\epsilon=\epsilon_c\simeq0.7003$, but the SCGF itself is regular
there. In this appendix we
(i) describe what happens at $\epsilon_c$ in terms of the zeros of
${\cal K}$,
(ii) explain the logic of the analytic continuation,
(iii) give the continued formulas and their numerical solution, and
(iv) derive from them the positive large-deviation tail.

The mechanism is analogous to several analytic continuations which
appear in exact KPZ large-deviation problems. In the periodic
one-layer problem, the stationary large-deviation function is
continued beyond the endpoint of its principal parametrization by
passing to another sheet \cite{appert1999universal}. A systematic
Riemann-surface interpretation of such continuations for periodic KPZ
fluctuations was developed in \cite{prolhac2020riemann,prolhac2024kpz}.
Closely related branch changes occur in the short-time KPZ problem
\cite{ledoussal2016shorttime,krajenbrink2017brownian}, while in the
inverse-scattering formulation they are associated with discrete
spectral contributions (solitons) \cite{krajenbrink2021inverse,krajenbrink2022flat}.
The mathematical analysis of the weak-noise variational problem in
\cite{tsai2023integrability} also emphasizes the role of the poles of
the scattering data. We use these works only as a guide: the
continuation below is derived directly from our Wiener--Hopf
representation.

\subsection{The endpoint of the principal sheet}

\begin{figure}[t]
\centering
\includegraphics[width=0.98\linewidth]{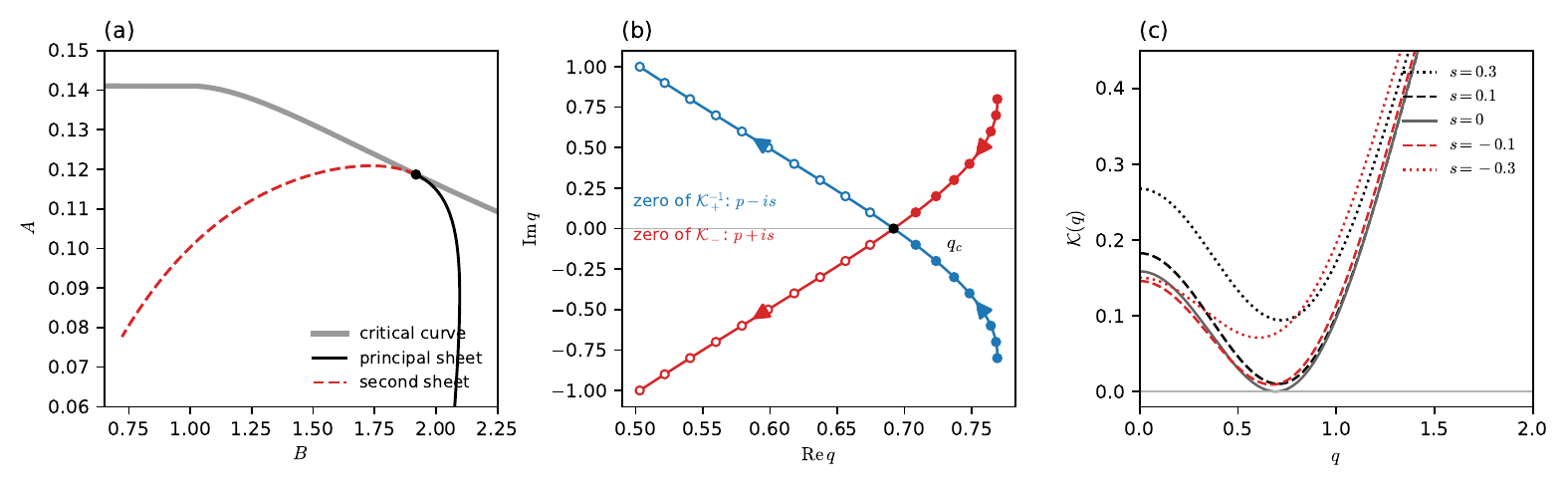}
\caption{The branch point and the change of sheet. (a) The physical branch in the $(B,A)$ plane: principal sheet
(solid, from $A=0$, $B=2$ outside the frame) and second sheet (dashed); it touches the critical curve
(grey: $A=A_*(B)$ for $B>1$, and $A=1/(4\sqrt\pi)$ for $B\le1$, where the minimum of ${\cal K}$ is at $q=0$)
tangentially at $(B_c,A_c)$ (dot) and returns into the region where ${\cal K}>0$ on the real axis. (b) Trajectories of the zeros of
${\cal K}$ near $+q_c$ along the branch, for $-1\le s\le0.8$ (the zeros near $-q_c$ follow by $q\to-q^*$): the zero
of ${\cal K}_+^{-1}$ (blue, at $p-is$) and the zero of ${\cal K}_-$ (red, at $p+is$). Filled symbols: principal
sheet; open symbols: second sheet; arrows point towards increasing $\epsilon$. (c) ${\cal K}(q)$ on the real axis at
points of the branch with $s=0.3,\,0.1,\,0,\,-0.1,\,-0.3$: the minimum touches zero at $s=0$ and lifts again on the second
sheet.}
\label{fig:contZerosNew}
\end{figure}

The entire procedure is based on the calculation of the integrals (\ref{eq:fhatAB}, \ref{eq:hhatAB}, \ref{eq:ghatAB}), which are well-defined as long as the function $\mathcal{K}(q)$ is positive on the real axis, so that $\log{\cal K}$ is real. It is convenient to consider the problem in the plane $(A,B) \in \mathbb{R}^2$, where the whole problem is defined. The condition \eqref{eq:fgeqABNew} determines a curve in this plane parameterized by the value of $\epsilon$ (see principal sheet curve -- solid black -- in Fig.~\ref{fig:contZerosNew}a).

For real $A$ and $B$, the function $\mathcal{K}(q)$ is even and real on the real axis. Its complex zeros therefore come in quartets
\be
\{\,q_0,\ q_0^*,\ -q_0,\ -q_0^*\,\},
\qquad
q_0=p+i\kappa,\quad p,\kappa>0,
\label{eq:contQuartetNew}
\ee
two in the upper half-plane ($q_0$ and $-q_0^*$) and two in the lower
half-plane ($q_0^*$ and $-q_0$). The zero $q_0$ in the first quadrant is
given explicitly by
\be
q_0
=
\left[
-\frac{1}{B}
-
W_{-1}\left(
-\frac{e^{-1/B}}{4\sqrt{\pi}\,AB}
\right)
\right]^{1/2},
\label{eq:contExplicitzNew}
\ee
which follows from writing ${\cal K}(q_0)=0$ as $u e^{-u}=e^{-1/B}/(4\sqrt\pi AB)$
with $u=q_0^2+1/B$.

On the branch connected to $(A=0,B=2)$, one has $A>0$ and $B$ remains larger
than one up to the singular point. For fixed $B>1$, and $q\in \mathbb{R}$ the function
$(1+Bq^2)e^{-q^2}$ has two symmetric maxima at
\be
q=\pm q_*(B),
\qquad
q_*^2(B)=1-\frac{1}{B}.
\label{eq:contqstarNew}
\ee
The first value of $A$ for which
$\mathcal{K}(q)$ vanishes on the real axis is then attained when $q_0 = q_\ast(B)$ and determines the critical curve in Fig.~\ref{fig:contZerosNew}a, parameterized as
\be
A_*(B)
=
\frac{e^{1-1/B}}{4\sqrt{\pi}\,B}.
\label{eq:contAstarNew}
\ee
At $A=A_*(B)$, the pair $(q_0,q_0^*)$ collides at $+q_*$ and, simultaneously, the pair $(-q_0,-q_0^*)$ collides at $-q_*$: the four simple zeros merge pairwise into two real double zeros.

The limiting values of the principal sheet are obtained by intersecting the
curve \eqref{eq:contAstarNew} with the self-consistency equation
$\hat f=\hat g$. Solving this equation directly on the critical curve,
with the logarithmic singularities at the double zeros treated
explicitly, gives
\be
A_c\simeq0.118672,
\qquad
B_c\simeq1.918303,
\qquad
q_c\simeq0.691886,
\label{eq:contCriticalABqNew}
\ee
and correspondingly
\be
\epsilon_c\simeq0.700325,
\qquad
{\cal E}_c\simeq1.927791.
\label{eq:contCriticalepsENew}
\ee

\paragraph{Collision with the real axis.}

It is convenient
to measure the distance to the critical curve by
\be
\delta=1-\frac{A}{A_*(B)} .
\label{eq:contdeltaNew}
\ee
Writing $q=q_*+x$, one finds
${\cal K}(q)=\delta+2q_*^2x^2+O(\delta x^2,x^3)$, so that the zeros close
to $+q_*$ are at $q_*\pm i\kappa$ with
\be
\kappa^2=\frac{\delta}{2q_*^2}+O(\delta^{3/2}),
\label{eq:contkappalocalNew}
\ee
and the same holds near $-q_*$ by parity. For $A>A_*(B)$ the zeros are
real, ${\cal K}<0$ on two symmetric intervals, $\log{\cal K}$ is no longer
real and the formulas of Sec.~\ref{sec:parametricSCGF} lose their meaning.
Figure~\ref{fig:contZerosNew}(b) shows the zeros approaching the real axis
along the physical branch.

From the numerical solution on the principal sheet, approaching the branch point ($A\to A_c^-$) and using as parameter the imaginary part $\kappa$ of the zero, Eq.~\eqref{eq:contQuartetNew}, we obtain
\bea
A(\kappa)&=&A_c-0.02565\,\kappa+O(\kappa^2),
\qquad
B(\kappa)=B_c+0.866\,\kappa+O(\kappa^2),
\nn\\
\epsilon(\kappa)&=&\epsilon_c-1.084\,\kappa+O(\kappa^2),
\qquad
{\cal E}(\kappa)={\cal E}_c-1.585\,\kappa+O(\kappa^2),
\label{eq:contLocalNumericsNew}
\eea
so that ${\cal E}'(\epsilon_c)\simeq1.46$. The linear coefficients of $A$
and $B$ satisfy the tangency condition
\be
\left.\frac{dA}{d\kappa}\right|_{\kappa=0^+}
=
A_*'(B_c)
\left.\frac{dB}{d\kappa}\right|_{\kappa=0^+},
\label{eq:contTangencyNew}
\ee
as visible in Fig.~\ref{fig:contZerosNew}a. Correspondingly, $\delta=2q_c^2\kappa^2+O(\kappa^3)$. So the natural hypothesis for the analytic continuation is that Eqs.~\eqref{eq:contLocalNumericsNew} are continued to negative values of this parameter (which we shall call $s$ below, since $\kappa>0$ by definition): $\delta$ remains positive, and the physical curve remains below the critical curve (see Fig.~\ref{fig:contZerosNew}a).

\subsection{Separating the colliding zeros}
It is convenient to factorize explicitly
\be
{\cal K}(q)=\tilde{\cal K}_{\rm reg}(q)\,\tilde P(q),
\qquad
\tilde P(q)
=
\frac{(q^2-q_0^2)(q^2-q_0^{*2})}{(q^2+d^2)^2}
=
\frac{\left[(q-p)^2+\kappa^2\right]\left[(q+p)^2+\kappa^2\right]}{(q^2+d^2)^2},
\label{eq:contPtildeNew}
\ee
where $d>0$ is an arbitrary constant, introduced only so that
$\tilde P\to1$ at infinity; nothing below depends on it. The two factors
have the following properties.
\begin{itemize}
\item $\tilde P$ is an analytic function of $(A,B)$, also on the critical
curve. Indeed, it depends on $(A,B)$ only through
\be
p=\frac{q_0+q_0^*}{2}
\quad\hbox{and}\quad
\kappa^2=-\frac{(q_0-q_0^*)^2}{4},
\label{eq:contpkappaAnalyticNew}
\ee
which are symmetric functions of the colliding zeros $q_0$, $q_0^*$, and
symmetric functions of colliding zeros are analytic. In contrast, $\kappa=(q_0-q_0^*)/(2i)$, which is
antisymmetric under $q_0\leftrightarrow q_0^*$, is not: $\kappa\propto\sqrt\delta$,
see \eqref{eq:contkappalocalNew}, with $\delta$ given in Eq.~\eqref{eq:contdeltaNew}.
\item Consequently $\tilde{\cal K}_{\rm reg}={\cal K}/\tilde P$ is also
analytic in $(A,B)$. On the real axis it is positive and it does not
vanish in a strip of half-width smaller than $d$ around it, uniformly close
to $(A_c,B_c)$: the
colliding zeros have been removed, and the other zeros of ${\cal K}$
(given by the other branches of the Lambert function in
\eqref{eq:contExplicitzNew}) stay at a finite distance from the real axis.
\end{itemize}
All the difficulty is therefore in $\tilde P$, which is an explicit
rational function.

\paragraph{Factorizing the two pieces.}
$\tilde{\cal K}_{\rm reg}$ has a standard Wiener--Hopf factorization 
\be 
\tilde{\cal K}_{\rm reg}=(\tilde{\cal K}_{\rm reg})_-/(\tilde{\cal K}_{\rm reg})_+ ,
\ee
given by Cauchy integrals of $\log\tilde{\cal K}_{\rm reg}$ along the real
axis as in \eqref{eq:Kpmexplicit}. Since $\log\tilde{\cal K}_{\rm reg}$ is
regular on the real axis, its factors are analytic in $(A,B)$, across the
critical curve as well. $\tilde P$ can instead be factorized by hand.
Calling $a$ and $b$ the zeros close to $+q_*$ which are assigned to
${\cal K}_+^{-1}$ and to ${\cal K}_-$, respectively (by parity, $-b$ and $-a$
are those close to $-q_*$),
\be
\tilde P_+^{-1}(q)=\frac{(q-a)(q+b)}{(q+id)^2},
\qquad
\tilde P_-(q)=\frac{(q-b)(q+a)}{(q-id)^2},
\qquad
\tilde P_-\,\tilde P_+^{-1}=\tilde P ,
\label{eq:contPtildeFactorsNew}
\ee
and
\be
{\cal K}_+^{-1}=(\tilde{\cal K}_{\rm reg})_+^{-1}\,\tilde P_+^{-1},
\qquad
{\cal K}_-=(\tilde{\cal K}_{\rm reg})_-\,\tilde P_- .
\label{eq:contKpmProductNew}
\ee
The double poles of $\tilde P$ at $q=\pm id$ are harmless. They are double
zeros of $\tilde{\cal K}_{\rm reg}$, which its factorization assigns to
$(\tilde{\cal K}_{\rm reg})_-$ (at $+id$) and to $(\tilde{\cal K}_{\rm reg})_+^{-1}$
(at $-id$), exactly as the factorization of ${\cal K}$ distributes its
own zeros away from the real axis. In \eqref{eq:contKpmProductNew} they cancel
the poles of $\tilde P_-$ and $\tilde P_+^{-1}$, which lie in the half-plane
where each factor is not required to be analytic. Changing $d$ thus moves
a factor $(q+id)^2$ from one piece to the other, and leaves ${\cal K}_\pm$
unchanged. The factorization of
Sec.~\ref{sec:parametricSCGF} requires ${\cal K}_+^{-1}$ (${\cal K}_-$) to be
non-vanishing in the upper (lower) half-plane, which fixes
\be
\hbox{principal sheet:}\qquad
a=q_0^*=p-i\kappa,
\qquad
b=q_0=p+i\kappa,
\qquad
\kappa>0,
\label{eq:contabPrincipalNew}
\ee
i.e.\ the zeros of ${\cal K}$ in the lower half-plane belong to
${\cal K}_+^{-1}$ and those in the upper half-plane to ${\cal K}_-$:
\be
{\cal K}_+^{-1}(q_0^*)={\cal K}_+^{-1}(-q_0)=0,
\qquad
{\cal K}_-(q_0)={\cal K}_-(-q_0^*)=0 .
\label{eq:contZeroAssignmentINew}
\ee
With this condition
the factorization is unique.

\paragraph{The critical curve is a square-root branch line of the functionals.}
In \eqref{eq:contKpmProductNew} every ingredient is analytic in $(A,B)$
except one: the choice \eqref{eq:contabPrincipalNew}, which requires the
positive root $\kappa=+\sqrt{\kappa^2}$. Through ${\cal K}_+$, this square
root enters the three functionals.
It is useful to see it explicitly for $\hat f$ and $\hat h$.
Expanding
$1/(k-q)$ in \eqref{eq:WHprojections},
\be
\log{\cal K}_+(q)
=
\frac{2i\hat f}{q}
-\frac{2i\hat h}{q^3}
+O(q^{-5}),
\qquad
|q|\to\infty,\ {\rm Im}\,q>0 .
\label{eq:contLogKplusNew}
\ee
At $q=iX$ this is the large-$X$ expansion
\eqref{eq:DBidentityomega}--\eqref{eq:omegaDerWH} which led to
\eqref{eq:fWH} and \eqref{eq:hWH}.
The
coefficients coming from $(\tilde{\cal K}_{\rm reg})_+$ are analytic in
$(A,B)$; we call $\hat f_{\rm reg}$, $\hat h_{\rm reg}$ their contributions.
Those coming from the rational factor follow from expanding
$\log(1-a/q)+\log(1+b/q)-2\log(1+id/q)$ in \eqref{eq:contPtildeFactorsNew},
whose odd coefficients involve only $b^n-a^n$, which equals
$2i\,{\rm Im}\,q_0^n$ with \eqref{eq:contabPrincipalNew}. One finds, exactly,
\begin{subequations}
\label{eq:contfhSingularNew}
\bea
&\hat f(A,B)=\hat f_{\rm reg}(A,B)+d-\kappa,
\label{eq:contfSingularNew}
\\
&\hat h(A,B)=\hat h_{\rm reg}(A,B)+\frac{d^3}{3}+\frac13{\rm Im}\,q_0^3
=\hat h_{\rm reg}(A,B)+\frac{d^3}{3}+p^2\kappa-\frac{\kappa^3}{3}.
\label{eq:conthSingularNew}
\eea
\end{subequations}
Since $p$ is analytic and $\kappa\propto\sqrt\delta$, the terms odd in
$\kappa$ are the non-analytic part of $\hat f$ and $\hat h$.

It is instructive to see how the real-axis integral \eqref{eq:fhatAB},
$\hat f=-\frac{1}{4\pi}\int dq\,\log{\cal K}$, which depends on $\kappa$
only through $\kappa^2$, produces the term $-\kappa$. Inserting
\eqref{eq:contPtildeNew} and shifting $q\to q\pm p$ in the contributions of
the two pairs of zeros, one needs
\be
\int_{\mathbb R}dx\,
\left[
\log(x^2+\kappa^2)-\log(x^2+d^2)
\right]
=
2\pi\left(|\kappa|-d\right),
\label{eq:contBasicLogIntegralNew}
\ee
which follows by integrating by parts (the boundary terms vanish): the
left-hand side equals $\int dx\,[2\kappa^2/(x^2+\kappa^2)-2d^2/(x^2+d^2)]$.
The result depends only on $\kappa^2$, but it is not an analytic function
of $\kappa^2$. The mechanism is that
$\partial_\kappa\log(x^2+\kappa^2)=2\kappa/(x^2+\kappa^2)$ is a Poisson kernel,
which tends to $2\pi\,{\rm sign}(\kappa)\,\delta(x)$ as $\kappa\to0$: a
zero which approaches the real axis leaves a contribution which depends
on the side from which it comes. The real-axis integral always sees the
zero ``from the side of the principal sheet'', which is why it gives
$|\kappa|$.

The important point is that the right-hand side is linear in the
square root of the distance to the collision. Its analytic continuation
is therefore naturally written in terms of the signed coordinate $s$
\be
s=+\kappa
\quad\hbox{on the principal sheet},
\qquad
s=-\kappa
\quad\hbox{on the continued sheet}.
\label{eq:contSignedCoordinateDefNew}
\ee
The difference between
the two determinations of the integral \eqref{eq:contBasicLogIntegralNew} is
therefore
\be
\left(2\pi s\right)_{\rm II}
-
\left(2\pi s\right)_{\rm I}
=
-4\pi\kappa.
\label{eq:contLogJumpOneMinimumNew}
\ee 
In terms of this signed coordinate, the zeros assigned to the
two Wiener--Hopf factors are
\be
a(s)=p-is,
\qquad
b(s)=p+is,
\qquad
\delta=2q_c^2s^2+O(s^3),
\label{eq:contSignedCoordinateNew}
\ee
i.e.\ ${\cal K}_+^{-1}(p-is)=0$ and ${\cal K}_-(p+is)=0$, and symmetrically
at $-p\mp is$. For $s=+\kappa>0$ this is the principal sheet
\eqref{eq:contabPrincipalNew}. Now $a$ and $b$ are analytic in $s$, and by
\eqref{eq:contKpmProductNew} so are ${\cal K}_\pm$ and the three functionals:
\eqref{eq:contfhSingularNew} becomes
$\hat f=\hat f_{\rm reg}+d-s$ and
$\hat h=\hat h_{\rm reg}+d^3/3+p^2s-s^3/3$. Nothing
prevents us from following these functions to $s<0$: this is the
analytic continuation. When $s$ changes sign, the zero of ${\cal K}_+^{-1}$
crosses the real axis at $q_c$ and moves into the upper half-plane,
while the zero of ${\cal K}_-$ moves into the lower half-plane
(Fig.~\ref{fig:contZerosNew}(b)). The set of zeros of ${\cal K}$ does not
cross the real axis: it touches it at $s=0$ and returns into the complex
plane, $\kappa=|s|$. What changes is to which factor each zero belongs:
for $s=-\kappa<0$ one has $a=q_0$ and $b=q_0^*$. We thus set
\be
\left({\cal K}_+^{\rm II}\right)^{-1}
=
(\tilde{\cal K}_{\rm reg})_+^{-1}\,\frac{(q-q_0)(q+q_0^*)}{(q+id)^2},
\qquad
{\cal K}_+^{-1}
=
(\tilde{\cal K}_{\rm reg})_+^{-1}\,\frac{(q-q_0^*)(q+q_0)}{(q+id)^2},
\ee
both evaluated at the same $(A,B)$. Their ratio does not depend on $d$
nor on $\tilde{\cal K}_{\rm reg}$:
\be
Q(q)
=
\frac{(q-q_0)(q+q_0^*)}{(q-q_0^*)(q+q_0)},
\label{eq:contQdefNew}
\ee
and
\be
\left({\cal K}_+^{\rm II}\right)^{-1}=Q\,{\cal K}_+^{-1},
\qquad
{\cal K}_-^{\rm II}=Q^{-1}\,{\cal K}_- .
\label{eq:contKpmIINew}
\ee

For $s=-\kappa<0$ we thus have, at the same $(A,B)$ and with the same
four zeros \eqref{eq:contQuartetNew}, $a=q_0$ and $b=q_0^*$:
\be
\hbox{second sheet:}\qquad
{\cal K}_-^{\rm II}(q_0^*)={\cal K}_-^{\rm II}(-q_0)=0,
\qquad
\left({\cal K}_+^{\rm II}\right)^{-1}(q_0)=\left({\cal K}_+^{\rm II}\right)^{-1}(-q_0^*)=0 .
\label{eq:contZeroAssignmentIINew}
\ee
We can now replace $\mathcal{K}_\pm \to \mathcal{K}_\pm^{\rm II}$ in $\hat f, \hat h, \hat g$. For the functions $\hat f, \hat h$
their relevant form on the second sheet can be read directly from Eqs.~\eqref{eq:contfhSingularNew}, flipping the terms odd in $\kappa$. Thus, one gets
\bea
&\hat f_{\rm II}(A,B)
=
\hat f(A,B)+2\kappa\;,
\label{eq:contfIINew}
\\
&\hat h_{\rm II}(A,B)-\hat h(A,B)
=
-2p^2\kappa+\frac{2}{3}\kappa^3.
\label{eq:conthJumpExplicitNew}
\eea
Since
\be
{\rm Im}\,q_0^3
=
3p^2\kappa-\kappa^3,
\ee
this last equation can also be written compactly as
\be
\hat h_{\rm II}(A,B)
=
\hat h(A,B)
-
\frac{2}{3}{\rm Im}\,q_0^3.
\label{eq:conthIINew}
\ee
Equations \eqref{eq:contfIINew} and \eqref{eq:conthIINew} are also the
counterpart, in the present representation, of the extra square-root
and $3/2$-power terms generated when one crosses the branch cut in the
one-layer polylogarithmic problem.

For real $q$,
\be
|Q(q)|=1,
\qquad
Q(-q)=Q(q)^*,
\label{eq:contQpropertiesNew}
\ee
so that $Q=e^{i\varphi}$, with
\be
\varphi(q)=\arg Q(q)=2\arg(q-q_0)+2\arg(q+q_0^*)
\pmod{2\pi}.
\label{eq:contphizNew}
\ee
Note that ${\cal K}_+^{\rm II}$ now has poles at $q_0$ and $-q_0^*$: it is
analytic only for ${\rm Im}\,q>\kappa$, which is enough for the
large-$q$ expansion \eqref{eq:contLogKplusNew}. This is the analogue, in the
present Wiener--Hopf language, of following the moving zeros or poles
when changing Riemann sheet, as in
Refs.~\cite{appert1999universal,prolhac2020riemann,prolhac2024kpz}.

For $\hat g$, we observe that $1-H_{A,B}=\frac12(1+{\cal K})$ and that, on
the real axis, ${\cal K}_+^{-1}=\sqrt{{\cal K}}\,e^{i\Omega_{A,B}}$ by
\eqref{eq:omegastarHilbert} and the definition of $\Omega_{A,B}$. The
denominator of \eqref{eq:ghatAB} is then
\be
1-H_{A,B}(q)+\sqrt{{\cal K}(q)}\cos\Omega_{A,B}(q)
=
\frac12\left|1+{\cal K}_+^{-1}(q)\right|^2 .
\label{eq:contgDenominatorNew}
\ee
Thus, replacing $\mathcal{K}_+ \to \mathcal{K}_+^{\rm II}$, the denominator
becomes
\be
D_{\rm II}(q)
=
\frac12\left|1+Q(q)\sqrt{{\cal K}(q)}\,e^{i\Omega_{A,B}(q)}\right|^2
=
1-H_{A,B}(q)+\sqrt{{\cal K}(q)}\cos\left[\Omega_{A,B}(q)+\varphi(q)\right],
\label{eq:contDIINew}
\ee
and
\be
\hat g_{\rm II}(A,B)
=
AB
\int_{\mathbb R}\frac{dq}{2\pi}\,
\frac{
8\sqrt{\pi}\,q^2e^{-q^2}
}{
D_{\rm II}(q)
}.
\label{eq:contgIINew}
\ee

The physical second-sheet solution is therefore defined by
\be
\hat f_{\rm II}(A,B)
=
\hat g_{\rm II}(A,B)
=
\epsilon,
\label{eq:contSelfConsistencyIINew}
\ee
with
\be
{\cal E}(\epsilon)
=
-4\hat h_{\rm II}(A,B).
\label{eq:contEIINew}
\ee
At the collision, $\kappa=0$ and the two determinations coincide.

 As a check of the continuation formulas, we verified numerically that $\hat f_{\rm II}+\hat f$,
$\hat h_{\rm II}+\hat h$ and $\hat g_{\rm II}+\hat g$ are analytic in $A$ at $A_c$, while the differences are
proportional to $\sqrt{A_c-A}$, as required for the two branches of a square-root branch point, and that the same
construction applied to $K=1-\xi e^{-q^2}$ (a single pair of zeros $q_0=\pm i\sqrt{-\log\xi}$, colliding at the
origin for $\xi\to1$) reproduces the polylogarithm monodromy \eqref{eq:mainN1SecondSheet} of the one-polymer problem.

\subsection{Numerical implementation}
\label{app:contNumericsNew}

\paragraph{Quadrature.}
For given $(A,B)$, all integrals are cut at $|q|\le9$ and evaluated with a
composite Gauss--Legendre rule whose panels are exponentially denser near $\pm p$, where $\log{\cal K}$ varies on the scale $\kappa$ (and
is logarithmically singular at the branch point). The phase is computed
at the quadrature nodes by subtracting the singularity,
\be
\dashint dk\,\frac{\ell(k)}{q-k}=\int dk\,\frac{\ell(k)-\ell(q)e^{q^2-k^2}}{q-k}+2\sqrt\pi\,\ell(q)\,e^{q^2}D(q),
\qquad \ell=\log{\cal K},
\ee
with $D(q)=e^{-q^2}\int_0^qe^{t^2}dt$ the Dawson function; the first
integrand is regular at $k=q$, where it equals $-\ell'(q)-2q\ell(q)$.
Close to a zero, $1-2H_{A,B}$ suffers from cancellations and is evaluated in
extended precision. With these precautions the functionals are obtained
to $\sim10^{-12}$ uniformly up to the branch point.

\paragraph{Principal sheet.}
For fixed $A$, $\hat f=\hat g$ is solved for $B$ by the secant method,
starting from \eqref{eq:BASeries} at small $A$ and from the previous
solution afterwards; the grid is refined at $\pm q_*(B)$. At small $A$
this reproduces the series \eqref{eq:BASeries}.
Close to $A_c$, however, $A$ becomes a poor parameter.

\paragraph{Parametrization by the zero.}
The physical branch is a curve in the parameter plane, and computing it
means choosing a coordinate along it and solving the single equation
$\hat f=\hat g$ for the remaining unknown. On the principal sheet the
coordinate is $A$ and the unknown is $B$. Near the branch point this is
not adequate: in the variables $(A,B)$ the functionals contain
$\kappa\propto\sqrt\delta$, their derivatives diverge on the critical
curve, and the sheet is not specified by $(A,B)$ alone. Moreover, $A$ is
not monotonic along the branch beyond $\epsilon_c$ (see below).

We therefore change coordinates in the parameter plane, from $(A,B)$
to the position of the zero, $q_0=p+i\kappa$. Below the critical curve
this is a one-to-one change of variables: \eqref{eq:contExplicitzNew} gives
$q_0$ from $(A,B)$, and conversely, separating real and imaginary parts
of ${\cal K}(q_0)=0$,
\be
B=-\frac{{\rm Im}\,e^{-q_0^2}}{{\rm Im}\,(q_0^2e^{-q_0^2})},
\qquad
A=\frac{1}{4\sqrt\pi\,{\rm Re}\left[(1+Bq_0^2)e^{-q_0^2}\right]} .
\label{eq:contABfromzNew}
\ee
In these coordinates the critical curve is simply $\kappa=0$. Replacing
$\kappa$ by the signed coordinate $s$ of \eqref{eq:contSignedCoordinateNew},
the pair $(p,s)$ covers the region below the critical curve twice: $(p,s)$
and $(p,-s)$ give the same $(A,B)$ by \eqref{eq:contABfromzNew} (with
$q_0=p+i|s|$), and differ only in the sheet, principal for $s>0$ and
second for $s<0$.

To follow the branch, $s$ is used as the coordinate and $p$ as the
unknown: for each $s$ one solves $\hat f=\hat g$ for $p$ if $s>0$, and
$\hat f_{\rm II}=\hat g_{\rm II}$ if $s<0$, starting from the solution at
the previous value of $s$. This is a one-dimensional root search, well
conditioned through $s=0$, where $B=1/(1-p^2)$, $A=A_*(B)$ and the
solution is the branch point \eqref{eq:contCriticalABqNew}. Along the branch
$s$ decreases monotonically as $\epsilon$ increases. On the principal
sheet the two parametrizations, by $A$ and by $s$, agree to $10^{-10}$.

\begin{figure}[t]
\centering
\includegraphics[width=0.98\linewidth]{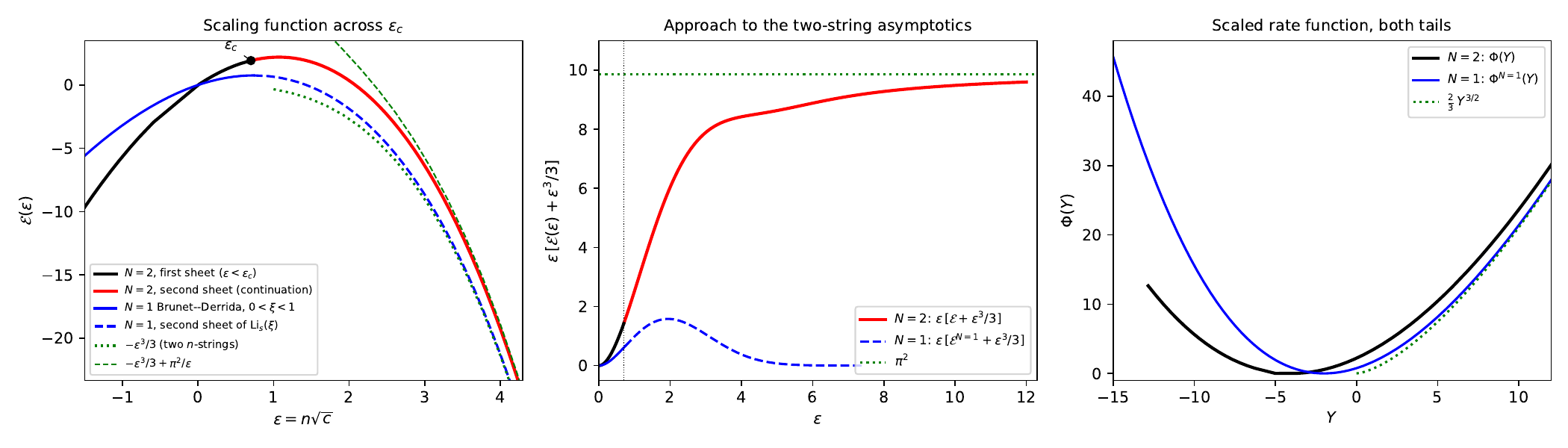}
\caption{Left: the scaling function ${\cal E}(\epsilon)$ across $\epsilon_c$: principal sheet (black), second
sheet (red), with the Brunet--Derrida function on its first ($0<\xi<1$, solid blue) and second (dashed blue)
sheet, and the asymptotics \eqref{eq:contELargePositiveNew} (green). Middle: $\epsilon[{\cal E}(\epsilon)+\epsilon^3/3]$,
which tends to $\pi^2$ for two polymers and to $0$ for one polymer. Right: the scaled rate functions $\Phi(Y)$ and
$\Phi^{N=1}(Y)$ on both tails, with the asymptote $\frac23Y^{3/2}$.}
\label{fig:continuationNew}
\end{figure}

Some points of the second sheet are given in Table~\ref{tab:continuationAcNew}.
Two features are worth stressing. First, $\epsilon$ increases
monotonically through $\epsilon_c$, whereas $A$ does not: just beyond
the collision $A$ increases slightly, reaches a shallow maximum
($A_{\max}=1.0187\,A_c$ at $\kappa=0.160$), then decreases, crossing $A_c$
again at $\kappa\simeq0.37$. Thus $A$ is not a global coordinate on the
physical branch. Second, as the continuation is followed further, $A$
and $B$ decrease while $\kappa$ grows, which points to the large-$\kappa$
regime analyzed below. The SCGF initially continues to increase beyond
the branch point, reaching a maximum ${\cal E}_{\max}=2.2023$ at
$\epsilon=1.072$ ($\kappa\simeq0.305$), before decreasing and crossing zero
at $\epsilon=2.081$. The branch point is therefore not an extremum of the
two-polymer SCGF. Since the maximum of ${\cal E}$ corresponds to $Y=0$ in
the Legendre transform \eqref{eq:mainScaledRateFunction}, the whole upper
tail $Y>0$ of the free-energy distribution is controlled by the second
sheet (Fig.~\ref{fig:continuationNew}).

\begin{table}[t]
\centering
\begin{tabular}{c|cccccc}
$\kappa$ & $p$ & $A$ & $B$ & $\delta$ & $\epsilon$ & ${\cal E}$\\
\hline
$0$     & $0.69189$ & $0.11867$ & $1.9183$ & $0$      & $0.70033$ & $1.92779$\\
$0.105$ & $0.67348$ & $0.12053$ & $1.8156$ & $0.0100$ & $0.81878$ & $2.07405$\\
$0.187$ & $0.65849$ & $0.12087$ & $1.7210$ & $0.0300$ & $0.91753$ & $2.15431$\\
$0.365$ & $0.62472$ & $0.11847$ & $1.4867$ & $0.0999$ & $1.14952$ & $2.19001$\\
$0.563$ & $0.58627$ & $0.11084$ & $1.2142$ & $0.2002$ & $1.43332$ & $1.93161$\\
$0.753$ & $0.54951$ & $0.09874$ & $0.9750$ & $0.2997$ & $1.72802$ & $1.29370$\\
$0.963$ & $0.50992$ & $0.08108$ & $0.7558$ & $0.3998$ & $2.07565$ & $0.02736$
\end{tabular}
\caption{Points on the second sheet, $s=-\kappa$; $q_0=p+i\kappa$ is the first-quadrant zero of ${\cal K}$ and
$\delta=1-A/A_*(B)$. The first row is the branch point.}
\label{tab:continuationAcNew}
\end{table}

\paragraph{Large $\kappa$.}
For $\kappa\gtrsim1.5$ the denominator $D_{\rm II}$ develops the two
quasipoles at $q=\pm\rho$, $\rho=|q_0|$, analyzed in the next subsection, with
$D_{\rm II}\simeq h_\rho^2/2$ exponentially small (down to $10^{-30}$ at
$\kappa=6$), which a fixed grid cannot resolve: a window around each
quasipole is integrated exactly in the variable $\theta$ defined by
$\Omega_{A,B}+\varphi-(2k+1)\pi=\sqrt{a/b}\,\tan\theta$, with
$a=H^2/(1-H+\sqrt{\cal K})$ and $b=\sqrt{\cal K}$ evaluated at the quasipole,
which turns the Lorentzian peak into a smooth function of $\theta$
(Gauss--Legendre with $400$ nodes; the result is stable under changes of
the window and agrees to $10^{-7}$ with adaptive quadrature where the
latter still converges). The branch was followed in this way up to
$\epsilon=12$ (${\cal E}=-575.2$, $\kappa=6$); the tabulated values of
$(\epsilon,{\cal E},A,B,p,\kappa)$ are available as supplementary data.

\subsection{Large-\texorpdfstring{$\kappa$}{kappa} solution of the second-sheet equation}
\label{app:contLargeKappaNew}

We now derive the far-tail asymptotics. Write
\be
q_0=p+i\kappa,
\qquad
\rho=|q_0|=\sqrt{p^2+\kappa^2},
\qquad
X=B\rho^2.
\label{eq:contRhoXNew}
\ee
The numerical continuation suggests $\kappa\to+\infty$,
$p/\kappa\to0$, and $A\to0$. We shall see that these properties
follow self-consistently from the continued equations.

A crucial property of the factor \eqref{eq:contQdefNew} is
\be
Q(\rho)=Q(-\rho)=-1.
\label{eq:contQrhoNew}
\ee
Moreover,
\be
\left.
\frac{d}{dq}\arg Q(q)
\right|_{q=\rho}
=
\frac{2}{\kappa}.
\label{eq:contQphaseDerivativeNew}
\ee
The principal phase $\Omega_{A,B}$ is exponentially small
in the regime derived below. It shifts the position of the
quasipole by an exponentially small amount, which does not
affect its integrated weight to the algebraic orders in
$1/\kappa$ considered here.
Let
\be
h_\rho
:=
H_{A,B}(\rho)
=
2\sqrt{\pi}\,A(1+X)e^{-\rho^2}.
\label{eq:conthrNew}
\ee
Near the shifted quasipole position $q_{\rm p}\simeq\rho$, the denominator \eqref{eq:contDIINew} therefore takes
the Lorentzian form
\be
D_{\rm II}(q)
\simeq
\frac{h_\rho^2}{2}
+
\frac12
\left(\frac{2}{\kappa}\right)^2
(q-q_{\rm p})^2.
\label{eq:contQuasipoleDenNew}
\ee
An identical quasipole occurs at $q=-\rho$. Although the two peaks
become exponentially narrow, their integrated contribution remains
finite.

The numerator of \eqref{eq:contgIINew} at $q=\rho$ is
\be
8\sqrt{\pi}\,AB\rho^2e^{-\rho^2}
=
8\sqrt{\pi}\,AXe^{-\rho^2}.
\label{eq:contQuasipoleNumNew}
\ee
Using \eqref{eq:contQuasipoleDenNew}, the contribution of one quasipole
to $\hat g_{\rm II}$ is
\be
\frac{2X}{1+X}\,\kappa,
\label{eq:contOneQuasipoleNew}
\ee
and the two symmetric quasipoles give
\be
\hat g_{\rm II}
\simeq
\frac{4X}{1+X}\,\kappa.
\label{eq:contgLargeKappaNew}
\ee

On the other hand, since $A$ is small,
\be
\hat f(A,B)
=
A\left(1+\frac{B}{2}\right)
+O(A^2),
\label{eq:contfSmallANew}
\ee
and the continued first equation gives
\be
\epsilon
=
2\kappa+O(A).
\label{eq:contepsLargeKappa1New}
\ee
The self-consistency condition
$\hat g_{\rm II}=\hat f_{\rm II}$ therefore imposes
\be
\frac{4X}{1+X}\,\kappa
=
2\kappa+O(A),
\ee
hence
\be
B\rho^2
=
1+O\left(\frac{A}{\kappa}\right).
\label{eq:contBrhoNew}
\ee
In particular, to all algebraic orders in $1/\kappa$ one may set
$B\rho^2=1$.

We can now return to the zero equation
$K_{A,B}(q_0)=0$. Using $B=1/\rho^2$ gives
\be
4\sqrt{\pi}\,A
\left(1+\frac{q_0^2}{\rho^2}\right)
e^{-q_0^2}
=
1,
\label{eq:contZeroLargeKappaNew}
\ee
and
\be
1+\frac{q_0^2}{\rho^2}
=
\frac{2p\,q_0}{\rho^2}.
\label{eq:contIdentityzNew}
\ee
Separating phase and modulus in \eqref{eq:contZeroLargeKappaNew},
and choosing the branch continuously reached from the first collision,
one obtains
\be
2p\kappa
=
\arg q_0
=
\arctan\left(\frac{\kappa}{p}\right),
\label{eq:contPhaseEquationNew}
\ee
and
\be
A
=
\frac{\rho}{8\sqrt{\pi}\,p}
e^{-\kappa^2+p^2}.
\label{eq:contAModulusNew}
\ee
Equation \eqref{eq:contPhaseEquationNew} gives
\be
p
=
\frac{\pi}{4\kappa}
-
\frac{\pi}{8\kappa^3}
+
\frac{\pi}{16\kappa^5}
+
O(\kappa^{-7}).
\label{eq:contpAsymptNew}
\ee
Combining this with \eqref{eq:contAModulusNew} and
\eqref{eq:contBrhoNew}, we find
\be
A
=
\frac{\kappa^2}{2\pi^{3/2}}
e^{-\kappa^2}
\left[
1+
\frac{1}{\kappa^2}
\left(
\frac12+\frac{\pi^2}{16}
\right)
+
O(\kappa^{-4})
\right],
\label{eq:contAAsymptNew}
\ee
and
\be
B
=
\frac{1}{\kappa^2}
-
\frac{\pi^2}{16\kappa^6}
+
O(\kappa^{-8}).
\label{eq:contBAsymptNew}
\ee
Thus the second sheet indeed returns to $A\to0^+$, but now with
$B\to0$ and with a zero moving arbitrarily far into the upper
half-plane. This closes the large-$\kappa$ solution
self-consistently.

\subsection{Cubic SCGF and the positive large-deviation tail}
\label{app:contCubicNew}

Since $A$ is exponentially small,
\be
\hat f(A,B)=O(\kappa^2e^{-\kappa^2}),
\qquad
\hat h(A,B)=O(\kappa^2e^{-\kappa^2}).
\label{eq:contfhExpSmallNew}
\ee
Equation \eqref{eq:contfIINew} therefore gives
\be
\epsilon
=
2\kappa
+
O(\kappa^2e^{-\kappa^2}).
\label{eq:contepsLargeKappaNew}
\ee
Furthermore,
\be
{\rm Im}\,q_0^3
=
3p^2\kappa-\kappa^3.
\label{eq:contImz3New}
\ee
Using \eqref{eq:conthIINew} and
${\cal E}=-4\hat h_{\rm II}$ gives
\bea
{\cal E}
&=&
-\frac{8}{3}\kappa^3
+
8p^2\kappa
+
O(\kappa^2e^{-\kappa^2})
\nn\\
&=&
-\frac{8}{3}\kappa^3
+
\frac{\pi^2}{2\kappa}
-
\frac{\pi^2}{2\kappa^3}
+
O(\kappa^{-5}).
\label{eq:contELargeKappaNew}
\eea
Eliminating $\kappa$ through
\eqref{eq:contepsLargeKappaNew}, we obtain
\be
{\cal E}(\epsilon)
=
-\frac{\epsilon^3}{3}
+
\frac{\pi^2}{\epsilon}
-
\frac{4\pi^2}{\epsilon^3}
+
O(\epsilon^{-5}),
\qquad
\epsilon\to+\infty.
\label{eq:contELargePositiveNew}
\ee
The numerical continuation confirms both coefficients: $\epsilon[{\cal E}+\epsilon^3/3]=8.885$, $9.281$, $9.487$, $9.601$ at $\epsilon=6,8,10,12$; a fit on $\epsilon\ge6$ gives $9.883$ for the constant ($\pi^2=9.870$) and $-38.7$ for the coefficient of $\epsilon^{-2}$ when the constant is fixed to $\pi^2$ ($-4\pi^2=-39.5$), and \eqref{eq:contELargePositiveNew} reproduces the data to $2\cdot10^{-2}$ at $\epsilon=6$ and $5\cdot10^{-4}$ at $\epsilon=12$. 

The structure of \eqref{eq:contELargePositiveNew} has a simple
interpretation in terms of the Bethe strings. Indeed, using
\eqref{eq:mainEnergyScaling}, the leading cubic term gives
\be
E(n,L)
\simeq
-\frac{c^2}{3L^2}\,n(n^2-1),
\label{eq:contTwoStringEnergyNew}
\ee
which is twice the $n$-string energy of the one-polymer problem.
It is therefore naturally interpreted as two asymptotically
non-interacting $n$-strings. By contrast, a single $2n$-string would
give a cubic contribution $-\frac43\epsilon^3$ to the scaling
function and is excluded by the $(n,n)$ replica sector.

The first finite-size correction also admits a simple interpretation:
through \eqref{eq:mainEnergyScaling},
$\pi^2/\epsilon$ gives $\pi^2/(nL^2)=E(1,L)/n$. Since the mass of an
$n$-string is proportional to $n$, this is naturally identified with
the lowest kinetic energy of the relative motion of the two composite
strings on the ring. This two-string picture is the replica
counterpart of the two spatially separated favorable corridors of
App.~\ref{app:rightTailOF}. For $N=1$ there is only one string, and
the corresponding correction is instead exponentially small, as in
\eqref{eq:mainN1RightTail}.

The scaled rate function is defined in
\eqref{eq:mainScaledRateFunction} by
\be
\Phi(Y)
=
\max_\epsilon
\left[
{\cal E}(\epsilon)+\epsilon Y
\right].
\ee
At leading order, \eqref{eq:contELargePositiveNew} gives
$Y\simeq\epsilon^2$, and therefore, including the first subleading term
\be
\Phi(Y)
=
\frac{2}{3}Y^{3/2}
+
\frac{\pi^2}{\sqrt{Y}}
+
O(Y^{-3/2}).
\label{eq:contPositiveTailCorrectionNew}
\ee

The exponent $3/2$ is the characteristic positive-tail exponent in
several KPZ large-deviation problems, but the present derivation is a
finite-volume two-layer calculation and should not be identified with
the short-time infinite-line problem. What is common is the analytic
mechanism: the physical continuation acquires a contribution from
singularities which have crossed between analytic factors. In the
inverse-scattering description of short-time KPZ this role is played
by discrete scattering data \cite{krajenbrink2021inverse,krajenbrink2022flat};
here it is encoded by the factor $Q$.

\subsection{Other sheets and branch selection}

The phase equation \eqref{eq:contPhaseEquationNew} admits further
determinations,
\be
2p\kappa=\arg q_0+2\pi m,
\qquad m\in\mathbb Z,
\label{eq:contOtherSheetsPhaseNew}
\ee
which, for fixed $m\geq0$ and $p>0$, give
\be
p_m\simeq\frac{(4m+1)\pi}{4\kappa}.
\label{eq:contpmNew}
\ee
This suggests a hierarchy of additional sheets, reminiscent of the
analytically continued excited-state sheets of periodic KPZ problems
\cite{prolhac2020riemann,prolhac2024kpz}, although we do not attempt
to identify them with specific excited states of the nested Bethe
equations. One may also formally continue only one of the two
collisions at $\pm q_c$, but this breaks the relation
$Q(-q)=Q(q)^*$ and hence leaves the real, parity-symmetric
Wiener--Hopf sector considered here. The simultaneous continuation of
the two zeros is therefore the natural continuation of the
ground-state branch. This contrasts with situations such as the
Brownian short-time KPZ problem, where symmetric and symmetry-broken
real saddles may coexist
\cite{krajenbrink2017brownian,krajenbrink2022flat}.

In summary, the principal parametrization reaches a square-root
endpoint at $A=A_c$, while the ground-state SCGF remains regular.
The second Wiener--Hopf sheet \eqref{eq:contKpmIINew} continues it to
larger positive $\epsilon$. Along this branch $A$ eventually returns
to zero and the zero $q_0=p+i\kappa$ moves to $\kappa\to\infty$.
The resulting quasipoles enforce $B|q_0|^2\to1$, leading to the cubic
asymptotic \eqref{eq:contELargePositiveNew} and hence to the right tail
\eqref{eq:contPositiveTailCorrectionNew}.

\section{Optimal-fluctuation interpretation of the right tail}
\label{app:rightTailOF}

We give here an optimal-fluctuation interpretation of the right-tail
relation between the one- and two-polymer problems.

For a single polymer, the $3/2$ exponent can already be understood
from a simple corridor argument. Consider a favorable disorder
fluctuation, persistent over a time $t$, with transverse width $\ell$
and amplitude $U$. The free-energy excess is of the form
$\Delta H\simeq U-a/\ell^2$, where the second term is the confinement
cost, while for Gaussian disorder the action per unit time is
proportional to $\ell U^2$. Minimization at fixed $\Delta H$ gives
$\ell\sim(\Delta H)^{-1/2}$ and $U\sim\Delta H$, hence
\be
-\frac1t\log P\sim(\Delta H)^{3/2}.
\label{eq:appCorridorRightTail}
\ee
The full optimal-fluctuation or weak-noise treatment is more
elaborate, since it determines the complete spacetime optimal history
\cite{kolokolov2007optimal,kolokolov2009explicit,meerson2016large,
krajenbrink2021inverse,gaudreaulamarre2023kpz,tsai2023integrability}.

For the long-time mechanism considered here, a particularly simple
formulation is obtained by assuming that the favorable corridor is
stationary in the longitudinal direction. The polymer growth rate is
then controlled by the bound states of a one-dimensional random
Schr\"odinger operator. This optimal-fluctuation problem was studied
directly by Smith \cite{smith2024groundstate}, both on the infinite
line and for a finite periodic system. In the deep tail the periodic
optimal fluctuation becomes strongly localized and approaches the
infinite-line soliton.

In the normalization convenient here, let
\be
{\mathsf h}_V=-\partial_x^2+V(x),
\qquad
{\mathsf h}_V\psi_j=-\kappa_j^2\psi_j,
\qquad
\int dx\,\psi_i\psi_j=\delta_{ij},
\label{eq:appSchrodingerOF}
\ee
and let $S[V]=\frac12\int dx\,V(x)^2$ denote the Gaussian disorder
cost. For $N$ non-crossing polymers the fermionic representation
implies that the relevant gain is the sum of the $N$ occupied
one-particle binding energies,
$\Delta_N=\sum_{j=1}^N\kappa_j^2$.

Minimizing $S[V]$ at fixed $\Delta_N$, and using the
Hellmann--Feynman formula, gives
\be
V(x)=-g\sum_{j=1}^N\psi_j(x)^2.
\label{eq:appOptimalPotentialN}
\ee
Inserting this relation back into the Schr\"odinger equations gives a
system of coupled focusing nonlinear Schr\"odinger equations. Such
NLS Euler--Lagrange equations also arise in the weak-noise
variational formulation of the SHE/KPZ equation
\cite{tsai2023integrability} and in finite-rank spectral optimization
\cite{frankgontierlewin2025}.

For $N=1$ one recovers
\be
\psi(x)\propto\sech(\kappa x),
\qquad
V(x)=-2\kappa^2\sech^2(\kappa x),
\qquad
S_1=\frac83\kappa^3,
\label{eq:appOneSoliton}
\ee
so that $\Delta_1=\kappa^2$ gives
$S_1\propto\Delta_1^{3/2}$. This is the deep-tail localized solution
of the periodic problem studied in Ref.~\cite{smith2024groundstate}.

For two polymers the same variational argument gives
\be
-\psi_j''
-g\left(\psi_1^2+\psi_2^2\right)\psi_j
=
-\kappa_j^2\psi_j,
\qquad
j=1,2,
\label{eq:appCoupledNLS}
\ee
with $\psi_1$ and $\psi_2$ orthogonal. The sharp one-dimensional
trace, or Lieb--Thirring, bound gives
\be
S[V]\geq
\frac83\left(\kappa_1^3+\kappa_2^3\right),
\label{eq:appTraceBound}
\ee
with the corresponding optimizers related to reflectionless KdV
multi-soliton potentials \cite{frankgontierlewin2025}. At fixed
$\Delta_2=\kappa_1^2+\kappa_2^2$, convexity shows that the lowest cost
is approached for equal sharing,
$\kappa_1^2\simeq\kappa_2^2\simeq\Delta_2/2$, and hence
\be
S_2(\Delta_2)
\geq
2S_1\left(\frac{\Delta_2}{2}\right).
\label{eq:appEqualSharing}
\ee

Since two bound states of a one-dimensional scalar Schr\"odinger
operator cannot be exactly degenerate, equality is approached through
a minimizing sequence in which two identical one-soliton wells move
far apart,
\be
V_R(x)=
-2\kappa^2\sech^2[\kappa(x-R/2)]
-2\kappa^2\sech^2[\kappa(x+R/2)],
\qquad
R\to\infty,
\qquad
\kappa^2=\frac{\Delta_2}{2}.
\label{eq:appTwoSeparatedSolitons}
\ee
The two lowest levels then become exponentially close to
$-\kappa^2$, while the overlap of the two wells vanishes and their
disorder costs add. Therefore
\be
S_2(\Delta_2)
=
2S_1\left(\frac{\Delta_2}{2}\right).
\label{eq:appRightTailCostRelation}
\ee

Thus the optimal right-tail fluctuation for two non-crossing polymers
consists asymptotically of two spatially separated favorable
corridors, each carrying one polymer and producing half of the total
excess free energy. Since the $N=2$ bulk free energy is twice the
$N=1$ one, this gives, in the corresponding large-$L$ right-tail
regime,
\be
\Phi_L(H)
\simeq
2\,\Phi_L^{N=1}\left(\frac{H}{2}\right),
\label{eq:appRightTailPhysicalRelation}
\ee
in agreement with \eqref{eq:mainRightTailPhysicalComparison}.
Equivalently, using \eqref{eq:mainPhiScaling} and
\eqref{eq:mainPhiScalingDB}, the leading scaled right tails coincide,
$\Phi(Y)\simeq\Phi^{N=1}(Y)\simeq\frac23Y^{3/2}$.

\end{document}